\RequirePackage{ifxetex}
\RequirePackage{ifluatex}
\newif\ifxeorlua
\ifxetex\xeorluatrue\fi
\ifluatex\xeorluatrue\fi

\ifxeorlua

\RequirePackage{expl3}
\ExplSyntaxOn

\ExplSyntaxOff
\else
\RequirePackage{expl3}
\ExplSyntaxOn
\pdf_version_gset:n{1.5}
\ExplSyntaxOff
\fi

\makeatletter
\newcommand{\disablepackage}[2]{%
  \disable@package@load{#1}{#2}%
}
\newcommand{\reenablepackage}[1]{%
  \reenable@package@load{#1}%
}
\makeatother
\ifxeorlua
\disablepackage{transparent}{}
\fi

\documentclass[nomenclature, english, bibtex]{kththesis}

\usepackage[style=ieee,citestyle=numeric-comp]{biblatex}
\ifxeorlua\relax
\else
\usepackage[scaled]{helvet}
\fi

\usepackage[force-eol=true]{scontents}              
\usepackage{pgffor}                 

\usepackage{xurl}                

\PassOptionsToPackage{dvipsnames, svgnames, table}{xcolor}
\usepackage{xcolor}

\usepackage[normalem]{ulem}
\usepackage{soul}
\usepackage{xspace}
\usepackage{braket}

\usepackage[locale=US]{siunitx}

\usepackage{balance}
\usepackage{stmaryrd}
\usepackage{booktabs}
\usepackage{graphicx}	        
\usepackage{multirow}	        
\usepackage{tabularx}		    
\usepackage{mathtools}
\usepackage{algorithm} 
\usepackage{algorithmic}  
\usepackage{amsmath}
\usepackage[linesnumbered,ruled,vlined,algo2e]{algorithm2e}
\usepackage{subfigure}  
\usepackage{optidef}
\usepackage{float}		        
\usepackage{pifont}

\usepackage{rotating}	    	
\usepackage{array}		        
\usepackage{mdwlist}            
\usepackage{setspace}           

\usepackage{enumitem}           

\usepackage{listings}		    

\usepackage{bytefield}          

\usepackage[obeyFinal]{todonotes}
\usepackage{notoccite} 

\usepackage{perpage}
\usepackage[perpage,symbol]{footmisc} 

\usepackage{tikz}
\usepackage{colortbl}
\usetikzlibrary{arrows,decorations.pathmorphing,backgrounds,fit,positioning, decorations.pathreplacing, calc,shapes, patterns}
\usepackage{pgfmath}	

\usepackage{bm} 

\usepackage{fvextra}
\usepackage{csquotes}               
\usepackage{placeins}

\usepackage{comment}  
\usepackage{refcount}   

\usepackage{eso-pic}
\usepackage[absolute,overlay]{textpos}

\usepackage{xparse}  

\ifnomenclature
\usepackage[nocfg]{nomencl}  
\fi

\usepackage{longtable}  
\usepackage{lscape}     
\usepackage{needspace}  
\usepackage{metalogo}   

\usepackage{etoolbox}
\renewcommand{\bfseries}{\fontseries{b}\selectfont}
\robustify\bfseries
\newrobustcmd{\B}{\bfseries}

\ifluatex\empty
\else
\usepackage{morewrites}
\fi

\usepackage{svg}
\usepackage{pgfplots}
\usepackage{pgfplotstable}
\pgfplotsset{compat=1.12}

    \tikzset{
        hatch distance/.store in=\hatchdistance,
        hatch distance=10pt,
        hatch thickness/.store in=\hatchthickness,
        hatch thickness=2pt
    }

    \makeatletter
    \pgfdeclarepatternformonly[\hatchdistance,\hatchthickness]{flexible hatch}
    {\pgfqpoint{0pt}{0pt}}
    {\pgfqpoint{\hatchdistance}{\hatchdistance}}
    {\pgfpoint{\hatchdistance-1pt}{\hatchdistance-1pt}}%
    {
        \pgfsetcolor{\tikz@pattern@color}
        \pgfsetlinewidth{\hatchthickness}
        \pgfpathmoveto{\pgfqpoint{0pt}{0pt}}
        \pgfpathlineto{\pgfqpoint{\hatchdistance}{\hatchdistance}}
        \pgfusepath{stroke}
    }
\makeatother

\definecolor{color1bg}{HTML}{1954a6}
\colorlet{color1bgFill}{color1bg!30!white}
\colorlet{color1bgDarkFill}{color1bg!90!white}

\definecolor{color2bg}{HTML}{24a0d8}
\colorlet{color2bgFill}{color2bg!30!white}
\colorlet{color2bgDarkFill}{color2bg!90!white}

\definecolor{color3bg}{HTML}{d85497}
\colorlet{color3bgFill}{color3bg!30!white}
\colorlet{color3bgDarkFill}{color3bg!90!white}

\definecolor{color4bg}{HTML}{b0c92b}
\colorlet{color4bgFill}{color4bg!30!white}
\colorlet{color4bgDarkFill}{color4bg!90!white}

\definecolor{color5bg}{HTML}{63666a}
\colorlet{color5bgFill}{color5bg!30!white}
\colorlet{color5bgDarkFill}{color5bg!90!white}

\pgfplotscreateplotcyclelist{rustcolors}{
  {color1bg,mark=none, pattern=
  {crosshatch}
  ,pattern color=color1bg},
  {color2bg,mark=none, pattern={vertical lines}, pattern color=color2bg},{color3bg,mark=none, pattern={horizontal lines}, pattern color=color3bg},{color4bg,mark=none, pattern={north east lines}, pattern color=color4bg},{color5bg,fill=color5bg,mark=none},
  {black,fill=gray,mark=none}
}

\pgfplotsset{cycle list name=rustcolors}
\pgfplotsset{/pgfplots/bar cycle list/.style={/pgfplots/cycle list name={rustcolors}}}

\usepackage{makecell}

\usepackage{pgf-pie}

\usetikzlibrary{tikzmark}

\ifluatex
\usepackage{luatexja-fontspec}
\setmainjfont{IPAexMincho} 
\fi

\input{lib/kthcolors}

\makeatletter
\newcommand{\DeclareLatinAbbrev}[2]{%
  \DeclareRobustCommand{#1}{%
  \@ifnextchar\cite{\textit{#2.\,}}{
    \@ifnextchar{.}{\textit{#2}}{%
      \@ifnextchar{,}{\textit{#2.}}{%
        \@ifnextchar{!}{\textit{#2.}}{%
          \@ifnextchar{?}{\textit{#2.}}{%
            \@ifnextchar{)}{\textit{#2.}}{%
              {\textit{#2.,\ }}}}}}}}}%
}
\makeatother
\DeclareLatinAbbrev{\eg}{e.g}
\DeclareLatinAbbrev{\Eg}{E.g}
\DeclareLatinAbbrev{\ie}{i.e}
\DeclareLatinAbbrev{\Ie}{I.e}
\DeclareLatinAbbrev{\etc}{etc}
\DeclareLatinAbbrev{\etal}{et~al}

\definecolor{ForestGreen} {RGB}{34,  139,  34}
\definecolor{HeraldRed2}   {rgb}{0.81, 0.12, 0.15}

\newcommand{\refscolor} {blue}
\newcommand{\linkscolor}{HeraldRed2}
\newcommand{\urlscolor} {ForestGreen}

\ExplSyntaxOn
\newcommand\typestoredx[2]{\expandafter\__scontents_typestored_internal:nn\expandafter{#1} {#2}}
\ExplSyntaxOff
\makeatletter
\let\verbatimsc\@undefined
\let\endverbatimsc\@undefined
\lst@AddToHook{Init}{\hyphenpenalty=50\relax}
\makeatother

\lstnewenvironment{verbatimsc}
    {
    \lstset{%
        basicstyle=\ttfamily\tiny,
        backgroundcolor=\color{white},
        columns=[l]fixed,
        language=[LaTeX]TeX,
        keywordstyle=\color{red},
        breaklines=true,                 
        breakatwhitespace=true,          
        breakindent=0em,
        frame=none,                     
        postbreak={}                    
    }
}{}

\lstdefinestyle{[LaTeX]TeX}{
morekeywords={begin, todo, textbf, textit, texttt}
}

\newcolumntype{L}[1]{>{\raggedright\let\newline\\\arraybackslash\hspace{0pt}}p{#1}}

\usepackage[plainpages=false]{hyperref}
\usepackage[all]{hypcap}	

\usepackage[acronym, style=super, section=section, nonumberlist, nomain,
nopostdot]{glossaries}
\usepackage[]{glossaries-extra}
\ifinswedish
\fi

\newglossary[tlg]{readme}{tld}{tdn}{README acronyms}

\usepackage{doi}
\usepackage[capitalise]{cleveref}         
\usepackage{hyperxmp}           

\usepackage{attachfile2}

\makeglossaries

\ifxeorlua
\setabbreviationstyle[acronym]{long-short}
\newacronym[longplural={Debugging Information Entities}]{DIE}{DIE}{Debugging Information Entity}
\newacronym[shortplural={OSes}, firstplural={operating systems (OSes)}]{OS}{OS}{operating system}

\newacronym{IQL}{IQL}{Independent Q‑Learning}

\newacronym[first={NVIDIA OpenSHMEM Library (NVSHMEM\texttrademark)}]{NVSHMEM}{NVSHMEM}{NVIDIA OpenSHMEM Library}

\newacronym{KTH}{KTH}{KTH Royal Institute of Technology}

\newacronym{LAN}{LAN}{Local Area Network}
\newacronym{VM}{VM}{virtual machine}
\newacronym{WiFi}{Wi‑Fi}{Wireless Fidelity}

\newacronym{WLAN}{WLAN}{Wireless Local Area Network}
\newacronym{UN}{UN}{United Nations}
\newacronym{SDG}{SDG}{Sustainable Development Goal}

\newacronym{gcd}{GCD}{Greatest Common Divisor}

\newacronym{W7-X}{W7-X}{Wendelstein 7-X}
\newacronym{MHD}{MHD}{Magnetohydrodynamics}
\newacronym{VMEC}{VMEC}{Variational Moments Equilibrium Code}
\newacronym{ECRH}{ECRH}{Electron Resonance Cyclotron Heating}
\newacronym{NBI}{NBI}{Neutral Beam Injection}
\newacronym{TS}{TS}{Thomson Scattering}

\newacronym{LCFS}{LCFS}{Last Closed Flux Surface}

\newacronym{ODEs}{ODEs}{Ordinary Differential Equations}

\newacronym{ITG}{ITG}{Ion Temperature Gradient}
\newacronym{ETG}{ETG}{Electron Temperature Gradient}
\newacronym{TEM}{TEM}{Trapped Electron Mode}

\newacronym{ML}{ML}{Machine Learning}
\newacronym{CG}{CG}{Critical Gradient}

\newacronym{ISS04}{ISS04}{International Stellarator Scaling 2004}

\newacronym{XICS}{XICS}{X-ray Imaging Crystal Spectrometer}
\newacronym{CXRS}{CXRS}{Charge Exchange Recombination Spectroscopy}

\newacronym{GB}{GB}{Gyro-Bohm}
\else
\setabbreviationstyle[acronym]{long-short}
\newacronym[longplural={Debugging Information Entities}]{DIE}{DIE}{Debugging Information Entity}
\newacronym[shortplural={OSes}, firstplural={operating systems (OSes)}]{OS}{OS}{operating system}

\newacronym{IQL}{IQL}{Independent Q^^e2^^80^^91Learning}

\newacronym{KTH}{KTH}{KTH Royal Institute of Technology}

\newacronym{LAN}{LAN}{Local Area Network}
\newacronym{VM}{VM}{virtual machine}
\newacronym{WiFi}{Wi^^e2^^80^^91Fi}{Wireless Fidelity}

\newacronym{WLAN}{WLAN}{Wireless Local Area Network}
\newacronym{UN}{UN}{United Nations}
\newacronym{SDG}{SDG}{Sustainable Development Goal}

\newacronym{W7-X}{W7-X}{Wendelstein 7-X}
\newacronym{MHD}{MHD}{Magnetohydrodynamics}
\newacronym{VMEC}{VMEC}{Variational Moments Equilibrium Code}
\newacronym{ECRH}{ECRH}{Electron Resonance Cyclotron Heating}
\newacronym{NBI}{NBI}{Neutral Beam Injection}
\newacronym{TS}{TS}{Thomson Scattering}

\newacronym{LCFS}{LCFS}{Last Closed Flux Surface}

\newacronym{ODEs}{ODEs}{Ordinary Differential Equations}

\newacronym{ITG}{ITG}{Ion Temperature Gradient}
\newacronym{ETG}{ETG}{Electron Temperature Gradient}
\newacronym{TEM}{TEM}{Trapped Electron Mode}

\newacronym{ML}{ML}{Machine Learning}
\newacronym{CG}{CG}{Critical Gradient}

\newacronym{ISS04}{ISS04}{International Stellarator Scaling 2004}

\newacronym{XICS}{XICS}{X-ray Imaging Crystal Spectrometer}
\newacronym{CXRS}{CXRS}{Charge Exchange Recombination Spectroscopy}

\newacronym{GB}{GB}{Gyro-Bohm}

\fi

\usepackage[separate-uncertainty = true]{siunitx} 
\let\svqty\qty
\usepackage{physics}
\let\qty\svqty
\renewcommand\vec{\mathbf}
\renewcommand\Vec{\boldsymbol}

\usepackage{isotope}
\usepackage{esint}
\newcommand{\reff}{r_{\mathrm{eff}}}

\newcommand{\me}{m_{\mathrm{e}}}
\newcommand{\nne}{n_{\mathrm{e}}}
\newcommand{\nni}{n_{\mathrm{i}}}
\newcommand{\mmi}{m_{\mathrm{i}}}
\newcommand{\Te}{T_{\mathrm{e}}}
\newcommand{\Ti}{T_{\mathrm{i}}}
\newcommand{\Qgb}{Q_{\mathrm{gB}}}
\newcommand{\Dgb}{D_{\mathrm{gB}}}
\newcommand{\aLT}{\frac{a}{L_\mathrm{T}}}
\newcommand{\aLn}{\frac{a}{L_\mathrm{n}}}
\newcommand{\alT}{a/L_\mathrm{T}}
\newcommand{\aln}{a/L_\mathrm{n}}
\newcommand{\alTcrit}{a/L_{\mathrm{T}_{\mathrm{crit}}}}
\newcommand{\nnl}{n_{\mathrm{l}}}
\newcommand{\aLTcrit}{\frac{a}{L_{\mathrm{T}_{\mathrm{crit}}}}}
\newcommand{\etacrit}{\eta_{\mathrm{crit}}}
\newcommand{\rhocrit}{\rho\textsubscript{crit}}

\newcommand{\tauiss}{\tau_{\mathrm{E}}^{\mathrm{ISS04}}}
\newcommand{\tauE}{\tau_{\mathrm{E}}}
\newcommand{\taugb}{\tau_{\mathrm{E}}^{\mathrm{gB}}}
\newcommand{\fren}{f_{\mathrm{ren}}}

\usepackage{stackengine}
\newcommand{\iotaslash}{%
  \stackengine{-.55ex}{$\iota$}{$\rule{0.6ex}{0.1ex}$}{O}{c}{F}{T}{S}%
}

\authorsLastname{Ricken}
\authorsFirstname{Jan Jonathan}
\email{ricken@kth.se}
\kthid{u1h4oz80} 
\authorsSchool{\schoolAcronym{EECS}}
\authorCityCountryDate{\MONTH\enspace\the\year}

\supervisorAsLastname{Jonsson}
\supervisorAsFirstname{Thomas}

\supervisorBsLastname{Plunk}
\supervisorBsFirstname{Gabriel}

\examinersLastname{Frassinetti}
\examinersFirstname{Lorenzo}
\examinersSchool{\schoolAcronym{EECS}}
\examinersDepartment{Fusion Science}

\hostcompany{Max-Planck-Institut für Plasmaphysik} 

\date{\today}

\courseCycle{2}
\courseCode{EA275X}
\courseCredits{30.0}

\programcode{TERFM}
\subjectArea{Electrical Engineering, specialising in Electromagnetics, Fusion and
Space Engineering}

\programcode{CDATE}
\courseCycle{2}
\courseCode{DA231X}
\courseCredits{30.0}
\degreeName{Degree of Master of Science in Engineering}
\subjectArea{Computer Science and Engineering}

\nationalsubjectcategories{10201, 10206}

\DeclareUnicodeCharacter{0196}{I}

\title{A Wendelstein-7X Profile Database for Informed Transport Studies}
\subtitle{}

\alttitle{En Wendelstein-7X Profildatabas för Informerade Transportstudier}
\altsubtitle{}

\presentationDateAndTimeISO{2025-06-23 16:00}
\presentationLanguage{eng}
\presentationRoom{via Zoom https://kth-se.zoom.us/j/67818155031}
\presentationAddress{Sten Velander room, Teknikringen 33}
\presentationCity{Stockholm}

\opponentsNames{Olle Sundberg}

\trita{TRITA -- EECS-EX}{2024:0000}

\makeatletter
\ifx\@subtitle\@empty
    \newcommand{\mytitle}{\@title}
\else
    \ifinswedish
        \newcommand{\mytitle}{\@title\xspace–\xspace\@subtitle}
    \else
        \newcommand{\mytitle}{\@title: \@subtitle}
    \fi
\fi
\makeatother

\makeatletter
\ifx\@altsubtitle\@empty\relax
    \newcommand{\myalttitle}{\@alttitle}
\else
    \ifinswedish
        \newcommand{\myalttitle}{\@alttitle: \@altsubtitle}
    \else
    \newcommand{\myalttitle}{\@alttitle\xspace–\xspace\@altsubtitle}
    \fi
    
\fi
\makeatother
\hypersetup{
     pdfsubject={\myalttitle}        
}

\ifinswedish
\XMPLangAlt{en}{pdfsubject={\myalttitle}}
\else
\XMPLangAlt{sv}{pdfsubject={\myalttitle}}
\fi

\ifinswedish
\hypersetup{%
    pdflang={sv},
    pdfmetalang={sv},
    pdftitle={\mytitle}        
}
\XMPLangAlt{en}{pdftitle={\myalttitle}}
\else
\hypersetup{%
    pdflang={en},
    pdfmetalang={en},
    pdftitle={\mytitle}        
}
\XMPLangAlt{sv}{pdftitle={\myalttitle}}
\fi

\makeatletter
\ltx@ifpackageloaded{hyperxmp}{
\ifx\@secondAuthorsLastname\@empty
\StrSubstitute{\@authorsLastname}{,}{\hyxmp@comma}[\@authorsLastnameXMP]
    \newcommand{\myauthor}{\xmpquote{\@authorsFirstname\space\@authorsLastnameXMP}} 
\else
\StrSubstitute{\@authorsLastname}{,}{\hyxmp@comma}[\@authorsLastnameXMP]
\StrSubstitute{\@secondAuthorsLastname}{,}{\hyxmp@comma}[\@secondAuthorsLastnameXMP]
    \newcommand{\myauthor}{\xmpquote{\@authorsFirstname\space\@authorsLastnameXMP},
\xmpquote{\@secondAuthorsFirstname\space\@secondAuthorsLastnameXMP}}
\fi
}{
\ifx\@secondAuthorsLastname\@empty
    \newcommand{\myauthor}{\@authorsFirstname\space\@authorsLastname} 
\else
    \newcommand{\myauthor}{\@authorsFirstname\space\@authorsLastname,
\space\@secondAuthorsFirstname\space\@secondAuthorsLastname}
\fi
}
\makeatother

\hypersetup{
     pdfauthor={\myauthor}      
}

\makeatletter
\ifx\@EnglishKeywords\@empty
    \ifx\@SwedishKeywords\@empty
        \newcommand{\mykeywords}{}
    \else
    \newcommand{\mykeywords}{\@SwedishKeywords}
    \fi
\else
    \ifx\@SwedishKeywords\@empty
        \newcommand{\mykeywords}{\@EnglishKeywords}
    \else
        \ifinswedish
            \newcommand{\mykeywords}{\@SwedishKeywords, \@EnglishKeywords}
        \else
            \newcommand{\mykeywords}{\@EnglishKeywords, \@SwedishKeywords}
        \fi
    \fi
\fi
\makeatother

\hypersetup{
     pdfkeywords={\mykeywords}        
}        

\makeatletter
\ifx\@secondkthid\@empty\relax
    \newcommand{\mykthids}{author: \@kthid}
\else
    \newcommand{\mykthids}{author: \@kthid,\xspace
    secondauthor: \@secondkthid}
\fi
\makeatother

\hypersetup{
     pdfcontactemail={\mykthids}        
}

\makeatletter
\ifinswedish
\hypersetup{
        pdfvolumenum={\@thesisSeries},        
        pdfissuenum={\@thesisSeriesNumber},
        pdfpublisher={Kungliga Tekniska högskolan (KTH)},
        pdfpubtype={report}
}
\else
\hypersetup{
        pdfvolumenum={\@thesisSeries},        
        pdfissuenum={\@thesisSeriesNumber},
        pdfpublisher={KTH Royal Institute of Technology},
        pdfpubtype={report}
}
\fi
\makeatother

\hypersetup{
	colorlinks  = true,
	breaklinks  = true,
	linkcolor   = \linkscolor,
	urlcolor    = \urlscolor,
	citecolor   = \refscolor,
	anchorcolor = black
}

\ifnomenclature

\renewcommand{\nompreamble}{The following symbols will be later used within the body of the thesis.}
\makenomenclature
\fi

\colorlet{punct}{red!60!black}
\definecolor{delim}{RGB}{20,105,176}
\definecolor{numb}{RGB}{106, 109, 32}
\definecolor{string}{RGB}{0, 0, 0}

\lstdefinelanguage{json}{
    numbers=none,
    numberstyle=\small,
    frame=none,
    rulecolor=\color{black},
    showspaces=false,
    showtabs=false,
    breaklines=true,
    postbreak=\raisebox{0ex}[0ex][0ex]{\ensuremath{\color{gray}\hookrightarrow\space}},
    breakatwhitespace=true,
    basicstyle=\ttfamily\small,
    extendedchars=false,
    upquote=true,
    morestring=[b]",
    stringstyle=\color{string},
    literate=
     *{0}{{{\color{numb}0}}}{1}
      {1}{{{\color{numb}1}}}{1}
      {2}{{{\color{numb}2}}}{1}
      {3}{{{\color{numb}3}}}{1}
      {4}{{{\color{numb}4}}}{1}
      {5}{{{\color{numb}5}}}{1}
      {6}{{{\color{numb}6}}}{1}
      {7}{{{\color{numb}7}}}{1}
      {8}{{{\color{numb}8}}}{1}
      {9}{{{\color{numb}9}}}{1}
      {:}{{{\color{punct}{:}}}}{1}
      {,}{{{\color{punct}{,}}}}{1}
      {\{}{{{\color{delim}{\{}}}}{1}
      {\}}{{{\color{delim}{\}}}}}{1}
      {[}{{{\color{delim}{[}}}}{1}
      {]}{{{\color{delim}{]}}}}{1}
      {’}{{\char13}}1,
}

\lstdefinelanguage{XML}
{
  basicstyle=\ttfamily\color{blue}\bfseries\small,
  morestring=[b]",
  morestring=[s]{>}{<},
  morecomment=[s]{<?}{?>},
  stringstyle=\color{black},
  identifierstyle=\color{blue},
  keywordstyle=\color{cyan},
  breaklines=true,
  postbreak=\raisebox{0ex}[0ex][0ex]{\ensuremath{\color{gray}\hookrightarrow\space}},
  breakatwhitespace=true,
  morekeywords={xmlns,version,type}
}

\makeatletter
\AtBeginDocument{\let\c@listing\c@lstlisting}
\makeatother
\usepackage{subfiles}

\begin{document}
%
\selectlanguage{english}

\pagenumbering{alph}
\kthcover
\clearpage\thispagestyle{empty}\mbox{} 
\titlepage

\bookinfopage

\frontmatter
\setcounter{page}{1}

\setlength{\parindent}{0pt}
\setlength{\parskip}{0.25\baselineskip}

\begin{abstract}
  \markboth{\abstractname}{}
\begin{scontents}[store-env=lang]
eng
\end{scontents}
\begin{scontents}[store-env=abstracts,print-env=true]
To advance the development of future fusion devices beyond Wendelstein 7-X (W7-X), a robust understanding of energy confinement time scaling is crucial. Universal temperature and density profile shapes are theoretically anticipated through a comparison of the empirical ISS04 confinement time scaling with predictions based on Gyro-Bohm heat flux scaling. This thesis investigates the presence of these universalities in W7-X experimental profiles and subsequently tests existing confinement time scalings against this data.

Experimental profiles are obtained from Thomson scattering diagnostics. Gradient length scales, defined as the radial derivative of the logarithmic quantity of the investigated profiles, are extracted from these profiles through fitting procedures. These scales serve as boundary conditions for subsequent gyrokinetic flux-tube simulations using the GX code. The simulated heat fluxes are then employed to reconstruct profiles by solving the power balance equation and density evolution.

A comparison of these reconstructed profiles with the averaged experimental profiles reveals that the predicted density profiles are steeper than those measured in the reference dataset. To investigate this discrepancy, additional discharges exhibiting more peaked density profiles, including both high-performance and low-power plasmas, are examined. Furthermore, the ratio of experimental temperature and density scale lengths is incorporated as an input into the simulations, and the resulting simulated heat fluxes are evaluated. 

Discrepancies between the simulated and measured profiles are discussed, with potential explanations encompassing particle sources and neoclassical transport mechanisms. This comprehensive approach aims to deepen our understanding of the critical role of profile shape and transport mechanisms in plasma confinement.

\end{scontents}

\subsection*{Keywords}
\begin{scontents}[store-env=keywords,print-env=true]
Wendelstein 7-X, Confinement time scaling, Gyro-Bohm heat flux, Thomson scattering diagnostics, Turbulent transport
\end{scontents}
\end{abstract}
\cleardoublepage
\babelpolyLangStart{swedish}
\begin{abstract}
    \markboth{\abstractname}{}
\begin{scontents}[store-env=lang]
swe
\end{scontents}
\begin{scontents}[store-env=abstracts,print-env=true]
För att driva utvecklingen av framtida fusionsreaktorer bortom Wendelstein 7-X (W7-X) krävs en gedigen förståelse för skalningen av energikonfineringstid. Universella former för temperatur- och densitetsprofiler härleds teoretiskt. Därefter görs en jämförelse av den empiriska ISS04-skalningen av konfineringstid med förutsägelser baserade på Gyro-Bohm värmeflödesskalning. Denna avhandling undersöker förekomsten av dessa universaliteter i W7-X experimentell data och testar därefter befintliga skalningar av konfineringstid mot denna data.

Experimentella profiler tas fram från Thomson-spridningsdiagnostik. Gradientlängdsskalor, definierade som den radiella derivatan av den logaritmiska kvantiteten för de undersökta profilerna, extraheras från dessa profiler genom anpassningsprocedurer. Dessa skalor utgör randvillkor för efterföljande gyrokinetiska flödesrörssimuleringar med GX-koden. De simulerade värmeflödena används sedan för att rekonstruera profiler genom att lösa effektbalansekvationen och densitetsutvecklingen.

En jämförelse av dessa rekonstruerade profiler med de medelvärdesbildade experimentella profilerna visar att de förutsagda densitetsprofilerna är brantare än de som mätts i referensdatauppsättningen. För att undersöka denna diskrepans analyseras ytterligare urladdningsfaser som uppvisar mer toppiga densitetsprofiler, inklusive både högpresterande och lågeffektsplasman. Vidare används förhållandet mellan experimentella temperatur- och densitetsskalor som indata i simuleringarna, och de resulterande simulerade värmeflödena utvärderas.

Avvikelser mellan de simulerade och uppmätta profilerna diskuteras, där möjliga förklaringar inkluderar partikelkällor och neoklassiska transportmekanismer. Denna övergripande strategi syftar till att fördjupa vår förståelse för den kritiska rollen av profilform och transportmekanismer i plasmainnesluting.
\end{scontents}
\subsection*{Nyckelord}
\begin{scontents}[store-env=keywords,print-env=true]
Wendelstein 7-X, Inneslutningsidsskalning, Gyro-Bohm värmeflöde, Thomson-spridningsdiagnostik, Turbulent transport
\end{scontents}
\end{abstract}
\babelpolyLangStop{swedish}

\cleardoublepage

\section*{Acknowledgments}
\markboth{Acknowledgments}{}

I am profoundly grateful for the invaluable guidance and support provided by my supervisor, Gabriel Plunk. Our regular meetings were instrumental not only in shaping this project but also in significantly deepening my understanding of plasma physics.

I would also like to extend my sincere gratitude to my second supervisor, Thomas Jonsson, at KTH. His insightful contributions and dedicated support throughout the years were highly appreciated and played a crucial role in my academic path.

Beyond my supervisors, I owe a debt of thanks to the many individuals who contributed to this thesis in various ways. Your efforts in proofreading, offering fresh perspectives, and, of course, providing welcome breaks for a game of table tennis, were truly invaluable.

Finally, I am thankful to my examiner for their flexibility in accommodating a presentation at relatively short notice and for ensuring the project remained on track.
\acknowlegmentssignature

\fancypagestyle{plain}{}
\renewcommand{\chaptermark}[1]{ \markboth{#1}{}} 
\cleardoublepage
\newglossarystyle{mylong}{%
  \setglossarystyle{long}%
  \renewenvironment{theglossary}%
     {\begin{longtable}[l]{@{}p{\dimexpr 2cm-\tabcolsep}p{0.8\hsize}}}
     {\end{longtable}}%
 }
\printglossary[style=mylong, type=\acronymtype, title={List of acronyms and abbreviations}]

\cleardoublepage
\tableofcontents
  \markboth{\contentsname}{}

\cleardoublepage


\mainmatter

\glsresetall
\renewcommand{\chaptermark}[1]{\markboth{#1}{}}
\selectlanguage{english}

\chapter{Introduction}

The pursuit of controlled nuclear fusion has been fundamentally guided by the need to achieve sufficiently high values of the triple product \textendash \ the product of plasma density $n$, temperature $T$, and energy confinement time $\tauE$. The Lawson criterion stipulates that for a self-sustaining fusion reaction, this triple product must exceed a critical threshold $n T \tauE > 3 \times 10^{21} \si{\per\cubic\m\kilo\eV\s}$ for deuterium-tritium reactions \cite{lawson_criteria_1957}. 

Whilst there are still challenges regarding the density limit and beta limit for future machines, the energy confinement time remains one of the primary limiting factors \cite{costley_fusion_2016}.
This fundamental constraint has driven the development of increasingly larger magnetic confinement devices, as confinement time scales favourably with machine volume $V$ \cite{yamada_characterization_2005}.

Stellarators, conceptualised by Lyman Spitzer in 1951, represent one of the earliest approaches to magnetic confinement fusion \cite{spitzer_stellarator_1958}. Unlike tokamaks, which possess toroidal symmetry and corresponding conserved quantities, stellarators lack such inherent symmetry, leading to more complex plasma confinement challenges. The initial successes of tokamaks, attributed to their symmetric design facilitating more straightforward plasma confinement, propelled them to the forefront of fusion research, culminating in the construction of ITER \cite{ITER2021}. However, stellarators offer distinct advantages for future fusion power plants: their intrinsic capability for steady-state operation eliminates the need for pulsed operations or reliance on bootstrap currents, and they are far less prone to current-driven disruptions that pose significant challenges for tokamaks.
\newpage
The \glsxtrfull{W7-X} stellarator exemplifies the potential of modern stellarator design. As the first neoclassically optimised stellarator, W7-X has demonstrated substantially reduced neoclassical transport, achieving plasma confinement comparable to that of tokamaks \cite{beidler_demonstration_2021}. 
This device employs a heating system consisting of \glsxtrfull{ECRH} as the primary heating method, delivering up to \SI{10}{\mega \W} of power through gyrotron-generated microwaves at \SI{140}{\giga \Hz}. Additionally, \glsxtrfull{NBI} can be used to provide supplementary heating and fuelling \cite{wolf_electron-cyclotron-resonance_2018}.

Advancing stellarator research necessitates a comprehensive understanding of how confinement and other critical physical properties scale with reactor size, magnetic field strength, and additional parameters. Developing accurate scaling laws and conducting simulations based on these laws are pivotal for informing the design of future reactors. However, this approach is challenged by the absence of reliable linear scaling relationships and the emergence of unforeseen effects that may not be accounted for in simplified models \cite{garcia-regana_turbulent_2021}.

If plasma transport in the confinement region is predominantly governed by turbulent processes, as extensive experimental evidence suggests for neoclassically optimised scenarios, then the underlying physics should follow \glsxtrfull{GB} scaling. This theoretical framework predicts that the heat flux scales as $Q \sim  T^{5/2}$ \cite{xanthopoulos_controlling_2014}.
Remarkably, the empirical \glsxtrfull{ISS04} \cite{yamada_characterization_2005} scaling law closely mirrors the dependencies predicted by Gyro-Bohm theory, suggesting that turbulent transport indeed dominates energy confinement in stellarators. This correspondence implies the existence of universal profile shapes and gradient scale lengths that should manifest across different stellarator configurations and operational regimes, as motivated in \cref{sec:scaling_laws}.

To characterise these universal profile shapes quantitatively, we define the temperature gradient length scale $L_\mathrm{T}$ as: $\alT = -a \dd{\ln{T}}/\dd{\reff} $, where $a$ denotes the minor radius and $\reff$ represents the effective normalised radius. This dimensionless parameter ($\alT$) encapsulates the characteristic scale length of temperature variations and serves as a crucial metric for comparing profile steepness across different plasma conditions and devices.

This thesis investigates the hypothesis that a scaling of the logarithmic temperature gradient exists at the outer plasma region, exhibiting only weak dependence on device configuration. The primary aim is to conduct a comprehensive study through the creation of an extensive profile database from W7-X \glsxtrfull{TS} measurements, representing the first large-scale experimental investigation of the temperature gradient scaling in W7-X. By systematically extracting these gradient lengths and comparing them with gyrokinetic simulations, we establish the foundation for understanding the existing profile shapes seen in the experiment. Whilst the development of a fully predictive framework remains a goal for future work, this thesis provides the essential experimental basis and initial validation necessary for such endeavours.

The thesis is structured as follows: \cref{sec:Background} provides the theoretical background, covering the W7-X device, TS diagnostics, gyrokinetic theory, and transport mechanisms. \cref{sec:Methods} details the methodological approach, including heat flux models, data filtering techniques, and profile fitting procedures. \cref{sec:Results} presents the analysis results from individual plasma discharges and the compiled database, including comparisons with gyrokinetic simulations. \cref{sec:Discussion} discusses the implications of these findings for understanding confinement time and transport scaling. Finally, \cref{sec:Conclusion} summarises the key conclusions and outlines directions for future research.
\chapter{Background} \label{sec:Background}

This chapter provides a concise introduction to W7-X and the Thomson scattering system as key diagnostic tools used to measure radial profiles of electron temperature and density.
An overview of transport processes is then presented, with a focus on turbulent transport mechanisms. Finally, a heuristic confinement time scaling based on the Gyro-Bohm scaling is contrasted with the empirical ISS04 scaling. From this comparison, the hypothesis of the thesis is then motivated.

\section{Wendelstein 7-X} \label{sec:W7X}

\begin{figure}[H]
    \centering
    \includegraphics[width =5.7cm]{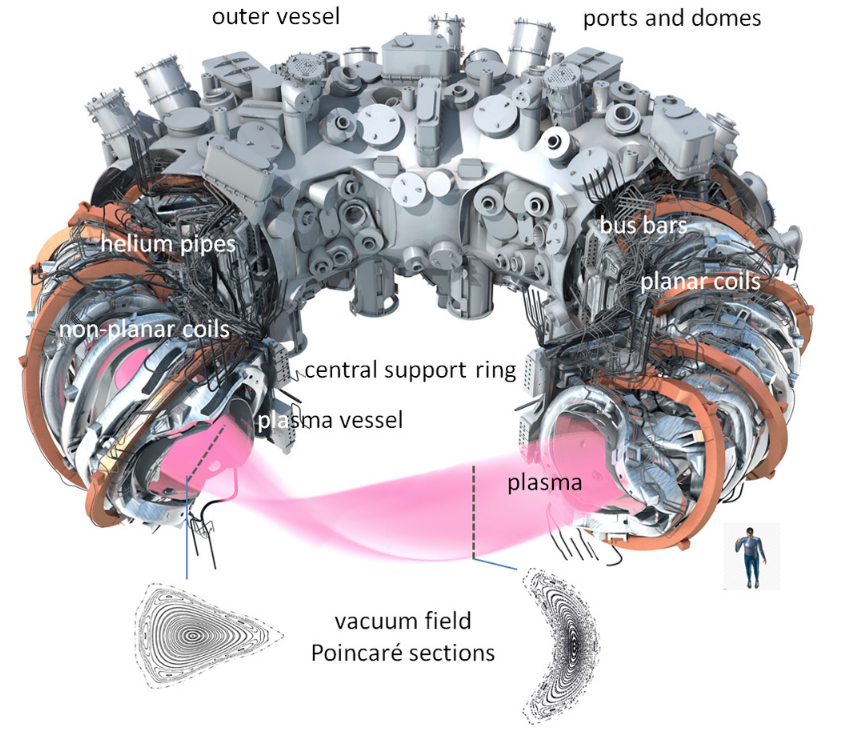}
    \caption[Illustration of W7-X.]{\textbf{Illustration of W7-X}. The plasma (pink) is contained in the vacuum vessel, which is surrounded by the coils, cooling and diagnostics. Further, two exemplary Poincaré plots of the magnetic field are shown. Taken from \cite{klinger_performance_2016}.}
\label{fig:Wendelstein_7X}
\end{figure}

W7-X, the latest in the Wendelstein stellarator series, is the first device optimised to reduce neoclassical transport. Located in Greifswald, Germany, it has been in operation since 2015. This five-field-period, quasi-isodynamic stellarator has a major radius of \SI{5.5}{\m} and an effective minor radius of \SI{0.514}{\m}, giving a plasma volume of approximately \SI{30}{\cubic\m} \cite{W7X_Einfuehrung}. An illustration of the device is shown in \cref{fig:Wendelstein_7X}.

\noindent The optimisation of W7-X's magnetic field configuration was achieved through the use of 50 non-planar and 20 planar superconducting modular coils. The design target of the magnetic equilibrium was to create a vacuum magnetic field that confines fusion plasmas while ensuring good nested magnetic surfaces, \glsxtrfull {MHD} equilibrium and stability, low neoclassical transport in the long-mean-free-path regime, and effective confinement of fast particles \cite{sunn_pedersen_key_2017}. The successful reduction of neoclassical energy transport in W7-X has been experimentally confirmed, validating the effectiveness of its optimised magnetic field configuration \cite{beidler_demonstration_2021}.

\section{Thomson Scattering} \label{sec:Thomson_Scattering}
\glsxtrfull{TS} describes the elastic scattering of electromagnetic radiation by free-charged particles, in the limit that the photon energy is much less than the rest mass energy of the particle ($h\nu \ll mc^2$). Given that the energy of a photon with wavelength $\lambda = \SI{1064}{\nano\meter}$ corresponds to approximately \SI{1.17}{\eV}, which is significantly smaller than the electron rest mass energy of \SI{511}{\kilo\electronvolt}, the Thomson scattering approximation is valid.

Ions are typically neglected in TS analyses due to their substantially greater mass compared to electrons, resulting in much lower accelerations under the same electromagnetic forces. However, it is noteworthy that ions are not entirely irrelevant; information about the ion population can, in principle, be inferred indirectly through their influence on the electron population \cite{hutchinson_scattering_2002}.

In high-temperature plasmas, electron thermal velocities can approach a significant fraction of the speed of light, necessitating a relativistic treatment of TS. Nonetheless, we focus on the non-relativistic incoherent TS process to establish a foundational understanding.

When an incident electromagnetic wave interacts with a free electron, the electric field component of the wave exerts a force on the electron, causing it to oscillate at the same frequency as the incident wave. This oscillation can be seen as a simple dipole antenna. Assuming that $\hat{\vec{s}}$ is the direction in which the radiation is detected, and $\Vec{\beta} = \vec{v}/c$, the scattered electric field is
\begin{equation}
    \vec{E}_{\text{s}} = \frac{r_{\text{e}}}{R} \hat{\vec{s}} \wedge \left( \hat{\vec{s}} \wedge \Dot{\Vec{\beta}} \right)
\end{equation}
with the classical electron radius $r_{\text{e}} =e^2/(4 \pi \epsilon_0 \me c^2)$ and $R$ denoting the (retarded) distance from the scattering electron to the observation point.
The scattered power per unit angle then takes the form
\begin{equation} 
\frac{\dd{P}}{\dd{\Omega\textsubscript{s}}} = r_{\text{e}}^2 \sin^2{\phi} \, c \epsilon_0 \abs{E\textsubscript{inc}}^2, 
\end{equation}
where $e$ denotes the elementary charge, $\epsilon_0$ the permittivity of free space, $\me$ represents the electron mass, $c$ is the speed of light, and $\phi$ is the scattering angle. Integrating over all solid angles and normalising by the incident power yields the total Thomson cross-section $\sigma_{\text{T}} = (8\pi/3) r_{\text{e}}^2$.

\noindent Thus far, we have only considered a single electron. When examining a plasma, we cannot simply sum over all electrons, as collective effects play a significant role. Moreover, electrons occupy different spatial locations, and as the wave travels with finite velocity, time retardation must be accounted for.

Performing the detailed calculation as outlined in \cite{naito_analytic_1993} yields the expression
\begin{equation}
    \frac{\dd{P}}{\dd{\Omega\textsubscript{s} \dd{\epsilon}}} = r_{\text{e}}^2 \int \langle S\textsubscript{i} \rangle \dd^3{r} \, S (\epsilon, \theta, 2\alpha).
    \label{eq:Thomson_Scattering_Spectrum}
\end{equation}
Here, $\langle S\textsubscript{i} \rangle$ denotes the mean Poynting vector, $\epsilon = (\lambda\textsubscript{s}-\lambda\textsubscript{i})/\lambda\textsubscript{i}$ represents the relative wavelength shift, $\alpha$ corresponds to a normalised inverse temperature, and $S (\epsilon, \theta, 2\alpha)$ is the spectral density function. The explicit form of $S$ is omitted from this discussion, as it does not substantively contribute to our current analysis; interested readers may consult \cite{naito_analytic_1993} for a comprehensive treatment.

\subsection{W7-X Thomson Scattering Diagnostic} \label{sec:W7X_Thomson_Scattering}
The TS diagnostic system at W7-X is described here as implemented during operational phase OP1.2 (2018), which forms the basis for the data analysed in this work. More recent developments and upgrades after this campaign are summarised towards the end of this section.

\begin{figure}[H]
    \centering
    \includegraphics[width = 8cm]{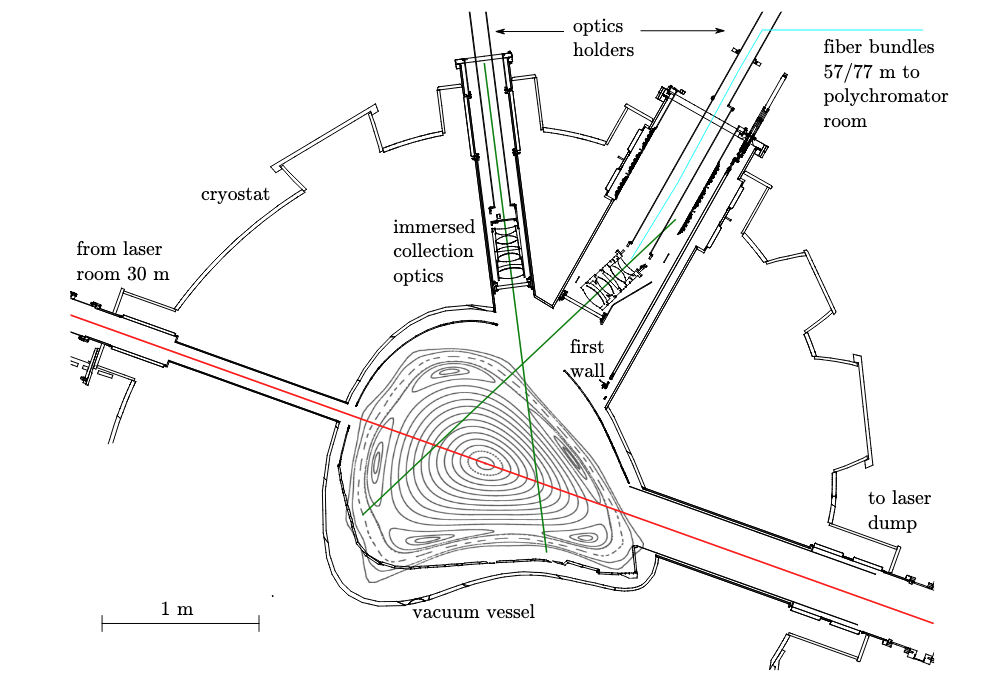}
\caption[Illustration of the TS system at W7-X.]{\textbf{Illustration of the TS system at W7-X}. Shown is an illustration of the beam path, entering from the inboard side (left) and exiting on the outboard side (right). The lines of sight for each observation optic are marked in blue and green. Taken from \cite{bozhenkov_thomson_2017}.}
\label{fig:Thomson_Scattering_System}
\end{figure}

The system is designed to measure radial profiles of electron temperature $\Te$ and density $\nne$ with high spatial and temporal resolution. It used three synchronised Nd:YAG lasers operating at \SI{1064}{\nm}, each with a repetition rate of \SI{30}{\Hz}, yielding an effective acquisition rate of up to \SI{90}{\Hz} \cite{bozhenkov_thomson_2017}. The laser beams are transported over a \SI{30}{\metre} optical path into the vessel, intersecting the plasma along a horizontal chord. An illustration of this is shown \cref{fig:Thomson_Scattering_System}.

Observation optics consist of two in-vessel telescopes mounted on the inboard and outboard sides of the vessel. Each telescope images half of the plasma cross-section onto arrays of fibre bundles, which define discrete scattering volumes. These fibres transmit the scattered light to filter-based polychromators that spectrally resolve the signal into five wavelength intervals. Silicon avalanche photodiodes are used as detectors. To optimise hardware usage, two scattering volumes are time-multiplexed per polychromator using a fibre delay line \cite{bozhenkov_thomson_2017}. 

Spectral calibration of the TS system is done using a pulsed supercontinuum light source, which provides a broad range of wavelengths to check how the polychromators respond across the spectrum. Absolute calibration, which determines the system’s overall sensitivity and allows measured signal intensities to be converted into quantitative electron density values, is carried out using Raman scattering in nitrogen \cite{pasch_thomson_2016}. To improve the accuracy and reliability of measurements, the optical alignment process was improved, and automatic systems were added to monitor the laser beam, ensuring it enters the system in a stable and repeatable way, which was implemented after OP 1.2 \cite{fuchert_novel_2022}. These upgrades also made it possible to create a new calibration method that works well even if the alignment isn’t perfect. This technique involves measuring how the system responds to different laser beam positions, then adjusting the calibration based on the actual beam path during each plasma shot. As a result, it greatly reduces errors in measuring electron density that can be caused by small misalignments between the laser and the observation optics \cite{fuchert_novel_2022}.

Data analysis is conducted using a Bayesian inference framework implemented in the Minerva software suite, which reconstructs $\Te$ and $\nne$ profiles with quantified uncertainties \cite{bozhenkov_thomson_2017}. The software has been extended to support multi-laser and burst-mode acquisitions, ensuring compatibility with evolving measurement configurations.

A more comprehensive system overview of the initial diagnostic system is available in \cite{pasch_thomson_2016}, and recent developments are detailed in \cite{fuchert_calibration_2024}.

 \section{Gyrokinetics}
 In this section, we provide a brief discussion of the gyro-averaged equations of motion, beginning with the single-particle motions and drifts in a strongly magnetised plasma.
 
\subsection{Single Particle Motion}
Before addressing the collective behaviour of many particles, it is instructive to begin with the simplest case: a single charged particle in a strong magnetic field $\vec{B}$, carrying charge $q$ and mass $m$. Further, we shall assume that the B-field is constant and aligned along the $z$-axis, thus $\vec{B} = B \hat{\vec{z}}$, and there is no background electric field $\vec{E}$.

The particle evolves according to the Lorentz force
\begin{equation}
m \Dot{\vec{v}} = q(\vec{E} + \vec{v} \times \vec{B}),
\label{eq:Lorentz_force}
\end{equation}
which in this case reduces the evolution of the velocity to a simple set of \glsxtrfull{ODEs}:
\begin{align}
\Dot{v_x} &= \frac{q B}{m} v_y = \Omega v_y \label{eq:Gyrp_Dvx}\\
\Dot{v_y} &= -\frac{q B}{m} v_x = -\Omega v_x \label{eq:Gyrp_Dvy}\\
\Dot{v_z} &= 0, \label{eq:Gyrp_Dvz}
\end{align}
where we define the gyro-frequency as $\Omega = qB/m$.

Solving this system and introducing the complex position variable $p = x + i y$, we find
\begin{equation}
    p = i\frac{v_\perp}{\Omega} \exp{(-i \Omega t) + \varphi} + \text{const.} ,
\end{equation}
where $v_\perp$ denotes the perpendicular velocity and $\varphi$ the initial gyro-phase. This represents uniform circular motion with radius
\begin{equation}
    \varrho_s = \frac{v_\perp m_s}{\abs{q_s} B},
    \label{eq:rho_s}
\end{equation}
known as the Larmor radius.

Adding a finite electric field introduces an additional term to \cref{eq:Gyrp_Dvx,eq:Gyrp_Dvy,eq:Gyrp_Dvz}, yielding a particular solution corresponding to the $\vec{E}\times\vec{B}$ drift. The homogeneous solution is unchanged, while the additional steady component of the velocity is
\begin{equation}
    \vec{v}_\mathrm{E} = \frac{\vec{E}\times \vec{B}}{B^2},
    \label{eq:E_cross_B_drift}
\end{equation}
which is perpendicular to both fields and independent of particle charge and mass \cite{chen_single-particle_2016}.

 The parallel velocity $\vec{v}_\parallel$ is not affected by the magnetic field and merely undergoes acceleration by the electric field. Here, parallel and perpendicular are defined, as they will be treated throughout the remainder of the text, with respect to the background magnetic field, unless otherwise specified.

 \Cref{eq:E_cross_B_drift} is not dependent on the charge nor the mass of the object, indicating how fundamental this drift is. By repeating the same calculation for an arbitrary force $\vec{F}$, instead of the $\vec{E}$, we obtain a drift reading as
 \begin{equation}
     \vec{v}_\mathrm{F} = \frac{\vec{F} \times \vec{B}}{q B^2}.
     \label{eq:general_force_drift}
 \end{equation}

\subsection{Gyrokinetic Vlasov Equation} \label{sec:Gyrokinetic_Vlasov_Equation}
To describe not just one particle, but a distribution function $f_s = f_s (\vec{r}, \vec{v}, t)$ in phase space, we use Vlasov equation 
\begin{equation}
    \partial_t f_s + \vec{v} \cdot \nabla f_s + \frac{q_s}{m_s} \left( \vec{E} + \vec{v} \times \vec{B} \right) =0.
\end{equation}
The magnetic and electric fields are defined by the vector potential $\vec{A}$ and scalar potential $\phi$ as:
\begin{equation}
    \vec{E} = - \nabla \phi - \partial_t \vec{A}, \quad \vec{B} = \nabla \times \vec{A} .
\end{equation}
This gives a full description of the development of the distribution function (neglecting collisions). To further analyse this equation, 
we split the distribution into an ensemble average  $f_{s0}=f$ and a fluctuating part with $\delta f_s$ with $\langle \delta f_s \rangle =0$. Assuming effectively a difference on time scale, such that the ensemble average behaves smoothly in time and space \cite{dwight_r_nicholson_introduction_1983}.

 To simplify the description of plasma dynamics in strongly magnetised fields, we transition to guiding centre coordinates. This involves averaging over the fast gyromotion, effectively removing the dependence on the gyrophase angle. 
To change to a gyro averaged formulation, we perform a coordinate transformation, namely the Catto transformation \cite{catto_generalized_1981}
\begin{equation}
    (\vec{r}, \vec{v}, t) \to (\vec{R}, \mu, \mathcal{E}, \vartheta, t) .
\end{equation}
The new coordinates are: the guiding centre position $\vec{R}$, the magnetic moment 
\begin{equation}
    \mu = \frac{m v_\perp^2}{2B},
    \label{eq:mag_momentum}
\end{equation} 
the total energy $\mathcal{E}$, and the gyro-phase $\vartheta$. In these variables, the gyrokinetic equation reads as
\begin{equation}
    \frac{\partial f}{\partial t } +  \frac{\partial \vec{R}}{\partial t } \frac{\partial f}{\partial \vec{R}}+  \frac{\partial \mu}{\partial t } \frac{\partial f}{\partial \mu} + \frac{\partial \mathcal{E}}{\partial t} \frac{\partial f}{\partial \mathcal{E} } + \frac{\partial \vartheta}{\partial t} \frac{\partial f}{\partial \vartheta } =0.
    \end{equation}
To perform a gyro average, one can now simply integrate over $\vartheta$
\begin{equation}
    \langle S (\vec{R}, \mu, \mathcal{E}, \vartheta, t) \rangle = \frac{1}{2 \pi} \int\limits_0^{2 \pi} S (\vec{R}, \mu, \mathcal{E}, \vartheta, t) \dd{\vartheta},
\end{equation}
where $S$ denotes a general variablem\cite{jan-peter_bahner_core_2022,catto_generalized_1981}.

\noindent Especially relevant is the gyro-averaging of the electric field $\vec{E}$. The field which the particle experiences depends on the gyro-motion and thus differs for different particles \cite{doi:10.1142/p015}. A more complete set of the equations discussed above is solved using the GX code, which is discussed in \cref{sec:GX_code}.

\section{Transport}
Transport in magnetically confined plasmas arises from two main mechanisms: collisional (neo-classical) processes and turbulence. In classical stellarators, neo-classical transport dominates because of unfavourable particle trajectories. Advances in magnetic field optimisation have significantly reduced this contribution in modern stellarators; in W7-X, for instance, turbulent transport now exceeds the neo-classical component \cite{beidler_demonstration_2021}.
Nevertheless, neo-classical transport remains important, both as a theoretical benchmark and as a contributor to the total heat flux. This section begins with an overview of collisional processes and then examines transport mechanisms through the lens of a random-walk model.

\noindent Regardless of the complexity of the transport model, we expect that the total amount of particles follows a density conservation equation 
\begin{equation}
    \partial_t \rho + \nabla \cdot \Vec{\Gamma} = S,
    \label{eq:conservation_eq}
\end{equation}
where $\rho$ denotes a density, $\Vec{\Gamma}$ the corresponding flux and $S$ illustrates a source term.
The evolution of the particle transport is determined by a diffusion part and an advection part, whereby we here solely focus on the diffusion part. This can be expressed by Fick's law
\begin{equation}
\Vec{\Gamma} = -D \nabla n ,
\label{eq:Fick_Gamma}
\end{equation}
with $D$ denoting the diffusion coefficient. 

\subsection{Collisions Frequencies}

In classical transport theory, the transport is driven by Coulomb collisions between charged particles and usually assumes a symmetrical, cylindrical, homogeneous magnetic field.
The collisions, in contrast to those in a neutral gas, are not head-on collisions but rather an accumulated effect over several long-distance interactions between many particles.

\noindent The frequency at which particles interact can be well described by the Coulomb collision frequency between electrons and ions, $\nu_{\text{ei}}$. The derivation of the corresponding characteristic collision time ($\tau_{\text{ei}} \sim \nu_{\text{ei}}^{-1}$) is standard and can be found in many plasma physics textbooks, including \cite{CollisionalTransportMagnetizedHelander}, whose specific formulation reads as
\begin{equation}
    \tau\textsubscript{ei} = \frac{12 \pi^{\frac{3}{2}} }{\sqrt{2}} \frac{\epsilon_0^2\sqrt{\me \Te^3  }}{n\textsubscript{i}Z^2 e^4 \ln{\Lambda}}.
\end{equation}
Here, $\ln{\Lambda}$ denotes the Coulomb logarithm, and the temperatures are specified in $\si{\eV}$. For ion-ion collisions, the corresponding collision time
\begin{equation}
    \tau\textsubscript{ii} = \sqrt{\frac{\mmi}{\me}} \frac{1}{Z^2} \tau\textsubscript{ei}.
\end{equation}
Assuming the plasma is strongly magnetised, meaning that the ion gyrofrequency is much larger than the collision frequency, the magnetic field rather than collisions governs the short–time–scale particle dynamics. We further require that plasma parameters $k$ vary only weakly across an ion Larmor radius $\varrho_\mathrm{i}$,
\begin{equation}
    \left( \frac{1}{k} \frac{\dd{k}}{\dd{x}} \right)^{-1} \gg \varrho\textsubscript{i}
\end{equation}
so that the plasma appears locally uniform to a gyrating particle \cite{CollisionalTransportMagnetizedHelander}.

\subsection{Classical Transport}

For the random walk model, we shall assume that between each collision, occurring at the characteristic time scale $\tau$, the particles move a radial distance of $\Delta x$ without a preferred direction. 
Taylor expanding the particle density around the pre-collision position $r_0$ and comparing the particle flux from $- \Delta x$ and $\Delta x$ gives the net flux
\begin{equation}
    \Gamma_\mathrm{r} = \Gamma_+ - \Gamma_- = - \frac{(\Delta x )^2}{2 \tau}  \frac{\dd{n}}{\dd{r}}\Bigr|_{r_0} .
\end{equation}
Comparing this expression to a Fickian process (\cref{eq:Fick_Gamma}) for the radial particle flux density $\Gamma_\mathrm{r}$ yields a diffusion coefficient for species $s$
\begin{equation}
    D^{\text{C}}_s = \frac{(\Delta x )^2}{2 \tau} \approx \frac{\varrho_{\text{s}}^2}{2 \tau\textsubscript{ss}}.
\end{equation}  
This diffusion is ambipolar, such that $D\textsubscript{C,e} = D\textsubscript{C,i}$ \cite{perbrunsellFUSIONPHYSICS}. 
The radial heat fluxes take a similar form as \cref{eq:Fick_Gamma}
\begin{equation}
    q_\mathrm{r}= -n \chi \frac{\dd{T}}{\dd{r}}.
    \label{eq:Fick_q}
\end{equation}
Here, $\chi$ denotes the heat diffusivity, which in the classical case is approximately $\chi^{\text{C}}_s \approx \sqrt{m_s/\me} D$, for the species $s$.

\subsection{Neoclassical Transport}
Neoclassical transport theory extends the above-discussed concepts to transport in a toroidal geometry. This includes the existence of an inhomogeneous magnetic field, which gives rise to an additional component in the transport.

In the high collisionality regime, guiding-centre orbits - the paths of the centres of rapid particle gyromotion - are constantly interrupted by collisions, which must be taken into account. This leads to an enhancement of the transport known as the Pfirsch-Schlüter transport.

The mean free path a particle can traverse before collision is measured by $\lambda = v\textsubscript{T}/\nu$, where $v\textsubscript{T}$ denotes the thermal velocity. Importantly, the guiding-centre orbits do not exactly coincide with flux surfaces. When particles collide, they change their direction along the magnetic field line, thereby altering their guiding-centre orbits. Consequently, the displacement $d$ of the guiding-centre orbit relative to the flux surface becomes the characteristic length scale that must be inserted in the random walk model \cite{jan-peter_bahner_core_2022}.

Comparing the diffusion coefficient for the Pfirsch-Schlüter transport with the classical estimate yields,
\begin{equation}
D^{\text{PS}} = \frac{1}{\iota^2} D^{\text{C}}.
\end{equation}
Here, $\iota$ represents the rotational transform, defined as:
\begin{equation}
\iota = \lim\limits_{m \to \infty} \frac{\theta_m}{2 \pi m } ,
\end{equation}
where $m$ denotes the number of toroidal transits and $\theta_m$ the corresponding poloidal angle \cite{spitzer_stellarator_1958}.
As the collisionality decreases in tokamaks, a distinct regime emerges, the so-called banana regime, which derives its name from the characteristic banana-shaped particle orbits.
In stellarators, the low collisionality regime is initially governed by the $1/\nu$-regime, which exhibits a diffusion coefficient proportional to
\begin{equation}
    D^{1/\nu} \sim \epsilon_{\text{h}}^{3/2} \frac{1}{\nu}.
\end{equation}
This relationship arises because the particles responsible for transport are trapped in local magnetic wells with a depth of $\delta B/B \sim \epsilon_{\text{h}}$ \cite{helander_stellarator_2012}.

At even lower collisionality, the $\sqrt{\nu}$-regime dominates, characterised by
\begin{equation}
    D^{\sqrt{\nu}} \sim \frac{v_{\text{D}}^2 \sqrt{\nu}}{\Omega_{\text{E}}^{3/2}}.
\end{equation}
Here, $\Omega\textsubscript{E} \sim E\textsubscript{r}/rB$ denotes the drift frequency caused by the $\vec{E} \times \vec{B}$-drift, and $v\textsubscript{D}$ represents the drift velocity. This is caused by the fact that electrons in general have much lower diffusion coefficients than ions \cite{helander_stellarator_2012}, and thus, as self-consistent local particle flux emerges, causing an ambipolar electric field \cite{jan-peter_bahner_core_2022}.

\subsection{Turbulent Transport} \label{sec:turbulent_transport}
In addition to neoclassical transport, the micro-instabilities in the plasma can drive plasma flows, further enhancing the overall transport. This turbulent transport, also known as anomalous transport, can massively influence plasma confinement \cite{plunk_theory_2009}.

To again use our random walk estimate of the diffusion, we start by considering at a fluctuating electric field giving rise to electrostatic turbulence. The electric field leads to a drift according to \cref{eq:E_cross_B_drift}, with a characteristic velocity 
\begin{equation}
    v\textsubscript{E} = \frac{\Delta x}{\tau} = - \frac{\nabla \Phi}{B},
\end{equation}
where $\Phi$ denotes the electric potential. The exact scaling of the  step size and saturation amplitude $e \Phi/T=k_0$ are difficult to obtain; the typical estimate is taken as $k_0=1/16$, yielding 
\begin{equation}
    D^{\text{Bohm}}_e = \frac{T}{16 e B},
\end{equation}
which has a far more significant scaling of the magnetic field and is known as Bohm diffusion \cite{CollisionalTransportMagnetizedHelander}.

In this picture, the characteristic step length is relatively large, determined by the $\vec{E}\times \vec{B}$ drift, which gives rise to Bohm diffusion. 
In contrast, for Gyro-Bohm diffusion, the characteristic step length is assumed to be much smaller, reflecting transport driven by microturbulent fluctuations. To analyse the heat fluxes arising from this anomalous transport, we can take a mixing-length estimate within the random walk model. In this approach, one characteristic length scale dominates the transport. Using the fluxes from \cref{eq:Fick_Gamma,eq:Fick_q} and assuming that
\begin{equation}
    D \sim \chi \sim \frac{(\Delta x)^2}{\Delta t},
\end{equation}
we obtain a Gyro-Bohm diffusivity
\begin{equation}
    \Dgb = - \varrho_i^2 v_{\mathrm{T}} L_{\mathrm{T}}^{-1},
\end{equation}
where the characteristic length scale is taken to be of the order of the Larmor radius, and the time step scales with the linear diamagnetic drift frequency $\omega \sim v_{\mathrm{T}}/L_{\mathrm{T}}$. Here $L_{\mathrm{T}}^{-1} = \dd{\ln{T}}/\dd{r}$ \cite{plunk_theory_2009}.

The Gyro-Bohm heat flux scaling can be defined in an analogous way. While several characteristic length scales are available,  including major radius $R_0$ and radius of curvature $R_{\text{c}}$, we follow the approach of \cite{xanthopoulos_controlling_2014} and normalise by the minor radius $a$,
\begin{equation}
    \Qgb= \frac{\varrho_\mathrm{i}^2}{a^2} c_\mathrm{i}  n_\mathrm{i}  T_\mathrm{i}.
    \label{eq:gyro_bohm}
\end{equation}

\subsubsection{Ion Temperature Gradient (ITG) Mode}
In magnetic confinement devices, turbulent transport is driven by several mechanisms, including the \glsxtrfull{ETG}, \glsxtrfull{TEM}, and \glsxtrfull{ITG} modes \cite{doi:10.1142/8362}. As demonstrated by 
\cite{plunk_stellarators_2019}, the ETG mode is less dominant in stellarators than in tokamaks, assuming $\Te \approx \Ti$. This temperature coupling is expected to be particularly relevant in future reactors and currently represents the primary transport driver in the outer plasma regions of W7-X \cite{beurskens_ion_2021,beurskens_confinement_2021}.

Physically, the ITG mode is an electrostatic drift–interchange instability driven by ion pressure gradients. When a small perturbation occurs in ion temperature or density, it induces a perturbed diamagnetic drift in regions of unfavourable curvature. This creates a pressure imbalance that drives an $\vec{E}\times \vec{B}$ flow, which subsequently reinforces the original perturbation \cite{horton_drift_1999}.

To analyse this mode comprehensively, we first examine the slab ITG mode before considering the additional effects arising from toroidal curvature. For the slab geometry, we assume that the electron bounce frequency $\omega_{\mathrm{be}} = \oint \dd{l}/v_\parallel$ significantly exceeds the ion diamagnetic drift frequency $\omega_\star$. Under a local treatment with constant perpendicular wavenumber $\vec{k}_\perp$, we obtain
\begin{equation}
    \omega_\star = \frac{k_\mathrm{y} \rho v_{\mathrm{T}}}{\sqrt{2}L_{\mathrm{n}} } ,
    \label{eq:omega_star}
\end{equation}
where $k_\mathrm{y} = k_\alpha B_0 (\dd{r}/\dd{\psi})/\sqrt{2}$ is the poloidal (binormal) wavenumber, $v_\mathrm{T}$ is the thermal velocity, and $L_\mathrm{n}$ is the density gradient length scale.

The magnetic drift frequency is given by
\begin{equation}
    \omega_{\mathrm{d}} = \vec{k}_\perp \cdot \vec{v_{\mathrm{d}}},
\end{equation}
where the drift velocity $\vec{v_{\mathrm{d}}}$ comprises components from \cref{eq:general_force_drift}. These include the magnetic curvature drift (with $\Vec{\kappa} = \vec{\hat{b}} \cdot \nabla \hat{b}$, where $\vec{\hat{b}} = \vec{B}/B$) and the gradient drift (with corresponding force $\vec{F} = - \mu \nabla B$), yielding
\begin{equation}
     \omega_{\mathrm{d}} = \vec{\hat{b}} \times \left( \mu \nabla \ln{B} + \frac{v_\parallel^2}{2 \Omega} \Vec{\kappa} \right) .
     \label{eq:omega_d}
\end{equation}
Here, $\mu$ denotes the magnetic moment as defined in \cref{eq:mag_momentum}.

For the slab ITG mode, where $\omega_{\text{d}} = 0$ and $\omega_\parallel \neq 0$, an instability criterion can be derived \cite{kadomtsev_turbulence_1995,plunk_collisionless_2014}. In the long-wavelength limit, this simplifies to
\begin{equation}
    k_{y\text{c}} =  \frac{2 k_\parallel L_\mathrm{n}}{\varrho_\mathrm{i}}   \sqrt{\frac{\tau (1+ \tau)}{\eta (\eta -2)}} ,
    \label{eq:kyc}
\end{equation}
where $\tau = \Ti/(Z \Te)$ represents the temperature ratio. To parametrise the ion temperature gradients, the gradient ratio is defined as:
\begin{equation}
    \eta = \frac{\partial_r \ln{\Ti}}{\partial_r \ln{\nni}} ,
    \label{eq:def_eta}
\end{equation}
where $r$ denotes a radial coordinate.
The expression of \cref{eq:kyc} clearly indicates a threshold for $\eta$, namely $\eta > 2$, for the slab ITG mode to be unstable. Below this threshold, the mode is linearly stable.  
In the limit of weak density gradients ($\eta \to \infty$), the temperature gradient length scale $L_{\mathrm{T}}$ becomes the governing parameter, consistent with the findings of \cite{rewoldt_collisional_1987}. This simplified model provides a fundamental understanding of how a sufficiently steep ion temperature gradient, relative to the density gradient, can drive an instability.

The slab model, while instructive, omits the crucial effects of magnetic curvature and gradient drifts, which are inherent to toroidal confinement devices such as stellarators. For the toroidal ITG mode, the assumption $\omega_\mathrm{d} = 0$ no longer holds and we must revert to \cref{eq:omega_d}. In the toroidal regime, magnetic drifts play a destabilising role, leading to a significantly lower instability threshold for $\eta$. As shown in \cite{biglari_toroidal_1989}, the threshold can be as low as $\eta > 2/3$. Thus, in toroidal plasmas, the susceptibility to ITG instabilities is enhanced compared to idealised slab configurations due to the presence of bad curvature. 

The growth rate ($\gamma$) of the ITG instability, which determines how quickly the perturbation amplifies, is a strong function of $\eta$. Theoretical models and gyrokinetic simulations consistently show that as $\eta$ increases above the critical gradient ($\etacrit$), the linear growth rate of the ITG mode rises sharply \cite{sandberg_finite_2007}. Consequently, a higher $\eta$ results in a more vigorous instability and, therefore, a greater level of turbulent fluctuations. This enhanced linear growth typically translates into increased turbulent transport once the instability reaches its nonlinear saturation phase.

This nonlinear consideration is essential to determine the actual level of turbulent transport, beyond the initial onset and amplification. Furthermore, the interplay between $\eta$ and stabilising mechanisms such as zonal flows and magnetic geometry is also critical, but lies outside the scope of this work. Interested readers may refer to \cite{hegna_theory_2018,tiwari_zonal_2025}.


\section{Equilibrium and Transport Solvers}
To analyse the equilibria and magnetic geometry of the used shots, we use the \glsxtrfull{VMEC} code to analyse the magnetic field geometry. The main tool used to compute the radial heat fluxes for different gradient length scales is the GX gyrokinetic solver.

\subsection{VMEC}
VMEC is commonly used in the field of stellarator optimisation. This code is tasked with computing a three-dimensional MHD equilibrium by minimising the energy functional
\begin{equation}
    W = \int_{\Omega_\mathrm{p}} \left( \frac{1}{2 \mu_0} B^2 +p \right) \dd{V}
\end{equation}
over the (toroidal) domain $\Omega_\mathrm{p}$. The non-trivial solutions are prescribed by magnetic flux surfaces $\Phi$ \cite{hirshman_steepestdescent_1983,hirshman_three-dimensional_1986}.

The radial direction is discretised into $n\textsubscript{s}$ flux surfaces, which define the normalised toroidal flux
\begin{equation}
    s_i = \frac{\Phi_i}{\Phi\textsubscript{LCFS}} \, \, \text{with} \, \, i \in [0, n\textsubscript{s}-1],
\end{equation}
with the boundary set by the \glsxtrfull{LCFS}.
This normalisation allows us to define an effective radius 
\begin{equation}
    \reff = a \sqrt{s},
\end{equation}
where $a$ represents an averaged minor radius. Typically, the normalised radius $\rho$ is used as a radial variable with 
\begin{equation}
    \rho = \reff/a = \sqrt{s}.
\end{equation}
For each experimental shot, the corresponding VMEC equilibrium is computed, and the measured volume is mapped onto a radial coordinate. One of the quantities required to solve the transport equations is \texttt{DVolDs}, the derivative of the plasma volume with respect to the flux coordinate $s$. To express this volume derivative in terms of the effective radius, we can write
\begin{equation}
    V^\prime = \frac{2}{a^2} \reff \cdot \texttt{DVolDs}.
\end{equation}

\begin{figure}[H]
    \centering
    \includegraphics[width = 10cm]{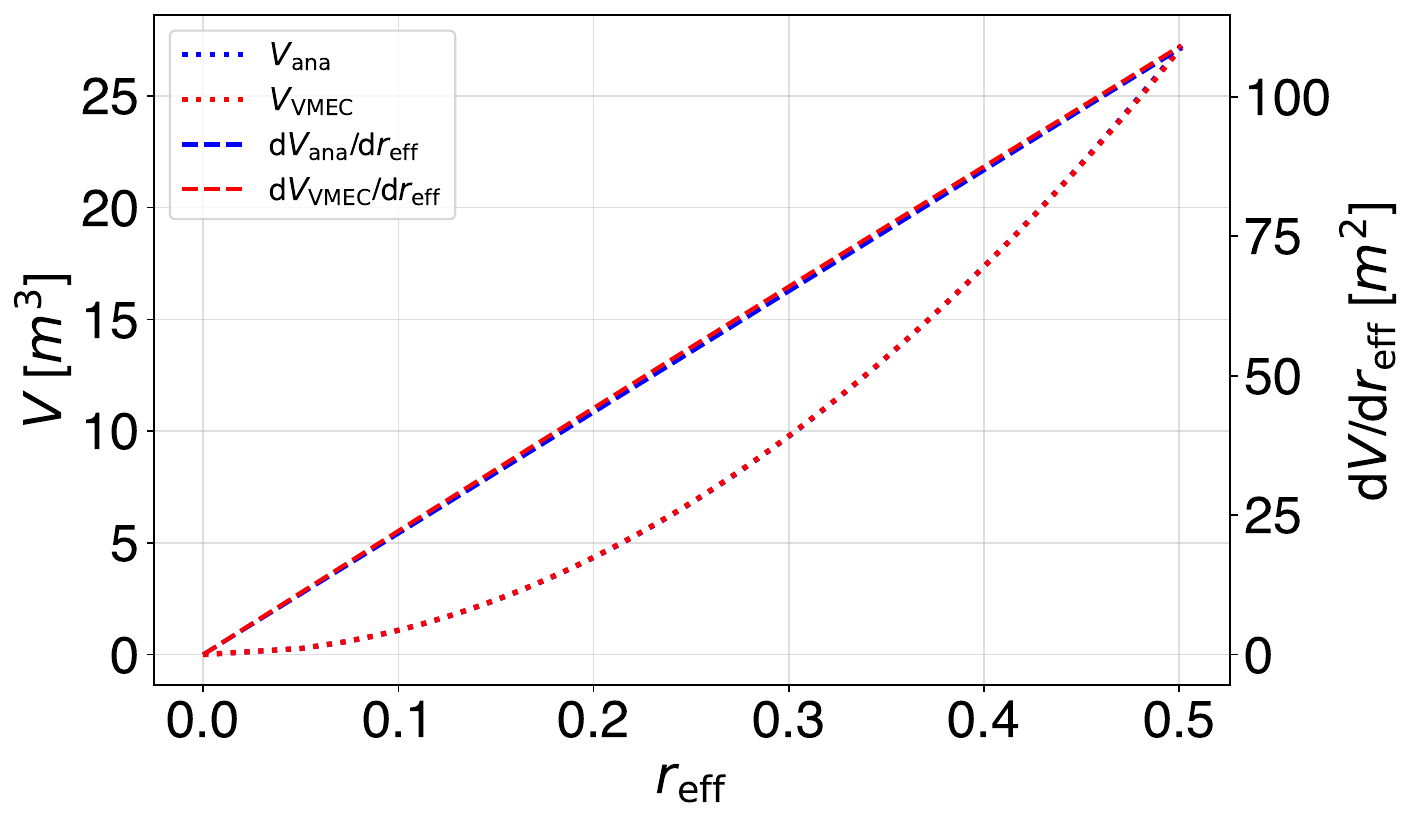}
    \caption[Volume derivative comparison.]{\textbf{Volume derivative comparison.} The blue curves show the volume and its derivative obtained from the VMEC calculation, while the red curves depict the analytical approximation.}
\label{fig:Volume_Derivative_Comparison}
\end{figure}

This volume derivative can be compared with a simple analytical expression by approximating W7-X as a torus with major radius $R_0$, minor radius $a$, and total volume $V$. In that case, 
\begin{equation}
    V^\prime_{\mathrm{ana}}(r) = 4 \pi^2  R_0  r .
\end{equation}
As shown in \cref{fig:Volume_Derivative_Comparison}, this approximation is sufficiently accurate for the subsequent theoretical derivations.

\subsection{GX Code} \label{sec:GX_code}
Extending on the discussion of \cref{sec:Gyrokinetic_Vlasov_Equation}, the GX code solves a system of non-linear gyrokinetic equations that are well suited to describe the low-frequency component in magnetised plasmas \cite{mandell_gx_2024}.

\noindent The length scales relevant for turbulent transport, as discussed in \cref{sec:turbulent_transport}, are much smaller than the macroscopic size of the machine. 
Traditionally, resolving these fine scales necessitated extremely fine grids, rendering simulations computationally intensive. GX addresses this challenge by employing a fully pseudo-spectral approach, utilising Fourier methods in physical space and Laguerre–Hermite polynomials \cite{mandell_laguerrehermite_2018} in velocity space. This formulation allows for spectrally accurate evaluations of derivatives and efficient handling of the velocity-space dynamics, effectively reducing the dimensionality of the problem without compromising physical fidelity. Moreover, the GX code is optimised for GPU architectures, leveraging their parallel processing capabilities to achieve substantial performance gains over conventional CPU-based simulations \cite{mandell_gx_2024}.

The GX simulations are used to compute the turbulent heat fluxes by performing flux-tube simulations at various radial positions. The net particle flux can either be enforced to vanish ($\Gamma = 0$) or be determined as an output. In the cases considered here, we focus on calculating the heat flux for radial positions with relevant $\alT$ values for both cases. And then interpolate between the set of simulations to get a full $Q(\rho, \alT)$ as discussed in \cref{sec:Sparse_Heat_Model}.

\section{Scaling Laws} \label{sec:scaling_laws}
The energy confinement time is defined as
\begin{equation}
\tau_{\text{E}} = \frac{W}{P} ,
\label{eq:tau_E_def}
\end{equation}
where $W$ is the stored plasma energy, and $P$ denotes the power loss. In steady-state conditions, the input power balances the power losses, yielding $P_{\text{in}} = P_{\text{loss}}  =P$. The power can be expressed as the product of heat flux $Q$ and surface area $A$, leading to $P = Q A$.

Approximating the plasma geometry as a torus, the surface area scales as $A \sim a R_0$.
The heat flux $Q$ is expected to scale with the Gyro-Bohm flux defined in \cref{eq:gyro_bohm},
\begin{equation}
Q = \hat{Q} \times \Qgb ,
\label{eq:heat_flux_model}
\end{equation}
where $\hat{Q}$ represents a dimensionless function independent of control parameters, leading to $Q \sim \Qgb$.

The stored energy $W$ over volume $V$, under the assumptions $\nni = \nne =n$ and $\Te = \Ti =T$, is
\begin{equation}
W/V = \frac{3}{2 V}  \int n T \dd{V} \sim \langle n T \rangle \sim n_{\text{bc}} T_{\text{bc}}.
\end{equation}
Here, $\langle n T \rangle$ denotes the volume-averaged plasma pressure $p= nT$, whilst $n_{\text{bc}}$ and $T_{\text{bc}}$ represent the plasma density and temperature at a specified radial (boundary condition) position. The first step follows directly from the definition, whilst the second assumes a universal shape for the density and temperature profiles, enabling the volume average to be related to values at a fixed radial position. This assumption is a priori not obvious and will be justified in the following.

Solving the Gyro-Bohm heat flux expression (\cref{eq:gyro_bohm}) for the temperature yields
\begin{equation}
T_{\text{bc}} = \left( \Qgb \sqrt{\frac{e}{\mmi}} a^2 B^2 \frac{1}{n_{\text{bc}}} \right)^{\frac{2}{5}}.
\label{eq:T_bc_gyro_Bohm}
\end{equation}
Substituting these expressions into \cref{eq:tau_E_def} and employing the analytical approximation for the volume, we obtain
\begin{align}
\tau_{\text{E}} &\sim \frac{n_{\text{bc}} V}{P  }  \left( \Qgb \sqrt{\frac{e}{\mmi}} a^2 B^2 \frac{1}{n_{\text{bc}}} \right)^{\frac{2}{5}} \\
&\sim  a^{2.4} R_0^{0.6} P^{-0.6} n_{\text{bc}}^{0.6} B^{0.8} .
\label{eq:tau_E_prediction}
\end{align}

 A more rigorous calculation utilising Gyro-Bohm scaling, as performed in \cite{stroth_stellarator-tokamak_2020}, yields
\begin{equation}
\tau_{\text{GB}} = 0.089 \, a_{\text{eff}}^{2.2} R_0^{0.8} P^{-0.6} \overline{n}_{\text{e}}^{0.6} B^{0.8} \iotaslash_{2/3}^{0.6},
\label{eq:tau_gb}
\end{equation}
where $\overline{n}_{\text{e}}$ denotes the line-averaged electron density, and $\iotaslash_{2/3}$ represents the reduced rotational transform at $r/a = 2/3$.

Remarkably, this theoretical derivation exhibits strong similarity to the empirical \glsxtrfull {ISS04} \cite{yamada_characterization_2005}
\begin{equation}
\tau_{\text{ISS04}} = 0.134 \, a^{2.28} R_0^{0.64} P^{-0.61} \overline{n}_{\text{e}}^{0.54} B^{0.84} \iotaslash_{2/3}^{0.41}.
\label{eq:tauISS04}
\end{equation}
This correspondence between the theoretical Gyro-Bohm scaling and the empirical ISS04 law is particularly striking when comparing the exponents, especially in the version of \cref{eq:tau_E_prediction}. The ISS04 scaling law was derived empirically from confinement time measurements across multiple stellarator experiments and has become a standard benchmark. In practice, a renormalisation factor $f_{\text{ren}}$ is typically introduced to account for configuration-specific effects on confinement properties
\begin{equation}
    \tau_{\text{E}} = f_{\text{ren}} \tau_{\text{ISS04}} .
    \label{eq:fren}
\end{equation}

Future fusion reactors must achieve sufficiently high Q-factors (i.e., fusion gain) to ensure economically viable electricity production. This requirement drives operation in parameter regimes that are fundamentally different, particularly in terms of size, from those of current experimental devices. 

Various scaling approaches can be employed to predict the performance of new machines and enhance understanding of existing experiments.
Since turbulence constitutes a primary energy loss mechanism in stellarators, the Gyro-Bohm scaling provides a crucial framework for understanding confinement trends in future reactors \cite{banon_navarro_assessing_2024}.

\section{Thesis Hypothesis}
Building upon the established connection between theoretical Gyro-Bohm scaling and empirical observations of global energy confinement time, this thesis proposes a focused investigation into the fundamental local scaling properties of the outer plasma regions. While the remarkable correspondence observed at a global level suggests underlying universal scaling relationships, a direct, local examination is crucial to validate the microscopic transport mechanisms responsible for these macroscopic behaviours and to enhance predictive capabilities for future fusion devices. The central hypothesis of this thesis can be formulated as:

\textit{A constant scaling of the logarithmic temperature derivative exists in the outer plasma region, which exhibits only weak dependence on device configurations. This local behaviour is consistent with Gyro-Bohm scaling across individual flux surfaces. 
Evidence of a universal edge profile shape, or refined scaling laws describing it, offers a more reliable basis for predicting confinement using realistic boundary conditions in the transport equations.}

This investigation shall be systematically pursued through three distinct objectives:

\begin{enumerate}
\item Conduct a comprehensive analysis of existing Thomson Scattering diagnostic data to extract robust estimates of the temperature gradient scaling length. These derived quantities will subsequently serve as essential input parameters for calculating heat fluxes through gyrokinetic simulations.
\item Investigate the validity of derived scaling following from the Gyro-Bohm definition, like the scaling of the temperature with power.
\item Use the heat fluxes calculated from direct numerical simulations of turbulence with the GX code to reconstruct temperature and density profiles, thereby verifying the internal consistency of the proposed model framework.
\end{enumerate}

Furthermore, the findings are discussed, and a final conclusion of the hypothesis is drawn.
\chapter{Methods} \label{sec:Methods}

\section{Heat Flux Models} \label{sec:Heat_Flux_Models}

We are interested in the conservation of energy, in analogy to the conservation of density as stated in \cref{eq:conservation_eq}. Let the (kinetic) energy density be denoted by $e_{\mathrm{kin}}$ and the energy flux by $\vec{q}$. While the conservation law applies in three dimensions, we are in this case only concerned with the radial component. 

\noindent To express the equation solely as a function of radius, the toroidal and poloidal coordinates are averaged out by performing a flux-surface average over all quantities \cite{wappl_web_2024}. The non-trivial term is the divergence of the heat flux, which can be written as

\begin{equation}
\begin{split}
    \langle \nabla \cdot q \rangle &= \frac{\iiint \vec{q} \, \dd^3{r}}{\iiint \dd^3{r}} \\
    &= \lim_{ \Delta \rho \to 0}   \frac{1}{\Delta V (\Delta \rho)} \left[ \oiint_{\rho + \Delta \rho} \vec{q} \cdot \dd{\vec{S}} - \oiint_{\rho} \vec{q} \cdot \dd{\vec{S}} \right] \\
    &= \lim_{ \Delta \rho \to 0}  \frac{1}{V'(\rho) \Delta \rho} \left[ \oiint_{\rho + \Delta \rho} \frac{\vec{q} \cdot \nabla \rho}{\abs{\nabla\rho}} \dd{S} - \oiint_{\rho} \frac{\vec{q} \cdot \nabla \rho}{\abs{ \nabla \rho}} \dd{S} \right] \\
    &= \lim_{ \Delta \rho \to 0} \left(  \frac{1}{V'(\rho) \Delta \rho}  \frac{\partial}{\partial \rho}\left[ \oiint_{\rho} \frac{\vec{q} \cdot \nabla \rho}{\abs{\nabla \rho}} \dd{S} \right] \Delta \rho + \mathcal{O}\left( (\nabla \rho)^2 \right) \right) \\
    &= \frac{1}{V^\prime} \frac{\partial}{\partial \rho} \left[ V^\prime (\rho) Q(\rho) \right],
\end{split}
\end{equation}
where the flux-surface averaged heat flux is defined as
\begin{equation}
    Q(\rho) = \frac{1}{V'} \oiint_{\rho} \frac{\vec{q} \cdot \nabla \rho}{\abs{\nabla \rho}} \dd{S} .
\end{equation}

\noindent Identifying the source term with the power $P$, the power balance equation becomes
\begin{equation}
   \partial_t e_{\mathrm{kin}}  = \frac{1}{V'(\rho)}\frac{\dd{}}{\dd{\rho}}[V'(\rho)Q(\rho)] - P(\rho). 
\end{equation}
In steady state, the partial time derivative vanishes, yielding
\begin{equation}
       P(\rho)  = \frac{1}{V'(\rho)}\frac{\dd{}}{\dd{\rho}}[V'(\rho)Q(\rho)]. 
       \label{eq:Power_Balance_eq}
\end{equation}

\noindent The heat flux is expected to scale with the Gyro-Bohm heat flux, as defined in \cref{eq:gyro_bohm} and \cref{eq:heat_flux_model}.
The function $\hat{Q}$ is determined by the specific heat flux model under consideration. Two examples used in this thesis are explained in the following sections.

\subsection{Critical Gradient Model}\label{sec:CG-Model}
\begin{figure}[H]
    \centering
    \includegraphics[width = 10cm]{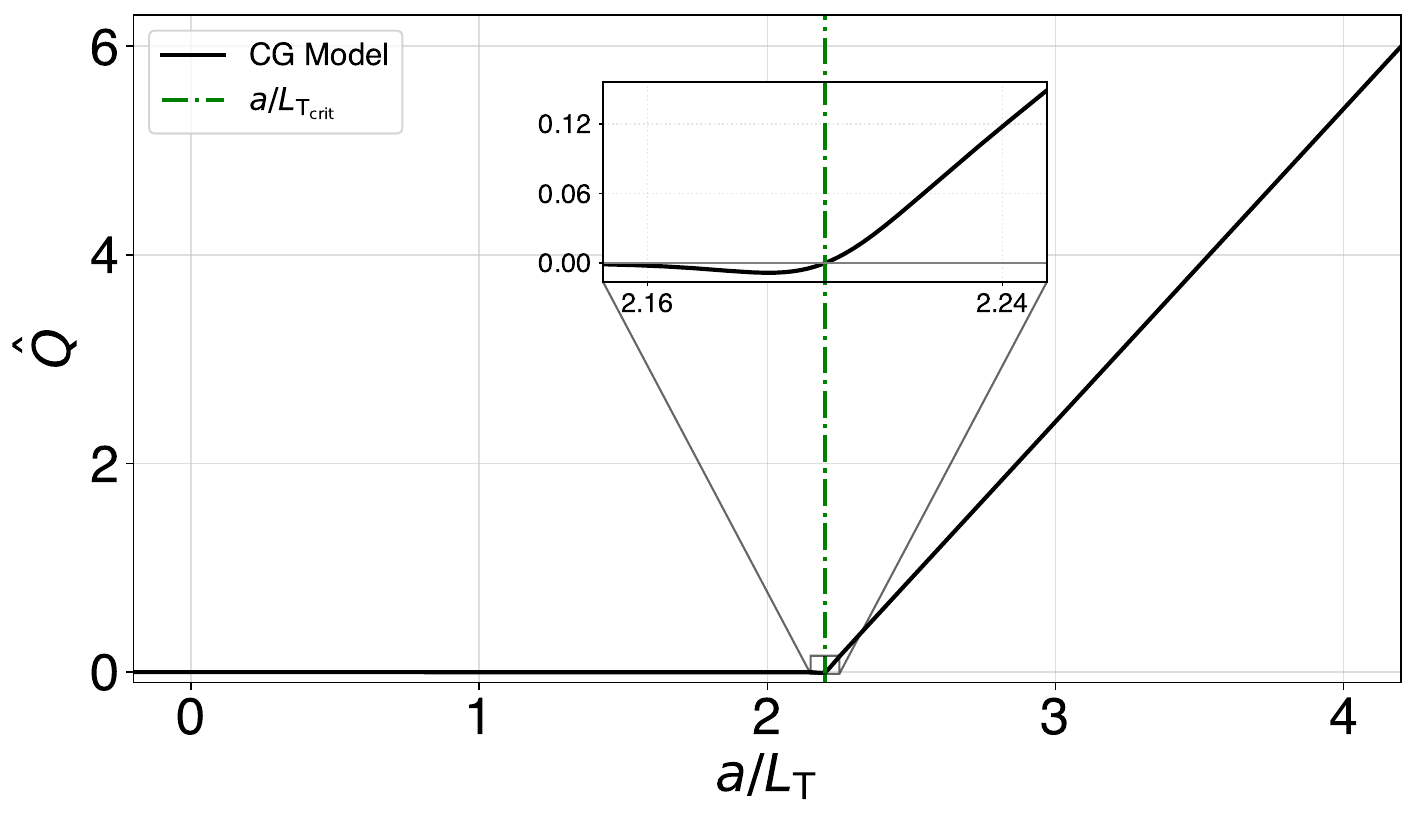}
    \caption[CG model illustration.]{\textbf{CG model illustration}. The normalised heat flux, with critical gradient $\alTcrit = 2.2$ and $\hat{C} = 3$ as predicted by \cref{eq:simple_Qhat_Model}, is shown. The transition region around $\alTcrit$ is highlighted.}
\label{fig:CG_Model_Illustration}
\end{figure}

Assuming we are interested only in the global power deposition $\mathsf{P}$, for example, the power deposited by ECRH heating, rather than detailed radial heating profiles, we can integrate over the volume to obtain a simplified equation. This integration is reasonable as we expect the heat absorption to occur at positions further inward than the region we are interested in describing with our model:
\begin{equation}
    \mathsf{P} = Q V^\prime .
\end{equation}
This means that our heat flux model is determined by:
\begin{equation}
    Q = \frac{\mathsf{P}}{V^\prime}.
\end{equation}
The critical heat flux model suggests that there is a critical gradient scaling length $\alTcrit$ below which there is no heat flux, followed by a linear increase thereafter.
This motivates a form of $\hat{Q}$ which reads as:
\begin{equation}
    \hat{Q} = C \times \text{H} \left( \aLT - \aLTcrit \right) \times \left( \aLT - \aLTcrit \right).
    \label{eq:simple_Qhat_Model}
\end{equation}
To ensure a smooth numerical implementation, the Heaviside function $H(x)$ is approximated by,
\begin{equation}
    H(x) = \frac{1}{2} (1 + \tanh(\alpha x)) , \quad \alpha = 50,
    \label{eq:Heaviside_Numerical}
\end{equation}
where the original function is recovered in the limit $\alpha \to \infty$. 
The heat flux model of \cref{eq:simple_Qhat_Model} with the transition governed by \cref{eq:Heaviside_Numerical} is illustrated in \cref{fig:CG_Model_Illustration}.

\subsection{Radially Dependent Sparse Model} \label{sec:Sparse_Heat_Model}
In contrast to the previously described approach, this model incorporates radial dependence, allowing the critical gradient threshold to vary across different plasma radii. Rather than fitting directly to temperature data, this model is meant to interpolate between heat flux values obtained from gyro-kinetic simulations with exponential analytical continuation, ensuring smoothness at critical gradient thresholds.

For each radial position $\rho_i$, we determine critical parameters through a linear fit to the first two data points,
\begin{equation} 
Q_h^{\text{linear}}(x, \rho_i) = a_i + b_i x .
\label{eq:fullrad_linear}
\end{equation}
From this fit, we derive the local critical gradient
\begin{equation} 
 Q_h^{\text{linear}}(x_0(\rho_i), \rho_i) = 0 \Rightarrow x_0(\rho_i) = -\frac{a_i}{b_i} .
\label{eq:fullrad_x0}
\end{equation}
These individual critical gradients are stored to form a radial profile
\begin{equation} 
\vec{x_0} = \{(\rho_1, x_0(\rho_1)), (\rho_2, x_0(\rho_2)), \ldots, (\rho_n, x_0(\rho_n))\}.
\label{eq:fullrad_x0data}
\end{equation}
Additionally, we identify and record the maximum gradient observed in the experimental data as $x_1(\rho_i)$ for each radial location. An adaptive fitting approach that initially implements a quadratic function but reverts to a linear model when necessary is employed, reading as
\begin{equation} \label{eq:sparse_fit}
Q_h^{\text{fit}}(x, \rho_i) = \alpha_i + \beta_i x + \gamma_i x^2.
\end{equation}
When the quadratic coefficient would produce non-physical behaviour, we transition to linear fitting \cite{plunk_profile_nodate}
\begin{equation} \label{eq:sparse_adapt}
 \gamma_i < 0 \Rightarrow Q_h^{\text{fit}}(x, \rho_i) = \alpha_i + \beta_i x, \quad \gamma_i = 0.
\end{equation}
The polynomial derivative at the critical gradient is
\begin{equation} \label{eq:sparse_poly_deriv}
\frac{\dd{Q_h^{\text{fit}}}}{\dd{x}}\bigg|_{x=x_0(\rho_i)} = \begin{cases}
\beta_i & \text{for linear fit,} \\
2\gamma_i x_0(\rho_i) + \beta_i & \text{for quadratic fit.}
\end{cases}
\end{equation}

The exponential analytical continuation for sub-threshold gradients is defined as
\begin{equation} \label{eq:sparse_exp_continuation}
Q_h(x, \rho_i) = \frac{\dd{Q_h^{\text{fit}}}}{\dd{x}}\bigg|_{x=x_0(\rho_i)} \cdot (x - x_0(\rho_i) + \varepsilon) \cdot \exp\left(\frac{x - x_0(\rho_i)}{\lambda}\right)
\end{equation}
for $x < x_0(\rho_i)$, where $\lambda$ is the characteristic length scale controlling the decay rate and $\varepsilon$ is a small regularisation parameter ensuring numerical stability.

For super-threshold gradients ($x \geq x_0(\rho_i)$), the polynomial fit is used directly
\begin{equation} \label{eq:sparse_superthreshold}
Q_h(x, \rho_i) = Q_h^{\text{fit}}(x, \rho_i) \quad \text{for} \quad x \geq x_0(\rho_i).
\end{equation}
 The complete radial-gradient data set is constructed by discretising the gradient space with step size $\Delta x$ up to a maximum gradient $x_{\text{max}}$, creating a comprehensive three-dimensional data structure
\begin{equation} \label{eq:sparse_3d_data}
\mathcal{D} = \{(\rho_i, x_j, Q_h(x_j, \rho_i)) : i \in [1,n_\rho], j \in [1,n_x]\}.
\end{equation}
 Boundary conditions are enforced by extending the data to $\rho = 0$ and $\rho = 1$ using the values from the nearest available radial positions.  The interpolation functions for critical parameters utilise linear interpolation across radial positions:
\begin{align} 
x_0(\rho) &= \mathcal{L}_1(\rho; \vec{x_0}), \label{eq:sparse_interp_x0} \\
x_1(\rho) &= \mathcal{L}_1(\rho; \vec{x_1}), \label{eq:sparse_interp_x1} \\
\rho_{\text{norm}}(\rho) &= \mathcal{L}_1(\rho; \vec{\rho_{\text{norm}}}) ,\label{eq:sparse_interp_rho}
\end{align}
where $\mathcal{L}_1$ denotes linear interpolation and the vectors represent the radial profiles of the respective quantities. 
The complete heat flux function integrates these components through two-dimensional linear interpolation:
\begin{equation} \label{eq:sparse_flux}
Q_h(x, \rho) = \mathcal{I}_{2D}(\mathcal{D}; \rho, x),
\end{equation}
where $\mathcal{I}_{2D}$ denotes two-dimensional linear interpolation over the complete data set $\mathcal{D}$, ensuring continuity and smoothness across both radial and gradient dimensions while maintaining the exponential analytical continuation properties at sub-threshold gradients.

\section{Data Filtering} \label{sec:Data_Filtering}
The experimental data for both the electron temperature and density are obtained through the TS diagnostics, as discussed in \cref{sec:W7X_Thomson_Scattering}.

\begin{figure}[H]
    \centering
    \includegraphics[width = 10cm]{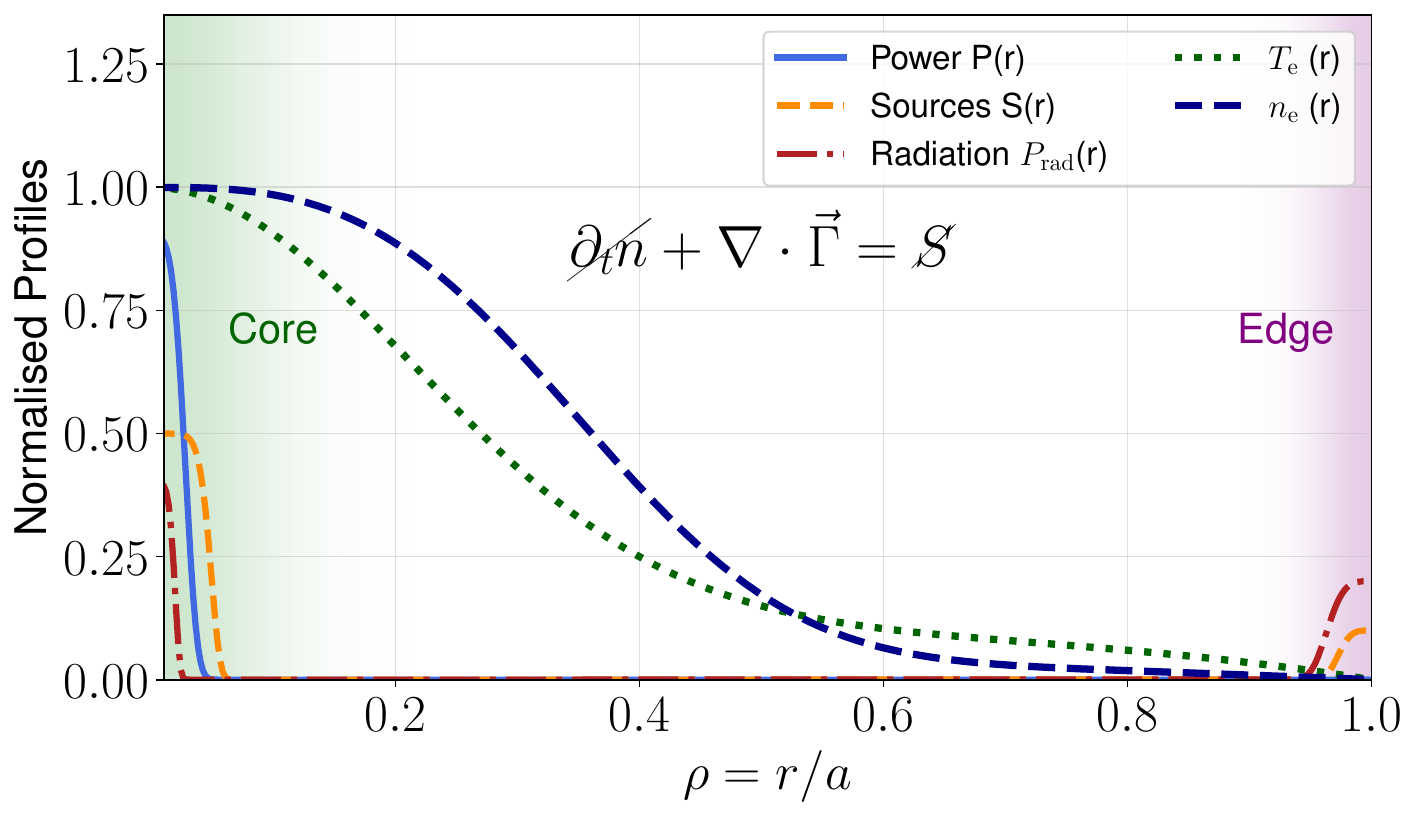}
    \caption[Radial experimental profiles.]{\textbf{Radial experimental profiles.} Illustration of heating power absorption, radiation and fuelling. The exact profile form and height are arbitrary. }
\label{fig:experimental_profiles}
\end{figure}

The typical radial profiles of temperature and density are illustrated in \cref{fig:experimental_profiles}. Here, the absorbed and radiated power, as well as fuelling are illustrated. 

To exclude the effects of heating and radiation, we shall restrict the radial range we are examining. 
In addition, the data must be filtered to ensure that the plasma is in steady state with net sources consistent with the assumptions of the heat-flux model, and to remove outliers most likely arising from measurement errors. 

Focusing on well-coupled, steady-state profiles is essential: only in such a consistent physical scenario can the steady-state power balance be meaningfully applied, as transient or weakly coupled conditions would obscure the discussed transport processes. This consistency is crucial for uncovering generic or even universal behaviour, since deviations from steady state would reflect localised or machine-specific effects.

\subsection{Steady-State Filter}

To determine the steady-state condition of the time shot, we compare the energy confinement time (\cref{eq:tau_E_def}), expressed as
\begin{equation}
    \tau_E = \frac{W_{\text{dia}}}{\mathsf{P}},
\end{equation}
where $W_{\text{dia}}$ is the diamagnetic energy, with the relative change of the energy. This leads to the criterion
\begin{equation}
   \hat{D}_1 = \frac{1}{W} \frac{\dd{W}}{\dd{t}} < C_1 \frac{1}{\tau_E},
    \label{eq:steady_state_filter_1}
\end{equation}
where $C_1$ is a coefficient. 
For a more stable numerical result and to avoid saddle points, which are not in steady state but are less affected by the previous criterion, we additionally enforce
\begin{equation}
     \hat{D}_2 = \frac{1}{W} \left( \frac{\dd^2{W}}{\dd{t^2}} \right) < C_2 \frac{1}{\tau_E^2}.
       \label{eq:steady_state_filter_2}
\end{equation}

\begin{figure}[H]
    \centering
    \includegraphics[width = 10cm]{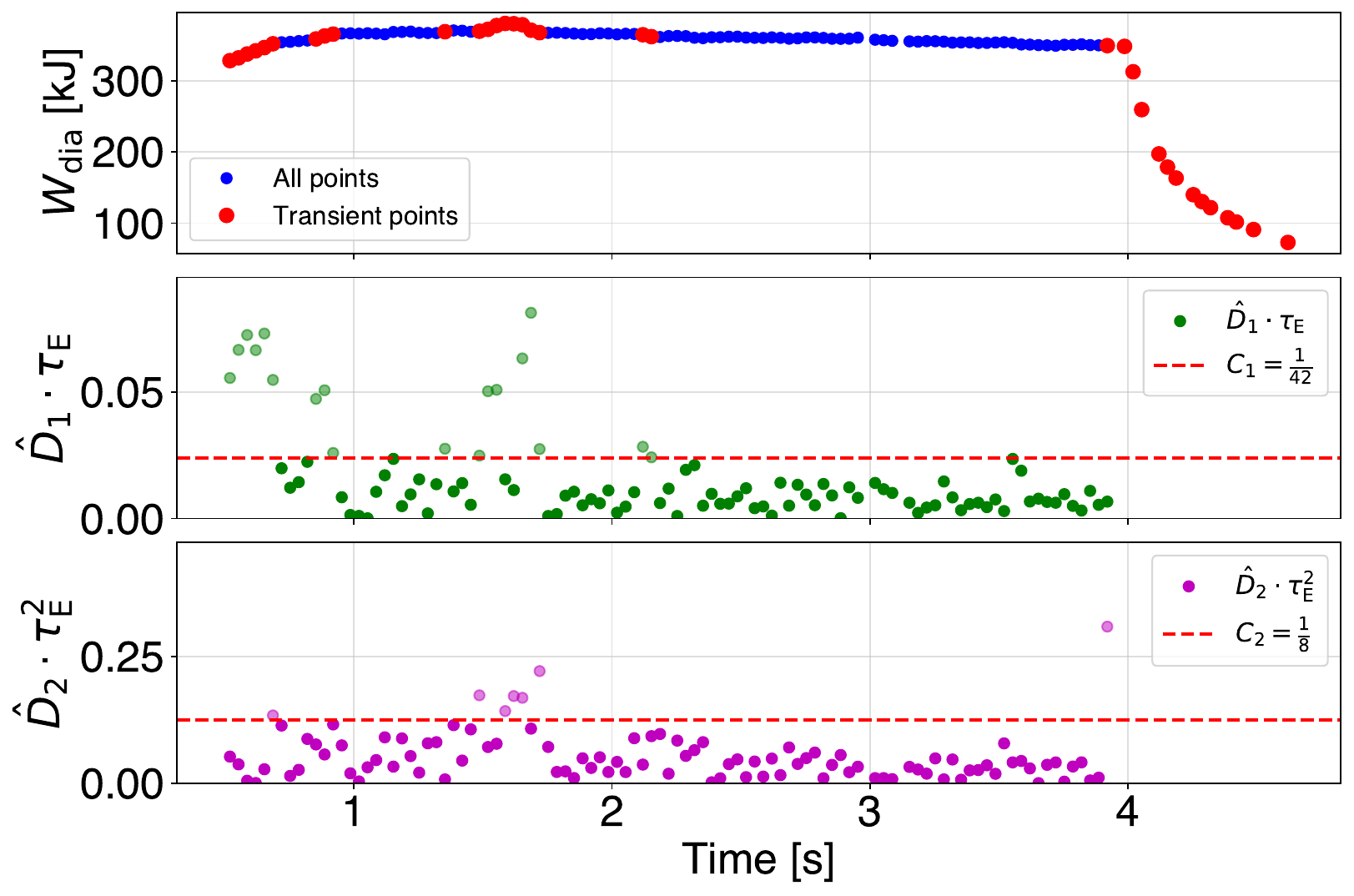}
    \caption[Steady-state filter.]{\textbf{Steady-state filter} The steady-state criterion \cref{eq:steady_state_filter_1} and \cref{eq:steady_state_filter_2} are illustrated. The coefficients used in this example are $C_1 =1/42$ and $C_2=1/8$. }
\label{fig:steady_state_filter}
\end{figure}
The values of the coefficients are not known a priori. Trial and error suggests approximate values of $C_1 = 1/42$ and $C_2 = 1/8$. However, the exact coefficients and their reasonable scaling can vary from shot to shot. Therefore, a more flexible approach is needed based on the data itself, to identify which regions can be considered in steady state. An example of this filtering process is shown in \cref{fig:steady_state_filter}.

\subsubsection{Peak Detection Method}
In the peak detection method, we identify and exclude regions corresponding to peaks in the first and second derivatives, as these represent rapid transitions rather than steady states. We calculate the normalised derivatives $\hat{D}_1$ and $\hat{D}_2$.

\begin{figure}[H]
    \centering
      \includegraphics[width = 12cm]{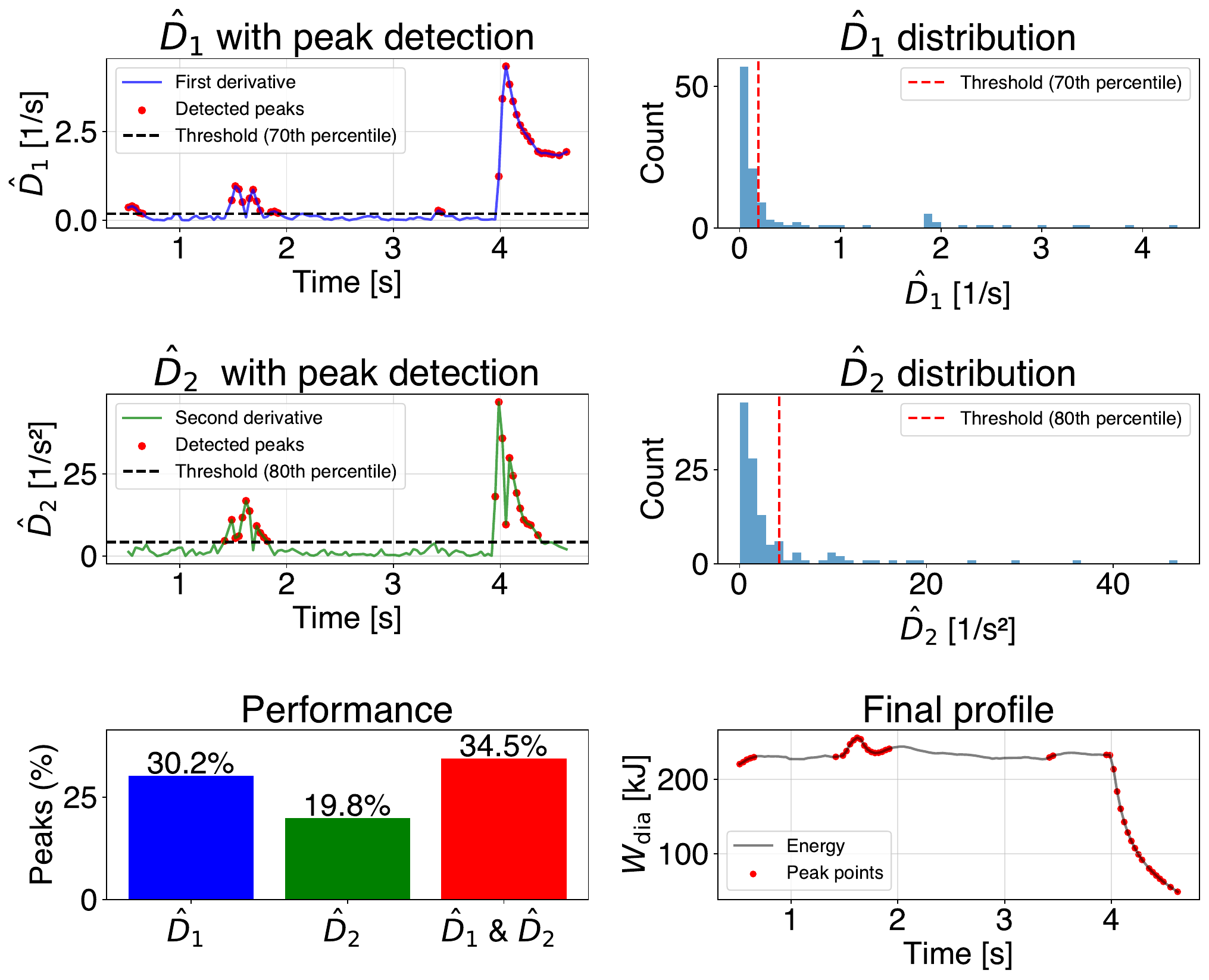}
    \caption[Steady-state filter: Peak detection.]{\textbf{Steady-state filter using peak detection.} The upper plot displays the first and second derivatives alongside a histogram of the data points. The lower plot illustrates the number of points captured by the first derivative filter and the second derivative filter. The combined filter is implemented as the logical AND of both individual filters. The lower right subplot shows the identified non-steady-state points.}
\label{fig:steady_state_filter_peak_detection}
\end{figure}
Instead of using fixed coefficients, we define the thresholds based on percentile values of the derivative magnitudes:
\begin{equation}
    \theta_1 = \text{Percentile}_{p_1}(|\hat{D}_1|),
    \label{eq:first_deriv_threshold},
\end{equation}
\begin{equation}
    \theta_2 = \text{Percentile}_{p_2}(|\hat{D}_2|),
    \label{eq:second_deriv_threshold}
\end{equation}
where $\hat{D}_1$ and $\hat{D}_2$ represent the first and second derivatives of the signal, respectively. In this work, the thresholds correspond to the 70th and 80th percentiles, chosen based on their performance across a representative set of discharges.
We identify peak regions where
\begin{equation}
    |\hat{D}_1| > \theta_1 \quad \text{or} \quad |\hat{D}_2| > \theta_2 .
    \label{eq:peak_regions}
\end{equation}
The steady-state regions are then defined as the complement of these peak regions:
\begin{equation}
    \text{Steady state} = \{t \mid |\hat{D}_1| \leq \theta_1 \: \text{and} \: |\hat{D}_2| \leq \theta_2\} .
    \label{eq:steady_state_regions}
\end{equation}
From these thresholds, we can derive the coefficients $C_1$ and $C_2$ by relating them to the energy confinement time:
\begin{equation}
    C_1 = \frac{\theta_1}{\overline{(1/\tau_E)}},
    \label{eq:C1_peak}
\end{equation}
\begin{equation}
    C_2 = \frac{\theta_2}{\overline{(1/\tau_E^2)}},
    \label{eq:C2_peak}
\end{equation}
allowing a comparison of this method with the previously manually set coefficients. Here, the overline denotes the median value.

\subsubsection{Moving Average Method}
The moving window approach identifies steady-state regions by analysing local variations in the diamagnetic energy. For each point in the time series, we calculate the local mean and standard deviation within a window of size $N$, where $W_j$ represents the diamagnetic energy values at time index $j$:
\begin{equation}
    \mu_i = \frac{1}{N}\sum_{j=i-N/2}^{i+N/2} W_j ,
    \label{eq:local_mean}
\end{equation}
\begin{equation}
    \sigma_i = \sqrt{\frac{1}{N}\sum_{j=i-N/2}^{i+N/2} (W_j - \mu_i)^2}.
    \label{eq:local_std}
\end{equation}
Here, the window size is set to $N=6$.
We then normalise the standard deviation by the local mean to obtain a dimensionless measure of the variation $\hat{\sigma}_i$.
A point is considered to be in steady state if this normalised standard deviation is below a specified threshold $\epsilon$
\begin{equation}
    \hat{\sigma}_i = \frac{\sigma_i}{\mu_i} < \epsilon.
    \label{eq:steady_state_window}
\end{equation}

\begin{figure}[H]
    \centering
    \includegraphics[width = 12cm]{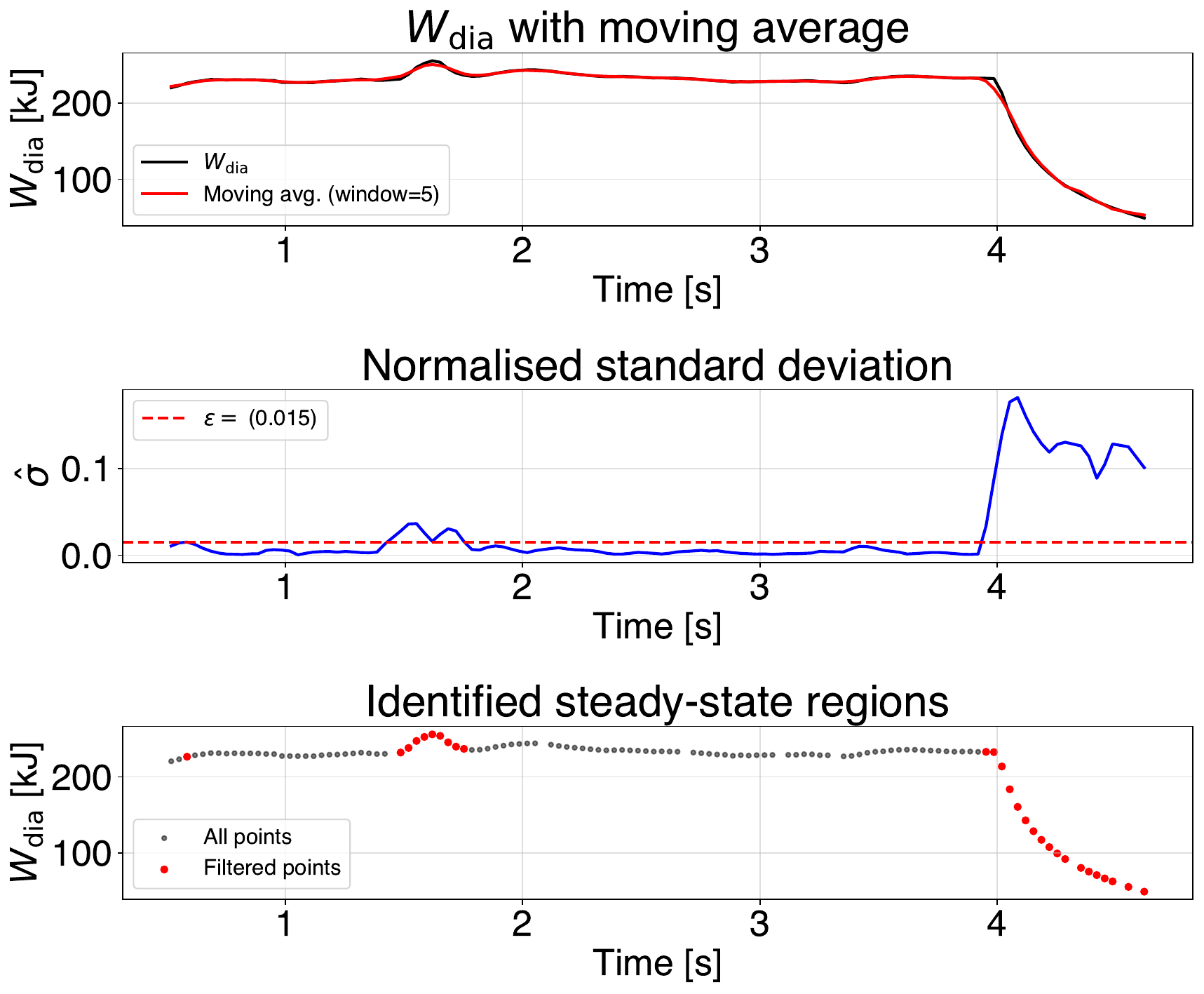}
    \caption[Steady-state filter: Window averaging.]{\textbf{Steady-state filter using window averaging.} The top plot shows the original signal together with its smoothed version; the middle plot displays the standard deviation with the selected cut-off; and the bottom plot illustrates the resulting filtered points.}
\label{fig:steady_state_filter_window_averaging}
\end{figure}
After identifying steady-state regions through this criterion, we compute the derivatives within these regions to determine appropriate coefficients. Instead of using median values as in the peak detection method, we use the 90th percentile:
\begin{equation}
    C_1 =  \text{Percentile}_{90}(\hat{D}_1 \, \tau_E),
    \label{eq:C1_window}
\end{equation}
\begin{equation}
    C_2 = \text{Percentile}_{90}(\hat{D}_2 \, \tau_E^2) .
    \label{eq:C2_window}
\end{equation}
An illustration of this approach is shown in \cref{fig:steady_state_filter_window_averaging}.  This approach has the advantage of being localised in time, making it more robust against slight changes and fluctuations in a flat $W_{\text{dia}}$ curve. 

\subsection{Density Transients Filtering}
In addition to requiring the vanishing of partial time derivatives under steady-state conditions, we must also ensure the absence of global particle sources and sinks, thereby maintaining an approximately constant integrated density. The integration of the radial density profile is proportional to the line interferometer measurements, as discussed in \cref{sec:Interferometry_Rescaling}. Consequently, we employ these measurement values directly rather than integrating the radial density profile, which exhibits greater noise levels.

\begin{figure}[H]
    \centering
    \includegraphics[width = 10cm]{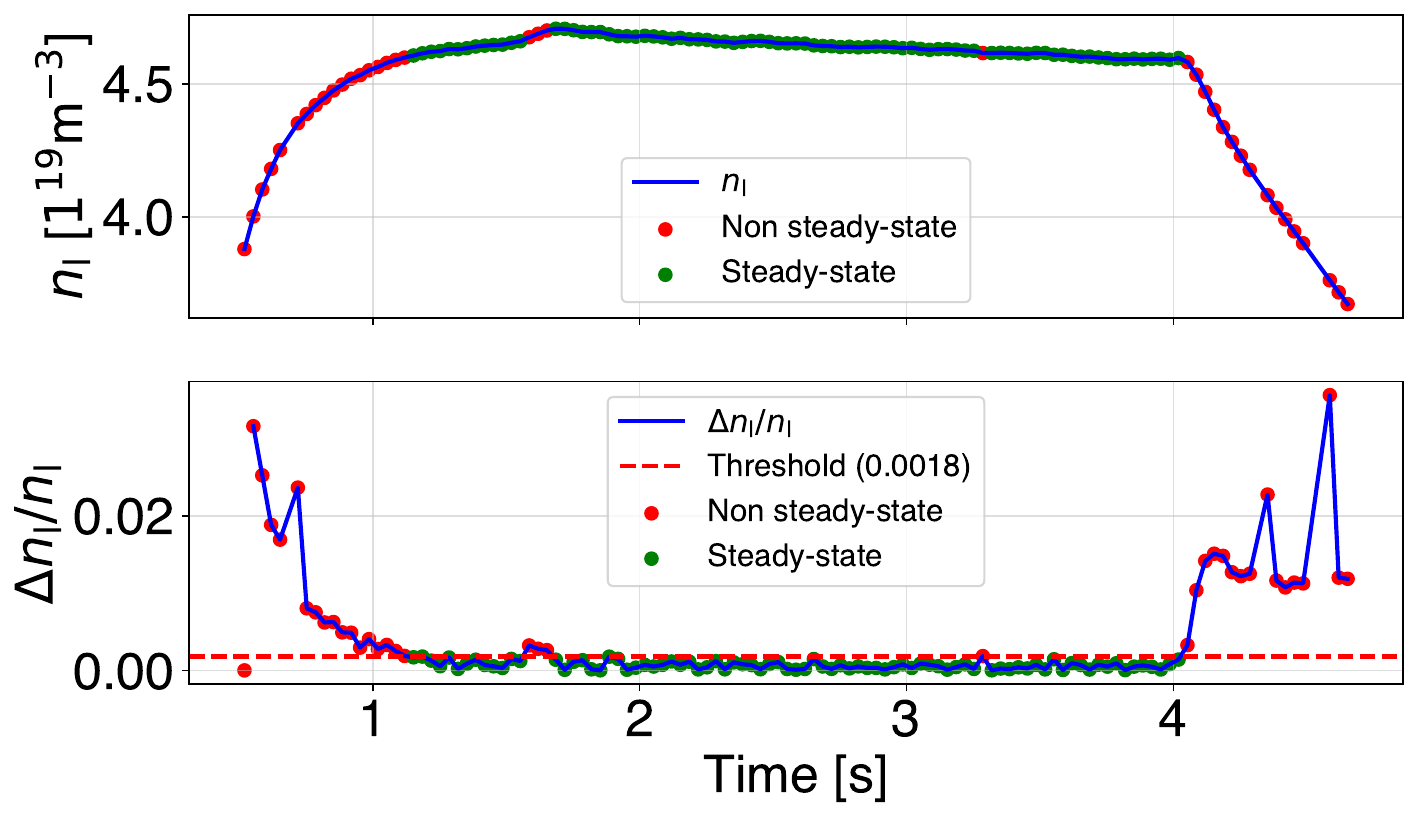}
    \caption[Integrated density filtering.]{\textbf{Filtering of the integrated density.} The upper subplot shows the integrated density with an indication of which points have been identified as steady state. The lower subplot shows the relative change as described in \cref{eq:steady_state_density} with the set threshold for this shot of $\epsilon = \SI{1.8}{\percent}$.}
\label{fig:density_filter_procedure}
\end{figure}

The filtering methodology is relatively straightforward. For each point in the time series of line integrated density measurements $\nnl$, we calculate the normalised point-to-point variation:
\begin{equation}
    \Delta \hat{\nnl}_i = \frac{|n_{\text{l}_i} - n_{\text{l}_{i-1}}|}{n_{\text{l}_{i-1}}},
    \label{eq:norm_density_change} 
\end{equation}
where $n_{\text{l}_i}$ represents the line integrated density at time index $i$.
A temporal point is classified as being in steady state if this normalised change remains below a specified threshold $\epsilon$:
\begin{equation}
    \Delta \hat{\nnl}_i < \epsilon .
    \label{eq:steady_state_density}
\end{equation}

All temporal points that exceed this threshold are classified as transient and subsequently excluded from the analysis. An illustration of this filtering procedure is presented in \cref{fig:density_filter_procedure}.

\subsection{Temperature Filtering}
Apart from already known faulty measurement locations, specific shots clearly display visible outliers. These outliers manifest as both strong temperature peaks and dips, which means a simple global temperature threshold cannot be set to exclude these points. To identify the steep gradients and consequently the unphysical points, we apply a Fourier transform to separate the underlying smooth profile from localised anomalies.

\begin{figure}[H]
    \centering
    \includegraphics[width = 12cm]{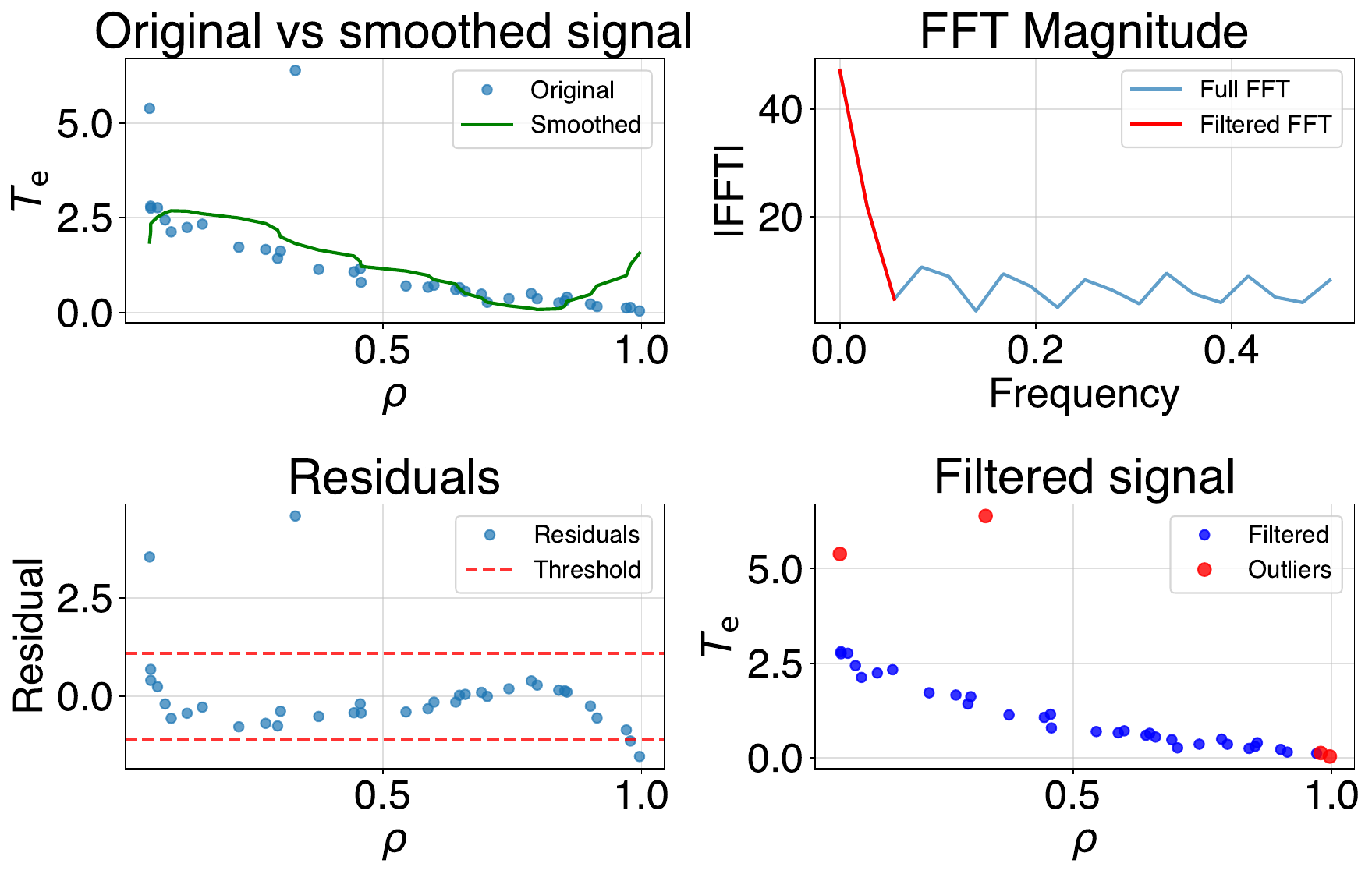}
\caption[Temperature filtering.]{\textbf{FFT filter of the temperature.} The threshold for the Fourier components was set to 10\%, meaning that in the smoothed profile, shown in the top-left subplot alongside the original data, only the lowest 10\% of the frequencies are retained. The threshold for filtering the residuals is likewise set to 10\%. The bottom-right subplot shows the filtered signal, with outliers marked as red crosses, indicating non-steady-state points that have been identified and removed from the analysis.}
\label{fig:FFT_temp_filter}
\end{figure}

\noindent The temperature data is first Fourier transformed:
\begin{equation}
\hat{T}_\mathrm{e}(k) = \mathcal{F}\{T_\mathrm{e}(r)\} ,
\end{equation}
and then a low-pass filter is applied: 
\begin{equation}
\hat{T}_{\text{e, filtered}}(k) = 
\begin{cases}
\hat{T}_\mathrm{e}(k), & \text{if } k < k_{\text{cutoff}} ,\\
0, & \text{otherwise},
\end{cases}
\end{equation}
where the cutoff index $k_{\text{cutoff}} = \max(3, N/10)$ retains approximately the lowest 10\% of frequency components, where $N$ is the number of radial points.
The smoothed profile is reconstructed through the inverse Fourier transform
\begin{equation}
T_{\text{e, smooth}}(r) = \mathcal{F}^{-1}\{\hat{T}_{\text{e, filtered}}(k)\}.
\end{equation}
This profile cannot be used directly, as it modifies all data points rather than only removing outliers. Instead, we calculate the residuals between the original and smoothed profiles:
\begin{equation}
\delta T_\mathrm{e}(r) = T_\mathrm{e}(r) - T_{\text{e, smooth}}(r).
\end{equation}
Outliers are identified where residuals exceed a threshold:
\begin{equation}
\text{outlier}(r_n) = 
\begin{cases}
\text{true}, & \text{if } |\delta T_\mathrm{e}(r_n)| > \alpha_\mathrm{T} \cdot \text{std}(\delta T_\mathrm{e}), \\
\text{false}, & \text{otherwise},
\end{cases}
\end{equation}
where $\alpha_\mathrm{T}$ is the threshold factor. In this analysis, it is set to $\alpha_\mathrm{T} = 1$.
This procedure is further visualised in \cref{fig:FFT_temp_filter}.

\section{Density Correction} \label{sec:density_correction}

The density measurement in TS data is in contrast to the temperature measurement absolute, making it difficult to obtain a consistent density measurement that agrees with other diagnostics.
To address this, the radial density profile is taken from the TS diagnostic, while the absolute density is determined using line-integrated interferometer data. Further, an attempt can be made to correct the form of the density profile using a machine-learning model that takes the shift of the laser position into account. 

\subsection{Interferometry Rescaling } \label{sec:Interferometry_Rescaling}
The interferometer beam paths are defined by entrance and exit points in 3D space, and 1001 equally spaced points along the paths are computed. The real-space distances of these points from the entrance are determined. The $\reff$ values are separated into inboard and outboard regions, and invalid data points are excluded using logical masks. The inboard and outboard data are then combined into a single $\reff$ array before fitting the polynomial function to the electron density profile using least-squares fitting.
The density is fitted with a sixth-order polynomial $n_\mathrm{e}(\reff)$, which is used to compute the line-integrated electron density along two lines of sight (LOS):
\begin{equation}
    I_1 = \int n_\mathrm{e}(r_{\text{eff,1}}) \, dr_{\text{real,1}},
\end{equation}
\begin{equation}
    I_2 = \int n_e(r_{\text{eff,2}}) \, dr_{\text{real,2}}.
\end{equation}
The total integral is then given by the sum of both contributions.   
Similarly, the total line-integral path length is 
\begin{equation}
    L = \int 1 \, dr_{\text{real,1}} + \int 1 \, dr_{\text{real,2}}.
\end{equation}
The final density estimate is obtained by averaging the integrals over the two LOS,  
$n_{\text{l}} = \frac{I}{2}$, where the factor of $1/2$ is conventional.  

The computed density is then rescaled based on the experimental measurement $n_{\text{l}}$:
\begin{equation}
    S = \frac{n_{\text{l}} }{n_{\text{l, tom}}}.
\end{equation}
The final scaled density is given by
\begin{equation}
    n_{\text{scaled}} = S \cdot n_{\text{raw}},
\end{equation}
where $n_{\text{raw}}$ denotes the unscaled density data \cite{golo_fuchert_thomson_nodate}.

\subsection{Machine Learning Profile Correction}\label{sec:ML_correction}
The interferometry data scales the absolute magnitude of the profile without altering the radial profile itself. However, the 2018 measurement data revealed a wide variety of profile shapes that could not be explained by physical mechanisms. One of the underlying reasons for these variations was identified as the movement of the laser beams themselves \cite{nelde_quantification_2023}. This movement relative to the observation line not only changes the magnitude of the measurement, but also alters the form of the profile itself. Importantly, the changes due to this movement are deterministic; thus, in principle, each laser position can be mapped to a specific profile form and vice versa. Therefore, if the laser position is known, the density profiles could be remapped to match the laser position of the initial calibration.

As there was no active laser tracking during the 2018 campaign, the laser position had to be inferred a posterior, for example with a \glsxtrfull {ML} algorithm.
This approach was implemented in Jule Frank's Bachelor Thesis \cite{JuleMLcorrection}, where normalised laser profiles of the density were used to train a neural network, more specifically an Autoencoder. 

The training of the encoder and decoder were conducted as separate steps, allowing the trained network to be subsequently applied to correct the density profiles. The general approach, in simplified terms, is that from the normalised profiles, three abstract dimensions were extracted, which can be understood as laser coordinates. The profiles can be sorted according to this set of coordinates and subsequently corrected by a coordinate transformation to the position which is believed to be in agreement with the original calibration. This optimal position is determined by selecting the profile which minimises a target function that rewards symmetry and smoothness \cite{JuleMLcorrection}.

\section{Temperature Coupling} \label{sec:Background_Temperature_Coupling}
To check whether the assumption that the ions and electron temperatures are well coupled, we either need to rely on direct measurement data or we need to compare characteristic time scales in the plasma. 
We do this by comparing the is the energy confinement time $\tauE$, which is an indication of how quickly energy leaves with the temperature equilibration time $\tau_{\text{eq}}$, measuring how quickly energy transfers between species.
A coupling parameter can then be defined as
\begin{equation}
    \zeta(r) = \frac{\tau_{\text{eq}}(r)}{\tauE(r)}.
    \label{eq:coupling_parameter}
\end{equation}
The smaller $\zeta$ is the stronger the coupling of the temperatures can be assumed.

Assuming the absence of a relative drift, the equilibration of temperature of different plasma species is described by
\begin{equation}
    \frac{\dd{T_\alpha}}{\dd{t}} = \sum_\beta \overline{\nu}_\epsilon^{\alpha/\beta} ( T_\beta - T_\alpha) .
\end{equation}
For ions and electrons with approximately the same temperature, $T_\mathrm{i} \approx T_\mathrm{e}= T$, the coefficient can be estimated as
\begin{equation}
    \frac{\overline{\nu}_\epsilon^{i/e}}{n_\mathrm{e}} = \frac{\overline{\nu}_\epsilon^{e/i}}{n_\mathrm{e}} = \num{3.2e-9}  \frac{Z^2 \lambda}{\mu T^{3/2}} .
\end{equation}
Here, $\lambda$ denotes the Coulomb Logarithm. For this case, it is approximated as $\lambda = 18$, which should, given the simple hydrogen electron plasma, be within reasonable bounds \cite{jd_huba_2013_2011}. The units of these equilibration coefficients are given in \si{\per\s}, which leads to the diffusion of the corresponding characteristic time scale
\begin{equation}
 \tau_{\text{eq}}(r) = \frac{1}{\overline{\nu}_\epsilon^{i/e}(r)}.
 \label{eq:tau_eq}
\end{equation}

To estimate a radial profile of the coupling, we require a local estimation of the energy confinement time, $\tau_{\text{E}}(r)$:
\begin{equation}
    \tauE(r) = \frac{r^2}{\chi_{\text{eff}}(r)},
    \label{eq:tau_E_def_local}
\end{equation}
where $r$ is the physical radius and $\chi_{\text{eff}}(r)$ is the effective thermal diffusivity \cite{taroni_global_1994}.  
The effective thermal diffusivity can be calculated from the heat flux and the temperature gradient:
\begin{equation}
    \chi_{\text{eff}}(r) = -\frac{q(r)}{\nne(r) \, \partial_r \Te(r)},
\end{equation}
where $q(r)$ is the heat flux density at radius $r$, $\nne(r)$ is the electron density, and $\partial_r \Te(r)$ is the electron temperature gradient.

The power crossing a flux surface at radius r equals the total heating power deposited within that radius
\begin{align}
    P(r)  &= \int\limits_0^r p_{\text{ECRH}} (r') 2 \pi r' 2 \pi R_0 \dd{r'} \\
    &= P_{\text{ECRH}} .
\end{align}

In the last step, we assumed that the ECRH heating power can be approximated as a delta function at the magnetic axis, where $P_{\text{ECRH}}$ is the total ECRH power and $R_0$ is the major radius of the torus.

This leads to a heat flux density at radius $r$
\begin{equation}
    q(r) = \frac{P_{\text{ECRH}}}{4\pi^2 R_0 r}.
    \label{eq:heat_flux}
\end{equation}

Combining these equations, the effective thermal diffusivity becomes
\begin{equation}
    \chi_{\text{eff}}(r) = \frac{P_{\text{ECRH}}}{4 \pi^2 R_0 r \nne(r)  |\partial_r \Te(r)|}.
    \label{eq:chi_eff}
\end{equation}

With both $\tau_{\text{eq}}(r)$ and $\tau_{\text{E}}(r)$ calculated, we can determine the coupling parameter $\zeta(r)$ at each radius. This allows us to predict the ratio of ion to electron temperature, assuming that all the ECRH heating is assumed by the electrons alone \cite{boozer_required_2022}
\begin{equation}
    \frac{T_\mathrm{i}(r)}{T_\mathrm{e}(r)} = \frac{\tauE(r)}{\tauE(r) + \frac{2}{3} \tau_{\text{eq}}(r)}.
    \label{eq:temp_ratio}
\end{equation}

\section{Profile Fitting}
The temperature is expected to be a smooth function that approaches zero towards the edge of the device. 
There is a variety of functions that can be used to fit the data, both parametric and non-parametric functions. Further, the function is fitted to measurement data, thus we need to define a loss function that can be minimised to find the optimal temperature form. 
This section addresses this challenge and introduces several fit functions.

\subsection{Fitting Functions} \label{sec:Methods_Fitting_Functions}
\subsubsection{Analytical Functions}
The easiest and most straightforward mathematical function to implement is a polynomial of order N
\begin{equation}
    f(\reff) =\sum\limits_{i=0}^N c_i \reff^i,
\end{equation}
where there are $N+1$ fit coefficients to be determined. There is an ongoing debate in the field regarding whether only symmetric terms should be retained, or whether the linear term should be excluded (setting $c_1 = 0$), such that the gradient at the magnetic axis vanishes \cite{wappl_web_2024}. In our approach, we choose not to make either of these assumptions and instead retain all terms in the expansion.

\noindent An alternative symmetry assumption is introduced when fitting a truncated Gaussian, which takes the form:
\begin{equation}
    f(\reff) = c_0 \frac{\exp{- \left( \frac{\reff}{c_1}\right)^2} - \exp{- \left( \frac{1}{c_1} \right)^2}}{1-\exp{- \left( \frac{1}{c_1} \right)^2}} +c_2,
\end{equation}
where the offset $c_2$ can be optionally set to introduce a third fitting parameter. 

Of course, there are several other prominent fitting options, like a two-power fit, with a form of $f(\reff) = c_0 (1-(\reff/a)^{c_1})^{c_2}$, or a LOESS fit, which is based on local regression.

\subsubsection{Gaussian Process Regression}
In addition to mathematical functions, we also implement Gaussian Process Regression (GPR) using the \texttt{scikit-learn} package in Python \cite{Scitkit_package}.  
GPR provides a non-parametric approach to profile fitting, defining a distribution over functions directly from the data:
\begin{equation}
f(\vec{x}) \sim \mathcal{GP}(m(\vec{x}), k(\vec{x}, \vec{x}')),
\end{equation}
where $m(\vec{x})$ is the mean function and $k(\vec{x}, \vec{x}')$ the covariance function \cite{rasmussen_gaussian_2005}.

\noindent We employ the Radial Basis Function (RBF) kernel combined with a white-noise kernel:
\begin{equation}
k(\vec{x}, \vec{x}') = \exp\left(-\frac{||\vec{x} - \vec{x}'||^2}{2l^2}\right) + \sigma_n^2\delta(\vec{x}, \vec{x}'),
\end{equation}
where $l$ is the length scale parameter controlling smoothness and $\sigma_n^2$ accounts for measurement noise \cite{GaussianProcesses_kernel_scikit}.
For numerical stability, we standardise the inputs as
\begin{equation}
\vec{x}_{\text{scaled}} = \frac{\vec{x} - \mu_{\vec{x}}}{\sigma_{\vec{x}}}, \quad \vec{y}_{\text{scaled}} = \frac{\vec{y} - \mu_{\vec{y}}}{\sigma_{\vec{y}}} .
\end{equation}

The kernel parameters are optimised by minimising the negative log marginal likelihood using the Nelder-Mead algorithm. After fitting, the final prediction is transformed back to the original scale.

To enforce physically meaningful monotonic profiles, we apply a cubic \texttt{UnivariateSpline} 
with smoothing factor $s$:
\begin{equation}
\min \sum_{i=1}^{m} |y_i - \hat{f}(x_i)|^2 + s \int \hat{f}''(x)^2 dx.
\end{equation}
This smoothing approach balances fidelity to the data whilst ensuring profile smoothness.

\noindent Finally, the profile gradient is computed via numerical differentiation, implemented using the \texttt{numpy.gradient} function. 

\subsubsection{Piecewise Profile Model} \label{sec:Piecewise_Model}
Experimental data \textendash \ particularly fits to L-mode discharges in the TCV tokamak \cite{sauter_non-stiffness_2014} \textendash \ together with theoretical considerations, indicate that the temperature profile is approximately linear in the outer plasma region, but exhibits a stronger non-linear increase towards the core. To capture this behaviour, we implement a piecewise model: a linear fit is applied in the outer region, while a third-order polynomial describes the inner region, with the transition occurring at a critical radius $\rhocrit$.
To ensure physical consistency, we impose continuity in the temperature profile at the transition point:
\begin{equation}
   \lim_{\rho \nearrow \rhocrit} T(\rho) = \lim_{\rho \searrow \rhocrit} T (\rho).
\end{equation}
Additionally, we enforce gradient continuity for a smooth temperature profile:
\begin{equation}
   \lim_{\rho \nearrow \rhocrit}  \partial_\rho T(\rho) = \lim_{\rho \searrow \rhocrit} \partial_\rho T (\rho).
\end{equation}
We further impose continuity of the gradient scale-length derivative, which, although more restrictive, still results in well-behaved fits:
\begin{equation}
   \lim_{\rho \nearrow \rhocrit}  \partial_\rho \frac{T^\prime (\rho)}{T(\rho)} = \lim_{\rho \searrow \rhocrit} \partial_\rho \frac{T^\prime (\rho)}{T(\rho)}.
\end{equation}

\noindent The outer region temperature is modelled linearly as
\begin{equation}
    T\textsubscript{outer} (\rho) = T_1 + \mu (1-\rho),
    \label{eq:lin_temp_prediction}
\end{equation}
while the inner region is represented by a polynomial of order $N \geq 3$:
\begin{equation}
    T\textsubscript{inner} (\rho)= \sum\limits_{i=0}^N c_i (\rho - \rhocrit)^i.
\end{equation}
The transformation of the radial variable leads to the vanishing of higher-order terms at the transition point, which is not only beneficial for specifying the set of boundary conditions but also improves numerical stability.    
Applying the three continuity conditions yields:
\begin{align}
    \mu &= - c_1  \\
    T_1  &= c_0 - \mu  (1 - \rhocrit) \\
    c_2 &= 0 .
\end{align}

\noindent This formulation results in an effective model with $N+4-m$ parameters, where $m$ denotes the number of enforced conditions.

\subsection{Objective Function}
The most straightforward objective function is arguably the least squares fit function:
\begin{equation}
    S = \sum\limits_{i=0}^n r_i^2 =\sum\limits_{i=0}^n \left( y_i - f (x_i, \vec{\xi})\right)^2,
    \label{eq:Least_Squares}
\end{equation}
where $x_i$ represents the independent variable, $y_i$ the measured value, and $\vec{\xi}$ encompasses all fitting parameters. The optimal parameter set is determined by minimising the residual $r_i$, which effectively minimises the squared distances between the fitted function and the observed measurement data. However, when dealing with noisy data containing potential outliers, this approach becomes suboptimal.

\noindent To enhance robustness against outliers, we employ the Huber loss function:
\begin{equation}
L_{\delta}(r) = 
\begin{cases}
\frac{1}{2}r^2, & \text{if } |r| \leq \delta, \\
\delta|r| - \frac{1}{2}\delta^2, & \text{otherwise}.
\end{cases}
\end{equation}

\noindent The parameter $\delta$ controls the transition between quadratic and linear behaviour and can be customised. In this implementation, we adopt a constant value of $\delta = 1.345$. For small residuals ($|r| \leq \delta$), the Huber loss is identical to the least squares approach in \cref{eq:Least_Squares}. For larger residuals ($|r| > \delta$), it transitions to a linear penalty, significantly reducing the influence of outliers on the overall fit \cite{HuberLoss}.

\noindent For improved numerical stability, we incorporate a regularisation term:
\begin{equation}
    \mathcal{R}_{\text{params}} = \lambda (\vec{\xi} \cdot \vec{\xi}),
\end{equation}
with the regularisation parameter set to $\lambda = \num{1e-4}$. This L2 regularisation penalises large parameter values by adding the sum of the squares of all fitting parameters, preventing the solution from being dominated by extreme values.

\noindent To prevent overfitting and ensure smoothness in the fitted function, we introduce a gradient penalty term:
\begin{equation}
    \mathcal{G}_{\text{smooth}} = 
    \begin{cases}
        \gamma \sum\limits_{i=1}^{m-1} \left( \hat{y}_{i+1} - \hat{y}_i \right)^2, & \text{if } m > 3, \\
        0, & \text{otherwise},
    \end{cases}
\end{equation}
\noindent To prevent over-fitting and ensure smoothness in the fitted function, we introduce a gradient penalty term:
\begin{equation}
    \mathcal{G}_{\text{smooth}} = 
    \begin{cases}
        \gamma \sum\limits_{i=1}^{m-1} \left( \hat{y}_{i+1} - \hat{y}_i \right)^2, & \text{if } m > 3, \\
        0, & \text{otherwise},
    \end{cases}
\end{equation}
where $\hat{y}_i$ represents the predicted values $f(x_i, \vec{\xi})$ after filtering out any invalid predictions, $m$ is the number of valid predictions, and $\gamma = 0.01\lambda$ is the gradient penalty coefficient. This term penalises rapid changes in the gradient of the fitted function, promoting smoother solutions. The constraint $m > 3$ ensures that the gradient penalty is only applied when sufficient valid data points exist to meaningfully evaluate the smoothness.

Further, to account for measurement errors, we incorporate a weighting scheme in which data points with larger uncertainties contribute less to the overall loss. For each data point with an associated error $\sigma_i$, we define a weight:
\begin{equation}
    w_i = \frac{1}{0.1 + \sigma_i^2}.
\end{equation}
This approach assigns higher weights to more reliable measurements (with smaller errors) and lower weights to less reliable ones.

\noindent The total objective function can thus be expressed as:
\begin{equation}
    \mathcal{L}_{\text{total}} = \alpha \cdot \frac{\sum\limits_{i=1}^{m} w_i L_{\delta}(r_i)}{\sum\limits_{i=1}^{m} w_i} + \beta \cdot \mathcal{R}_{\text{params}} + \omega \cdot \mathcal{G}_{\text{smooth}} ,
    \label{eq:total_loss_function}
\end{equation}
where $\alpha = 5$, $\beta = 0.5$, and $\omega = 0.1$ are weighting factors that balance the influence of each component. These factors are chosen by trial and error to produce stable and reliable fits, given the data set. 

\subsection{Fitting of the CG Model}

For each time trace, the total heating power $P$, the volume derivative $V^\prime$, the magnetic field $B$, and the density profile $\nne$ are known. Thus, the power balance equation of \cref{eq:Power_Balance_eq} is fully determined. 

\begin{figure}[H]
    \centering
    \includegraphics[width = 13cm]{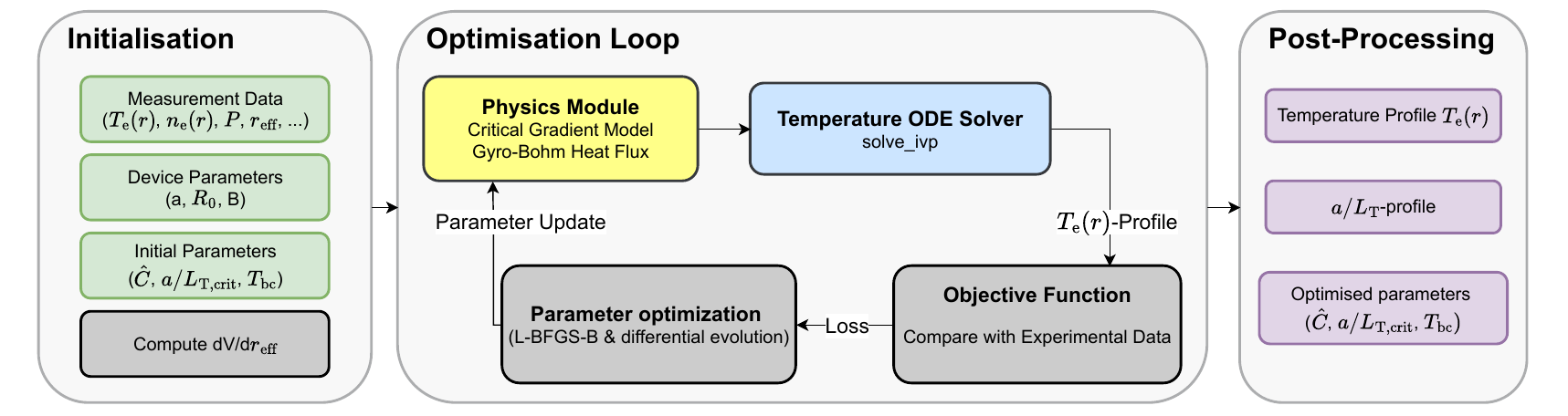}
    \caption[CG-Model fitting workflow.]{\textbf{CG-Model fitting workflow}. The fitting procedure is divided into three phases. In the initial phase, the data is loaded, and all required quantities are calculated. In the second phase, the heat fluxes are calculated (yellow field) and the resulting ODE is solved (blue field), with the solution subsequently compared against the experimental temperature profile. The temperature prediction is refined iteratively until the target loss falls below the specified threshold. The final temperature prediction can then be analysed further in the third phase.
    The green fields denote input data, the grey help function, and the post-processing is indicated by purple fields.}
\label{fig:CG_Fit_Workflow}
\end{figure}

The gradient length scale is calculated from the power balance equation. To facilitate numerical implementation, we split the terms and first determine the gradient length scale without the critical threshold constraint:
\begin{equation}
   \left(\aLT\right)_{\text{raw}}  =   \aLTcrit + \frac{P}{\hat{C} \Qgb V^\prime}.
\end{equation}
The effective gradient length scale is then obtained by applying the critical gradient model with the Heaviside function:
\begin{equation}
   \left(\aLT\right)_{\text{eff}}  =   \aLTcrit + \text{H}\left[\left(\aLT\right)_{\text{raw}} - \aLTcrit\right] \left(\left(\aLT\right)_{\text{raw}}- \aLTcrit\right).
\end{equation}
This leads to an ODE for the temperature profile:
\begin{equation}
        \frac{\dd{T}}{\dd{r}} = -\frac{T}{a}    \left(\aLT\right)_{\text{eff}},
        \label{eq:CG_model_dTdr}
\end{equation}
which extends the fitting procedure to three parameters: the two model parameters $\hat{C}$ and $\alTcrit$, along with the boundary condition $T_{\text{bc}}$ required for solving the ODE. The workflow of solving this and matching it to the experiment is illustrated in \cref{fig:CG_Fit_Workflow}.

For each set of trial parameters, we first calculate the Gyro-Bohm heat flux based on experimental density and magnetic field data, as formulated in \cref{eq:gyro_bohm}. Using this heat flux, we then determine the temperature gradient according to the critical gradient model, as stated in \cref{eq:CG_model_dTdr}, incorporating the Heaviside function from \cref{eq:Heaviside_Numerical}.

This ODE can either be solved as a boundary value problem using the \texttt{solve\_bvp} function from SciPy, with the boundary condition $T(r_{\text{outer}}) = T_{\text{bc}}$ applied at the outer edge of the fitting region.
Alternatively, what has proven to be faster as an initial value problem, where the boundary condition is used as an initial value and the solution is then integrated from $r_{\text{outer}}$ to $r_{\text{inner}}$ using the \texttt{solve\_ivp} function from SciPy.

The predicted temperature profile is then compared with experimental measurements using a modified version of the objective function from \cref{eq:total_loss_function}, with the gradient smoothness factor removed.

An external estimate $x_{\text{guess}}$ of the critical gradient can optionally be incorporated as a regularisation term; in the present implementation, however, this regularisation term is not employed and is therefore omitted from the objective function.

To constrain the parameter space to physically meaningful values and enhance optimisation efficiency, we impose bounds on the parameters:
\begin{equation}
1 \leq \hat{C} \leq 10, \quad 0.8 \leq \aLTcrit \leq 4, \quad 0.01 \leq T_{\text{bc}} \leq 1.0.
\end{equation}
The temperature boundary values reflect typical edge temperatures observed in the experimental data, whilst the $\hat{C}$ and $\alTcrit$ ranges encompass values observed in early fit results.

The fitting process employs a multi-stage optimisation approach to overcome the computational challenges posed by the ODE-based temperature prediction.
The initial estimate for $T_{\text{bc}}$ is obtained from a polynomial fit to the experimental data, which proves more reliable when the fitting region does not coincide with edge data points or when these points exhibit strong fluctuations.

A diverse set of starting points is generated around estimates that have proven successful in previous analyses. For each initial guess, we apply a rapid local optimisation using the \texttt{L-BFGS-B} algorithm. If the resulting loss remains above a predefined threshold, we perform a global optimisation attempt using \texttt{differential\_evolution}. The best solution from these stages undergoes final high-precision refinement with tighter convergence tolerances.

\section{Result Analysis}

\subsection{Correlation Analysis}
To analyse dependencies between parameters, we employ both the Pearson and Spearman rank correlation coefficients. The Pearson coefficient quantifies linear relationships and is defined as
\begin{equation}
    r = \frac{\sum\limits_{i=1}^{n} (X_i - \bar{X})(Y_i - \bar{Y})}{\sqrt{\sum_{i=1}^{n} (X_i - \bar{X})^2 \sum\limits_{i=1}^{n} (Y_i - \bar{Y})^2}},
\end{equation}
where $X_i$ and $Y_i$ represent individual fit results, and $\bar{X}$ and $\bar{Y}$ denote their respective means. This can be expressed more concisely using covariance:
\begin{equation}
    r = \frac{\text{Cov}(X,Y)}{\sigma_X \sigma_Y},
\end{equation}
with standard deviations calculated as
\begin{equation}
    \sigma_X = \sqrt{\frac{1}{n} \sum\limits_{i=1}^{n} (X_i - \bar{X})^2}.
\end{equation}

\noindent When fitted parameters exhibit non-linear but monotonic relationships, Spearman's rank correlation provides greater sensitivity. It is calculated as the Pearson correlation of rank values:
\begin{equation}
r_s = \frac{\sum\limits_{i=1}^{n} (R(X_i) - \bar{R}_X)(R(Y_i) - \bar{R}_Y)}{\sqrt{\sum_{i=1}^{n} (R(X_i) - \bar{R}_X)^2 \sum\limits_{i=1}^{n} (R(Y_i) - \bar{R}_Y)^2}},
\end{equation}
where $R(X_i)$ and $R(Y_i)$ represent the ranks of the fit parameter values.
The Spearman coefficient helps identify consistent monotonic trends when direct linear correlations are less evident in the optimised parameter sets \cite{winter_comparing_2016}.

\subsection{Theoretical \texorpdfstring{$\alT$}  P Predictions} \label{sec:Theoretical_aLT_predictions}
Near the edge, the temperature is expected to scale linearly. A simple form of the temperature profile is given in \cref{eq:lin_temp_prediction}.
This leads to a gradient length scale of
\begin{equation}
    \aLT = \frac{\mu}{T_1 + \mu (1-\rho)} = \frac{1}{\alpha + (1-\rho)},
    \label{eq:linear_alT_prediction}
\end{equation}
but a priori the physical meaning of the radial variable $\rho$ is not defined. 
Thus, we can define the radial variable to be a second-order coordinate transformation of the $\rho$, in the form 
\begin{equation}
    \hat{\rho} = \alpha + \beta (1-\rho) + \gamma (1-\rho)^2.
\end{equation}
This leads then to a proposed form
\begin{equation}
    \aLT = \frac{1}{\alpha + (1-\rho) + \gamma (1-\rho)^2}.
    \label{eq:higher_order_alT_prediction}
\end{equation}
The coefficients $\alpha$ and $\gamma$ can then be fitted to the experimental data.

\section{Profile Solver}
This section presents the formulation and implementation of a numerical solver that solves the power balance equation and the $\etacrit$-model to infer, from given heat flux data, the radial temperature and density profiles.  In the following sections, this implementation will simply be referred to as the profile solver. 

\begin{figure}[H]
    \centering
    \includegraphics[width = 13cm]{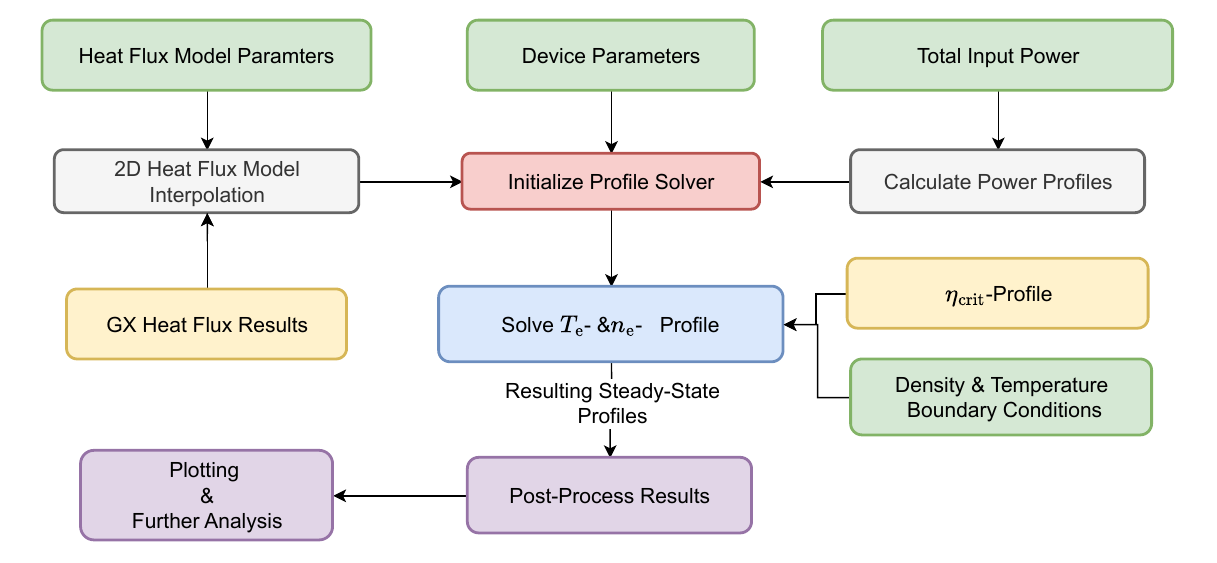}
    \caption[Profile solver workflow.]{\textbf{Profile solver workflow.} The first step is the initialisation of the model (red field) with the input data (e.g. initial temperature and density) marked in green. The set of ODEs is then solved to obtain the corresponding $\Te$ and $\nne$ profiles, which can subsequently be further processed.}
\label{fig:Profile_Solver_Workflow}
\end{figure}

The simplified structure and typical workflow of the solver are illustrated in \cref{fig:Profile_Solver_Workflow}

\subsection{Heating Profile} \label{sec:methods_heating_profiles}
In a real fusion reactor, a significant fraction of the heating is expected to come from $\alpha$-particle heating \cite{windsor_alpha-particle_1999}. However, in the case of W7-X, it is sufficient to prescribe an external power profile, as the plasma consists solely of hydrogen and does not reach fusion-relevant temperatures.

In our model, we consider two main heating methods: ECRH and NBI, as well as losses through radiation, denoted here as $P_{\text{RAD}}$. 

For the ECRH heating, we use a Gaussian distribution of the form:
\begin{equation}
P_{\text{ECRH}}(\rho) = \frac{P_{\text{ECRH,total}}}{N_{\text{ECRH}}} \exp\left(-\frac{(\rho - \rho_{\text{in}})^2}{\rho_{\text{w}}^2}\right) \, a^3 \, A(\rho) ,
\end{equation}
where $P_{\text{ECRH,total}}$ is the total injected ECRH power, $\rho_{\text{in}}$ denotes the normalised deposition radius, and $\rho_{\text{w}}$ the width.
$N_{\text{ECRH}}$ is a normalisation factor ensuring that the volume integral equals $P_{\text{ECRH,total}}$:
\begin{equation}
N_{\text{ECRH}} = a^3 \int_0^1 \exp\left(-\frac{(\rho - \rho_{\text{in}})^2}{\rho_{\text{w}}^2}\right) A(\rho) \, \dd{\rho} .
\end{equation}

The position of the deposition radius generally depends on plasma parameters such as the plasma pressure $\beta$ \cite{wolf_electron-cyclotron-resonance_2018}. In the analysed shots, we assume $\beta$ is low and place the Gaussian near the magnetic axis.

For NBI heating and radiation losses, the profile form is less well defined. Currently, two profile shapes are implemented: uniform heating and a radially decreasing profile. For NBI, this assumption is reasonable, as discussed in LHD experiments \cite{yamada_impact_2003}.  

For uniform profiles, the power density is constant throughout the plasma volume:
\begin{equation}
P_{\text{NBI,uniform}}(\rho) = \frac{P_{\text{NBI,total}}}{V_{\text{total}}} \, a^3 \, A(\rho) ,
\end{equation}
where $V_{\text{total}}$ is the total plasma volume, given by:
\begin{equation}
V_{\text{total}} = a^3 \int_0^1 A(\rho) \, \dd{\rho} .
\end{equation}

For linear profiles, the power density decreases linearly from the plasma core to zero at the edge:
\begin{equation}
P_{\text{NBI,linear}}(\rho) = \frac{P_{\text{NBI,total}}}{N_{\text{linear}}} (1 - \rho) \, a^3 \, A(\rho) ,
\end{equation}
where $N_{\text{linear}}$ is the normalisation factor for the linear profile:
\begin{equation}
N_{\text{linear}} = a^3 \int_0^1 (1 - \rho) A(\rho) \, \dd{\rho} .
\end{equation}

Similarly, the radiation losses are represented as
\begin{equation}
P_{\text{RAD}}(\rho) = -\frac{P_{\text{RAD,total}}}{N_{\text{profile}}} f_{\text{profile}}(\rho) \, a^3 \, A(\rho) ,
\end{equation}
where $f_{\text{profile}}(\rho)$ is either 1 for uniform profiles or $(1 - \rho)$ for linear profiles, and $N_{\text{profile}}$ is the corresponding normalisation factor. The negative sign indicates that radiation represents a power loss.

This power loss is only considered for high-power scenarios and is neglected otherwise. For low density and low heating power, the radiation region is considered to be reasonably low. This can no longer be assumed for high-power shots \cite{zhang_plasma_2021}.

The total power profile is then given by the sum of all contributions:
\begin{equation}
P_{\text{total}}(\rho) = P_{\alpha}(\rho) + P_{\text{ECRH}}(\rho) + P_{\text{NBI}}(\rho) + P_{\text{RAD}}(\rho) .
\end{equation}

\subsection{Power Balance Equation}

The power balance is derived in \cref{sec:Heat_Flux_Models} and stated in \cref{eq:Power_Balance_eq}.
Substituting the Gyro-Bohm heat flux and model definition from \cref{eq:gyro_bohm,eq:heat_flux_model}, we obtain
\begin{equation}
    V^\prime(\rho) P(\rho) = \frac{\dd{}}{\dd{\rho}} \left[ A(\rho)  n(\rho)  e T_\mathrm{i}(\rho)  c_\mathrm{i}(\rho) \varrho_\mathrm{i}^2(\rho)  \hat{Q}\left(\aLT, \rho\right) \right],
    \label{eq:power_balance_full}
\end{equation}
where $T_\mathrm{i}(\rho)$ is expressed in joules. The factor $A(\rho)$ incorporates the minor radius and is defined as
\begin{equation}
A(\rho) = a^{-2} \frac{\dd{V}}{\dd{r}}.
\label{eq:aoa2}
\end{equation}

\subsection{\texorpdfstring{$\etacrit -$} MModel}\label{sec:crit_eta_model}

If the density or the $\eta$ profile is not externally fixed, it can be determined in the gyro-kinetic simulations using the $\eta_{\text{crit}}$ model. This model follows the definition from \cref{eq:def_eta}:
\begin{equation}
    \etacrit (\rho) = \frac{L_\mathrm{n} }{L_\mathrm{T}} = \frac{\aLT}{\aLn} = \frac{n}{n^\prime}  \frac{T^\prime}{T}.
    \label{eq:eta_crit}
\end{equation}

The term $\etacrit (\rho)$ is determined by enforcing that the particle flux ($\Gamma = 0$) vanishes. This constraint leads to a relationship between the temperature and density gradients:
\begin{equation}
    \frac{n^\prime}{n} = \frac{T^\prime}{T}  \frac{1}{\eta_{\text{crit}}(\rho)},
    \label{eq:density_gradient}
\end{equation}
or equivalently, in terms of the evolution of the density profile:
\begin{equation}
    \frac{\dd{n}}{\dd{\rho}} = \frac{\dd{T}}{\dd{\rho}}  \frac{1}{T}   \frac{1}{\etacrit(\rho)}  n.
    \label{eq:density_evolution}
\end{equation}

\subsection{Coupled ODE System} \label{sec:coupled_ODE_system}
The power balance equation and the $\etacrit$ model form a coupled system. The temperature profile affects the density profile through the $\etacrit$ model, while the density profile influences the heat transport and, consequently, the temperature evolution.
 
To express this coupling explicitly, we formulate a system of first-order differential equations. Let us define:
\begin{align}
y_1 = T, \quad
y_2 = \frac{\dd{T}}{\dd{\rho}}, \quad
y_3 = n.
\end{align}
The system then becomes:
\begin{align}
\frac{\dd{y_1}}{\dd{\rho}} &= y_2 \label{eq:ode_T}, \\
\frac{\dd{y_2}}{\dd{\rho}} &= \frac{d^2T}{d\rho^2} \label{eq:ode_Tprime} ,\\
\frac{\dd{y_3}}{\dd{\rho}} &= \frac{y_2}{y_1}  \frac{1}{\etacrit(\rho)}  y_3. \label{eq:ode_n}
\end{align}
From the power balance equation \cref{eq:power_balance_full}, we calculate the $\dd^2{T}/\dd{\rho^2}$.
The heat flux at any radial position depends on the temperature, its gradient, and the density:
\begin{align}
Q(\rho, T, T', n) &= n(\rho)  T(\rho)  c_{\text{i}}(\rho) \varrho_{\text{i}}^2(\rho) \hat{Q} \left(\aLT, \rho \right) \\
&= \frac{m^{\frac{1}{2}}}{e^2 B^2} n (\rho) T^{\frac{5}{2}} \hat{Q}\left(\aLT, \rho \right) .
\label{eq:heat_flux_calculation}
\end{align}
To determine $\dd^2{T}/\dd{\rho^2}$ from the power balance equation, we differentiate \cref{eq:heat_flux_calculation} with respect to $\rho$:
\begin{equation}
\begin{split}
\frac{\dd{}}{\dd{\rho}}[A(\rho)Q(\rho, T, T', n)] &= V'(\rho)P(\rho). \\
\end{split}
\label{eq:power_balance_differentiated}
\end{equation}
Applying the chain rule gives
\begin{equation}
\begin{split}
\frac{\dd{}}{\dd{\rho}}[A(\rho)Q(\rho, T, T', n)]  &= \frac{\dd{ A(\rho)}}{\dd{\rho}}Q + A(\rho)\frac{\partial Q}{\partial \rho} + A(\rho)\frac{\partial Q}{\partial T}T' \\
&+ A(\rho)\frac{\partial Q}{\partial T'}T'' + A(\rho)\frac{\partial Q}{\partial n}n' .
\end{split}
\label{eq:chain_rule_expanded}
\end{equation}
Substituting $n' = \frac{T'}{T} \cdot \frac{n}{\etacrit(\rho)} $ from \cref{eq:density_evolution} and solving for $T''$ leads to

\begin{equation}
\begin{split}
T'' &= \frac{1}{A(\rho)\frac{\partial Q}{\partial T'}} \left( a^2 A(\rho)P(\rho) - \frac{\dd{ A(\rho)}}{\dd{\rho}}Q - A(\rho)\frac{\partial Q}{\partial \rho} \right. \\
&\left. - A(\rho)\frac{\partial Q}{\partial T}T' - A(\rho)\frac{\partial Q}{\partial n}\frac{T'}{T}\frac{1}{\eta_{\text{crit}}(\rho)}n \right) .
\end{split}
\label{eq:second_derivative_expanded}
\end{equation}

\noindent The partial derivatives in \cref{eq:second_derivative_expanded} with respect to the radial coordinate are calculated numerically. For the initial step:
\begin{align}
\frac{\partial Q}{\partial \rho} &\approx \frac{Q(\rho + h_\rho, T, T', n) - Q(\rho, T, T', n)}{h_\rho} .\label{eq:partial_rho}
\end{align}
while for subsequent steps, the central difference method is employed for improved stability:
\begin{equation}
    \frac{\partial Q}{\partial \rho} \approx \frac{Q(\rho + h_\rho, T, T', n) - Q(\rho - h_\rho, T, T', n)}{2 h_\rho}.
\end{equation}
The same procedure applies to $A (\rho)$. Since we use a parametrised volume derivative, it cannot be differentiated analytically. For improved stability, we use adaptive step sizes for $h_\rho$:
\begin{equation}
    h_\rho = \min{\left\{ \num{1e-5}, 0.001  \rho \right\}}.
\end{equation}

The remaining partial derivatives can be expressed analytically as
\begin{align}
    \frac{\partial Q}{\partial n} &= \frac{Q}{n}\\
    \frac{\partial Q}{\partial T} &= \frac{5}{2} \frac{Q}{T} + \frac{m^{\frac{1}{2}}}{e^2 B^2} n (\rho) T^{\frac{5}{2}} \frac{\partial \hat{Q}\left(\aLT, \rho \right)}{\partial T} \\
        \frac{\partial Q}{\partial T^\prime} &= \frac{m^{\frac{1}{2}}}{e^2 B^2} n (\rho) T^{\frac{5}{2}} \frac{\partial \hat{Q}\left(\aLT, \rho \right)}{\partial T^\prime} .
\end{align}
The derivatives of $\hat{Q}$ can be obtained from the fitting conditions at each radius by differentiating \cref{eq:sparse_fit,eq:sparse_adapt}:
\begin{align}
    \frac{\partial \hat{Q} \left(-\frac{T^\prime}{T}, \rho_i \right)}{\partial T^\prime} &=  - \beta \frac{1}{T} + 2 \gamma_i \frac{T^\prime}{T^2} , \\
    \frac{\partial \hat{Q} \left(-\frac{T^\prime}{T}, \rho_i \right)}{\partial T} &=   \beta \frac{T^\prime}{T^2} - 2  \gamma_i\frac{(T^\prime)^2}{T^3}. 
\end{align}

The system is then solved numerically using the \texttt{scipy.solve\_ivp} solver, which integrates the final set of equations:
\begin{align}
\frac{\dd{T}}{\dd{\rho}} &= T' , \label{eq:final_ode_T} \\
\frac{\dd{T'}}{\dd{\rho}} &= \frac{a^2 A P - \frac{\dd{ A}}{\dd{\rho}}Q - A\frac{\partial Q}{\partial \rho} - A\frac{\partial Q}{\partial T}T' - A\frac{\partial Q}{\partial n}\frac{T'}{T}\frac{1}{\etacrit}n}{A\frac{\partial Q}{\partial T'}}, \label{eq:final_ode_Tprime} \\
\frac{\dd{n}}{\dd{\rho}} &= \frac{T'}{T}  \frac{n}{\etacrit} .
 \label{eq:final_ode_n}
\end{align}
The boundary conditions are defined as:
\begin{align}
T(\rho_{\text{min}}) &= T_0 , \label{eq:bc_T} \\
T'(\rho_{\text{min}}) &= -\,x_0\,T_0 , \label{eq:bc_Tprime} \\
n(\rho_{\text{min}}) &= n_0 . \label{eq:bc_n}
\end{align}
Here, $x_0$ denotes the critical normalised temperature gradient as defined in \cref{eq:fullrad_x0data}. 
The second condition explicitly specifies the initial temperature gradient, requiring it to match the critical gradient at $\rho_{\text{min}}$, scaled by the on-axis temperature $T_0$.  
This formulation reflects the physical constraint that the heat flux must vanish at the magnetic axis.

\chapter{Results}\label{sec:Results}

This section is structured as follows. First, we analyse a single time trace in \cref{sec:Results_Shot_Analysis}, where the prescribed filtering methods are tested, and the discussed fits and loss functions are examined. 

Next, we extract trends and behaviours across the entire database in \cref{sec:Results_Database_Analysis}. Here, the term "database" refers exclusively to shots from OP.1.2 (2018) that were heated solely by ECRH. This section also analyses the behaviour of the temperature and confinement times.

The results from the GX simulation, together with the operation of the profile solver, are presented in \cref{sec:Results_Reconstructed}. 
A comparison between the experimental profiles, the simulations, and the reconstructed profiles is provided in \cref{sec:Results_Comparison}. 

Finally, to broaden the scope of the analysis, we examine peaked density profiles from both high- and low-performance shots in \cref{sec:Peaked_Density_profile}.

\section{Individual Shot Analysis} \label{sec:Results_Shot_Analysis}

This section focuses on a detailed examination of a single time trace. We begin by outlining the filtering procedure and the correction applied to the density profile. The subsequent discussion considers the performance of the different fitting methods and compares their respective behaviours.

\subsection{Data Correction and Filtering} \label{sec:Data_Correction_and_Filtering}

\subsubsection{Steady-State Filter}
\begin{figure}[H]
    \centering
    \includegraphics[width = 10.5cm]{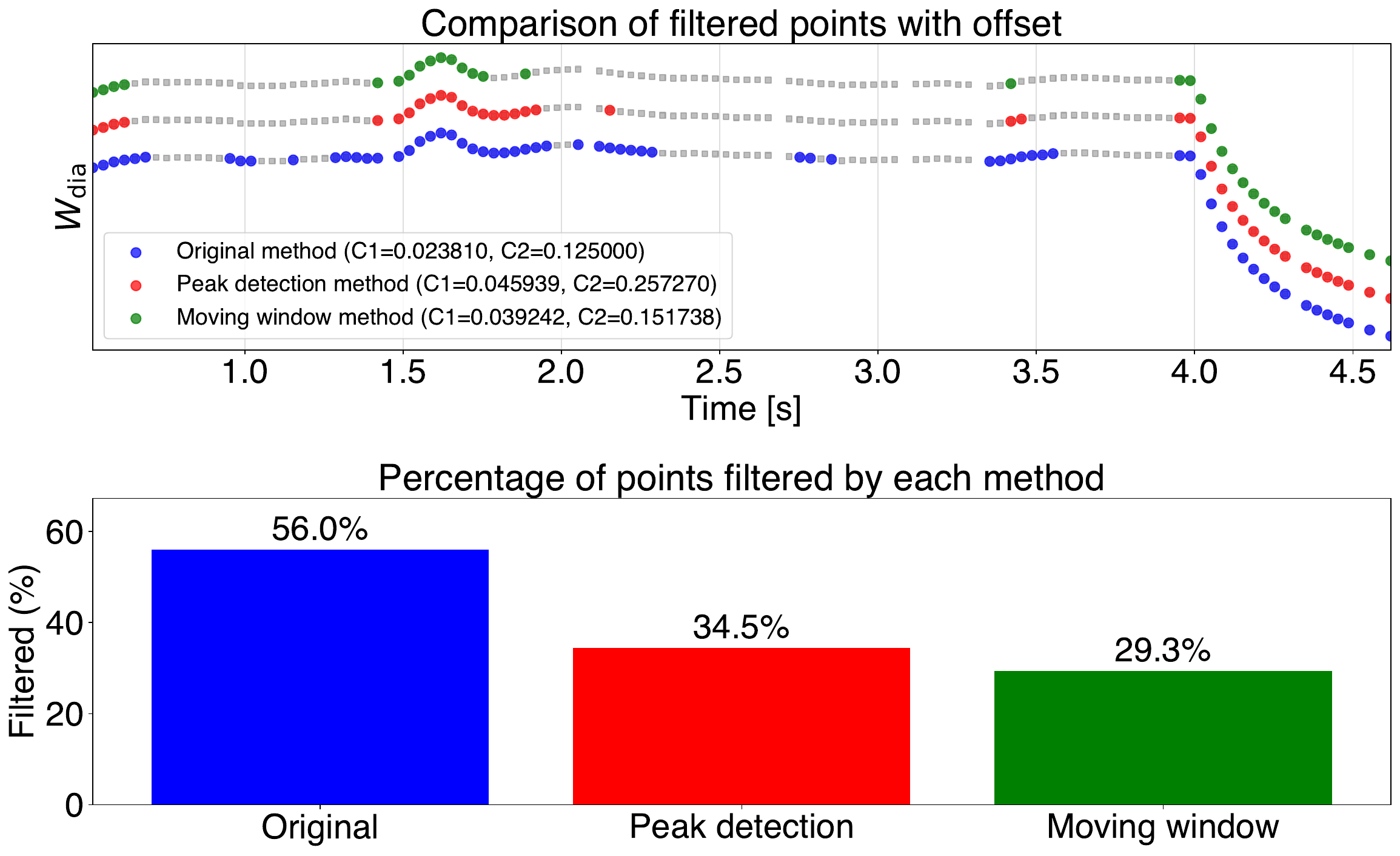}
    \caption[Steady-state filter: Window averaging]{\textbf{Comparison of different steady-state filters.} The upper three subplots display the three different filtering methods, whilst the bottom subplot shows the percentage of points identified as being in a non-steady state.}
\label{fig:steady_state_filter_window_overview}
\end{figure}

Given a diamagnetic energy trace, we have developed two data-driven approaches to determine the steady-state coefficients $C_1$ and $C_2$. 
A comparison of the points filtered out, as well as the total proportion of filtered points, is shown in \cref{fig:steady_state_filter_window_overview}.

The regions identified by all three methods are broadly similar. In this example, the peak detection and window averaging methods identify fewer transient points than the initial choice of $C_1$ and $C_2$. A comparison of the extracted scales, using the thresholds set by these two methods, is presented in \cref{tab:steady_state_filter_method_comparison}. 

\begin{table}[H]
    \centering
    \begin{tabular}{c|c|c|c}
         Method & $C_1$ & $C_2$ & $P_{\text{filtered}}$ [\si{\percent}]\\
         \hline
        Initial estimate & $\num{0.023810}$ & $\num{0.125000}$ & \num{35.3} \\
        Peak detection & $\num{0.036(28)}$ & $\num{0.13(13)}$ & \num{40.5} \\
        Window averaging & $\num{0.026(11)}$ & $\num{0.107(85)}$ & \num{53.8} \\
    \end{tabular}
    \caption[Steady-state filter method comparison.]{\textbf{Steady-state filter method comparison.} The table shows the estimated coefficients $C_1$ and $C_2$, and the percentage of total points filtered across the database.}
    \label{tab:steady_state_filter_method_comparison}
\end{table}

The results of this scan show that, although the resulting coefficients are similar, the percentage of filtered points differs significantly.  
The peak detection method struggles with small fluctuations, a limitation not observed with the window averaging method. However, compared to using constant coefficients, window averaging incurs a slightly higher computational cost when analysing the shots. As this increase is marginal, this method will therefore be used as the primary filter.

\subsubsection{Density Correction}

\begin{figure}[H]
\centering
\includegraphics[width=12cm]{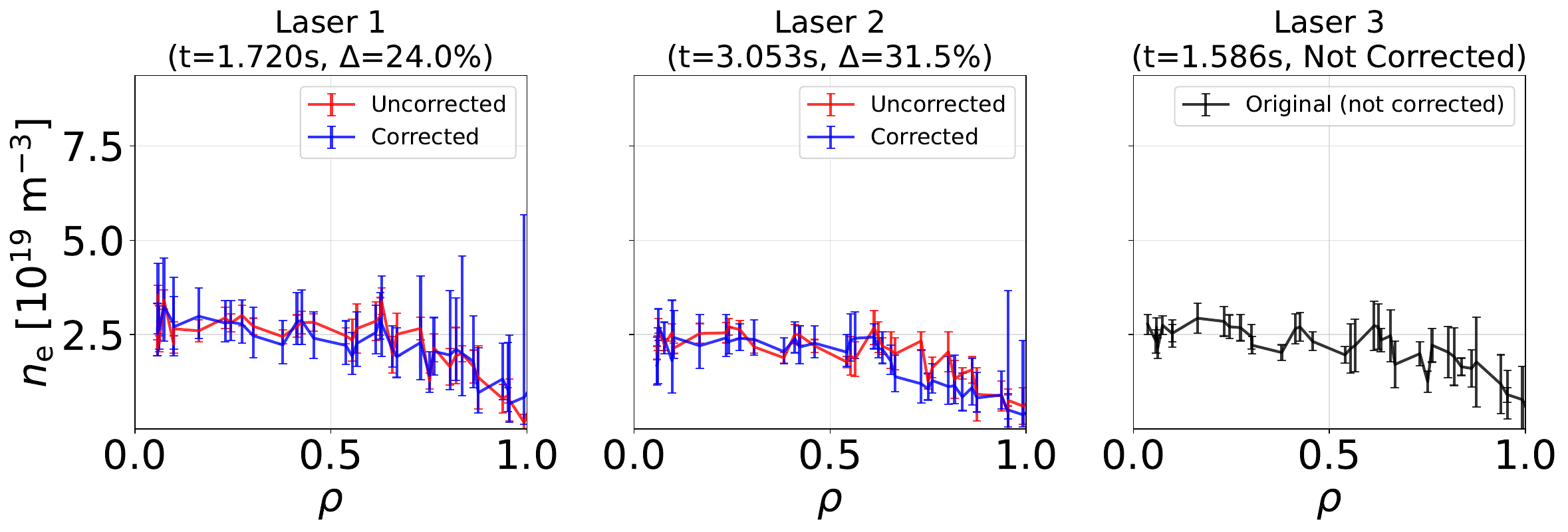}
\caption[ML density profile correction]{\textbf{ML density profile correction.} The original profiles are shown in red, and the profiles after the ML profile correction and interferometer rescaling are shown in blue. The three different lasers considered are indicated. Shown is the correction with the highest MSE value. Laser 3 has not been corrected.}
\label{fig:ML_Laser_Correction}
\end{figure}

The density must be corrected both in amplitude and in the profile itself, as discussed in \cref{sec:density_correction}. The first step is to correct the profile shape using the ML correction described in \cref{sec:ML_correction}.
In \cref{fig:ML_Laser_Correction}, the maximum correction for each laser number in a typical plasma shot is shown. Laser No.\ 3 is not corrected but is included here for completeness. This laser is excluded from the ML correction because, unlike the others, it does not share the same beam path; consequently, a correction based on its position would result in an unreliable adjustment.

\begin{figure}[H]
    \centering
    \subfigure[Raw data.]{%
        \label{fig:2D_Illustration_Density_Rescaling_raw}%
        \includegraphics[width=6.2cm]{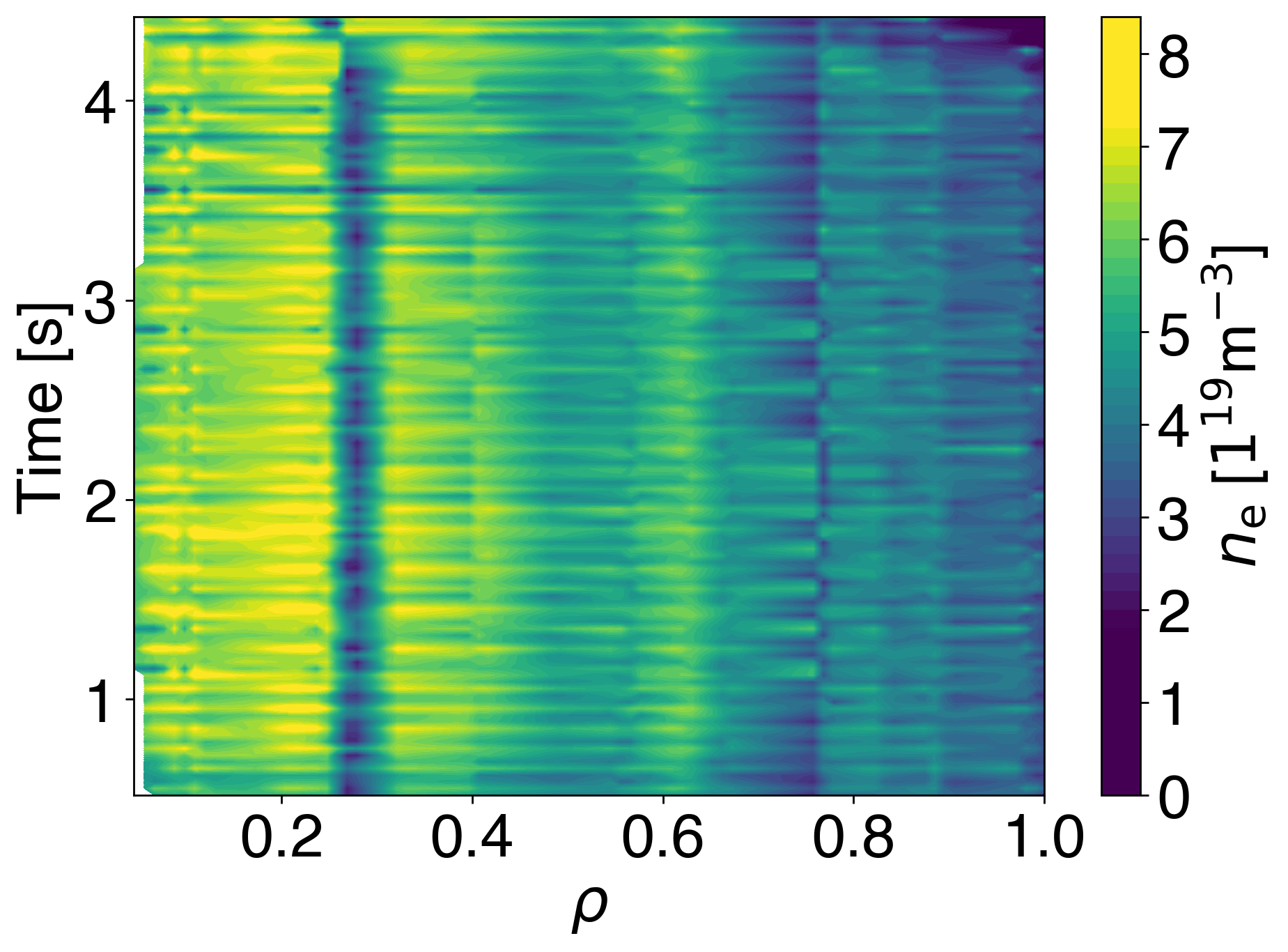}%
    }%
    \quad
    \subfigure[Filtered data.]{%
        \label{fig:2D_Illustration_Density_Rescaling_corrected}%
        \includegraphics[width=6.2cm]{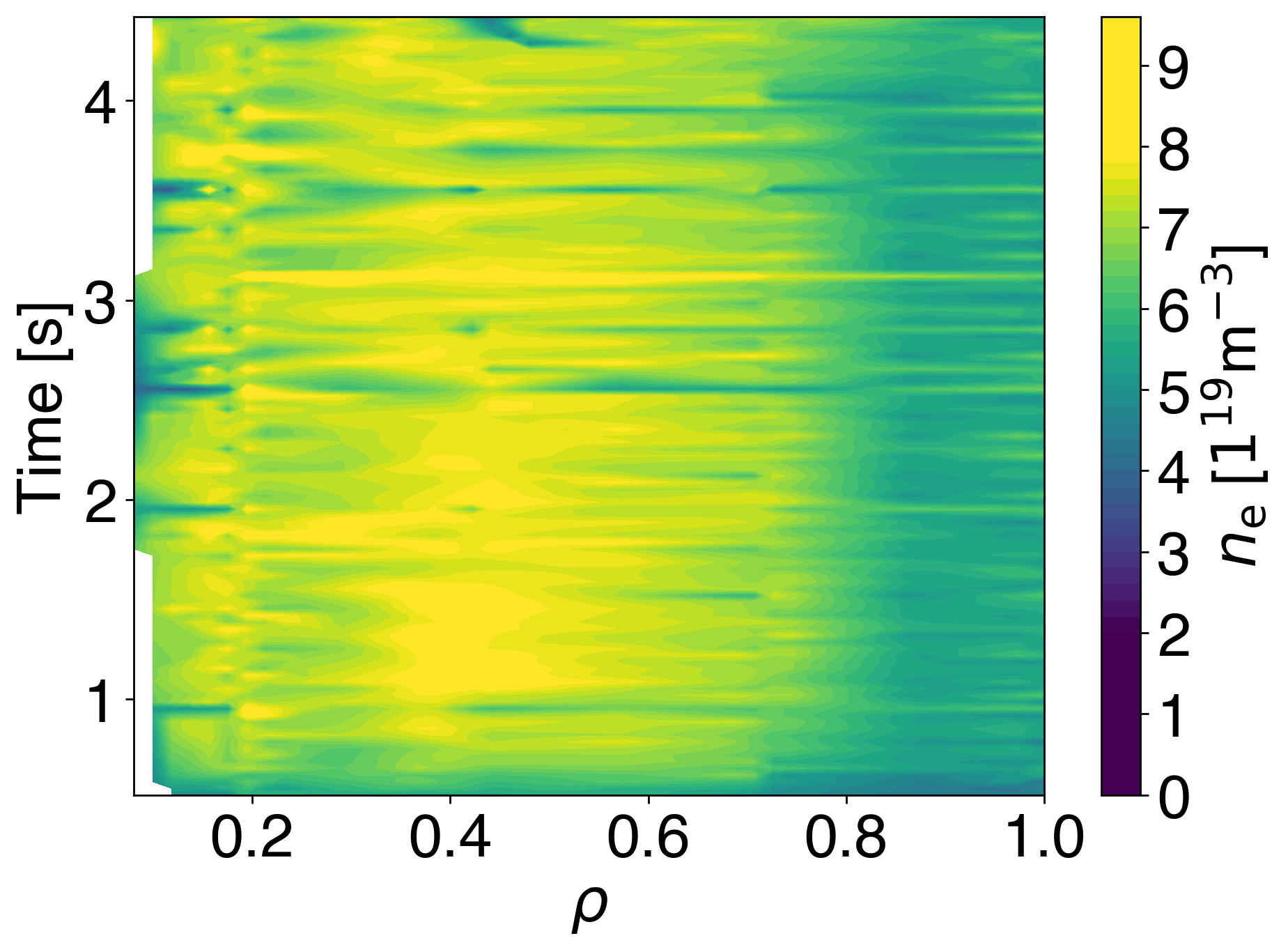}%
    }%
    \caption[Density rescaling]{\textbf{Density rescaling.} The plot in \textbf{a)} shows the density value as a function of $\rho$ and $t$ directly coming from the W7-X archive, and \textbf{b)} shows the corrected profiles. Shot No. \#20180925.025.}
    \label{fig:2D_Illustration_Density_Rescaling}
\end{figure}

During the ML correction, the magnitude of the density profile is not preserved; only the profile shape is adjusted. The magnitude is subsequently corrected via rescaling using the interferometer measurement, as outlined in \cref{sec:Interferometry_Rescaling}. A typical shot before and after this rescaling is shown in \cref{fig:2D_Illustration_Density_Rescaling}. 

\subsubsection{Temperature Filter}
The temperature values are not affected by the above-discussed correction. The temperature values are taken as correct, and we only filter certain outliers, which we attribute to errors in the measurement process.

\begin{figure}[H]
    \centering
    \subfigure[Raw data.]{%
        \label{fig:2D_Illustration_Temperature_Filtering_raw}%
        \includegraphics[width=6.2cm]{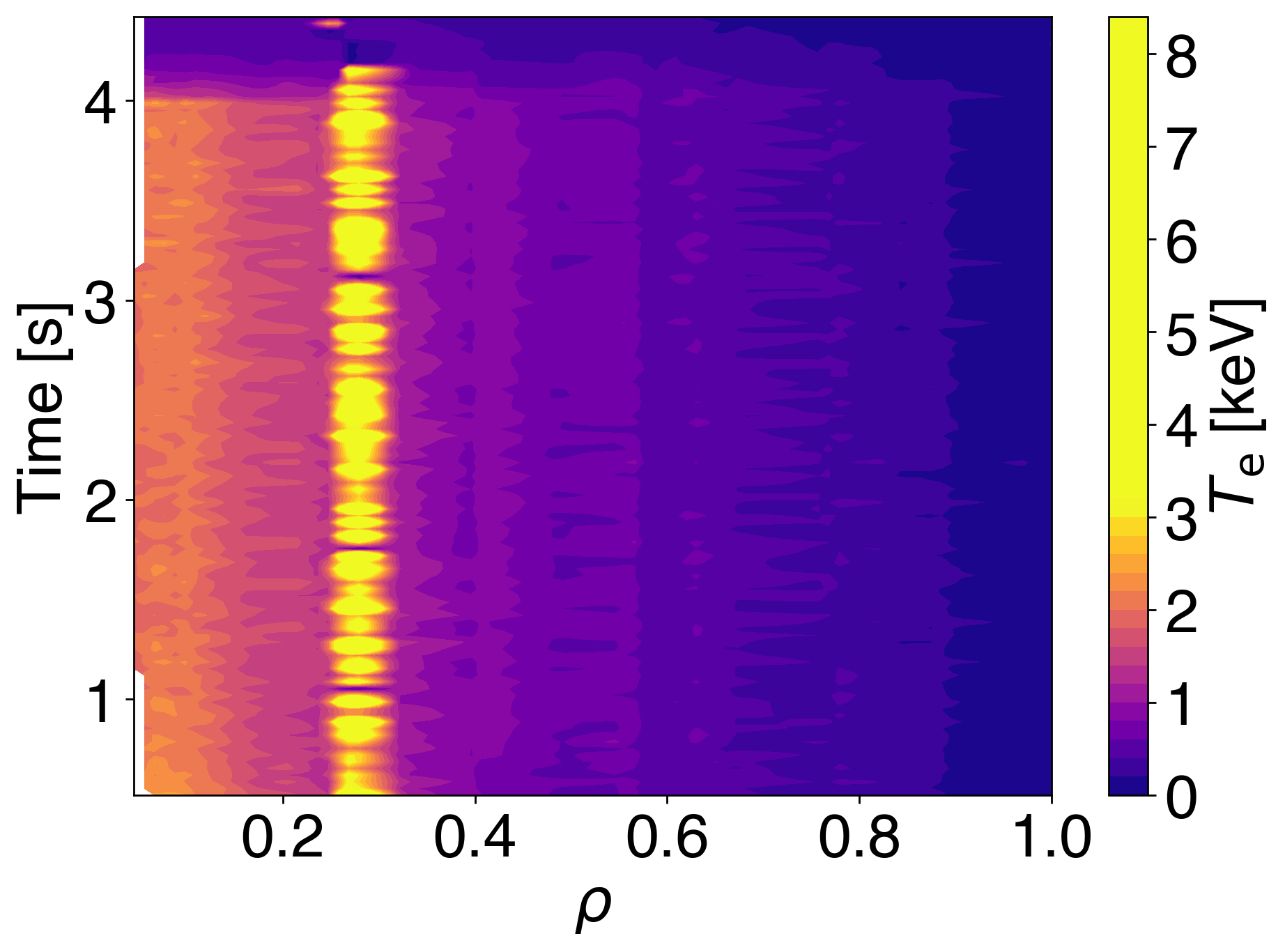}%
    }%
    \quad
    \subfigure[Filtered data.]{%
        \label{fig:2D_Illustration_Temperature_Filtering_filtered}%
        \includegraphics[width=6.2cm]{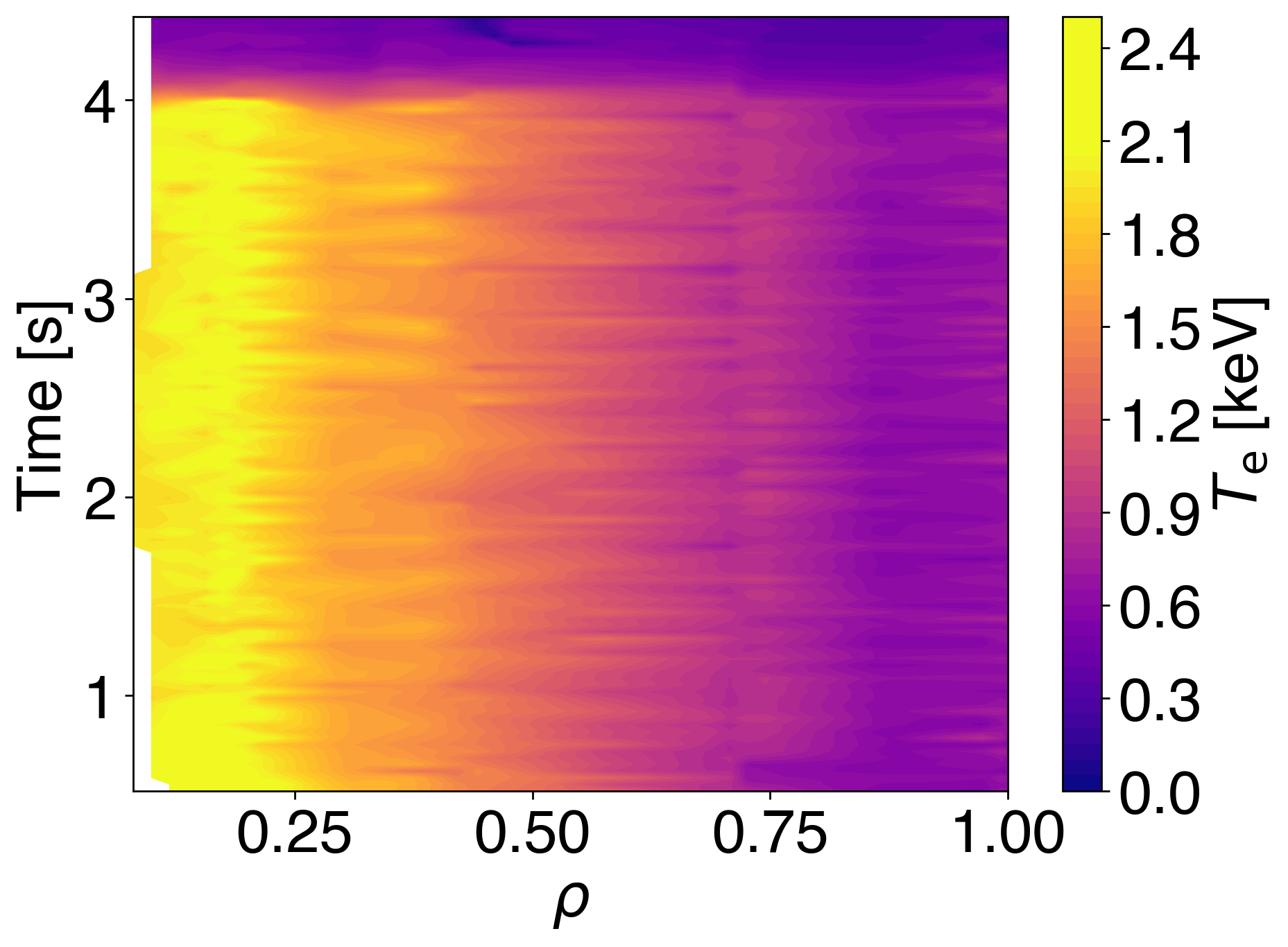}%
    }%
\caption[Illustration of FFT filtering]{\textbf{Illustration of FFT filtering.} The plot in \textbf{a)} shows the uncorrected data, where the temperature is displayed as a function of $\rho$ and $t$. In \textbf{b)}, the same data with the FFT filter applied is shown for comparison. Shot No. \#20180925.025.}
\label{fig:2D_Illustration_Temperature_Filtering}
\end{figure}

A typical timeline is illustrated in \cref{fig:FFT_temp_filter}. There appear to be mainly two types of temperature outliers. Firstly, faulty measurement volumes, where there is a clear temperature spike at a specific measurement volume that remains constant throughout all the shots. Secondly, there are outliers at the outer edge. These outliers are related to very small density measurements and indicate that the fitting process used to determine the temperature has failed.

The effect of the temperature filter throughout the shot is illustrated in \cref{fig:2D_Illustration_Temperature_Filtering}. The majority of the temperature values are indeed unchanged, but the high outliers have been filtered out successfully.

\subsubsection{Temperature Coupling} \label{sec:Result_Temperature_Coupling}
To determine an appropriate fitting range, we must first assess how well the ion and electron temperatures are coupled as a function of radius. As discussed in \cref{sec:Data_Filtering}, focusing on well-coupled, steady-state profiles ensures that the single steady-state power balance is applicable and that our analysis reflects generic transport behaviour rather than machine-specific artefacts. 

\begin{figure}[H]
    \centering
    \includegraphics[width = 10cm]{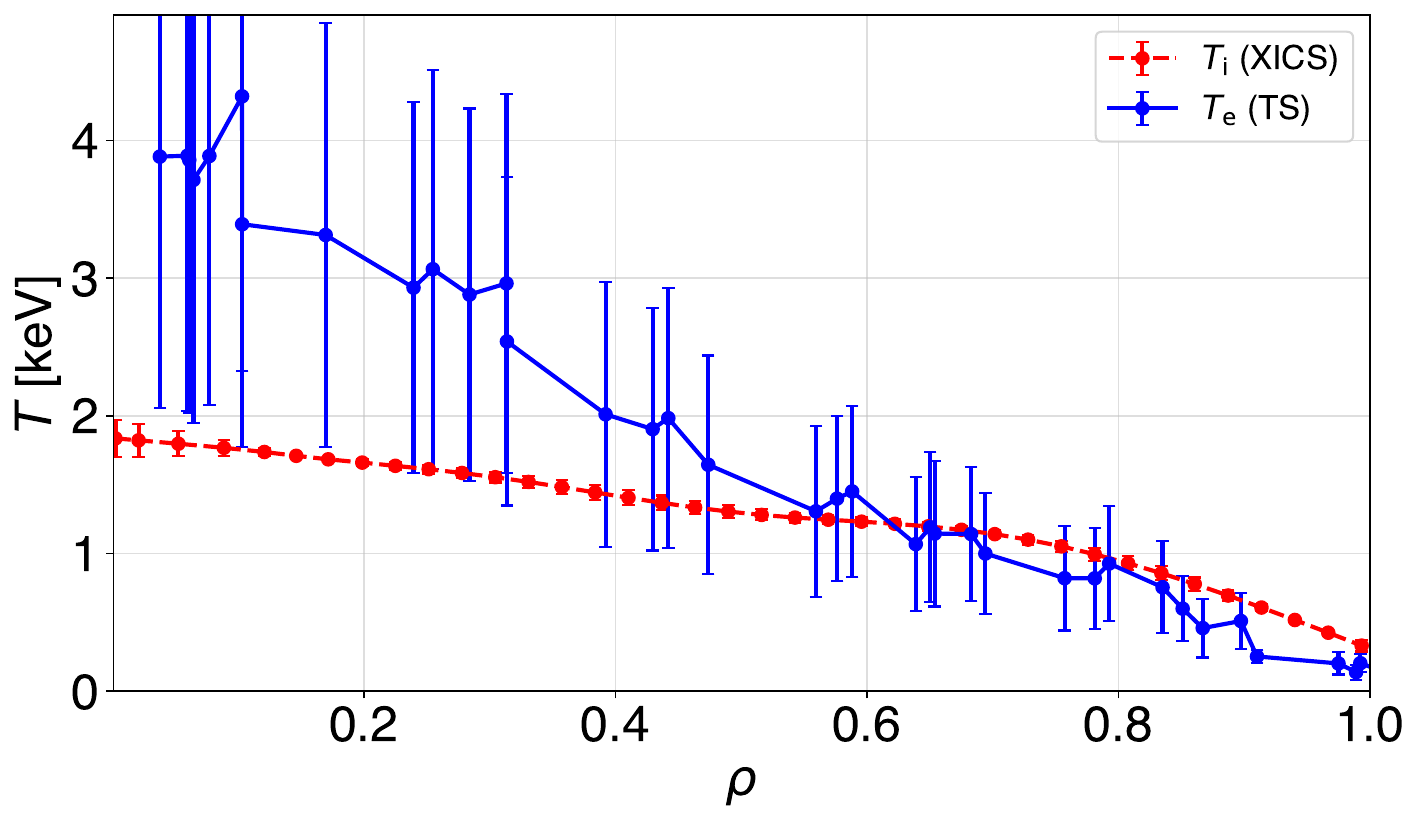}
    \caption[$\Ti$ and $\Te$-profiles.]{\textbf{$\mathbf{\Ti}$ and $\mathbf{\Te}$-profiles.} Typical averaged $\Te$ (TS) and $\Ti$ (XICS) profiles.  Shot No. \#20180927.021.}
\label{fig:coupling_comparison_Ti_Te}
\end{figure}

The most direct approach is to examine experimental temperature profiles.
These data can be obtained through \glsxtrfull{XICS}\cite{kring_situ_2018} or \glsxtrfull{CXRS}\cite{ford_charge_2020}. An exemplary shot, in which the electron temperature (from TS) is compared with the ion temperature (from XICS), is shown in \cref{fig:coupling_comparison_Ti_Te}.

\begin{figure}[H]
    \centering
    \includegraphics[width = 10cm]{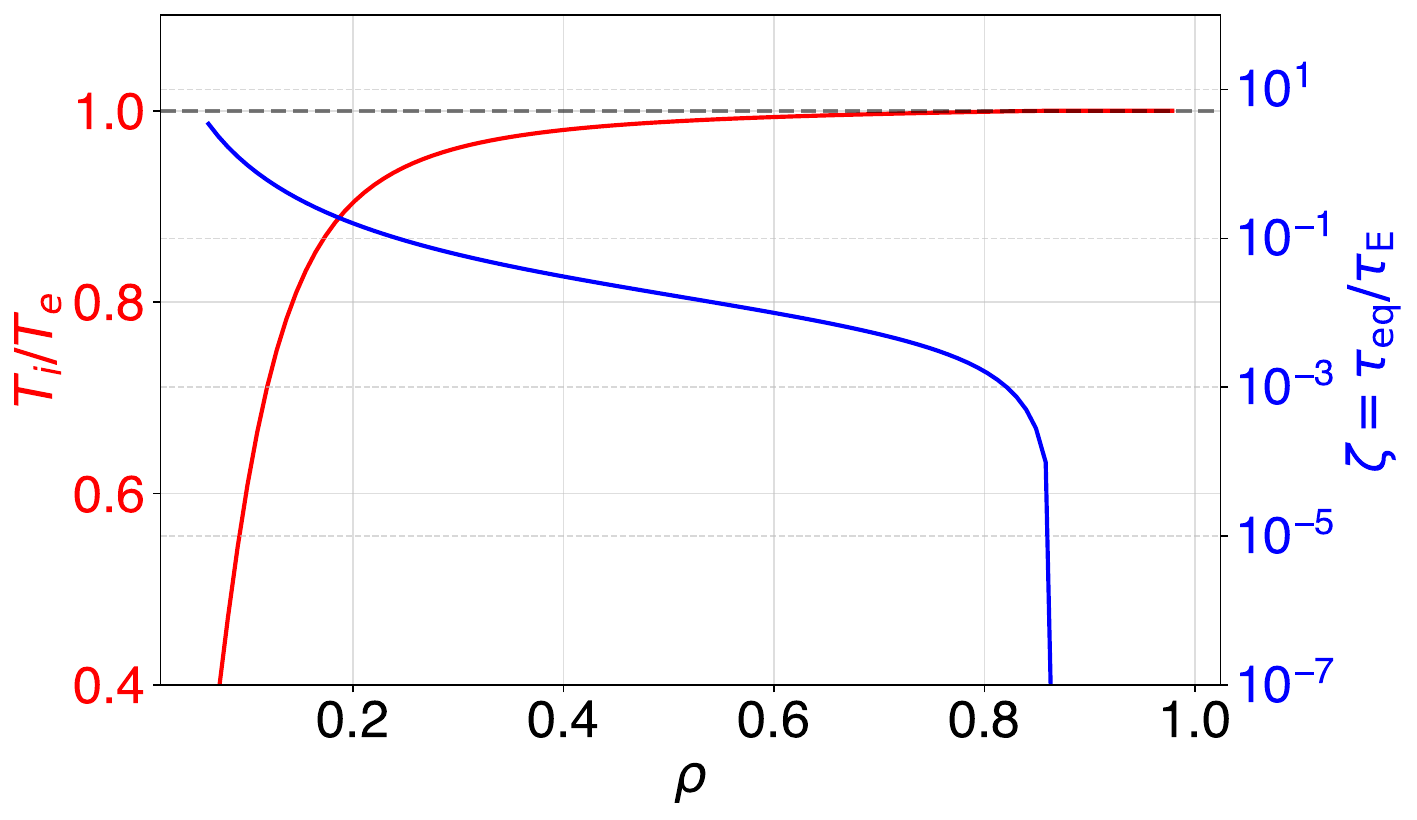}
    \caption[Coupling of $\Ti$ and $\Te$-profiles.]{\textbf{Coupling $\mathbf{\Ti}$ and $\mathbf{\Te}$-profiles.} The left axis (red) shows the coupling parameter as defined in \cref{eq:coupling_parameter}. The right axis (blue) shows the ion-electron temperature ratio. Shot No. \#20180927.021.}
\label{fig:prediction_coupling_comparison_Ti_Te}
\end{figure}

A different, more theoretical approach, as detailed in \cref{sec:Background_Temperature_Coupling}, instead compares characteristic time-scales. Using the parameters corresponding to the conditions in \cref{fig:coupling_comparison_Ti_Te}, we obtain the prediction shown in \cref{fig:prediction_coupling_comparison_Ti_Te}.

The general trend shows good agreement beyond a radius of $\rho \approx 0.1$. In the core region, however, such matching is not expected, as the equilibration is insufficient to keep the species strongly coupled. While W7-X cannot reach the high values of $\tau_E$ required to overcome this limitation, future devices are expected to do so.

\subsection{Data Fitting} 

\subsubsection{Fitting Loss Function}

Given a clamped temperature profile, the coupling of the ion and electron temperature is only reasonable to assume until a certain radius, as discussed in \cref{sec:Background_Temperature_Coupling,sec:Result_Temperature_Coupling}. This means that we have to limit the radial range in which we fit the temperature values. 

\begin{figure}[H]
    \centering
    \includegraphics[width = 10cm]{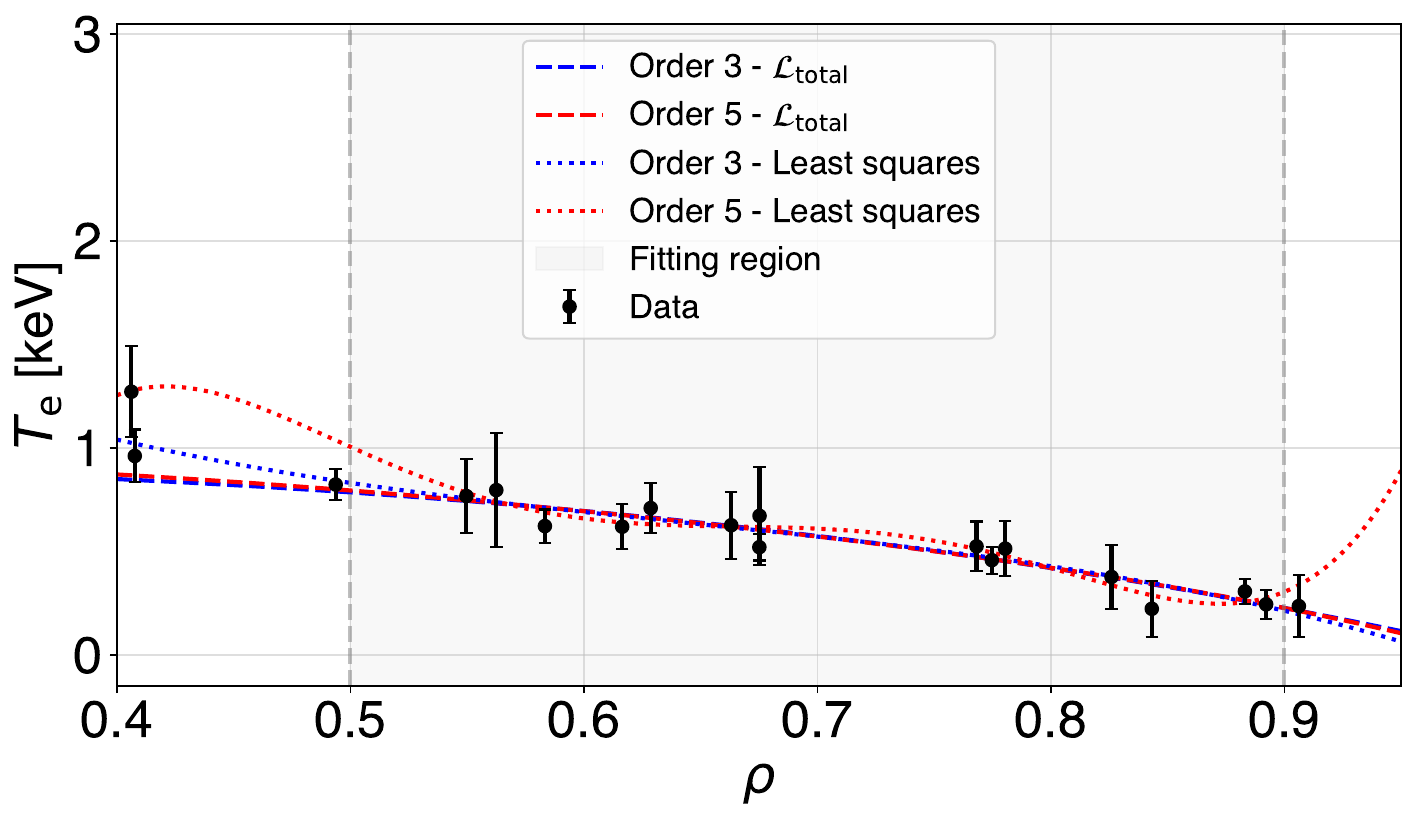}
    \caption[Comparison of loss functions.]{\textbf{Comparison of loss functions} To compare the least square fit with the result of the loss function defined in \cref{eq:total_loss_function}, we fit both a 3rd- (blue line) and 5th (red line) order polynomial to the outer region of the temperature profile. The least square fit is indicated by the dashed line style, and the extended loss function by the dotted line style. Shot No. \#20180925.012.}
\label{fig:comparison_target_loss}
\end{figure}

The target loss function of \cref{eq:total_loss_function} is closely related to a least square fit and should thus behave similarly, while showing a more robust behaviour regarding outliers and should stop higher order polynomials from steep gradients. A comparison of the fitting using the target loss function stated in \cref{eq:total_loss_function} is illustrated in \cref{fig:comparison_target_loss}.

From the different fitting methods discussed in \cref{sec:Methods_Fitting_Functions}, the truncated Gaussian fitting and two-power fitting have proven to be less reliable. As dictated by the underlying profile shape, they work well when the profile falls quickly enough after the core and then flattens out. As the density profiles and temperature profiles in the regions of interest are typically very flat. Thus, for the remainder of this section, we shall focus on the polynomial fits, the CG- and piecewise model.

\subsubsection{Piecewise Model}
\begin{figure}[H]
    \centering
    \includegraphics[width = 10cm]{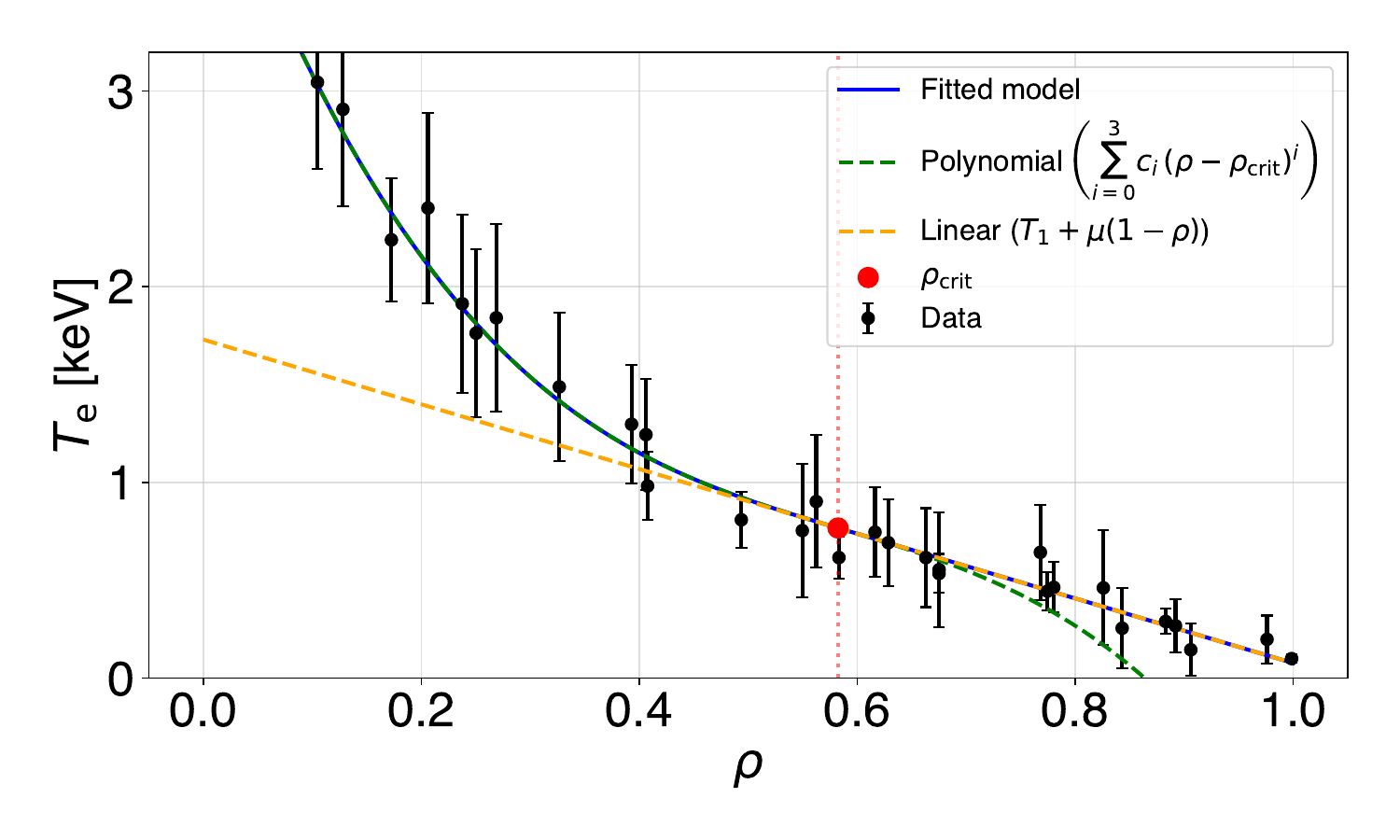}
    \caption[Fitting of piecewise model.]{\textbf{Piecewise model} Fit of the $\Te$ profile with a fixed $\rhocrit=0.6$ (red dot). The linear fit is shown in orange, the 3rd order polynomial in green, and the combined final model fit in blue. Shot No. \#20180925.012.}
\label{fig:piecewise_Model_single_Te_fit}
\end{figure}

The piecewise model allows for a set of different specifications. First, the number of constraints can be adjusted, and the transition point $\rhocrit$ can either be manually set or taken as a fitting parameter. 
One typical fit of the temperature with a fixed $\rhocrit$ and all three constraints ($T$, $T^\prime$, $\alT$) is illustrated in \cref{fig:piecewise_Model_single_Te_fit}.

The figure clearly shows that the outer part is well described by a linear fit, while the inner part has a steeper decrease. 
A more thorough analysis of the different constraints, placement and fitting of $\rhocrit$ is given in \cref{sec:Results_Database_Piecewise_Model}.

\subsubsection{CG Model Fit}
\begin{figure}[H]
    \centering
    \includegraphics[width = 12cm]{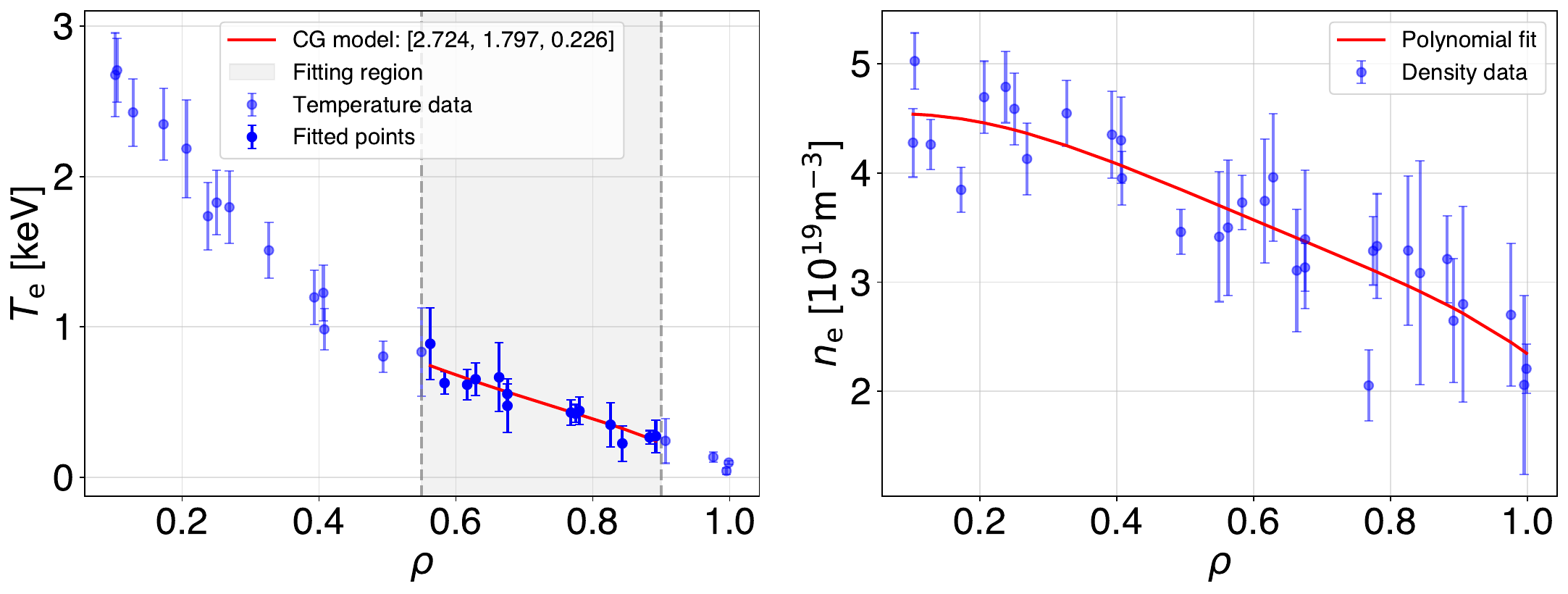}
    \caption[Fit of the CG model.]{\textbf{Fit of the CG model.} The left subplot shows the fitted temperature profile (red), optimised to match the data (blue). The right subplot shows the corresponding density profile and a 5th-order fit.
    Shot No.\ 20180925.012.}
\label{fig:Single_CG_Model_Fit}
\end{figure}

The temperature curve shown in \cref{fig:Single_CG_Model_Fit} (left) represents the temperature profile obtained as a solution to \cref{eq:CG_model_dTdr}.

In contrast to all other modules, the density profiles also directly influence the fitting. To obtain a smoother curve and facilitate the fitting procedure, a fit of the density that enters the GB heat flux is employed. This fit is illustrated in \cref{fig:Single_CG_Model_Fit} (right).

The fit is typically close to linear, but its exact form depends on the $\alTcrit$ value and, upon closer inspection, deviates from a purely linear behaviour. 

\subsubsection{Fit Model Comparison}

\begin{figure}[H]
    \centering
    \includegraphics[width = 12cm]{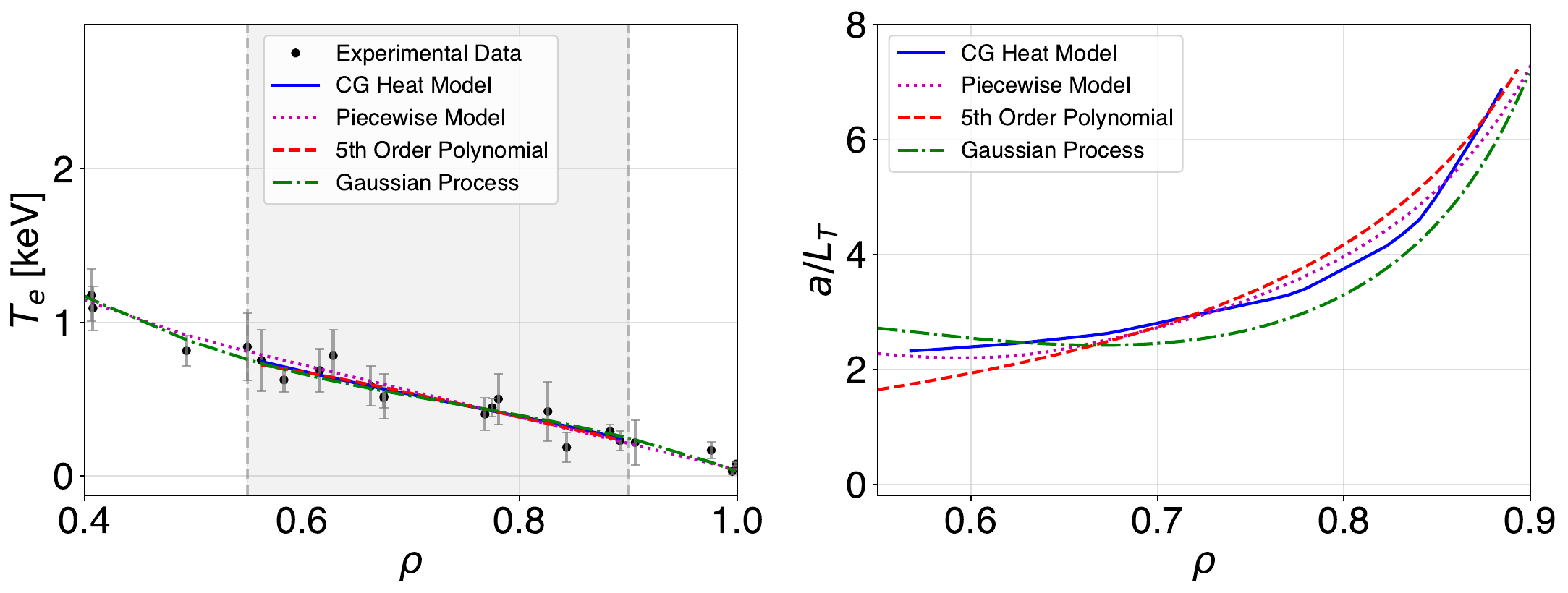}
    \caption[Fit of the temperature profile with several fitting functions.]{\textbf{Fit of the temperature profile with several fitting functions.} On the left-hand side, the fits are shown and on the right-hand side the corresponding $\alT$ values. Shot No. \#20180925.012. }
\label{fig:Fit_Comparison_Indv_Fit}
\end{figure}

A selection of the different fit models is shown in \cref{fig:Fit_Comparison_Indv_Fit}.
Important to note is that the piecewise model has to be fitted to the whole profile in order to capture the transition to a linear decrease. Also, the Gaussian regression only produces sensible results when more data is taken into account, as available in the fitting region, at least in the current implementation. A finer tuning may make it less sensitive to fluctuations in the measurement data. 

Even though the temperature fits shown on the right are quite similar, the resulting $\alT$ profiles exhibit a clear distinction. Thus, the differences observed arise not only from comparing profiles at different time points, but also from the use of different fitting methods.

\section{Database Analysis}  \label{sec:Results_Database_Analysis}

The database comprises selected shots from the OP 1.2 campaign (2018), with a total of 338 shots included. For each shot, the filters discussed above are applied, and several functions are fitted to the resulting data.
First, the renormalisation factor $f_{\text{ren}}$ from the ISS04 scaling is extracted in \cref{sec:Results_Confinment_Time}.
Then, the predicted Gyro-Bohm scaling of the temperature is examined in \cref{sec:Gyro_Bohm_Scaling_Temperature}.
Finally, the results of the individual fit functions are presented, followed by a discussion of the extracted gradient length scales.


\subsection{Confinement Time} \label{sec:Results_Confinment_Time}
\begin{figure}[H]
    \centering
    \includegraphics[width = 12cm]{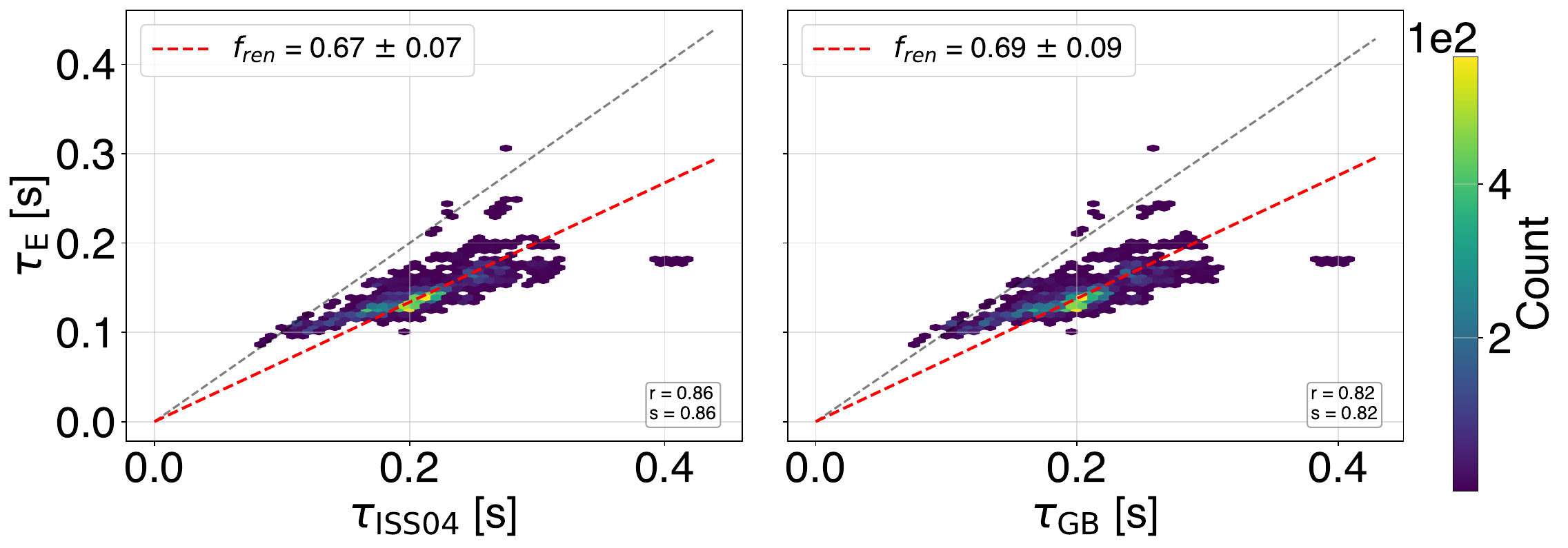}
    \caption[Comparison of $\tauE$ with $\tauiss$ and $\taugb$.]{\textbf{Comparison of $\mathbf{\tauE}$ with $\tauiss$ and $\taugb$.} The red dashed line represents the linear fit used to determine the normalisation factor between $\tauE$ and the confinement-time scalings. The colour bar indicates the number of data points contained in each hexagonal bin.}
\label{fig:Confinment_Time_Comparison}
\end{figure}

The confinement time at each point is calculated using \cref{eq:tau_E_def} and compared with the ISS04 and GB scalings given in \cref{eq:tauISS04,eq:tau_gb}. For the line-averaged density $\overline{n}$, a simple average over $\reff$ is taken. From this comparison, we extract the renormalisation factor $f_{\text{ren}}$ (\cref{eq:fren}). An example of the calculated confinement times for the standard configuration (as defined in \cite{breznsek_plasmasurface_2021}) is shown in \cref{fig:Confinment_Time_Comparison}. The iota profile is taken as $\iotaslash_{2/3} = 0.9$. Radiation losses are neglected, and only ECRH power is considered.  

A clear linear correlation is observed.  
For the ISS04 scaling, we obtain a correlation coefficient of $r = 0.86$ and a renormalisation factor of $f_{\text{ren, ISS04}}=\num{0.67(0.07)}$.  
For the Gyro-Bohm scaling, the result is similar, with $r = 0.82$ and $f_{\text{ren,GB}}=\num{0.69(0.09)}$.  
Both scalings therefore predict nearly identical trends, although the ISS04 scaling agrees slightly more closely with the data. 

\begin{figure}[H]
    \centering
    \includegraphics[width = 9cm]{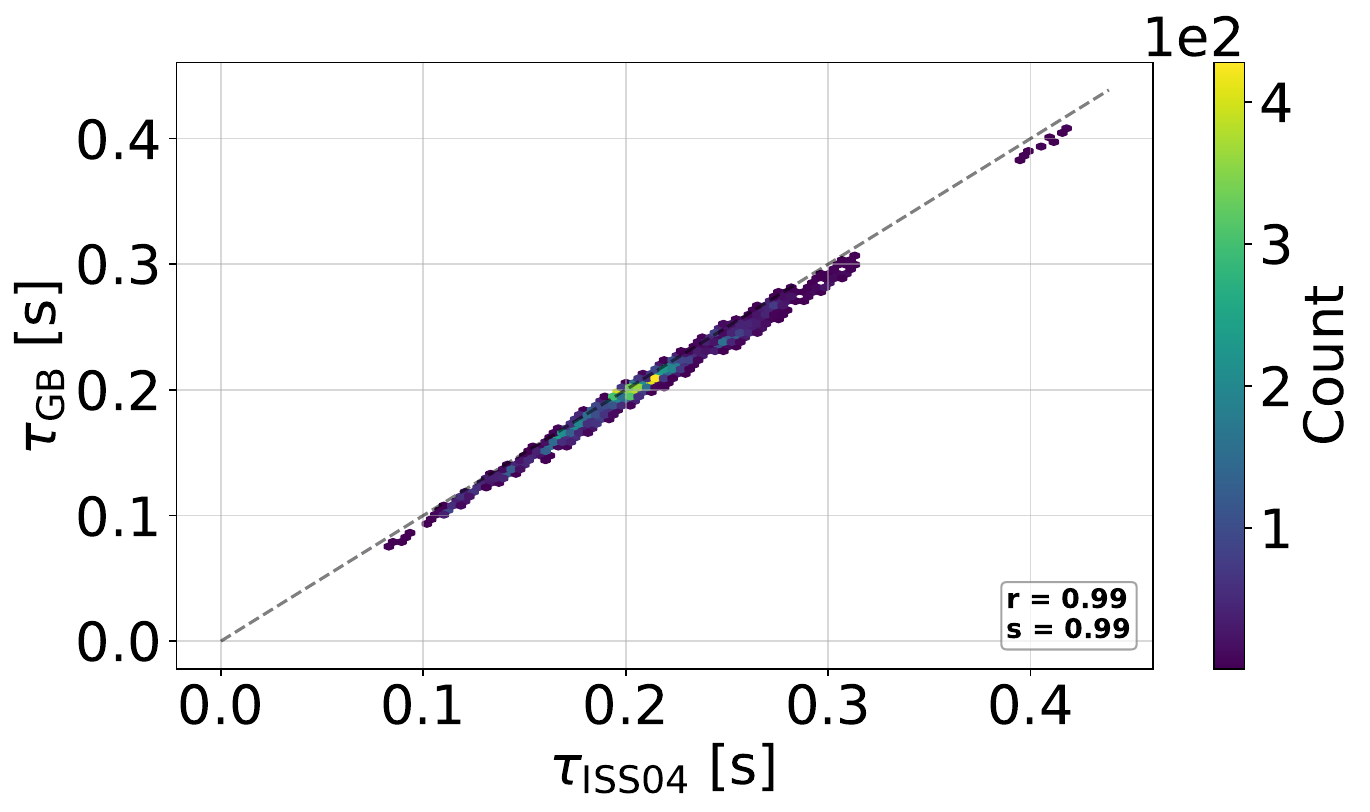}
    \caption[Comparison of $\tauiss$ with $\taugb$.]{\textbf{Comparison of $\tauiss$ with $\taugb$.} The grey line represents $x = y$, indicating perfect agreement between the two confinement-time scalings.}
\label{fig:Confinment_Time_scaling_Comparison}
\end{figure}

The similarity of the two confinement-time scalings is further illustrated in \cref{fig:Confinment_Time_scaling_Comparison}.

\subsection{Gyro-Bohm Temperature Scaling} \label{sec:Gyro_Bohm_Scaling_Temperature}
\begin{figure}[H]
    \centering
    \includegraphics[width = 12cm]{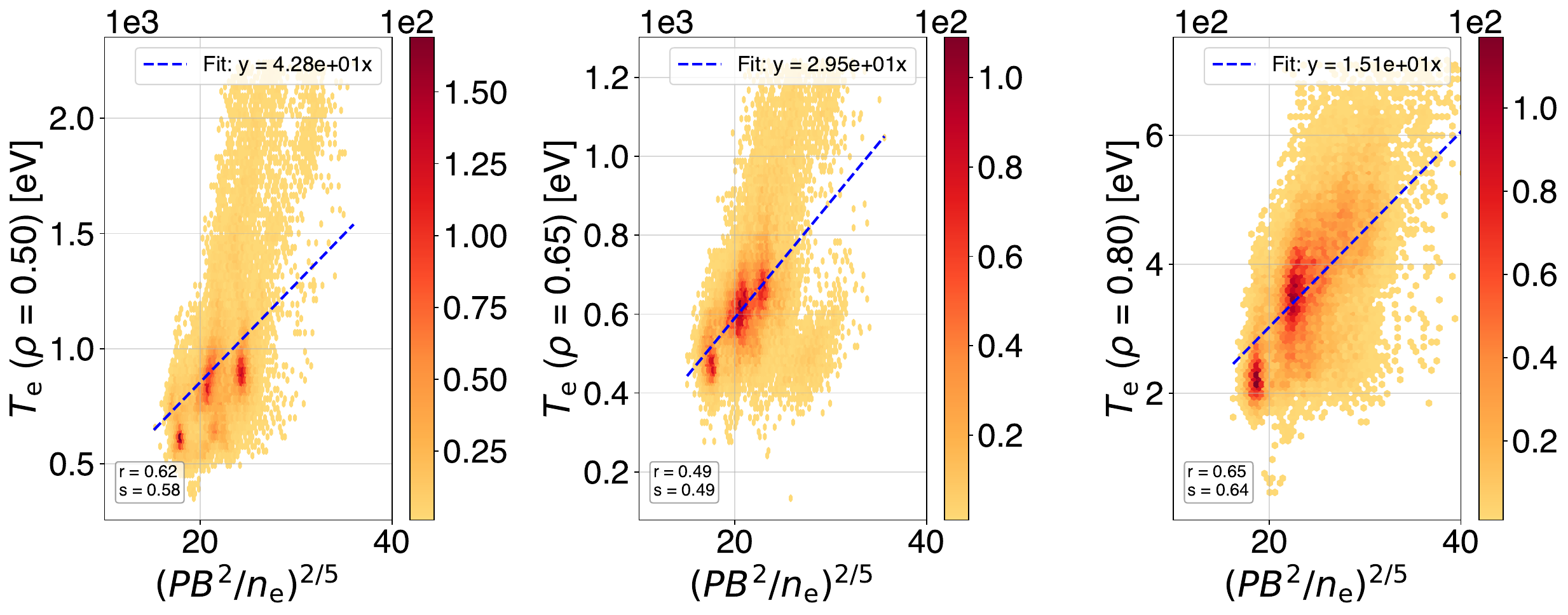}
    \caption[Temperature scaling based on Gyro-Bohm theory.]{\textbf{Temperature scaling based on Gyro-Bohm theory.} The relation in \cref{eq:T_Gyro_Bohm_Power_Scaling} is evaluated at three radial positions, $\rho \in [0.50, 0.60, 0.80]$.}
\label{fig:Temperature_Gyro_Bohm_Scaling}
\end{figure}

According to Gyro-Bohm theory, the temperature is expected to scale as 
\begin{equation}
    T \propto \left(\frac{P B^2}{n} \right)^{\frac{2}{5}},
    \label{eq:T_Gyro_Bohm_Power_Scaling}
\end{equation}
which is evaluated at three radial positions, as shown in \cref{fig:Temperature_Gyro_Bohm_Scaling}.

The results indicate that the temperature broadly follows the predicted scaling.
However, the Pearson correlation coefficient of $r = 0.5$–$0.6$ suggests that the relationship with the control parameters is not strictly linear and deviates from the ideal behaviour implied by \cref{eq:T_Gyro_Bohm_Power_Scaling}.

\subsection{Piecewise Model Fit} \label{sec:Results_Database_Piecewise_Model}
As described in \cref{sec:Piecewise_Model}, the piecewise model allows for several physical continuity constraints to be imposed at the transition point $\rhocrit$. The most natural condition is the continuity of the temperature itself, as there is no visible discontinuity in the measured temperature profiles. Enforcing continuity in the first derivative, and especially in the gradient scale length, is more difficult to justify physically. However, such constraints can improve the smoothness and numerical stability of the fit.

\begin{figure}[H]
    \centering
    \includegraphics[width = 10cm]{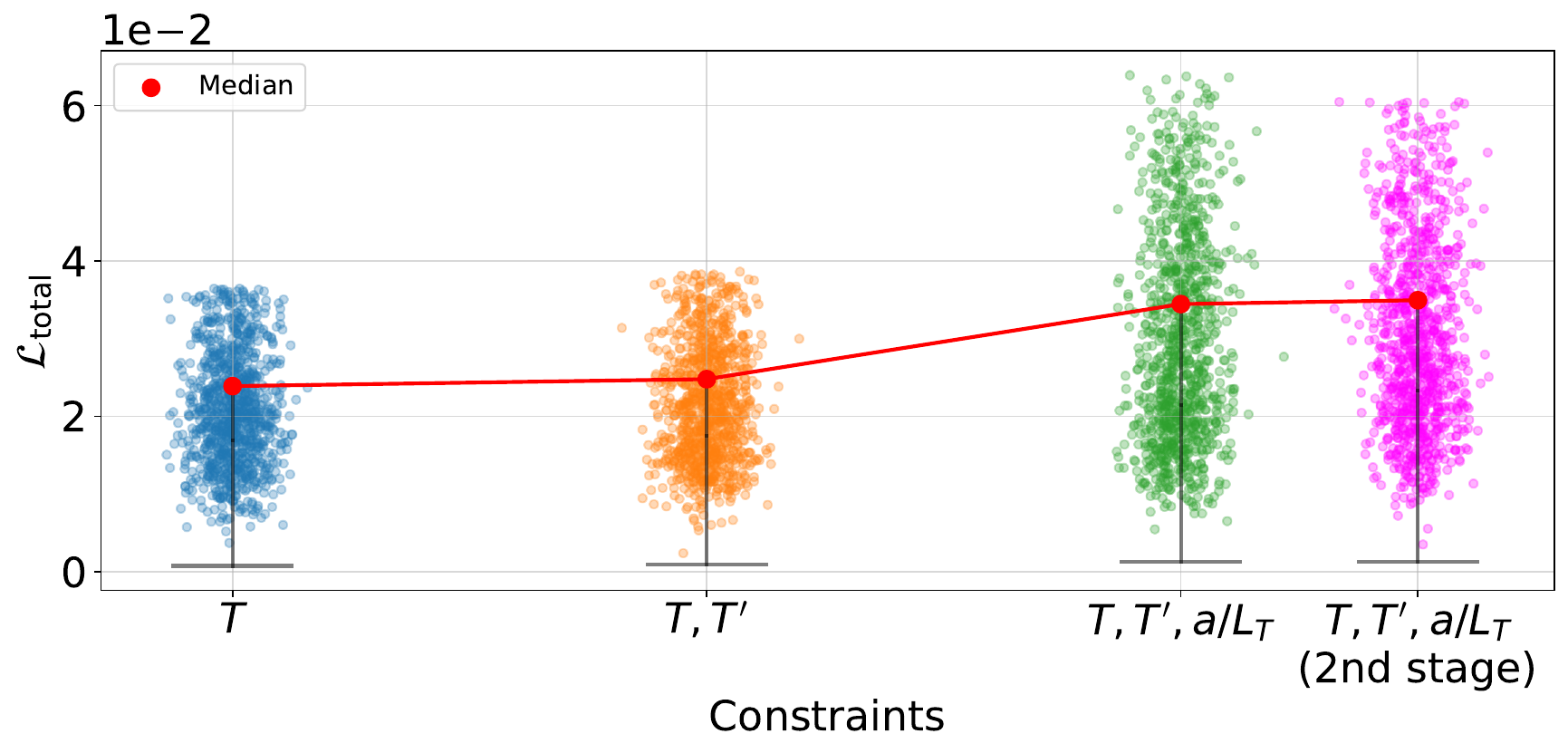}
    \caption[Piecewise model: Loss vs number of constraints.]{\textbf{Target loss function.} The first three configurations use a fixed transition point at $\rhocrit = 0.7$, while the fourth case applies a two-stage fit where $\rhocrit$ is optimised using a two-constraint model before applying the full three-constraint model.}
\label{fig:Target_Loss_Piecewise_Model_07}
\end{figure}

\Cref{fig:Target_Loss_Piecewise_Model_07} shows the mean total error as a function of the number of constraints applied. Each group corresponds to a different level of constraint imposed at a fixed transition point $\rhocrit = 0.7$, except for the fourth case, which employs a two-stage fitting process: first determining $\rhocrit$ with a two-constraint model, then applying all three constraints using the fitted value of $\rhocrit$.

To assess the sensitivity of the fit quality to the position of the transition, we repeated this analysis for $\rhocrit \in \{0.6, 0.7, 0.8\}$. The detailed numerical results for each case are listed in \cref{tab:loss_piecewise_model}. Generally, imposing more constraints leads to a modest increase in the target loss, but the fits remain well-behaved. The influence of $\rhocrit$ on the overall model performance appears mild, with no sharp degradation observed across the tested values.

\begin{table}[htbp]
    \centering
    \begin{tabular}{c|l|c|c|c}
        $\rhocrit$ & Constraints & Median & Q3 & Std. \\
        \hline
        0.6 & T               & 0.024 & 0.037 & 0.490 \\
        0.6 & T, T'           & 0.027 & 0.047 & 0.291 \\
        0.6 & T, T', a/L\textsubscript{T} & 0.043 & 0.077 & 0.312 \\
        \hline
        0.7 & T               & 0.024 & 0.038 & 0.385 \\
        0.7 & T, T'           & 0.025 & 0.040 & 0.210 \\
        0.7 & T, T', a/L\textsubscript{T} & 0.035 & 0.065 & 0.298 \\
        \hline
        0.8 & T               & 0.024 & 0.038 & 0.243 \\
        0.8 & T, T'           & 0.025 & 0.039 & 0.234 \\
        0.8 & T, T', a/L\textsubscript{T} & 0.031 & 0.056 & 0.328 \\
        \hline
    \end{tabular}
        \caption[Target loss of piecewise model for different $\rhocrit$.]{\textbf{Target loss of piecewise model for different $\mathbf{\rhocrit}$}. The table shows the median, upper quartile (Q3), and standard deviation (Std) of the target loss under various constraints.}
    \label{tab:loss_piecewise_model}
\end{table}
A further hypothesis is that the optimal transition point $\rhocrit$ depends on the heating power. Specifically, under ECRH, one would expect the transition point to shift outward with increasing power. 

\begin{figure}[H]
    \centering
    \includegraphics[width = 13cm]{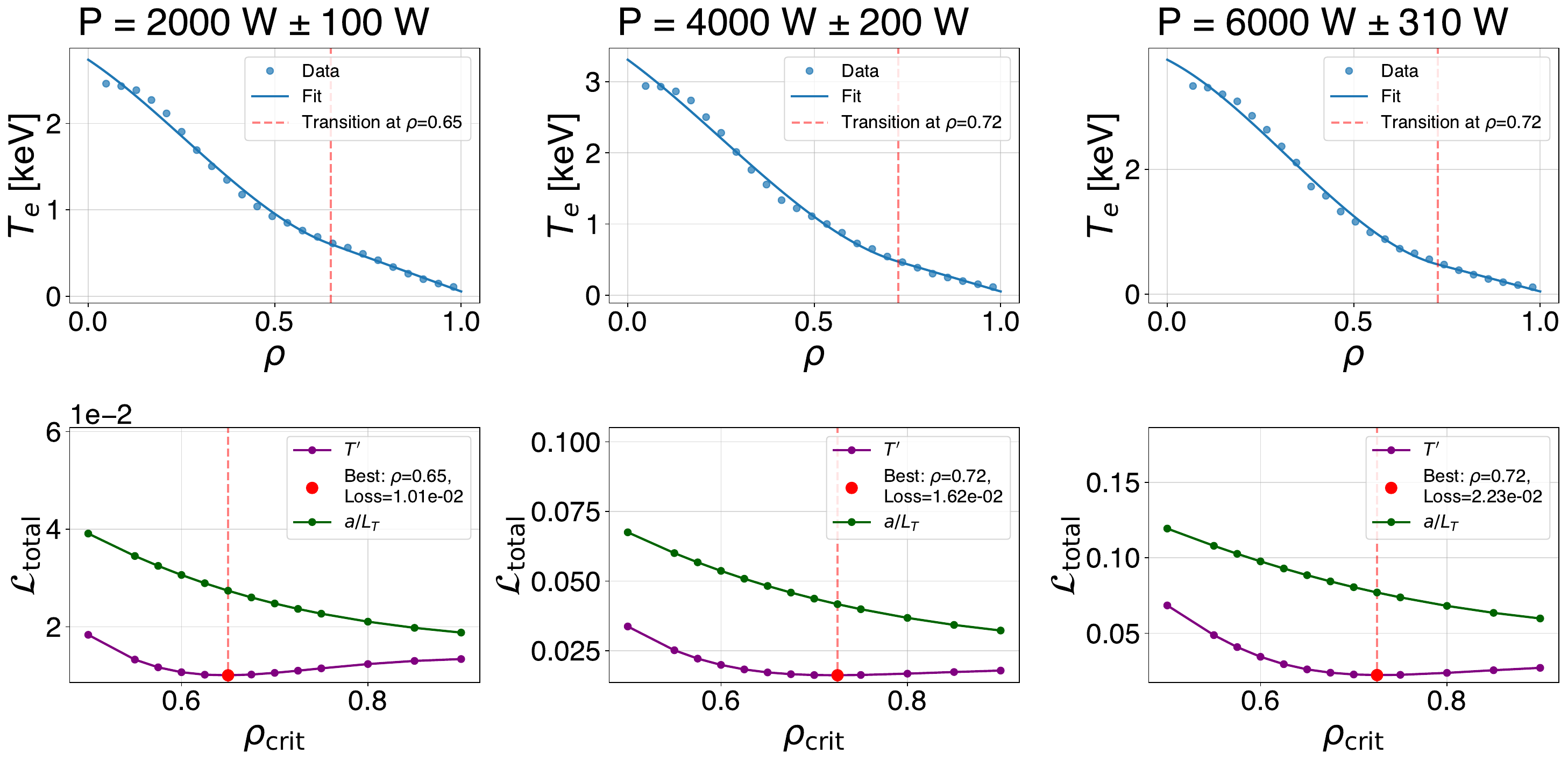}
    \caption[Piecewise model: Fit for different powers.]{\textbf{Piecewise model: Fit for different powers.} Median fitted profiles for three heating powers. An outward shift of the transition point with increasing power is observed.}
\label{fig:Piecewise_Power_Transition}
\end{figure}

\Cref{fig:Piecewise_Power_Transition} shows the median fitted temperature profiles for three different power levels: \SIlist{2;4;6}{\mega\W}. The transition point visibly moves to higher $\rho$ with increasing power, supporting the initial hypothesis.

\begin{figure}[H]
    \centering
    \includegraphics[width = 12cm]{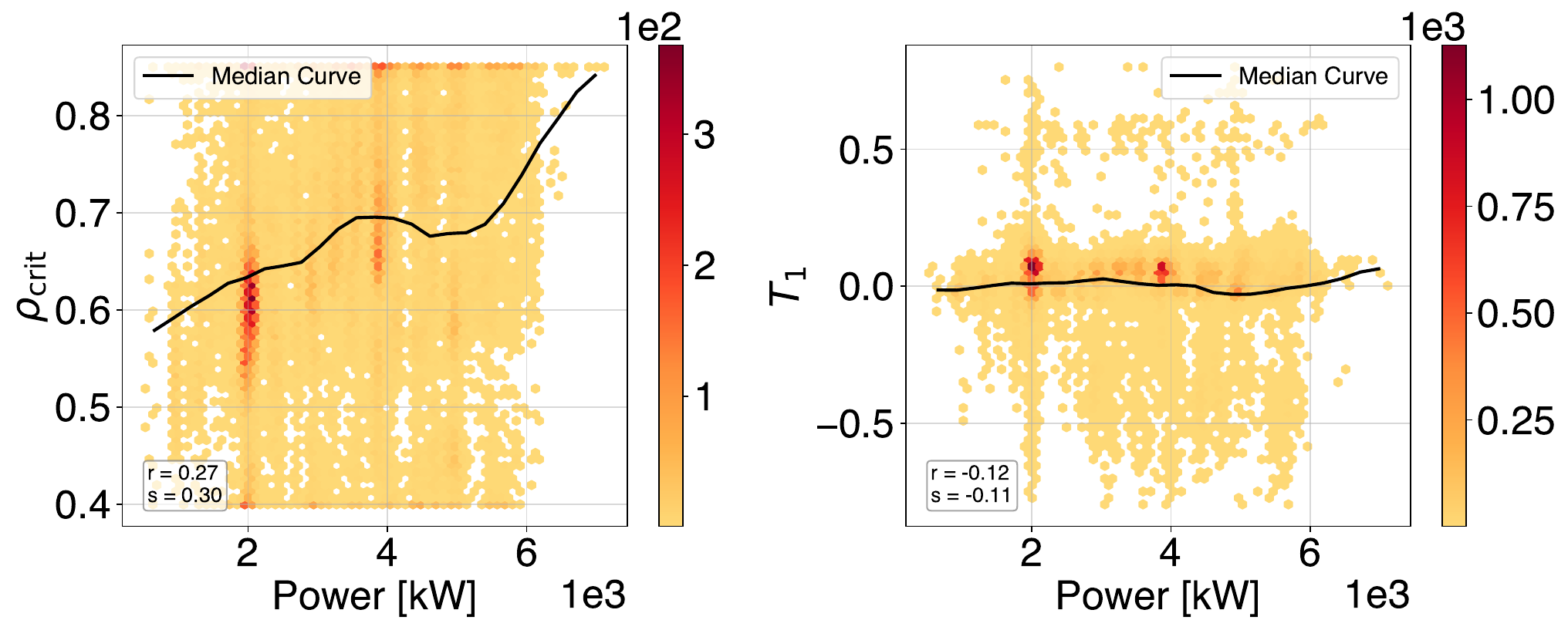}
    \caption[Piecewise model parameter correlation with power.]{\textbf{Piecewise model parameter correlation with power.} Transition point $\rhocrit$ shows a weak but visible correlation with heating power.}
\label{fig:Piecewise_Power_Correlation}
\end{figure}

This trend is further explored in \cref{fig:Piecewise_Power_Correlation}, where the fitted values of $\rhocrit$ are plotted against the heating power. A modest correlation is observed, with a Pearson coefficient of $r = 0.27$ and a Spearman rank correlation of $s = 0.30$, suggesting a statistically weak but consistent trend. The linear offset parameter $T_1$ fluctuates around zero, with no apparent dependence on power.

\subsection{CG Model Fit}
\begin{figure}[H]
    \centering
    \includegraphics[width = 10.5cm]{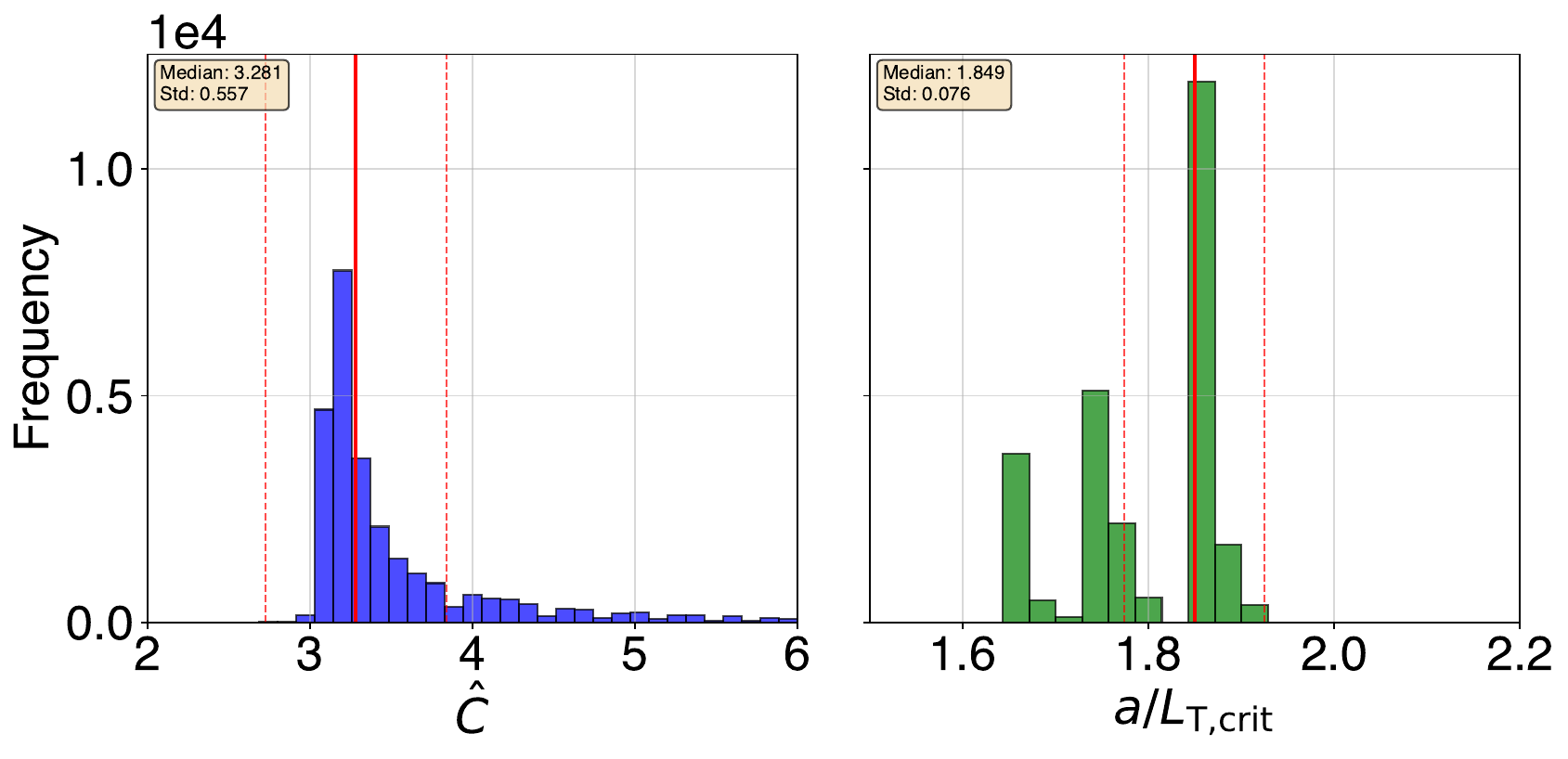}
    \caption[CG model parameter distribution.]{\textbf{CG model parameter distribution.} The left histogram shows the distribution of $\hat{C}$, while the right histogram shows the distribution of $\alTcrit$.}
\label{fig:CG_Model_Paramter_Distribution}
\end{figure}

The CG model involves three fitting parameters. Among these, the boundary temperature is essentially fixed by the experimental temperature at the outer edge of the fitting region, and the relevant correlations have already been discussed in \cref{sec:Gyro_Bohm_Scaling_Temperature}. The behaviour of the slope $\hat{C}$ and, in particular, of $\alTcrit$ is also of interest.
The overall distributions of $\hat{C}$ and $\alTcrit$ are shown in \cref{fig:CG_Model_Paramter_Distribution}.

\begin{figure}[H]
    \centering
    \includegraphics[width = 12cm]{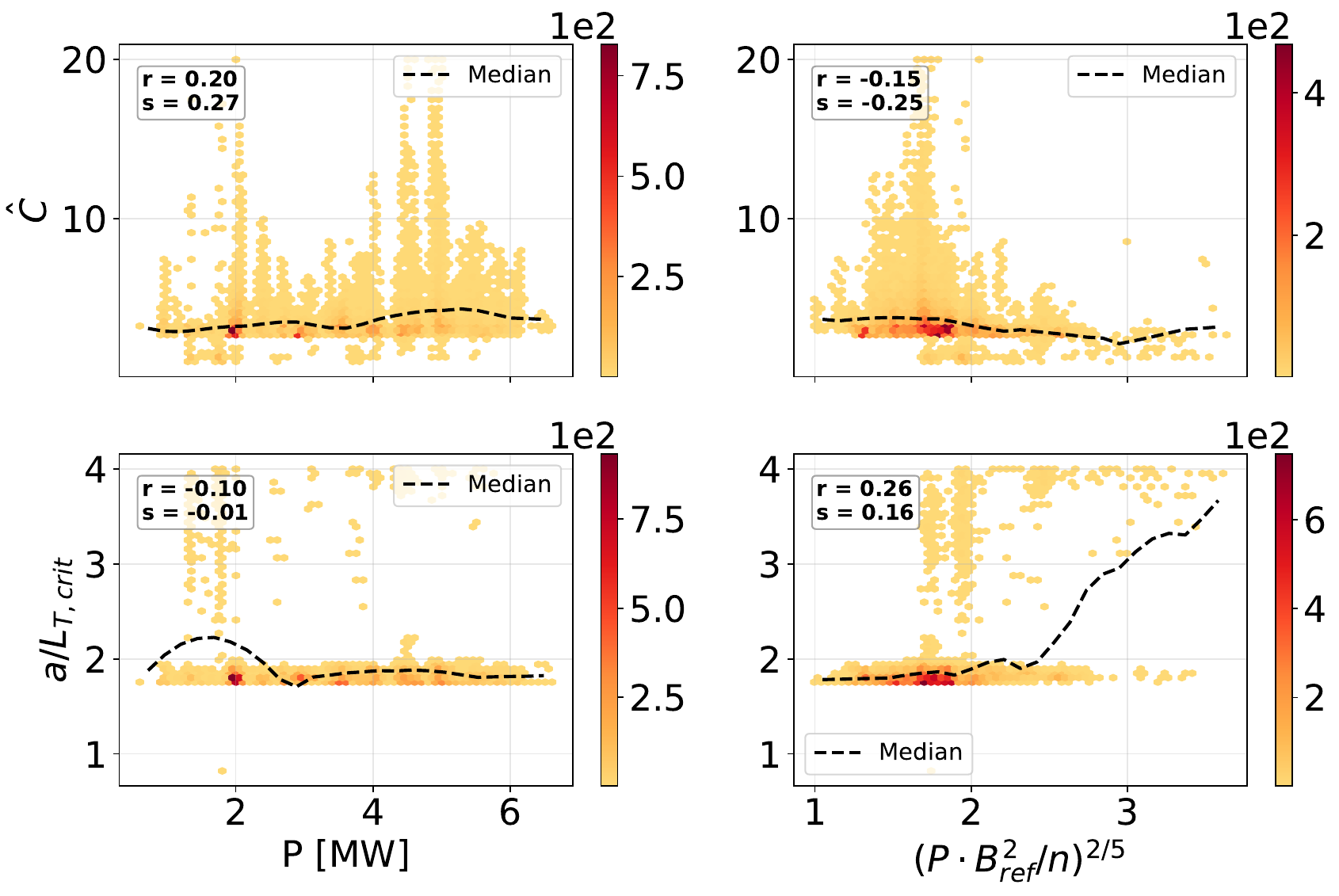}
    \caption[CG model parameter power correlation.]{\textbf{CG model power correlation.} The variation of the CG model parameters is shown both as a function of power (first column) and the GB scaling (second column). Additionally, the median value (black dashed) curve and the correlation coefficients are shown.}
\label{fig:CG_Model_Paramter_power_corr}
\end{figure}

The dependence of those parameters on power and the Gyro-Bohm scaling is shown in \cref{fig:CG_Model_Paramter_power_corr}. 

The results indicate that the variation of $\hat{C}$ is greater than that of $\alTcrit$. However, neither parameter exhibits a clear correlation with the power or with the GB scaling, providing further support for the central hypothesis of this thesis.

\subsection{Extracted Gradient Length Scales} \label{sec:Results_lenght_scale_profiles}
\begin{figure}[H]
    \centering
    \includegraphics[width = 13cm]{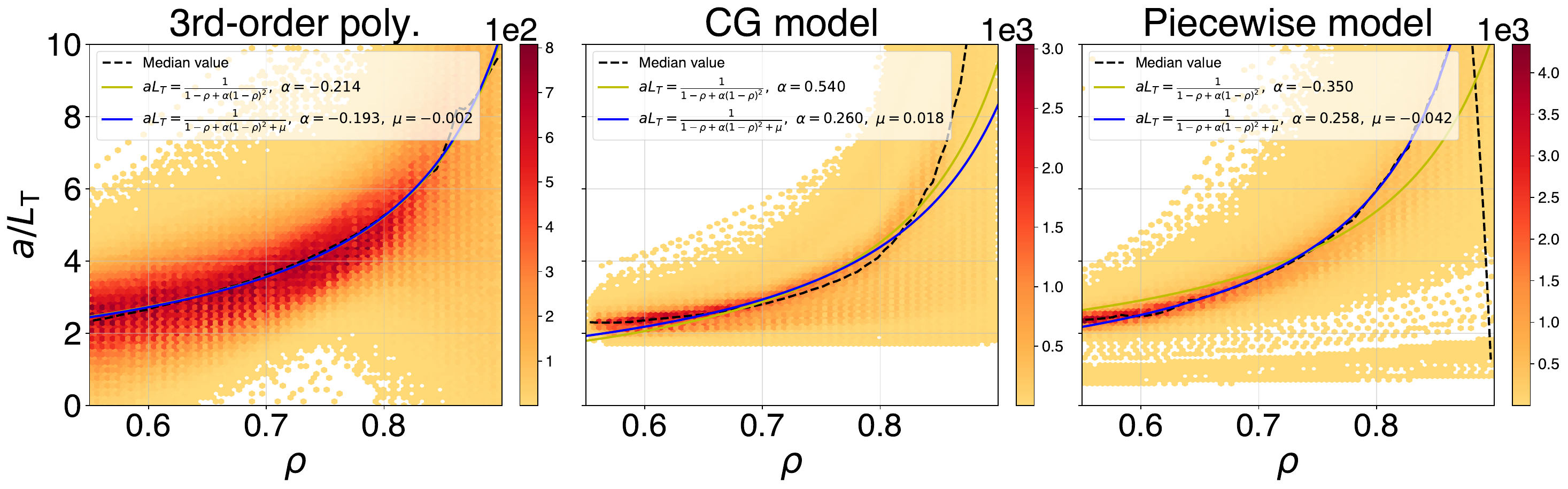}
    \caption[Comparison of fitting methods for database.]{\textbf{Comparison of fitting methods for $\alT$.} Shown are the $\alT$ profiles obtained using a third-order polynomial, the CG model, and a piecewise model.}
\label{fig:Database_Fit_Model_Comparison_Temperature}
\end{figure}

The main purpose of the database analysis is to extract reliable profiles of the gradient length scales. 
An extensive range of fitting methods is available, and a small selection for the $\alT$ profiles is shown in \cref{fig:Database_Fit_Model_Comparison_Temperature}.

To quantify the behaviour and provide an analytical form as input for the simulations, the theoretically anticipated $\alT$ curves, discussed in \cref{sec:Theoretical_aLT_predictions}, are fitted. The fit parameters are shown in the plot. 
All three methods reproduce the expected behaviour very well, with differences in the fits reflected in the corresponding parameters. While the spread of $\alT$ values is smaller at lower radii, it increases with radius. To obtain a consistent fit across the entire profile, the optimisation is performed on the median profiles, which are also shown in the plot.

\begin{figure}[H]
    \centering
    \includegraphics[width = 9cm]{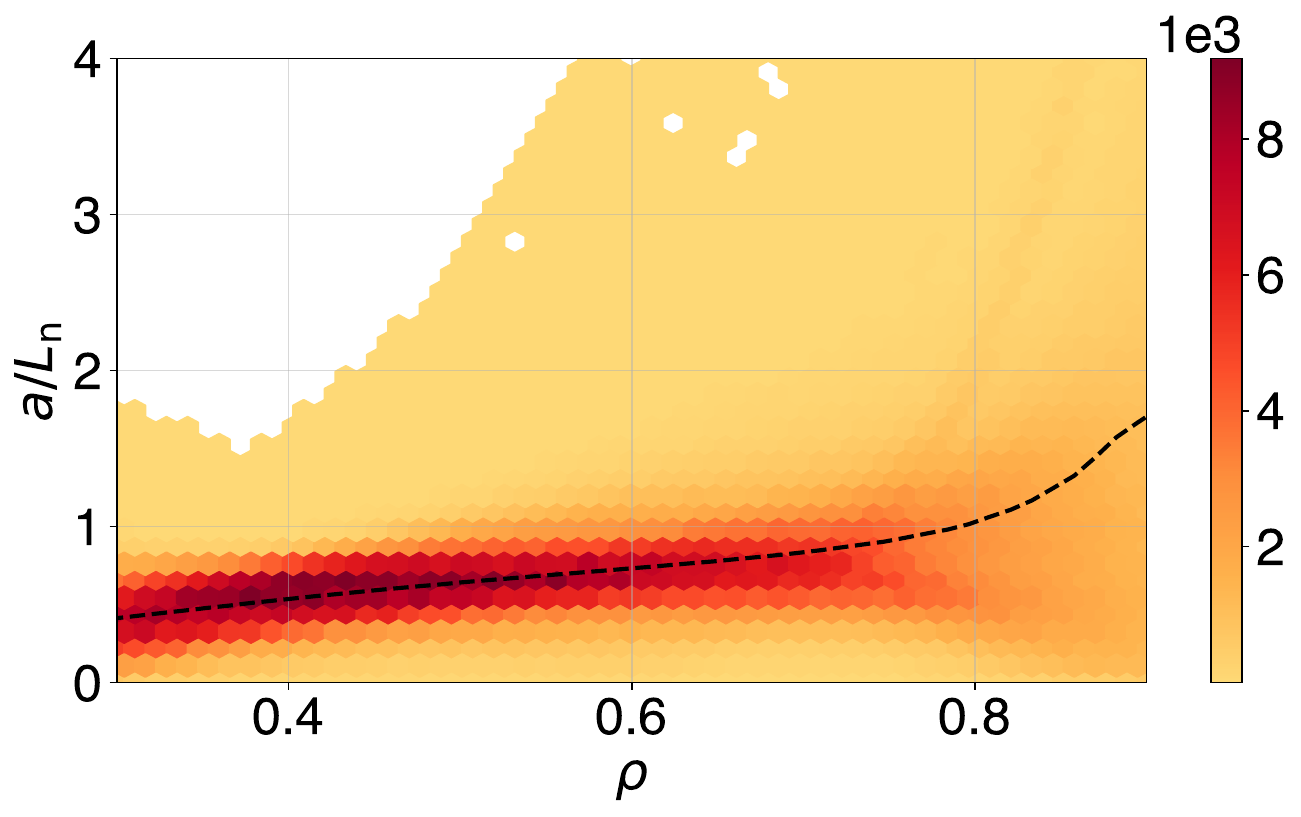}
    \caption[Density length scales.]{\textbf{Density length scales.} The plot visualises the density gradient length scales obtained through the fitting of a 5th-order polynomial.}
\label{fig:Database_Fit_Model_Comparison_Density}
\end{figure}

For the $\aln$ profiles, several fitting options remain. The most robust proved to be the 5th-order polynomial, providing the best overall fit quality. An overview is shown in \cref{fig:Database_Fit_Model_Comparison_Density}.

The majority of the profiles lie within $\aln \in [0.5,0.7]$, where this is a slight increase with radius. However, much slower compared to the $\alT$ profiles.

\begin{figure}[H]
    \centering
    \includegraphics[width = 12cm]{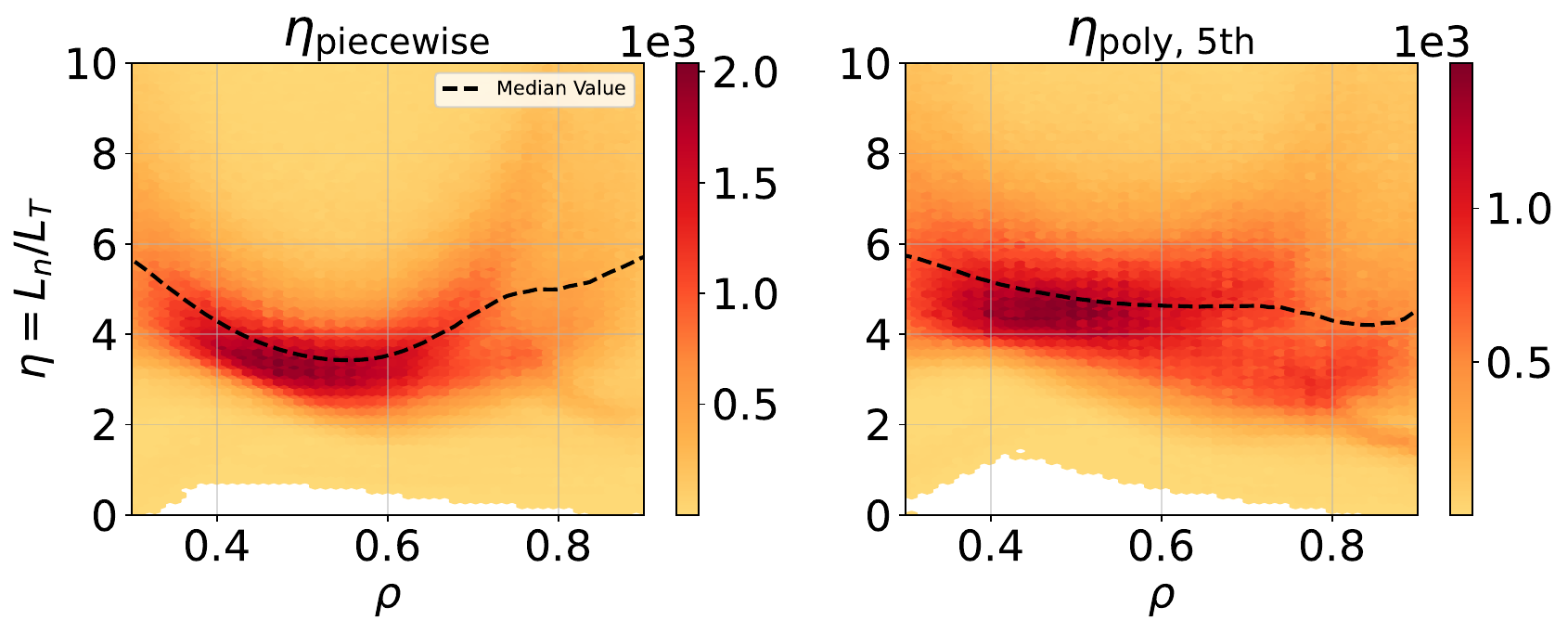}
    \caption[Comparison of fitting methods for database.]{\textbf{Comparison of fitting methods.} Shown are the $\alT$ profiles obtained using the third-order polynomial, CG model, and piecewise model.}
\label{fig:Database_Fit_eta}
\end{figure}

The two resulting $\eta$ profiles are calculated from the ratio of the two shown temperature gradient length scales and the density length scales, and are illustrated in \cref{fig:Database_Fit_eta}.

As expected, the spread in the data is larger, since uncertainties in both $\alT$ and $\aln$ contribute.  
The majority of profiles within the database have an $\eta$ value in the shown radial range between four and six.

\section{Reconstructed Profiles} \label{sec:Results_Reconstructed}

The GX Simulation results 
are discussed in \cref{sec:Results_GX_Simulation_Results}. With the known heat flux, and a given specified input power, as well as initial temperature and density value, we can recalculate the temperature curves as presented in \cref{sec:Results_Profile_Solver}.

\subsection{GX Simulation Results} \label{sec:Results_GX_Simulation_Results}
The simulation results were provided by P. Xanthopoulos and use the most common $\alT$ profile and VMEC equilibrium, as determined in the analysis discussed in \cref{sec:Results_lenght_scale_profiles}.

\begin{figure}[H]
    \centering
    \subfigure[Normalised heat flux.]{%
        \label{fig:GX_Heat_Flux}%
        \includegraphics[width=6.2cm]{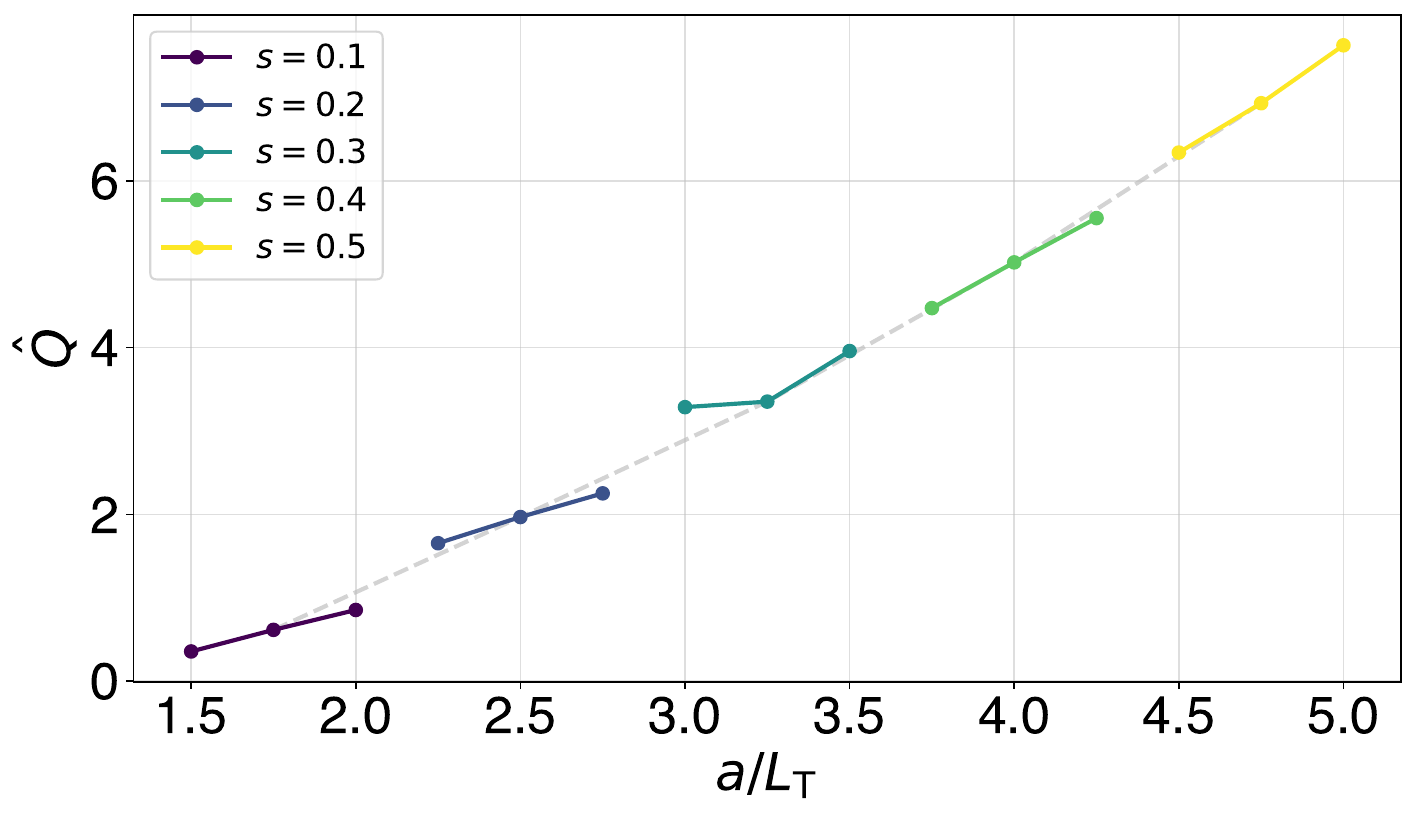}%
    }%
    \quad
    \subfigure[$\etacrit$-profile.]{%
        \label{fig:GX_etacrit}%
        \includegraphics[width=6.2cm]{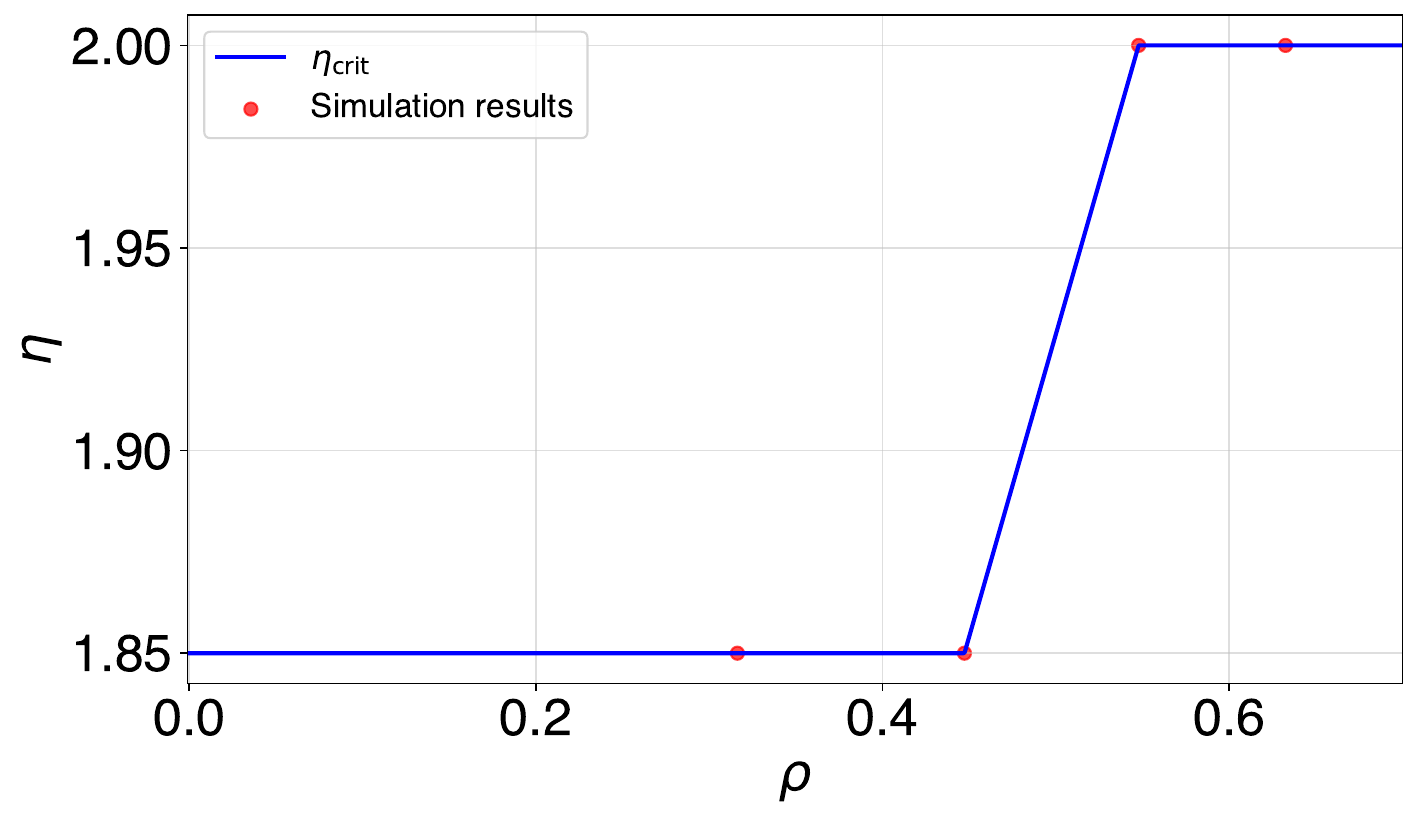}%
    }%
\caption[GX simulation results using the $\etacrit$ model.]{\textbf{GX simulation results using the $\mathbf{\etacrit}$ model.} The heat flux, normalised in Gyro-Bohm units, is shown in \textbf{a)}. The grey line connects the midpoints and serves as a guide to the eye. The resulting $\etacrit$ profile ensuring power balance is shown in \textbf{b)}.}
\label{fig:GX_Simulation_Results}
\end{figure}

\begin{figure}[H]
    \centering
    \subfigure[Normalised heat flux.]{%
        \label{fig:GX_Heat_Flux_set_eta}%
        \includegraphics[width=6.2cm]{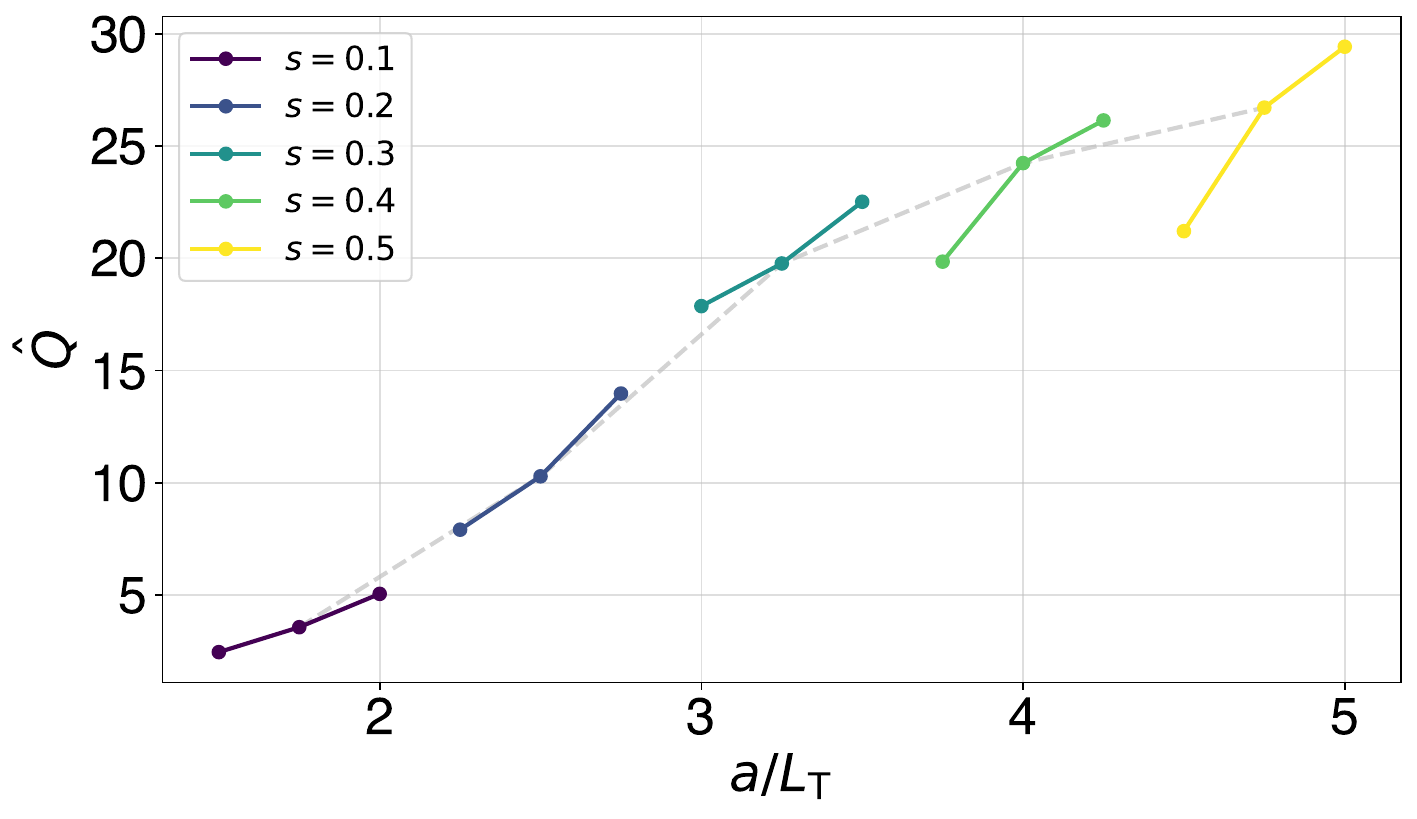}%
    }%
    \quad
    \subfigure[$\eta_\mathrm{exp}$-profile.]{%
        \label{fig:GX_eta_set}%
        \includegraphics[width=6.2cm]{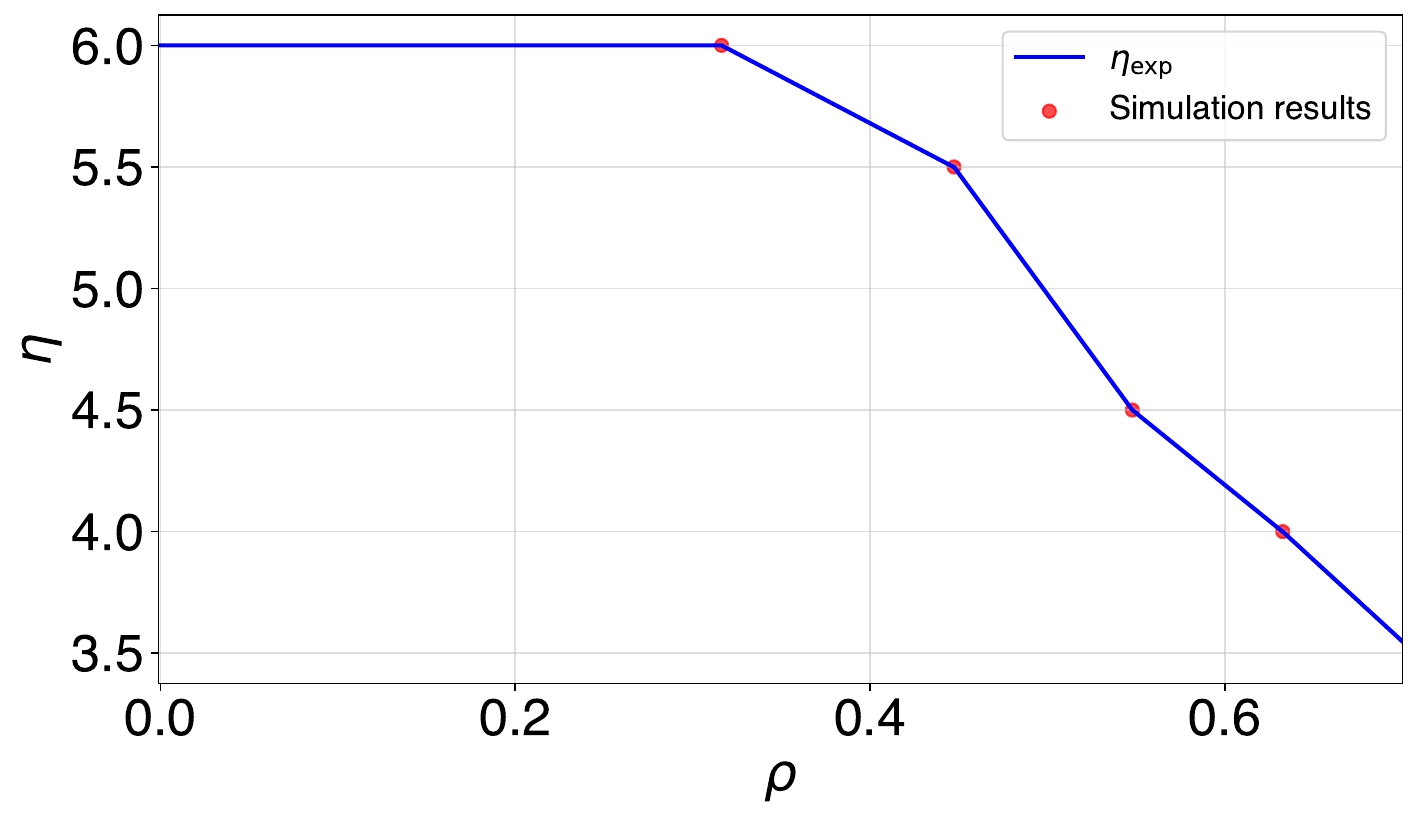}%
    }%
    \caption[GX simulation with set $\eta$-profile.]{\textbf{GX simulation with set $\eta$-profile.} The heat flux normalised in Gyro-Bohm units is shown in \textbf{a)}. The grey line connects the middle points and is a guide to the eye. In \textbf{b)} the underlying $\eta$ profile is shown.}
    \label{fig:GX_Simulation_Results_manuel_gradient}
\end{figure}

As expected, additional sources of flux play a role. A second round of simulations was therefore conducted, this time using the experimentally observed $\eta$ profile (\cref{fig:GX_eta_set}) as input and without enforcing the $\Gamma = 0$ criterion.  
The heat flux results from this simulation are shown in \cref{fig:GX_Heat_Flux_set_eta}.

Not only are the $\eta$ profile and shape very different, but also $\hat{Q}$ is much larger for the second case, demonstrating the strong influence of $\eta$ on regulating ITG drive. 
The individual $\alT$ simulation points for different radii do not lie longer on the same slope, but have a more tilted behaviour. 

\subsection{Radial Profiles} \label{sec:Results_Profile_Solver}

The shape of the resulting radial profiles is governed by the system of ODEs presented in \cref{sec:coupled_ODE_system}. While this system constrains the form of the solution, it still allows for some degrees of freedom, which must be specified to fully determine the profiles. These include the boundary conditions defined in \cref{eq:bc_T,eq:bc_Tprime,eq:bc_n}, as well as the spatial distribution of the heating power, described in \cref{sec:methods_heating_profiles}.

In particular, the deposition profile of ECRH, characterised by its width and peak position, is not known a priori and must be chosen. For the present case, the profile is set with a peak at $\rho_{\mathrm{in}} = 0.0 $ and a width of $ \rho_{\mathrm{w}} = 0.05 $. With these choices, the system is integrated radially outward, starting from the magnetic axis.

As the solver progresses along the radial coordinate $\rho $, it evaluates local values of the normalised temperature gradient $ \alT$, which in turn determines the local heat flux $ \hat{Q}(\rho, \alT)$. Through this dependency, the solution dynamically adapts to the imposed heating profile and boundary conditions, resulting in a self-consistent set of radial profiles.

\begin{figure}[H]
    \centering
    \includegraphics[width = 13cm]{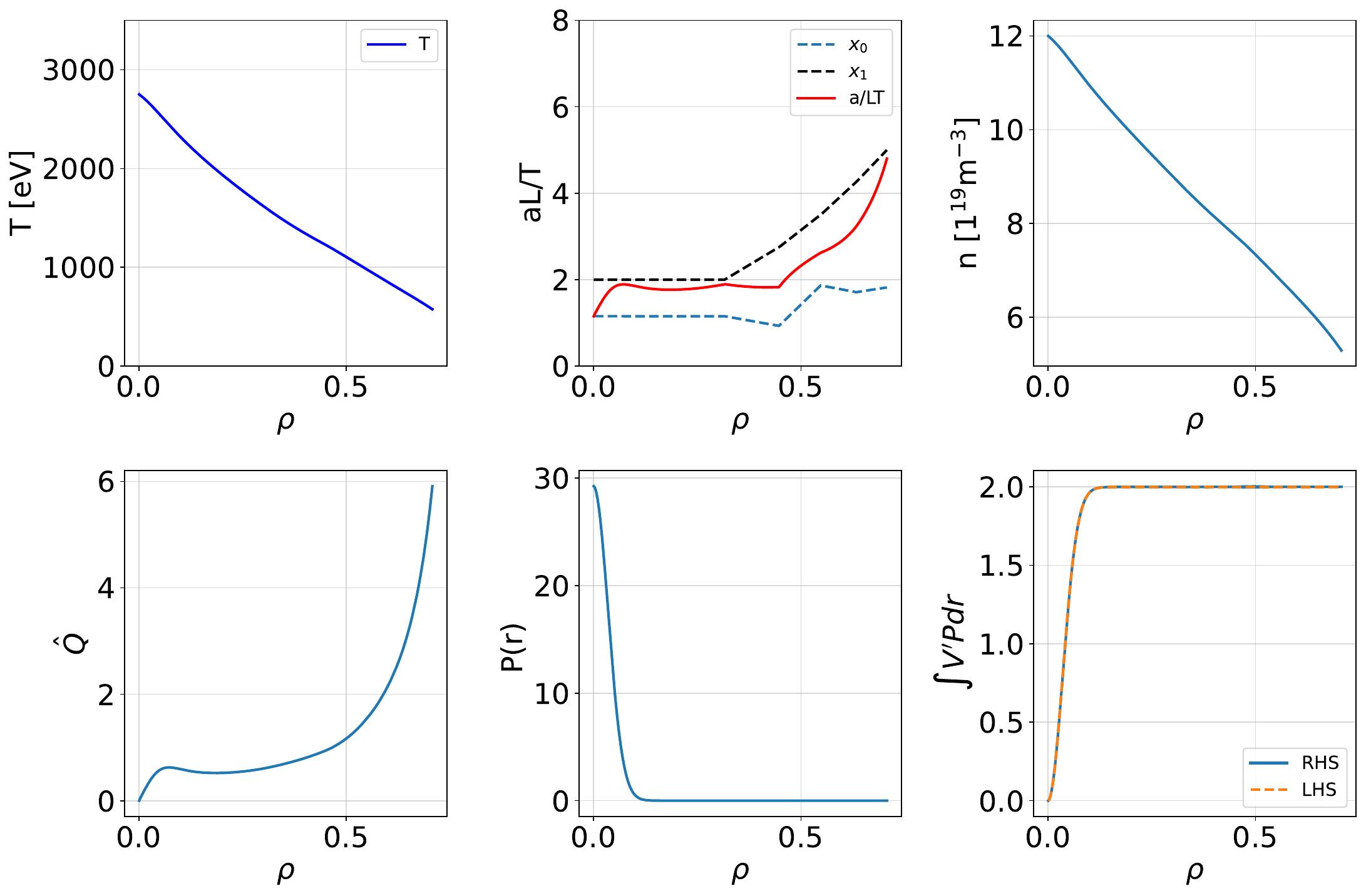}
    \caption[Diagnostics of the profile solver.]{\textbf{Diagnostics of the profile solver.} Overview of the key radial quantities computed by the profile solver. The first row shows the parameter profiles: temperature, gradient length (with $x_0$ and $x_1$ defined in \cref{eq:sparse_interp_x0} and \cref{eq:sparse_interp_x1}, respectively), and density. The second row displays the normalised heat flux $\hat{Q}$, the power profile, and the accumulated power. To assess convergence, both the LHS and RHS of \cref{eq:power_balance_full} are plotted.}
    \label{fig:Profile_Solver_Working_Illustration}
\end{figure}

An example of such a solution is visualised in \cref{fig:Profile_Solver_Working_Illustration}, showing the solver's diagnostics for a specific set of boundary conditions.

\section{Comparison} \label{sec:Results_Comparison}


\subsection{Heat Flux Comparison}
\begin{figure}[H]
    \centering
    \subfigure[$\hat{Q}$ for $\etacrit$.]{%
        \label{fig:Heat_Flux_Model_Comparison_eta_crit}%
        \includegraphics[width=6.2cm]{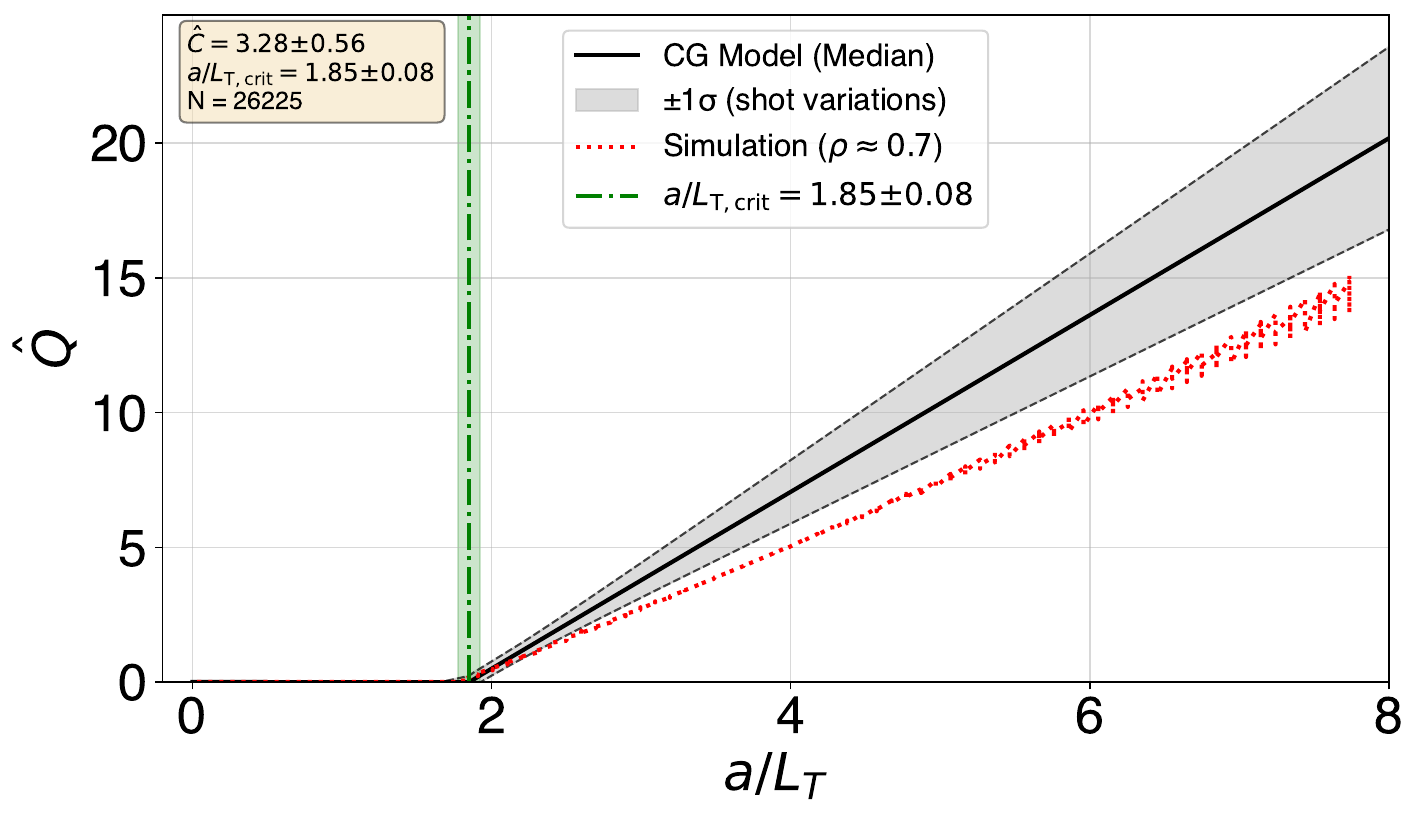}%
     }%
    \quad
    \subfigure[$\hat{Q}$ for $\eta_{\text{exp}}$.]{%
        \label{fig:Heat_Flux_Model_Comparison_eta_set}%
        \includegraphics[width=6.2cm]{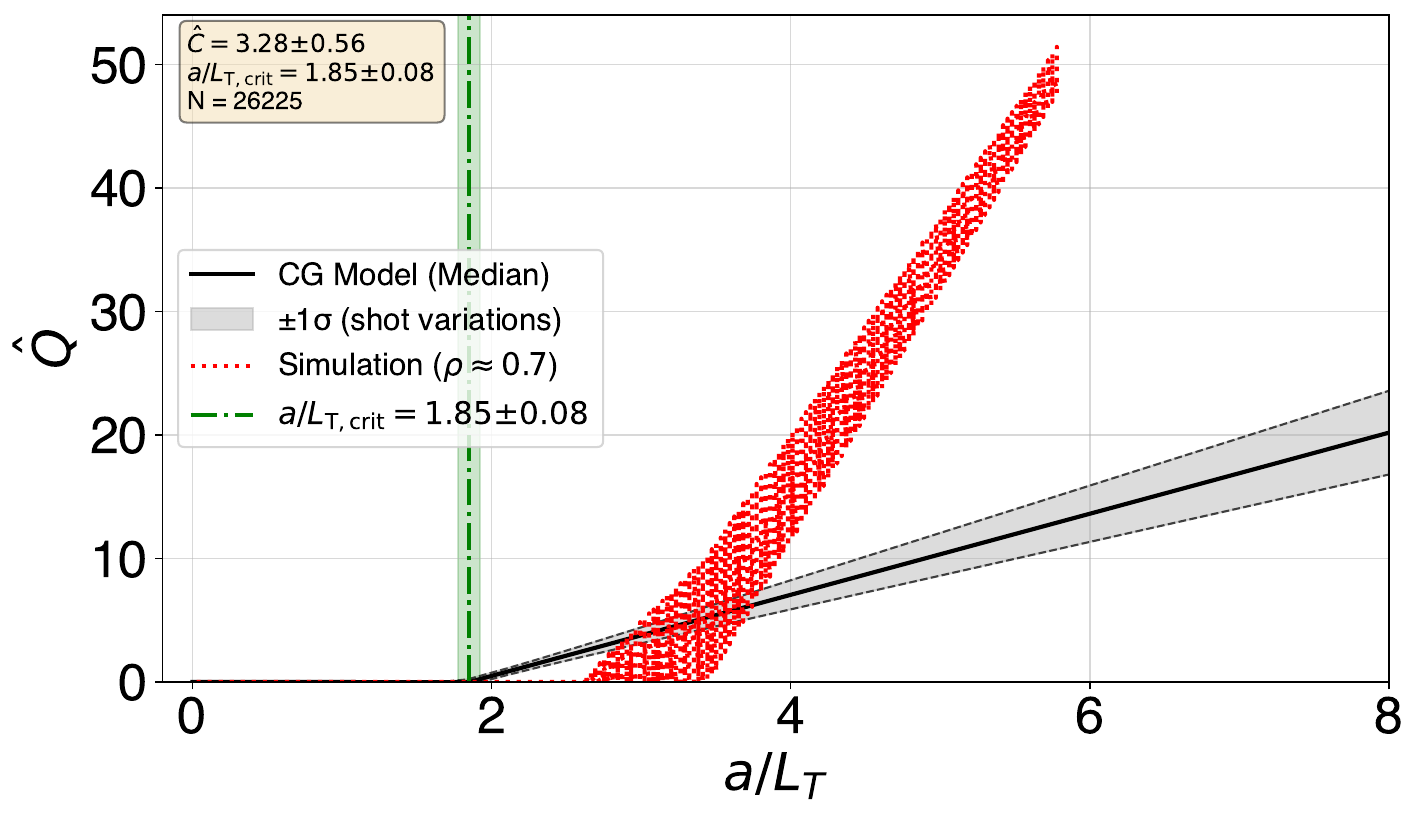}%
    }%
    \caption[Comparison of the normalised heat flux $\hat{Q}$.]{\textbf{Comparison of the normalised heat flux $\mathbf{\hat{Q}}$.} The heat flux from the GX simulation in a small area around $\rho=0.7$. The median heat flux, with one standard deviation indicated, is shown as a function of $\alT$. In \textbf{a)}, the heat flux for $\Gamma =0$, is shown and in \textbf{b)}, the fluxes resulting from the experimental $\eta$ profile.}
    \label{fig:Heat_Flux_Model_Comparison}
\end{figure}

The GX simulation provides a normalised heat flux as a direct function of both radial position and temperature: $\hat{Q}_{\text{GX}} = \hat{Q}(\rho, \alT)$. In contrast, the CG-Model fitted to experimental data (\cref{sec:CG-Model}) yields a temperature-dependent function only: $\hat{Q}_{\text{CG}} = \hat{Q}(\alT)$. This fundamental difference makes direct comparison challenging.
To address this, we compare the predicted heat flux at the outermost radial point from the GX model with the average heat flux from the CG-fit, which incorporates data points spanning $\rho \in [0.55-0.90]$. This approach is illustrated in \cref{fig:Heat_Flux_Model_Comparison}. Although the steep density profile yields fewer data points, the heat flux comparison follows a similar pattern, which is not shown in this plot.

The plot shows that for the first case, the heat fluxes are underestimating the experimental values, while the $\alTcrit$ value aligns very well. In the second case, the heat flux coming from the simulation is much higher and also the corresponding $\alTcrit$, which also does not reproduce what is seen in the experiment.

\subsection{Profile Comparison} \label{sec:Results_Database_Comparison}

\begin{figure}[H]
    \centering
    \subfigure[Profiles for $\etacrit$.]{%
        \label{fig:Direct_Profile_Comparison_2018_eta_crit}%
        \includegraphics[width=12.0cm]{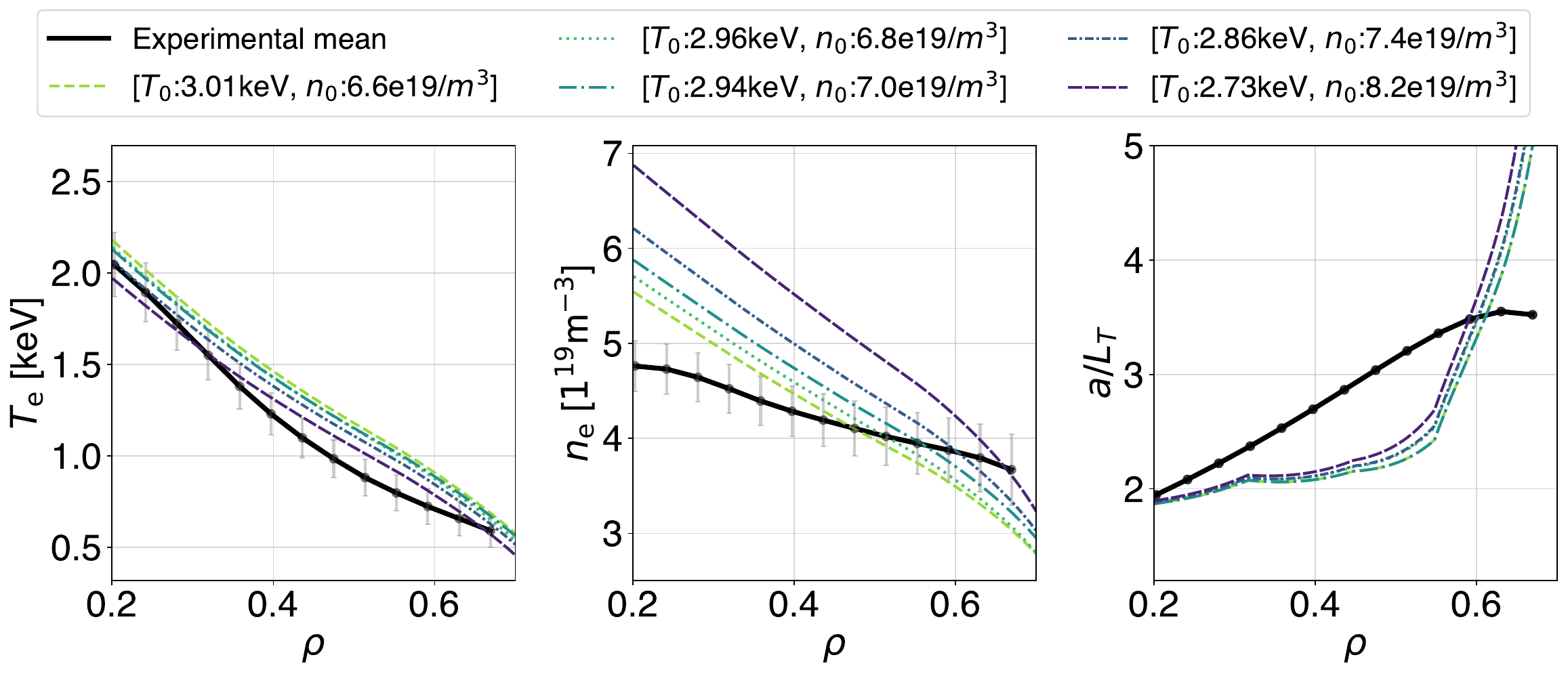}%
    }%
    \\
    \subfigure[Profiles for $\eta_{\text{exp}}$.]{%
        \label{fig:Direct_Profile_Comparison_2018_eta_set}%
        \includegraphics[width=12cm]{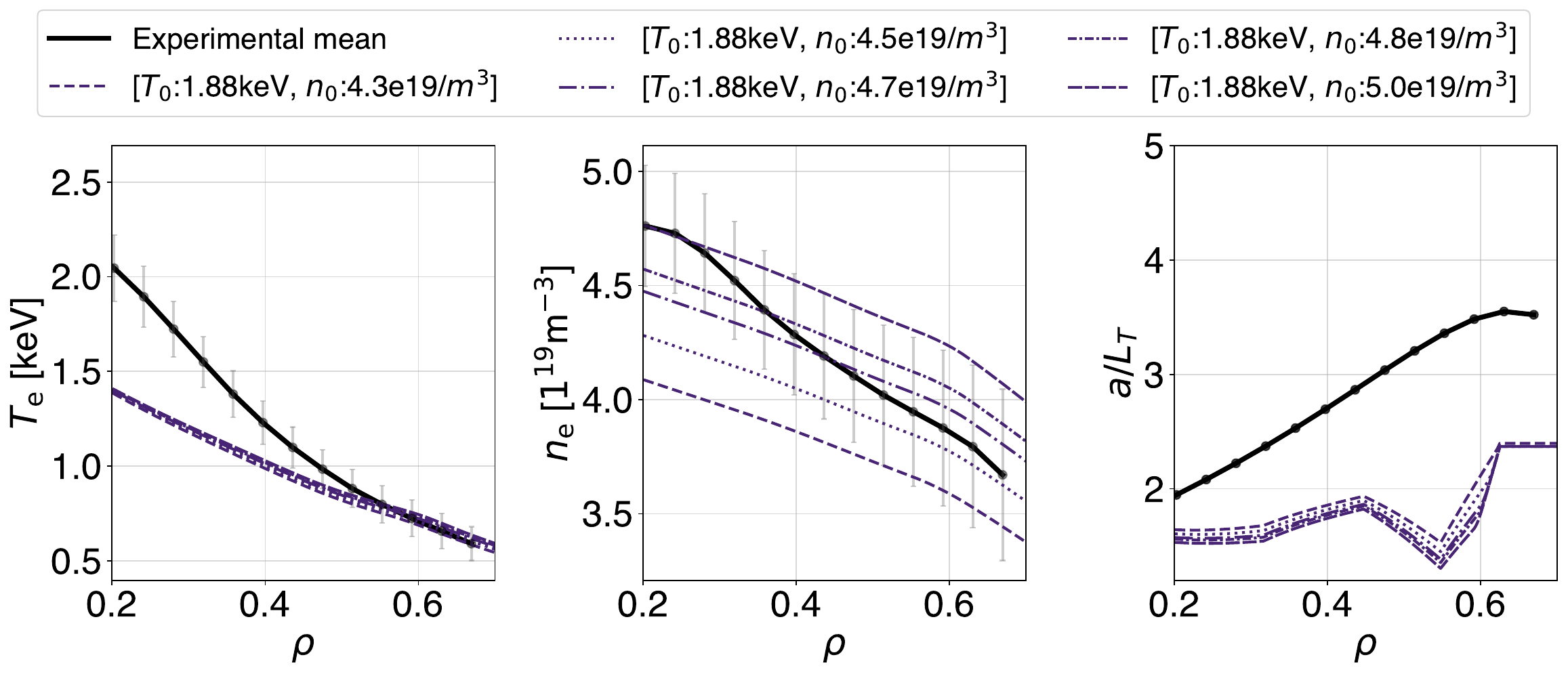}%
    }%
    \caption[Profile comparison of simulation results and averaged experimental profiles for \SI{2}{\mega \W}.]{\textbf{Profile comparison of simulation results and averaged experimental profiles for $\mathbf{\SI{2}{\mega \W}}$.} The experimental averaged profiles are indicated by the black solid lines with error bars, and are averaged over shots with $\SI{2(0.1)}{\mega\W}$. The simulation results for different initial densities and temperature values are shown. The first plot shows the temperature profiles, the second the density profiles and the last the radial $\alT$ profile. }
    \label{fig:Direct_Profile_Comparison_2018}
\end{figure}

By taking the heat flux data and scanning over different initial density and temperature values, we can attempt to reproduce the average temperature and densities observed in the experiment.
The comparison with the database data is shown in \cref{fig:Direct_Profile_Comparison_2018}. In this case, the total power is delivered via ECRH heating, and the experimental data are filtered for shots with $P = \SI{2.0(0.1)}{\mega \W}$ to allow a consistent comparison.
The figure presents a comparison using both the $\etacrit$ model in \cref{fig:Direct_Profile_Comparison_2018_eta_crit} and the experimental $\eta$ profile in \cref{fig:Direct_Profile_Comparison_2018_eta_set}. For the latter, the solver converges to values of the same magnitude; however, in both cases, the exact profile shape cannot be fully reproduced by the simulations. 

\begin{figure}[H]
    \centering
    \includegraphics[width = 12cm]{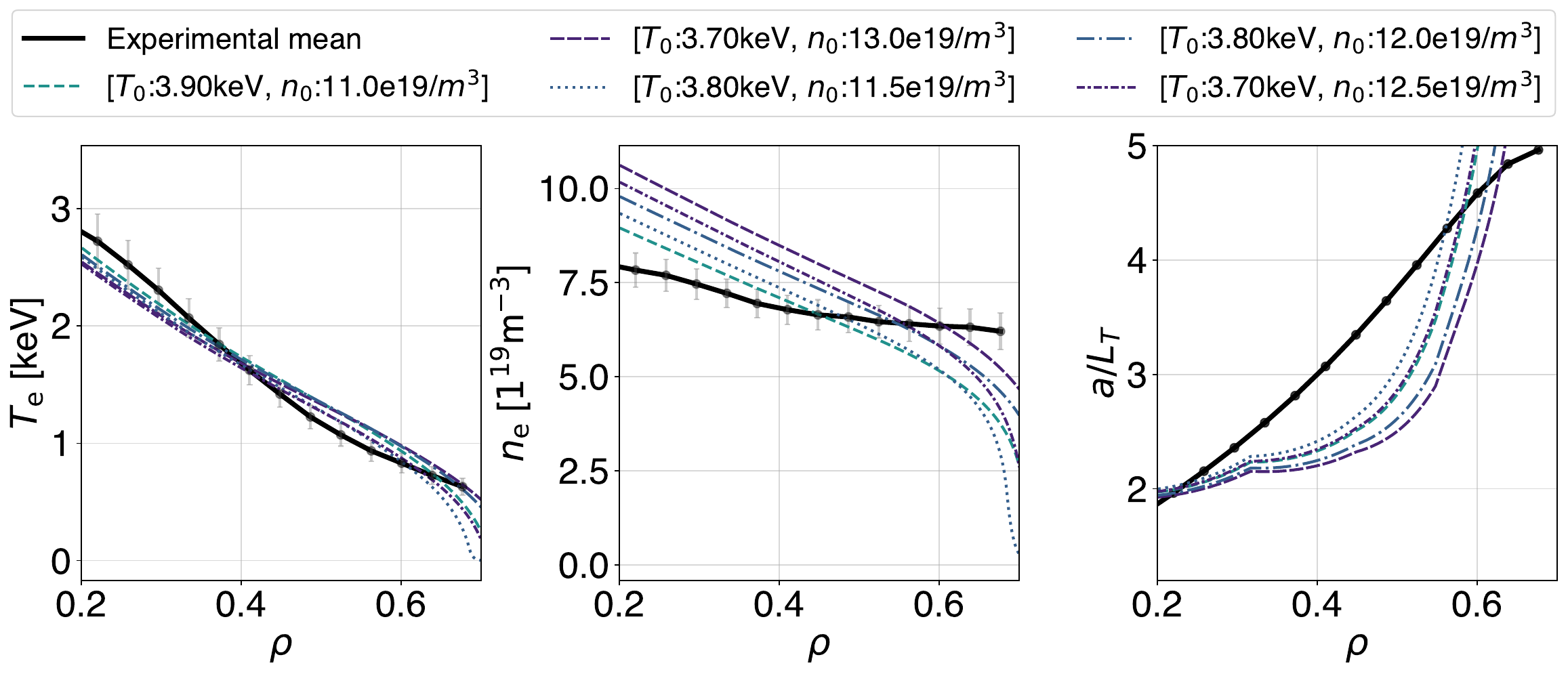}
    \caption[Profile comparison of simulation results and averaged experimental profiles for \SI{6}{\mega\W}.]{\textbf{Profile comparison of simulation results and averaged experimental profiles for \SI{6}{\mega\W}.} The experimental averaged profiles are indicated by the black solid lines with error bars, and are averaged over shots with $\SI{2(0.1)}{\mega\W}$.}
\label{fig:Direct_Profile_Comparison_2018_high_power}
\end{figure}

Further, another comparison is made for shots with $P = \SI{6.0(0.3)}{\mega \W}$ for the $\etacrit$ case. This comparison is shown in \cref{fig:Direct_Profile_Comparison_2018_high_power}.

\subsection{\texorpdfstring{$\eta -$  } CComparison}

As shown in the comparison of the profiles, the steepness of the temperature and density cannot be matched simultaneously. 
The ratio of the profile shape of both is quantified by the $\eta$-profile. 

\begin{figure}[H]
    \centering
    \includegraphics[width = 10cm]{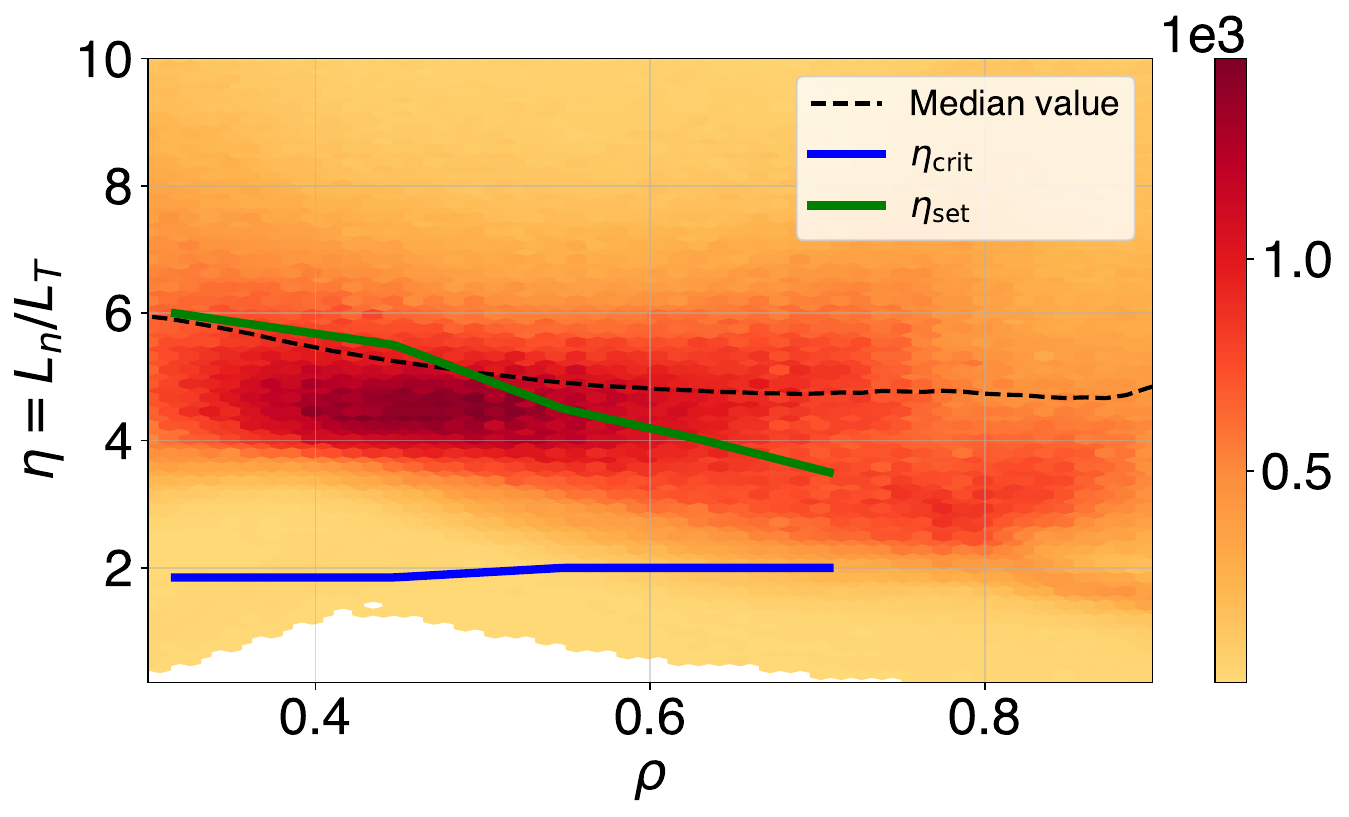}
    \caption[$\eta$-profile comparison.]{\textbf{$\eta$-profile comparison.} The experimental $\eta$ profiles are shown as a heat map, with the median curve indicated by the black dashed line. The obtained $\etacrit$ is marked in blue, and the input $\eta$ profile is also shown in blue.}
\label{fig:eta_profile_comparison}
\end{figure}

\section{Peaked Density Scenarios} \label{sec:Peaked_Density_profile}

\subsection{High Performance Shots} \label{sec:Results_Steep_Density_Profiles}

In recent years, several high-performance discharges have been observed. Unlike the shots considered for the database, these rely on either the addition of NBI or pellet injection, whilst maintaining a high power of ECRH. Given the complex dynamics of pellets, we shall restrict ourselves to cases where NBI has been combined with ECRH.

\begin{figure}[H]
    \centering
    \subfigure[Shot \#20230216.063.]{%
        \label{fig:Heating_Power20230216}%
        \includegraphics[width=6.2cm]{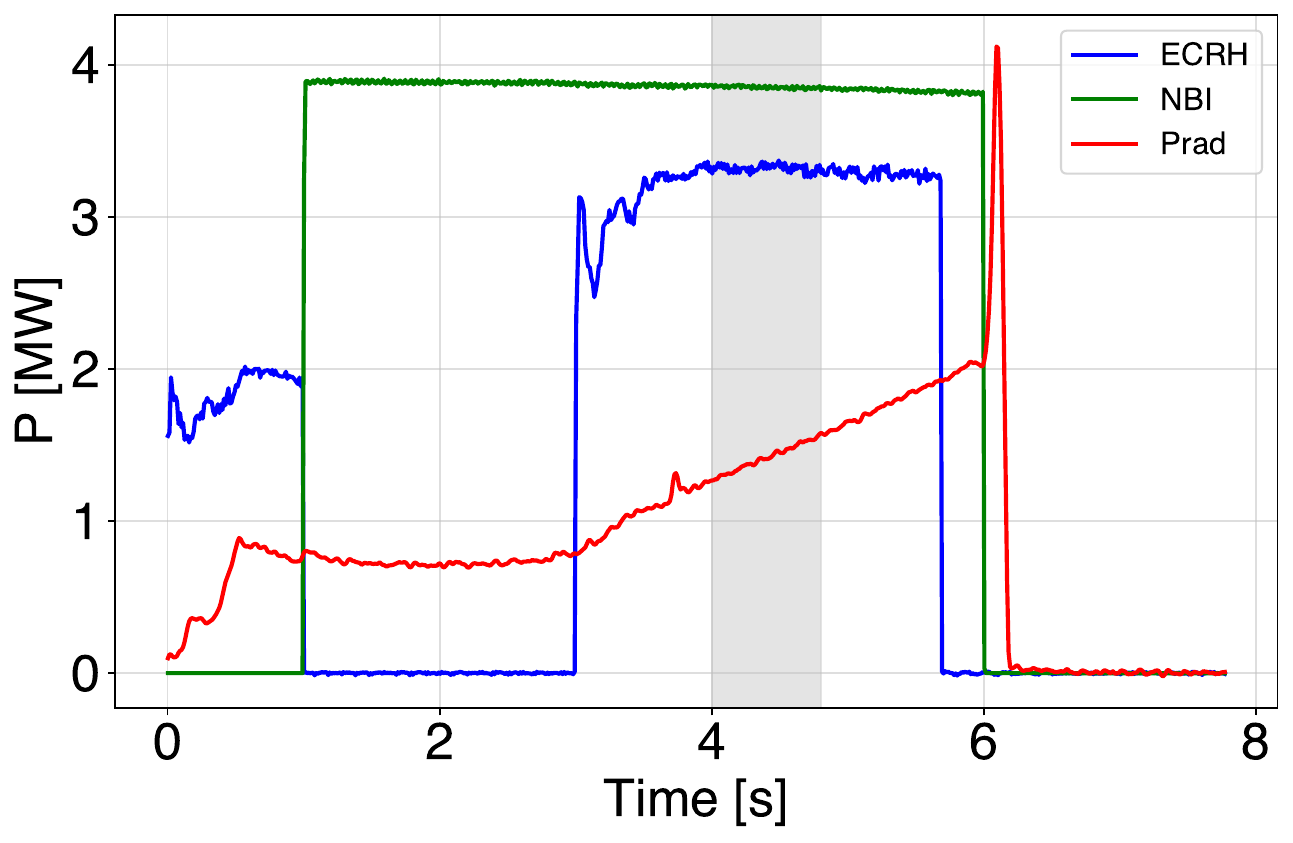}%
    }%
    \quad
    \subfigure[Shot \#20230323.034.]{%
        \label{fig:Heating_Power20230323}%
        \includegraphics[width=6.2cm]{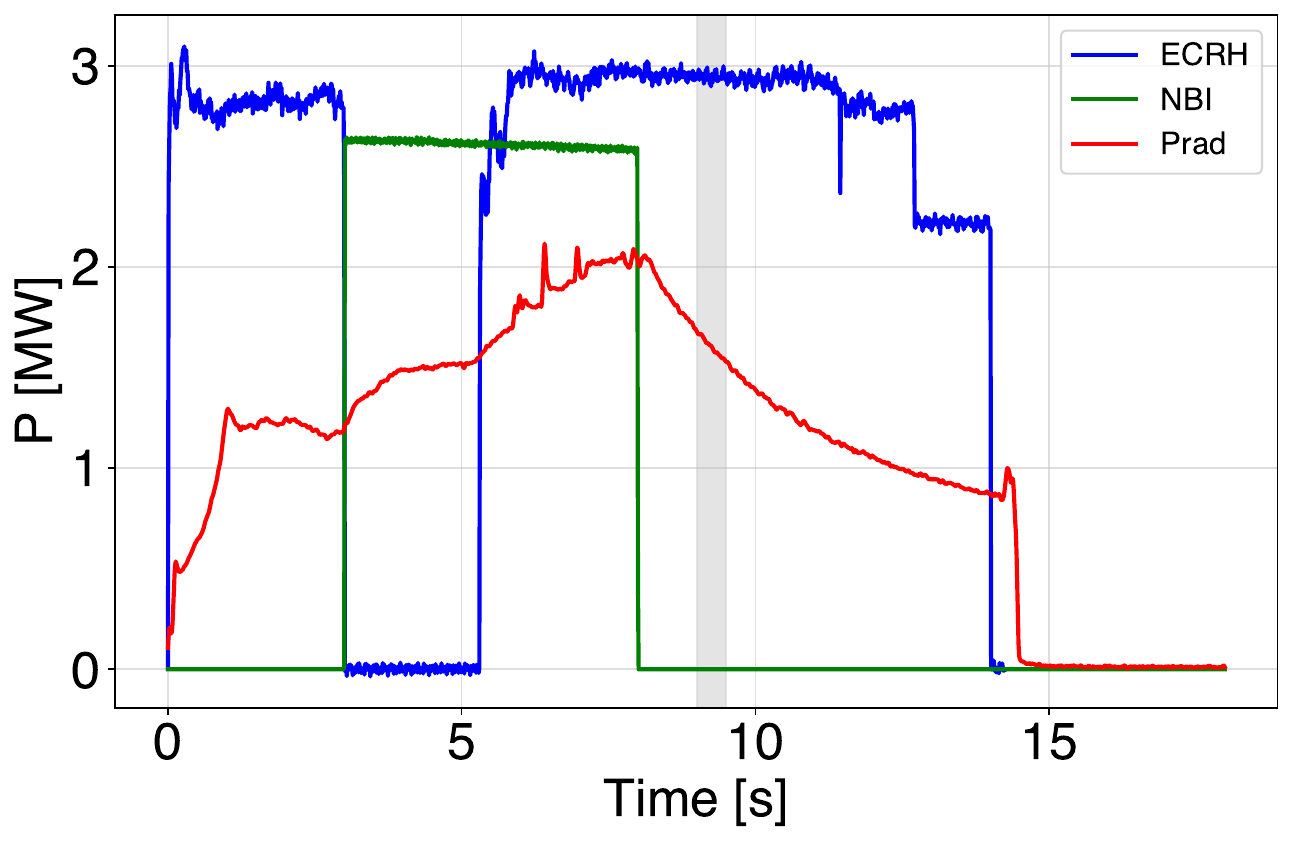}%
    }%
    \caption[Power profile of high performance shots.]{\textbf{Power profile of high-performance discharges.} The time traces of the different heating mechanisms are shown. The ECRH power is indicated in blue, the NBI in green, and the radiation losses in red. The grey array in the left subplot indicates a time range in which the full ECRH and NBI heating power is active. In the right subplot, a region just after the NBI has been switched off has been chosen.}
    \label{fig:Hig_Power_Power_Profile}
\end{figure}

The heating profiles for two of these discharges are illustrated in \cref{fig:Hig_Power_Power_Profile}.

\begin{figure}[H]
    \centering
    \includegraphics[width = 12cm]{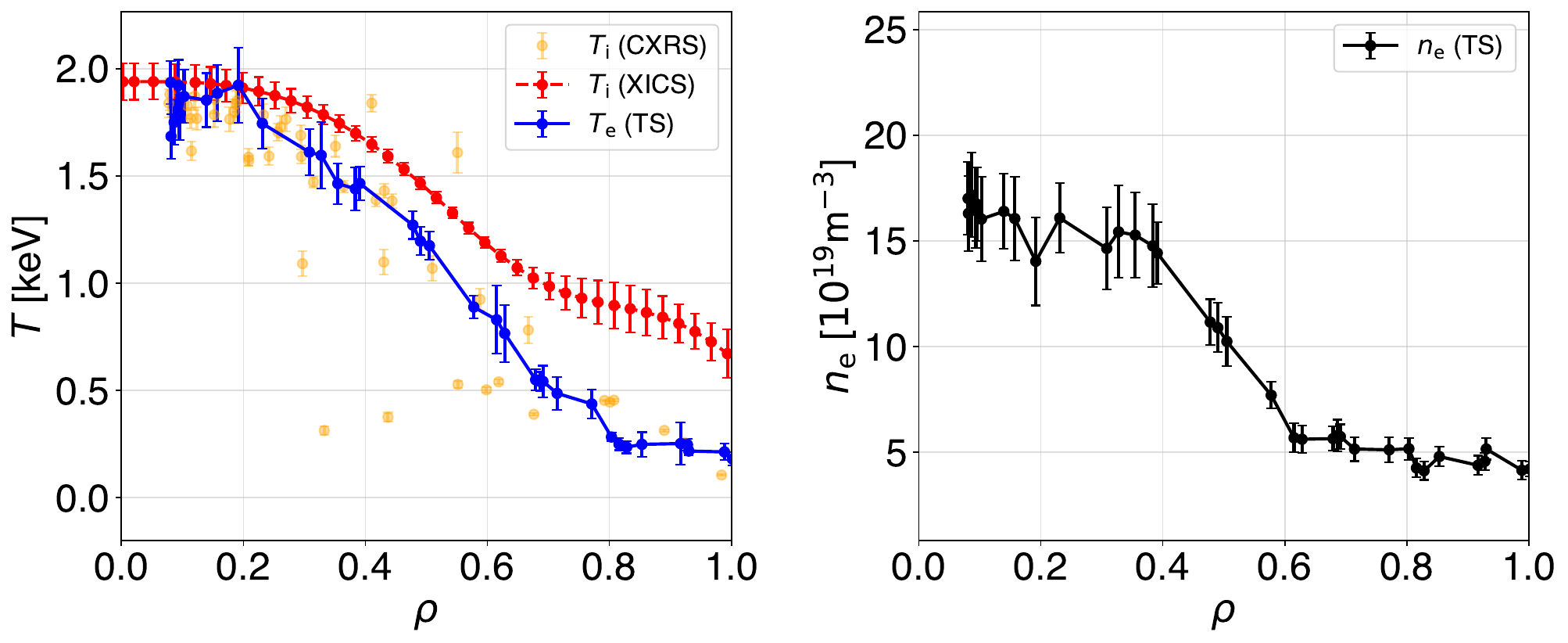}
    \caption[High-performance shot profiles.]{\textbf{High-performance shot profiles.} The left subplot shows the $\Te$ and $\Ti$ profiles. For the ion temperature, measurements from both XICS (red) and CXRS (orange) are shown. The right subplot shows the electron density. No reliable error bars are available for the displayed profiles. Shot No. \#20230216.063.}
\label{fig:High_Perfomance_Shot_Profiles}
\end{figure}

The result of this heating combination is a much higher density with a stronger gradient and an improved coupling of $\Ti$ and $\Te$. The averaged profiles for the region highlighted in \cref{fig:Heating_Power20230216} are shown in \cref{fig:High_Perfomance_Shot_Profiles}.

\begin{figure}[H]
    \centering
    \includegraphics[width = 12cm]{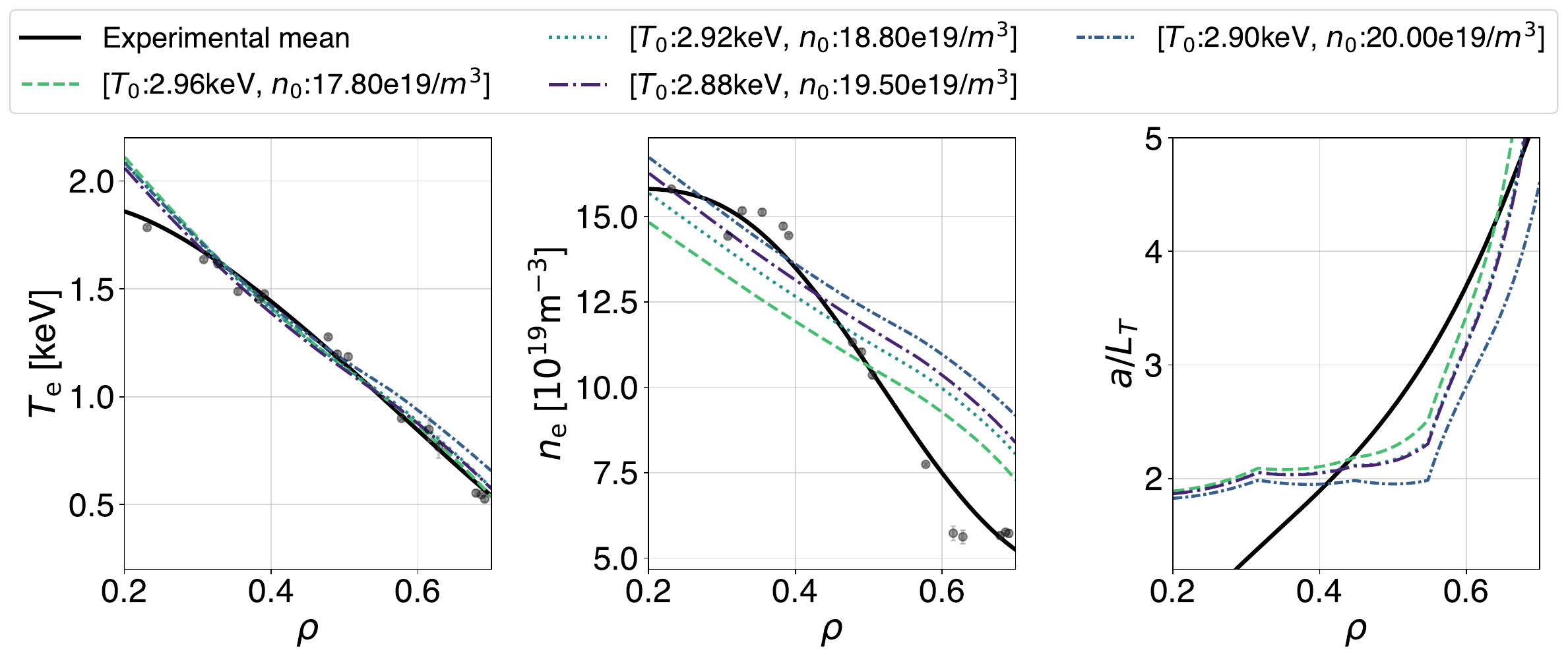}
    \caption[Profile comparison for high performance shots.]{\textbf{Profile comparison for a high-performance discharge.} The power is set to $P_{\mathrm{ECRH}}= \SI{3.34}{\mega \W}$, $P_{\mathrm{NBI}}= \SI{3.85}{\mega \W}$, $P_{\mathrm{RAD}}= \SI{1.32}{\mega \W}$. The experimental profile is indicated in black (Shot No. \#20230216.063).}
\label{fig:Direct_Profile_Comparison_High_Performance}
\end{figure}

Similarly to \cref{sec:Results_Database_Comparison}, we can compare this by also including the NBI and radiation in the heating profile, which is shown in \cref{fig:Direct_Profile_Comparison_High_Performance}.

Again, the magnitude and general trend can be reproduced; however, a consistent reproduction of the profiles is not expected as the $\eta$ values seen in this experimental discharge are lower than the $\etacrit$ value of the simulation.

The case shown in \cref{fig:Heating_Power20230323} would be better suited for comparison; however, the available measurement data are not reliable. We therefore consider a different type of discharge, in which a peaked density profile is achieved after the NBI has been switched off and only ECRH remains active at much lower power. This scenario is presented in \cref{sec:Low_Power_Discharges}.

\subsection{Low-Power Discharges} \label{sec:Low_Power_Discharges}

A peaking of the density profiles can be achieved with a lower power input as well. This was demonstrated in the most recent campaign OP2.2. Here, the density is increased by reducing the edge density.

\begin{figure}[H]
    \centering
    \includegraphics[width = 13cm]{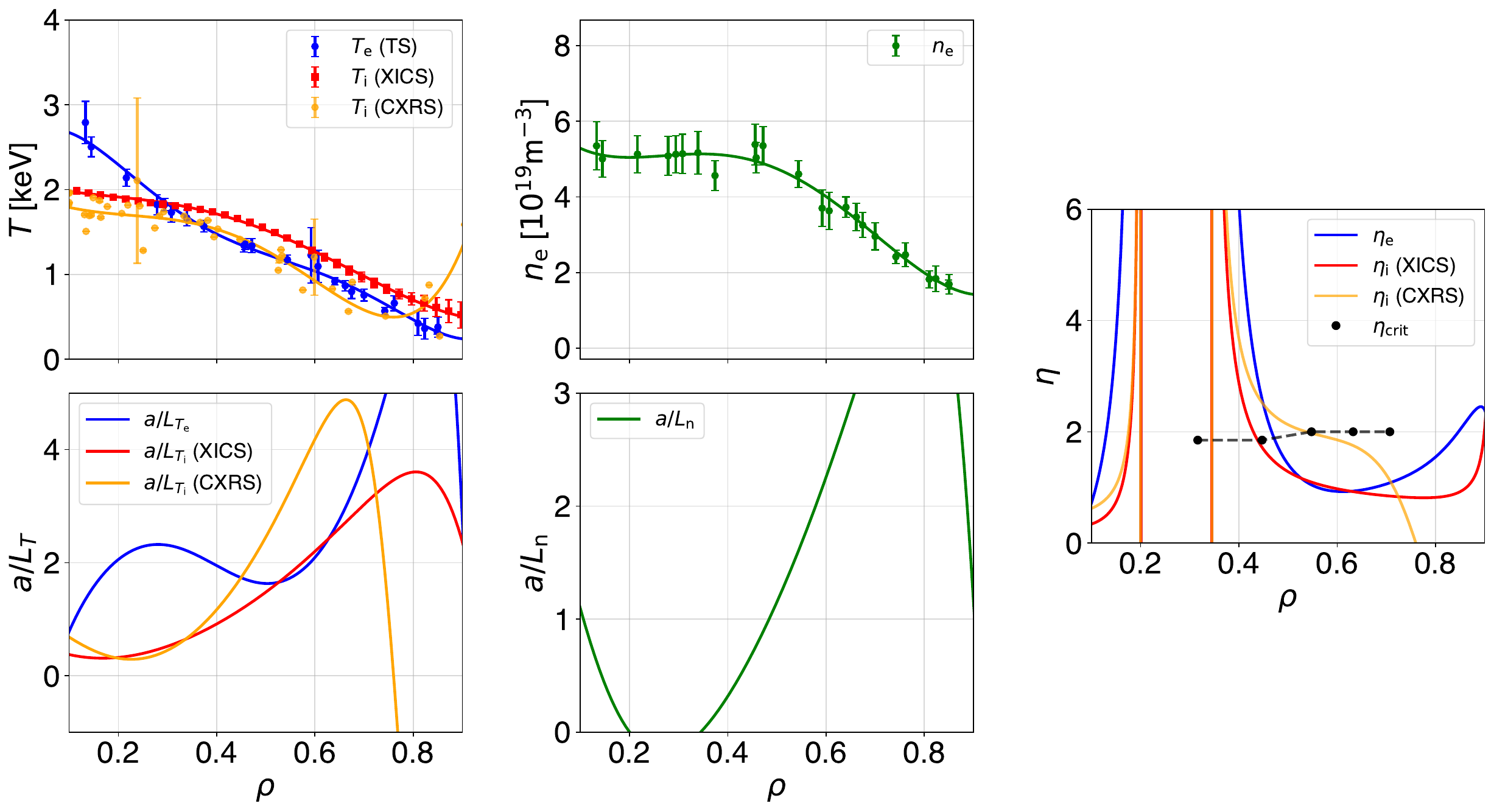}
    \caption[Profile overview of low power discharge.]{\textbf{Profile overview of a low-power discharge.} The upper subplots show the temperature and density profiles. Below are the corresponding gradient length scales, and the bottom plot shows the resulting $\eta$ profiles. Shot No. \#20250513.032.}
\label{fig:Low_Power_Shot_Overview}
\end{figure}

An overview of one of these discharges, including temperature, density, and gradient length scales, is shown in \cref{fig:Low_Power_Shot_Overview}.

\begin{figure}[H]
    \centering
    \includegraphics[width = 12cm]{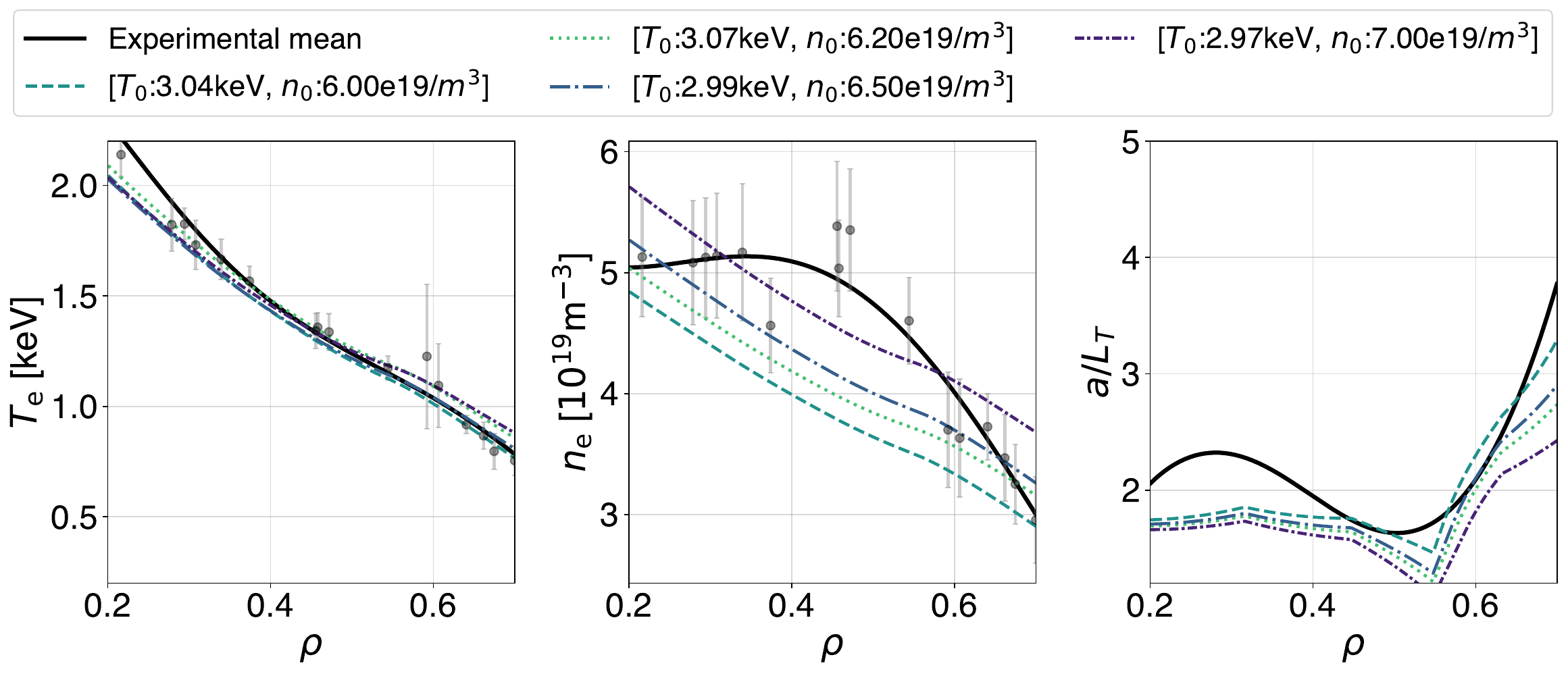}
    \caption[Profile comparison for low-power shots.]{\textbf{Profile comparison for low-power discharges.} Both the ion (XICS \& CXRS) and electron temperatures and their gradients are shown. The power is set to $P = \SI{1}{\mega \W}$..}
\label{fig:Direct_Profile_Comparison_Low_Performance}
\end{figure}

With these new profiles, another comparison can be performed, as shown in \cref{fig:Direct_Profile_Comparison_Low_Performance}. The advantage here is that only ECRH is active at the selected time point for comparison.

Once again, the temperature and density profiles cannot be exactly reproduced simultaneously.

\subsection{Improved Confinement Times}

As previously discussed, a steeper density profile should increase the confinement time and also increase $\tauE$ with respect to the confinement time scaling, thus $\fren$ is expected to increase.

To test this hypothesis, we calculate the confinement times of the high-performance and known edge density discharges. The results, with the parameters used, are summarised in \cref{tab:Improved_Confinment_Time}. For the high-power discharge (\cref{fig:Heating_Power20230216}), we also compare an earlier time before the ECRH has been turned on again, thus only having NBI as a power source.

\begin{table}[H]
    \centering
    \begin{tabular}{l | c c c}
        Parameter & 20230216.063 & 20230216.063 & 20250513.032 \\
         & [2.2-2.4 \si{s}] & [4.2-4.8 \si{s}] & [6.4-6.8 \si{s}] \\
        \hline
        $W\mathrm{dia}$ [\si{kJ}] & 785 & 1070 & 453 \\
        $P$ [\si{MW}] & 3.05 & 5.64 & 0.73 \\
        $\overline{n}$ [\si{10^{19} m^{-3}}] & 8.27 & 10.30 & 3.94 \\
        $B$ [\si{T}] & 2.470 & 2.430 & 2.480 \\
        $\iotaslash{2/3}$  & 1.010 & 1.010 & 0.930 \\
        $a$ [\si{m}] & 0.510 & 0.510 & 0.520 \\
        \hline 
        $\tauE$ [\si{s}] & 0.2574 & 0.1897 & 0.6205 \\
        $\tauiss$ [\si{s}] & 0.2923 & 0.2226 & 0.4750 \\
        $\taugb$ [\si{s}] & 0.2985 & 0.2318 & 0.4495 \\
        $f{\text{ren,ISS04}}$ & 0.881 & 0.852 & 1.306 \\
        $f{\text{ren,GB}}$ & 0.862 & 0.818 & 1.380 \\
    \end{tabular}
    \caption[Confinement time calculation for peaked density cases.]{\textbf{Confinement time calculation for peaked density cases.} The used parameters for the different shots and time intervals are presented, along with the calculated experimental and theoretical confinement times, and their respective renormalisation factors. The major radius is a device parameter and is constant in all shots with $R_0 = \SI{5.5}{\m}$.}
    \label{tab:Improved_Confinment_Time}
\end{table}

The confinement time calculations indicate that the diamagnetic energy is very high for the high-performance discharges; however, due to the large input power, the confinement time does not increase significantly, both for pure NBI and ECRH cases. In contrast, a substantial increase, where $\tau_E$ even exceeds the predicted confinement time, is observed for the lower-power discharge, for which $\fren > 1$.
\chapter{Discussion} \label{sec:Discussion}
The results presented in this thesis provide a compelling argument that certain universal features emerge in the profile shape within the outer region of the plasma. However, no definitive conclusion can yet be drawn. In this chapter, we first discuss the general data analysis and then the validation of the temperature and confinement time scalings, before turning to an analysis of the individual models, particularly the comparison between physics-informed fits and purely mathematical formulations. Subsequently, we examine potential factors contributing to the discrepancies between the simulated profiles and the averaged experimental profiles. This includes a closer consideration of the peaked density scenarios.

\section{Data Analysis}
In this section, we wish to highlight some of the main uncertainties related to the data analysis and their impact on the extraction of the respective profile slopes. An attempt was made to be as close to a steady-state condition as possible by filtering out time points which clearly violate this criterion. This data cleaning procedure was implemented to reduce the influence of transient effects on the subsequent profile fits. 

Furthermore, instead of fitting each discharge individually and then averaging the resulting slopes, we also tried averaging over consecutive shots and fitting, or averaging over all normalised profiles before applying the fitting procedure. 
These procedures, however, did not produce a clear advantage; in all cases, the resulting slopes fell within the range of uncertainty. 

The uncertainty in the density values is one of the biggest challenges. Even though it does not directly influence the temperature fits, it has a major impact on the $\eta$ profiles, as here also the $\aln$ term contributes. Profile correction attempts using an ML framework have been made, but with limited success, considering the effects on the whole database.

\section{Validation of Temperature and Confinement Time Scaling}
By comparing the energy loss time $\tauE$ with $\tauiss$, we obtain a renormalisation factor of $\fren=\num{0.67(0.07)}$, while for $\taugb$ the corresponding value is $\fren=\num{0.69(0.09)}$. These results are slightly lower than previous findings, such as those reported in \cite{fuchert_increasing_2020}, where a value of \num{0.75} was determined for the standard configuration. For other configurations, such as high-iota scenarios, somewhat different values are expected. Nevertheless, because $\iotaslash$ is close to unity, the dependence remains relatively weak.

A limiting factor is thought to be the clamping of ion temperatures at $\Ti \leq \SI{1.7}{\kilo \eV}$ \cite{beurskens_ion_2021}. This constraint has been partially alleviated in more recent experiments featuring peaked density profiles \cite{langenberg_achieving_2024}. The removal of this clamping is associated with the stabilisation of ITG turbulence by density profile shaping, which enables higher $\Ti$ and improved confinement \cite{grulke_overview_2024}.
 
The density values constitute one of the main uncertainties discussed throughout this thesis. In principle, one could utilise interferometric measurements and project them onto a radial axis to obtain the line-averaged density $\overline{n}$. The general linear trend, with a correlation coefficient of $r = 0.86$, still indicates good agreement between the two confinement time scalings and the measured $\tauE$.

The temperature scaling with input power is clearly evident. The scaling relation shown in \cref{eq:T_Gyro_Bohm_Power_Scaling} is observable and exhibits a correlation. Nonetheless, the data distribution suggests the presence of an additional offset, indicating that this scaling may not sufficiently capture the observed behaviour. It should also be noted that neither the radial dependence of the magnetic field nor that of the power deposition has been explicitly accounted for here. If both are treated as constants relative to the reference value used, the qualitative behaviour remains unchanged.

\section{Physics Informed Fitting Methods}
As discussed previously, several functional forms can be fitted to both the temperature and density profiles, each based on different assumptions regarding the underlying physics. In all cases, continuity and at least (local) second-order differentiability are assumed to enable the extraction of smooth gradient length scales.

The polynomial fits, by contrast, are mathematically agnostic, with no embedded physical assumptions. It is possible to enforce symmetry by setting the linear term to zero or by considering only even-order terms, as discussed in \cite{wappl_web_2024}. While diffusive transport theory suggests that the temperature gradient should vanish at the magnetic axis, thus justifying a zero linear coefficient, higher-order effects could, in principle, yield a finite gradient. A more detailed discussion of heat transport phenomena near the magnetic axis lies beyond the scope of this thesis, as we focus on the outer part of the plasma.
To minimise assumptions in the purely mathematical fit, all polynomial terms are retained.
A modification of the fitting behaviour is introduced by modifying the loss function to \cref{eq:total_loss_function} from a simple least squares approach. Here, we enforce that the changes in temperature are slow, meaning there are no steep gradients or oscillating terms expected.

The principal advantage of the piecewise model applied to the temperature profile is its capacity to capture the apparent linear transition beyond a certain radius, which is particularly visible in the case of flat density profiles. The behaviour for a different number of constraints has been thoroughly analysed. The decision for the final used model was to keep all three boundary conditions. The increase in total loss was expected. Furthermore, as the position of $\rhocrit$ is set arbitrarily, it is hard to justify why exactly at this position the profiles should behave in a non-smooth way and not in the rest of the plasma in the absence of any instabilities or abrupt turbulence or transport transitions. Thus, following Ockham’s razor, we decided to retain smooth profiles with the minimum number of assumptions.   

As electron and ion temperatures begin to decouple, the model effectively describes the non-linear rise in $\Te$ while also preserving the linear behaviour at the edge, as resulting in the simplified model given by \cref{eq:linear_alT_prediction}. However, introducing an additional offset term, such as in \cref{eq:higher_order_alT_prediction}, yields improved agreement with the experimental gradient length scales.

A similar piecewise model has been implemented by \cite{sauter_non-stiffness_2014}, where the edge transport has been studied for L-mode tokamak plasmas as found in TCV. The linear edge model beyond $\rhocrit$ is the same, although the inner region is to be treated as exponential. In the fits presented here, the transition region was varied between $\rhocrit \in [0.7;0.9]$, with a focus on 0.8, which is significantly further outside than in the fit of our model for the analysed TS data. Furthermore, the stiffness of profiles and the manifestation of pedestal-like behaviour were not really apparent in this context.

The motivation for fitting the CG model goes beyond the extraction of heat fluxes. It aims to provide a more physically consistent representation of the data, as the underlying equations describe a (simplified) steady-state plasma scenario. Nevertheless, when comparing the different models, the CG approach shows no significant performance advantage. The additional uncertainties inherent in the density profiles, combined with the increased computational complexity of solving an ODE during each optimisation step, result in no clear improvement in the final gradient length scales.

\section{Transport Mechanisms Beyond ITG}
The simulations presented in this thesis account only for turbulent transport. In experiments, particularly at the scale of W7-X, turbulence is the dominant contribution, though not the sole one, and its relative importance is scenario-dependent. Our analysis focuses on the outer plasma region, which is generally considered less sensitive to variations in heating schemes and device configurations. In \cite{fernando_validation_2025}, where heat fluxes are evaluated by coupling a gyrokinetic solver with a neoclassical solver, the neoclassical contribution to the total heat flux is found to be small; however, the associated particle fluxes may still play a significant role in the overall transport.

For the peaked density profiles, the resulting $\eta$ profiles are below the $\etacrit$ of the GX simulation. This indicates that there is no missing neoclassical heat transport; instead, there is a particle transport contribution. An outward particle flux would lower the $\eta$. In low-power discharges, this could be due to neutral particles that can penetrate far into the plasma.

In general, validating the predictions against experimental data using this framework is very challenging. Recently, see \cite{hofler_milestone_2025}, ASDEX Upgrade data have been successfully predicted, but the validation and prediction of temperature profiles remain an important task in understanding the underlying physics and how these mechanisms may evolve in future reactors.
\chapter{Conclusion and Future Work} \label{sec:Conclusion}

This thesis investigates the profile shape within a comprehensive database of Thomson scattering electron temperature and density profiles. From these radial profiles, the gradient length scales $\aln$ and $\alT$ are extracted through several fitting procedures and subsequently used for comparison with gyrokinetic simulations (GX).

The GX simulations used in this thesis were performed by collaborators and follow the profile shape extracted from the experiment. The resulting radial profiles underestimate the $\eta$ values observed in the experiment by a factor of two to three. The critical gradient $\alTcrit$ retrieved from the interpolation of these simulation results agrees within a standard deviation with the corresponding experimental estimates. The resulting heat fluxes follow the trend predicted by the CG model but remain lower in absolute terms. The opposite occurs when the experimental $\eta$ profile is used directly as an input for the simulations, in which case the predicted flux overestimates the measured flux, implying a non-zero net flux of turbulence.

Additionally, this work investigates two types of peaked density profiles. The first occurs in high-performance discharges, where NBI is combined with ECRH; in these cases, a net transport is introduced, which makes a rigorous, direct comparison difficult. The second involves low-performance discharges, where no NBI is present, allowing for a more reliable comparison.

With regard to the project goals of this thesis, the first step, developing a framework in Python for extracting and analysing the existing gradient length scales, was successfully completed. These length scales subsequently serve as a basis for the GX simulations, where the $\alT$ profiles were set to match the experimental data. The second step was also successfully implemented, and a weak but visible correlation with the Gyro-Bohm scaling was found. The third step constitutes a partial success; a profile solver was implemented in Python, which recalculates the parameter profiles based on the heat fluxes. 
However, the resulting fluxes and the corresponding temperature and density profiles do not fully match the average profiles measured in the experiment.

As for the main hypothesis, it could be demonstrated that there are universal scalings for the logarithmic temperature gradient, although with a larger scatter than initially expected. The results follow the expected power scaling, and a first validation of this scaling against the experimental data was performed. Further predictive studies remain for future work.

Additionally, validating and interpreting the profiles is a complex task, and there are numerous directions in which this work can be continued.

The most obvious subsequent step is to compare the particle fluxes from the GX simulations, where the experimental $\eta$ profile is used as an input, to a neoclassical estimate, for instance with Neotransp \cite{h_smith_neotransp_2022}. Coupling a neoclassical code to the gyrokinetic solver would enable a more accurate validation of the profiles in W7-X. However, this would significantly increase the complexity and computational cost of the simulations. Alternatively, a simplified particle source model could be implemented to account for the effects of NBI instead of being restricted to pure ECRH.

Furthermore, when running the simulations, it would be desirable to enable ratios of ion to electron temperature different from unity in order to facilitate a comparison between electron and ion profiles. Currently, the simulations are performed for only a single magnetic configuration; extending this to a different configuration, e.g. high-mirror configuration, could provide additional insight into the robustness of the results.

When reconstructing the profiles, it would be useful to extend the framework described in \cref{sec:coupled_ODE_system} to allow more flexible placement of boundary conditions, particularly for comparisons in the outer plasma region.

As shown, there are discharge scenarios with distinctly different transport behaviour, which may indicate regimes that are primarily turbulence-dominated and show closer agreement with the simulation results. In general, further extension of this work could involve an analysis of discharges covering a wider range of performance parameters, particularly those with significant variation in $\eta$.

Such an analysis would allow a more critical assessment of the validity of "universal" behaviour in the temperature gradient length scales at the plasma edge. Deviations from universality may become more pronounced in higher-performance regimes and could be associated with different underlying transport mechanisms. Expanding the dataset in this way would also provide a more robust basis for testing the proposed scalings across varied operational conditions.

\label{pg:lastPageofPreface}

\cleardoublepage

\listoffigures

\cleardoublepage

\listoftables

\printbibliography

@article{winter_comparing_2016,
	title = {Comparing the {Pearson} and {Spearman} {Correlation} {Coefficients} {Across} {Distributions} and {Sample} {Sizes}: {A} {Tutorial} {Using} {Simulations} and {Empirical} {Data}},
	volume = {21},
	issn = {1939-1463, 1082-989X},
	shorttitle = {Comparing the {Pearson} and {Spearman} {Correlation} {Coefficients} {Across} {Distributions} and {Sample} {Sizes}},
	url = {http://arxiv.org/abs/2408.15979},
	doi = {10.1037/met0000079},
	number = {3},
	urldate = {2025-09-06},
	journal = {Psychological Methods},
	author = {Winter, J. C. F. de and Gosling, S. D. and Potter, J.},
	month = sep,
	year = {2016},
	note = {arXiv:2408.15979 [stat]},
	pages = {273--290},
}

@article{grulke_overview_2024,
	title = {Overview of the first {Wendelstein} 7-{X} long pulse campaign with fully water-cooled plasma facing components},
	volume = {64},
	issn = {0029-5515},
	url = {https://dx.doi.org/10.1088/1741-4326/ad2f4d},
	doi = {10.1088/1741-4326/ad2f4d},
	language = {en},
	number = {11},
	urldate = {2025-08-22},
	journal = {Nuclear Fusion},
	author = {Grulke, O. and Albert, C. and Alcuson Belloso, J.A. and Aleynikov, P. and Aleynikova, K. and Alonso, A. and Anda, G. and Andreeva, T. and Arvanitou, M. and Ascasibar, E. and Aymerich, E. and Avramidis, K. and Bähner, J.-P. and Baek, S.-G. and Balden, M. and Baldzuhn, J. and Ballinger, S. and Banduch, M. and Bannmann, S. and Bañón Navarro, A. and Baylor, L. and Beidler, C.D. and Beurskens, M. and Biedermann, C. and Birkenmeier, G. and Bluhm, T. and Boeckenhoff, D. and Boeyaert, D. and Bold, D. and Borchardt, M. and Borodin, D. and Bosch, H.-S. and Bouvain, H. and Bozhenkov, S. and Bräuer, T. and Braune, H. and Brandt, C. and Brezinsek, S. and Brunner, K.J. and Büschel, C. and Bussiahn, R. and Buzás, A. and Buttenschoen, B. and Bykov, V. and Calvo, I. and Cappa, A. and Carovani, F. and Carralero, D. and Carls, A. and Carvalho, B. and Castaño-Bardawil, D. and Chaudhary, N. and Chelis, I. and Chen, S. and Cipciar, D. and Coenen, J.W. and Conway, G. and Cornelissen, M. and Corre, Y. and Costello, P. and Crombe, K. and Cseh, G. and Csillag, B. and Cu Castillo, H.I. and Czymek, G. and Damm, H. and Davies, R.J. and Day, C. and Degenkolbe, S. and De Wolf, R. and Dekeyser, W. and Demby, A. and Despontin, P. and Dhard, C.P. and Dinklage, A. and d’Isa, F.A. and Dittmar, T. and Dreval, M. and Drevlak, M. and Drews, P. and Droste, J. and Dunai, D. and Dyhring, C. and van Eeten, P. and Edlund, E. and Endler, M. and Ennis, D.A. and Escoto, F.J. and Espinosa, M.S. and Estrada, T. and Fehling, D. and Feuerstein, L. and Fellinger, J. and Feng, Y. and Fernando, D.L.C. and Fischer, S. and Flom, E.R. and Ford, O. and Fornal, T. and Frank, J. and Frerichs, H. and Fuchert, G. and Gantenbein, G. and Gao, Y. and Garcia, K. and García-Cortés, I. and García-Regaña, J.M. and Geiger, B. and Geiger, J. and Geissler, P. and Gerard, M. and Godino-Sedano, G. and Gonda, T. and González, A. and Goriaev, A. and Gradic, D. and Grahl, M. and Greuner, H. and Grigore, E. and Gruca, M. and Guerrero Arnaiz, J.F. and Haak, V. and van Ham, L. and Hammond, K. and Hamstra, B. and Han, X. and Hansen, S.K. and Harris, J. and Hartmann, D. and Hathiramani, D. and Hegedus, S. and Heinrich, S. and Helander, P. and Henke, F. and Henneberg, S. and Henschke, L. and Hirsch, M. and Hoefel, U. and Hoefler, K. and Hoermann, S. and Hollfeld, K.-P. and Holtz, A. and Höschen, D. and Houry, M. and Huang, J. and Huang, J. and Hubeny, M. and Hunger, K. and Hwangbo, D. and Ida, K. and Igitkhanov, Y. and Illy, S. and Ioannidis, Z. and Jablczynska, M. and Jablonski, S. and Jabłoński, B. and Jagielski, B. and Jakubowski, M. and Jelonnek, J. and Jenko, F. and Jin, J. and Johansson, A. and Jouniaux, G. and Kajita, S. and Kallmeyer, J.-P. and Kamionka, U. and Kasparek, W. and Kawan, C. and Kazakov, Ye. O. and Kenmochi, N. and Kernbichler, W. and Kharwandikar, A.K. and Khokhlov, M. and Killer, C. and Kirschner, A. and Kleiber, R. and Klepper, C.C. and Klinger, T. and Knauer, J. and Knieps, A. and Kobayashi, M. and Kocsis, G. and Kolesnichenko, Y. and Könies, A. and Kontula, J. and Kornejew, P. and Korteweg, S.A. and Koschinsky, J. and Koster, J. and Kovtun, Y. and Krämer-Flecken, A. and Krause, M. and Kremeyer, T. and Krier, L. and Kriete, D.M. and Krychowiak, M. and Ksia¸zek, I. and Kubkowska, M. and Kuczyński, M.D. and Kulla, D. and Kumar, A. and Kurki-Suonio, T. and Kuzmych, I. and Kwak, S. and Lancelotti, V. and Langenberg, A. and Laqua, H. and Laqua, H.P. and Larsen, M.R. and Lazerson, S. and Lechte, C. and Lee, B. and LeViness, A. and Lewerentz, M. and Liang, Y. and Liao, L. and Litnovsky, A. and Liu, J. and Loizu, J. and Lopez-Cansino, R. and Lopez Rodriguez, L.D. and Lorenz, A. and Lunsford, R. and Luo, Y. and Lutsenko, V. and Maaziz, N. and Machielsen, M. and Mackenbach, R. and Makowski, D. and Maragkoudakis, E. and Marchuk, O. and Markl, M. and Marsen, S. and Martínez, J. and Marushchenko, N. and Masuzaki, S. and Maurer, D.A. and Mayer, M. and McCarthy, K.J. and McNeely, P. and Medina Roque, D. and Meineke, J. and Meitner, S. and vaz Mendes, S. and Menzel-Barbara, A. and van Milligen, B. and Mishchenko, A. and Moiseenko, V. and Möller, A. and Möller, S. and Moseev, D. and Motojima, G. and Mulas, S. and Mulholland, P. and Nagel, M. and Nagy, D. and Narbutt, Y. and Naujoks, D. and Nelde, P. and Neu, R. and Neubauer, O. and Neuner, U. and Nicolai, D. and Nielsen, S. and Nührenberg, C. and Ochoukov, R. and Offermanns, G. and Ongena, J. and Oosterbeek, J.W. and Otte, M. and Pablant, N. and Panadero Alvarez, N. and Pandey, A. and Partesotti, G. and Pasch, E.A. and Pavlichenko, R. and Pawelec, E. and Pedersen, T.S. and Perseo, V. and Peterson, B. and Pisano, F. and Plaum, B. and Plunk, G. and Podavini, L. and Polei, N.S. and Poloskei, P. and Ponomarenko, S. and Pons-Villalonga, P. and Porkolab, M. and Proll, J. and Pueschel, M.J. and Puig Sitjes, A. and Ragona, R. and Rahbarnia, K. and Rasiński, M. and Rasmussen, J. and Refy, D. and Reimold, F. and Richou, M. and Riemann, J.S. and Riße, K. and de la Riva Villén, J. and Roberg-Clark, G. and Rodriguez, E. and Rohde, V. and Romazanov, J. and Romba, T. and Rondeshagen, D. and Rud, M. and Ruess, T. and Rummel, T. and Runov, A. and Ruset, C. and Rust, N. and Ryc, L. and Rzesnicki, T. and Salewski, M. and Sánchez, E. and Sanchis Sanchez, L. and Satheeswaran, G. and Schacht, J. and Scharff, E. and Schilling, J. and Schlisio, G. and Schmid, K. and Schmitt, J.C. and Schmitz, O. and Schneider, M. and Van Schoor, M. and Schröder, T. and Schroeder, R. and Schweer, B. and Sereda, S. and Shanahan, B. and Sias, G. and Simko, S. and Singh, L. and Siusko, Y. and Slaby, C. and Śle¸czka, M. and Smith, B.S. and Smith, D.R. and Smith, H. and Spolaore, M. and Spring, A. and Stange, T. and von Stechow, A. and Stepanov, I. and Stern, M. and Stroth, U. and Suzuki, Y. and Swee, C. and Syrocki, L. and Szabolics, T. and Szepesi, T. and Takacs, R. and Takahashi, H. and Tamura, N. and Tantos, C. and Terry, J. and Thiede, S. and Thienpondt, H. and Thomsen, H. and Thumm, M. and Thun, T. and Togo, S. and Tork, T. and Trimino Mora, H. and Tsikouras, A. and Turkin, Y. and Vano, L. and Varoutis, S. and Vecsei, M. and Velasco, J.L. and Verstraeten, M. and Vervier, M. and Viezzer, E. and Wagner, J. and Wang, E. and Wang, F. and Wappl, M. and Warmer, F. and Wegner, T. and Wei, Y. and Weir, G. and Wendler, N. and Wenzel, U. and White, A. and Wilms, F. and Windisch, T. and Winter, A. and Winters, V. and Wolf, R. and Wurden, G. and Xanthopoulos, P. and Xiang, H.M. and Xu, S. and Yamada, H. and Yang, J. and Yi, R. and Yokoyama, M. and Zamorski, B. and Zanini, M. and Zarnstorff, M. and Zhang, D. and Zhou, S. and Zhu, J. and Zimmermann, J. and Zocco, A. and Zoletnik, S.},
	month = aug,
	year = {2024},
	note = {Publisher: IOP Publishing},
	pages = {112002},
}

@article{breznsek_plasmasurface_2021,
	title = {Plasma–surface interaction in the stellarator {W7}-{X}: conclusions drawn from operation with graphite plasma-facing components},
	volume = {62},
	issn = {0029-5515},
	shorttitle = {Plasma–surface interaction in the stellarator {W7}-{X}},
	url = {https://dx.doi.org/10.1088/1741-4326/ac3508},
	doi = {10.1088/1741-4326/ac3508},
	language = {en},
	number = {1},
	urldate = {2025-08-22},
	journal = {Nuclear Fusion},
	author = {BrezƖnsek, S. and Dhard, C.P. and Jakubowski, M. and König, R. and Masuzaki, S. and Mayer, M. and Naujoks, D. and Romazanov, J. and Schmid, K. and Schmitz, O. and Zhao, D. and Balden, M. and Brakel, R. and Butterschoen, B. and Dittmar, T. and Drews, P. and Effenberg, F. and Elgeti, S. and Ford, O. and Fortuna-Zalesna, E. and Fuchert, G. and Gao, Y. and Goriaev, A. and Hakola, A. and Kremeyer, T. and Krychowiak, M. and Liang, Y. and Linsmeier, Ch. and Lunsford, R. and Motojima, G. and Neu, R. and Neubauer, O. and Oelmann, J. and Petersson, P. and Rasinski, M. and Rubel, M. and Sereda, S. and Sergienko, G. and Sunn Pedersen, T. and Vuoriheimo, T. and Wang, E. and Wauters, T. and Winters, V. and Zhao, M. and Yi, R. and Team, the W7-X.},
	month = dec,
	year = {2021},
	note = {Publisher: IOP Publishing},
	pages = {016006},
}

@book{doi:10.1142/8362,
	title = {Turbulent transport in magnetized plasmas},
	url = {https://www.worldscientific.com/doi/abs/10.1142/8362},
	publisher = {WORLD SCIENTIFIC},
	author = {Horton, Wendell},
	year = {2012},
	doi = {10.1142/8362},
	note = {tex.eprint: https://www.worldscientific.com/doi/pdf/10.1142/8362},
}

@book{doi:10.1142/p015,
	title = {Basic space plasma physics},
	url = {https://www.worldscientific.com/doi/abs/10.1142/p015},
	publisher = {PUBLISHED BY IMPERIAL COLLEGE PRESS and DISTRIBUTED BY WORLD SCIENTIFIC PUBLISHING CO.},
	author = {Baumjohann, Wolfgang and Treumann, Rudolf A},
	year = {1996},
	doi = {10.1142/p015},
	note = {tex.eprint: https://www.worldscientific.com/doi/pdf/10.1142/p015},
}

@misc{tiwari_zonal_2025,
	title = {Zonal flow suppression of turbulent transport in the optimized stellarators {W7}-{X} and {QSTK}},
	url = {http://arxiv.org/abs/2501.12722},
	doi = {10.48550/arXiv.2501.12722},
	urldate = {2025-08-12},
	publisher = {arXiv},
	author = {Tiwari, Abhishek and Das, Joydeep and Alageshan, Jaya Kumar and Roberg-Clark, Gareth and Plunk, Gabriel and Xanthopoulos, Pavlos and Sharma, Sarveshwar and Lin, Zhihong and Kuley, Animesh},
	month = aug,
	year = {2025},
	note = {arXiv:2501.12722 [physics]},
}

@article{hegna_theory_2018,
	title = {Theory of {ITG} turbulent saturation in stellarators: {Identifying} mechanisms to reduce turbulent transport},
	volume = {25},
	issn = {1070-664X},
	shorttitle = {Theory of {ITG} turbulent saturation in stellarators},
	url = {https://doi.org/10.1063/1.5018198},
	doi = {10.1063/1.5018198},
	number = {2},
	urldate = {2025-08-12},
	journal = {Physics of Plasmas},
	author = {Hegna, C. C. and Terry, P. W. and Faber, B. J.},
	month = feb,
	year = {2018},
	pages = {022511},
}

@article{sandberg_finite_2007,
	title = {Finite {Larmor} radius effects on the coupled trapped electron and ion temperature gradient modes},
	volume = {14},
	issn = {1070-664X},
	url = {https://doi.org/10.1063/1.2768938},
	doi = {10.1063/1.2768938},
	number = {9},
	urldate = {2025-08-12},
	journal = {Physics of Plasmas},
	author = {Sandberg, I. and Isliker, H. and Pavlenko, V. P.},
	month = sep,
	year = {2007},
	pages = {092504},
}

@article{ford_charge_2020,
	title = {Charge exchange recombination spectroscopy at {Wendelstein} 7-{X}},
	volume = {91},
	issn = {0034-6748},
	url = {https://doi.org/10.1063/1.5132936},
	doi = {10.1063/1.5132936},
	number = {2},
	urldate = {2025-06-30},
	journal = {Review of Scientific Instruments},
	author = {Ford, O. P. and Vanó, L. and Alonso, J. A. and Baldzuhn, J. and Beurskens, M. N. A. and Biedermann, C. and Bozhenkov, S. A. and Fuchert, G. and Geiger, B. and Hartmann, D. and Jaspers, R. J. E. and Kappatou, A. and Langenberg, A. and Lazerson, S. A. and McDermott, R. M. and McNeely, P. and Neelis, T. W. C. and Pablant, N. A. and Pasch, E. and Rust, N. and Schroeder, R. and Scott, E. R. and Smith, H. M. and Wegner, Th. and Kunkel, F. and Wolf, R. C. and {W7-X Team}},
	month = feb,
	year = {2020},
	pages = {023507},
}

@misc{h_smith_neotransp_2022,
	title = {{NEOTRANSP}},
	url = {https://gitlab.mpcdf.mpg.de/smithh/neotransp},
	publisher = {IPP},
	author = {H. Smith},
	year = {2022},
}

@article{sauter_non-stiffness_2014,
	title = {On the non-stiffness of edge transport in {L}-mode tokamak plasmasa)},
	volume = {21},
	issn = {1070-664X},
	url = {https://doi.org/10.1063/1.4876612},
	doi = {10.1063/1.4876612},
	number = {5},
	urldate = {2025-06-13},
	journal = {Physics of Plasmas},
	author = {Sauter, O. and Brunner, S. and Kim, D. and Merlo, G. and Behn, R. and Camenen, Y. and Coda, S. and Duval, B. P. and Federspiel, L. and Goodman, T. P. and Karpushov, A. and Merle, A. and Team, TCV},
	month = may,
	year = {2014},
	pages = {055906},
}

@article{hofler_milestone_2025,
	title = {Milestone in predicting core plasma turbulence: successful multi-channel validation of the gyrokinetic code {GENE}},
	volume = {16},
	copyright = {2025 The Author(s)},
	issn = {2041-1723},
	shorttitle = {Milestone in predicting core plasma turbulence},
	url = {https://www.nature.com/articles/s41467-025-56997-2},
	doi = {10.1038/s41467-025-56997-2},
	language = {en},
	number = {1},
	urldate = {2025-06-13},
	journal = {Nature Communications},
	author = {Höfler, Klara and Görler, Tobias and Happel, Tim and Lechte, Carsten and Molina, Pedro and Bergmann, Michael and Bielajew, Rachel and Conway, Garrard D. and David, Pierre and Denk, Severin S. and Fischer, Rainer and Hennequin, Pascale and Jenko, Frank and McDermott, Rachael M. and White, Anne E. and Stroth, Ulrich},
	month = mar,
	year = {2025},
	note = {Publisher: Nature Publishing Group},
	pages = {2558},
}

@article{klinger_performance_2016,
	title = {Performance and properties of the first plasmas of {Wendelstein} 7-{X}},
	volume = {59},
	issn = {0741-3335},
	url = {https://dx.doi.org/10.1088/0741-3335/59/1/014018},
	doi = {10.1088/0741-3335/59/1/014018},
	language = {en},
	number = {1},
	urldate = {2025-06-12},
	journal = {Plasma Physics and Controlled Fusion},
	author = {Klinger, T and Alonso, A and Bozhenkov, S and Burhenn, R and Dinklage, A and Fuchert, G and Geiger, J and Grulke, O and Langenberg, A and Hirsch, M and Kocsis, G and Knauer, J and Krämer-Flecken, A and Laqua, H and Lazerson, S and Landreman, M and Maaßberg, H and Marsen, S and Otte, M and Pablant, N and Pasch, E and Rahbarnia, K and Stange, T and Szepesi, T and Thomsen, H and Traverso, P and Velasco, J L and Wauters, T and Weir, G and Windisch, T and Team, The Wendelstein 7-X.},
	month = oct,
	year = {2016},
	note = {Publisher: IOP Publishing},
	pages = {014018},
}

@article{langenberg_achieving_2024,
	title = {Achieving stationary high performance plasmas at {Wendelstein} 7-{X}},
	volume = {31},
	issn = {1070-664X},
	url = {https://doi.org/10.1063/5.0199958},
	doi = {10.1063/5.0199958},
	number = {5},
	urldate = {2025-06-12},
	journal = {Physics of Plasmas},
	author = {Langenberg, A. and Warmer, F. and Fuchert, G. and Ford, O. and Bozhenkov, S. and Andreeva, T. and Lazerson, S. and Pablant, N. A. and Gonda, T. and Beurskens, M. N. A. and Brunner, K.-J. and Buttenschön, B. and Dinklage, A. and Hartmann, D. and Knauer, J. and Marchuk, O. and Pasch, E. and Reimold, F. and Stange, T. and Wegner, Th. and Grulke, O. and Wolf, R. C. and {W7-X Team}},
	month = may,
	year = {2024},
	pages = {052502},
}

@misc{fernando_validation_2025,
	title = {Validation of a {Comprehensive} {First}-{Principles}-{Based} {Framework} for {Predicting} the {Performance} of {Future} {Stellarators}},
	url = {http://arxiv.org/abs/2503.08943},
	doi = {10.48550/arXiv.2503.08943},
	language = {en},
	urldate = {2025-06-12},
	publisher = {arXiv},
	author = {Fernando, D. L. C. Agapito and Navarro, A. Bañón and Carralero, D. and Alonso, A. and Siena, A. Di and Velasco, J. L. and Wilms, F. and Merlo, G. and Jenko, F. and Bozhenkov, S. A. and Pasch, E. and Fuchert, G. and Brunner, K. J. and Knauer, J. and Langenberg, A. and Pablant, N. A. and Gonda, T. and Ford, O. and Vanó, L. and Windisch, T. and Estrada, T. and Maragkoudakis, E. and Team, the Wendelstein 7-X.},
	month = mar,
	year = {2025},
	note = {arXiv:2503.08943 [physics]},
}

@article{fuchert_increasing_2020,
	title = {Increasing the density in {Wendelstein} 7-{X}: benefits and limitations},
	volume = {60},
	issn = {0029-5515},
	shorttitle = {Increasing the density in {Wendelstein} 7-{X}},
	url = {https://dx.doi.org/10.1088/1741-4326/ab6d40},
	doi = {10.1088/1741-4326/ab6d40},
	language = {en},
	number = {3},
	urldate = {2025-06-12},
	journal = {Nuclear Fusion},
	author = {Fuchert, G. and Brunner, K.J. and Rahbarnia, K. and Stange, T. and Zhang, D. and Baldzuhn, J. and Bozhenkov, S.A. and Beidler, C.D. and Beurskens, M.N.A. and Brezinsek, S. and Burhenn, R. and Damm, H. and Dinklage, A. and Feng, Y. and Hacker, P. and Hirsch, M. and Kazakov, Y. and Knauer, J. and Langenberg, A. and Laqua, H.P. and Lazerson, S. and Pablant, N.A. and Pasch, E. and Reimold, F. and Sunn Pedersen, T. and Scott, E.R. and Warmer, F. and Winters, V.R. and Wolf, R.C. and Team, W7-X.},
	month = feb,
	year = {2020},
	note = {Publisher: IOP Publishing},
	pages = {036020},
}

@article{beurskens_confinement_2021,
	title = {Confinement in electron heated plasmas in {Wendelstein} 7-{X} and {ASDEX} {Upgrade}; the necessity to control turbulent transport},
	volume = {62},
	issn = {0029-5515},
	url = {https://dx.doi.org/10.1088/1741-4326/ac36f1},
	doi = {10.1088/1741-4326/ac36f1},
	language = {en},
	number = {1},
	urldate = {2025-05-30},
	journal = {Nuclear Fusion},
	author = {Beurskens, M.N.A. and Angioni, C. and Bozhenkov, S. A. and Ford, O. and Kiefer, C. and Xanthopoulos, P. and Turkin, Y. and Alcusón, J.A. and Baehner, J.P. and Beidler, C. and Birkenmeier, G. and Fable, E. and Fuchert, G. and Geiger, B. and Grulke, O. and Hirsch, M. and Jakubowski, M. and Laqua, H.P. and Langenberg, A. and Lazerson, S. and Pablant, N. and Reisner, M. and Schneider, P. and Scott, E.R. and Stange, T. and von Stechow, A. and Stober, J. and Stroth, U. and Wegner, Th. and Weir, G. and Zhang, D. and Zocco, A. and Wolf, R.C. and Zohm, H. and Team, the W7-X. and Team, the ASDEX Upgrade and Team, the EUROfusion MST1},
	month = dec,
	year = {2021},
	note = {Publisher: IOP Publishing},
	pages = {016015},
}

@article{wolf_electron-cyclotron-resonance_2018,
	title = {Electron-cyclotron-resonance heating in {Wendelstein} 7-{X}: {A} versatile heating and current-drive method and a tool for in-depth physics studies},
	volume = {61},
	issn = {0741-3335},
	shorttitle = {Electron-cyclotron-resonance heating in {Wendelstein} 7-{X}},
	url = {https://dx.doi.org/10.1088/1361-6587/aaeab2},
	doi = {10.1088/1361-6587/aaeab2},
	language = {en},
	number = {1},
	urldate = {2025-05-30},
	journal = {Plasma Physics and Controlled Fusion},
	author = {Wolf, R C and Bozhenkov, S and Dinklage, A and Fuchert, G and Kazakov, Y O and Laqua, H P and Marsen, S and Marushchenko, N B and Stange, T and Zanini, M and Abramovic, I and Alonso, A and Baldzuhn, J and Beurskens, M and Beidler, C D and Braune, H and Brunner, K J and Chaudhary, N and Damm, H and Drewelow, P and Gantenbein, G and Gao, Yu and Geiger, J and Hirsch, M and Höfel, U and Jakubowski, M and Jelonnek, J and Jensen, T and Kasparek, W and Knauer, J and Korsholm, S B and Langenberg, A and Lechte, C and Leipold, F and Mora, H Trimino and Neuner, U and Nielsen, S K and Moseev, D and Oosterbeek, H and Pablant, N and Pasch, E and Plaum, B and Pedersen, T Sunn and Sitjes, A Puig and Rahbarnia, K and Rasmussen, J and Salewski, M and Schilling, J and Scott, E and Stejner, M and Thomsen, H and Thumm, M and Turkin, Y and Wilde, F and Team, the Wendelstein 7-X.},
	month = nov,
	year = {2018},
	note = {Publisher: IOP Publishing},
	pages = {014037},
}

@article{costley_fusion_2016,
	title = {On the fusion triple product and fusion power gain of tokamak pilot plants and reactors},
	volume = {56},
	issn = {0029-5515},
	url = {https://dx.doi.org/10.1088/0029-5515/56/6/066003},
	doi = {10.1088/0029-5515/56/6/066003},
	language = {en},
	number = {6},
	urldate = {2025-05-30},
	journal = {Nuclear Fusion},
	author = {Costley, A.E.},
	month = apr,
	year = {2016},
	note = {Publisher: IOP Publishing},
	pages = {066003},
}

@misc{golo_fuchert_thomson_nodate,
	title = {Thomson {Density} {Data} {Rescaling}},
	author = {Golo Fuchert},
	note = {Method is based on private commincation with Golo Fuchert},
}

@misc{jd_huba_2013_2011,
	title = {2013 {NRL} {PLASMA} {FORMULARY}},
	publisher = {Naval Research Laboratory},
	author = {J.D. Huba},
	year = {2011},
}

@misc{plunk_profile_nodate,
	title = {Profile {Solver}},
	author = {Plunk, Gabriel G.},
	note = {Method is based on private communication with Gabriel Plunk.},
}

@article{rewoldt_collisional_1987,
	title = {Collisional effects on kinetic electromagnetic modes and associated quasilinear transport},
	volume = {30},
	issn = {0031-9171},
	url = {https://doi.org/10.1063/1.866332},
	doi = {10.1063/1.866332},
	number = {3},
	urldate = {2025-05-28},
	journal = {The Physics of Fluids},
	author = {Rewoldt, G. and Tang, W. M. and Hastie, R. J.},
	month = mar,
	year = {1987},
	pages = {807--817},
}

@incollection{kadomtsev_turbulence_1995,
	address = {Boston, MA},
	title = {Turbulence in {Toroidal} {Systems}},
	isbn = {978-1-4615-7793-5},
	url = {https://doi.org/10.1007/978-1-4615-7793-5_2},
	language = {en},
	urldate = {2025-05-28},
	booktitle = {Reviews of {Plasma} {Physics}: {Volume} 5},
	publisher = {Springer US},
	author = {Kadomtsev, B. B. and Pogutse, O. P.},
	editor = {Leontovich, M. A.},
	year = {1995},
	doi = {10.1007/978-1-4615-7793-5_2},
	pages = {249--400},
}

@article{biglari_toroidal_1989,
	title = {Toroidal ion‐pressure‐gradient‐driven drift instabilities and transport revisited},
	volume = {1},
	issn = {0899-8221},
	url = {https://doi.org/10.1063/1.859206},
	doi = {10.1063/1.859206},
	number = {1},
	urldate = {2025-05-28},
	journal = {Physics of Fluids B: Plasma Physics},
	author = {Biglari, H. and Diamond, P. H. and Rosenbluth, M. N.},
	month = jan,
	year = {1989},
	pages = {109--118},
}

@article{horton_drift_1999,
	title = {Drift waves and transport},
	volume = {71},
	url = {https://link.aps.org/doi/10.1103/RevModPhys.71.735},
	doi = {10.1103/RevModPhys.71.735},
	number = {3},
	urldate = {2025-05-27},
	journal = {Reviews of Modern Physics},
	author = {Horton, W.},
	month = apr,
	year = {1999},
	note = {Publisher: American Physical Society},
	pages = {735--778},
}

@article{plunk_stellarators_2019,
	title = {Stellarators {Resist} {Turbulent} {Transport} on the {Electron} {Larmor} {Scale}},
	volume = {122},
	url = {https://link.aps.org/doi/10.1103/PhysRevLett.122.035002},
	doi = {10.1103/PhysRevLett.122.035002},
	number = {3},
	urldate = {2025-05-27},
	journal = {Physical Review Letters},
	author = {Plunk, G. G. and Xanthopoulos, P. and Weir, G. M. and Bozhenkov, S. A. and Dinklage, A. and Fuchert, G. and Geiger, J. and Hirsch, M. and Hoefel, U. and Jakubowski, M. and Langenberg, A. and Pablant, N. and Pasch, E. and Stange, T. and Zhang, D. and W7-X Team, the},
	month = jan,
	year = {2019},
	note = {Publisher: American Physical Society},
	pages = {035002},
}

@article{lawson_criteria_1957,
	title = {Some {Criteria} for a {Power} {Producing} {Thermonuclear} {Reactor}},
	volume = {70},
	issn = {0370-1301},
	url = {https://dx.doi.org/10.1088/0370-1301/70/1/303},
	doi = {10.1088/0370-1301/70/1/303},
	language = {en},
	number = {1},
	urldate = {2025-05-24},
	journal = {Proceedings of the Physical Society. Section B},
	author = {Lawson, J. D.},
	month = jan,
	year = {1957},
	pages = {6},
}

@misc{boozer_required_2022,
	title = {Required toroidal confinement for fusion and omnigeneity},
	url = {https://arxiv.org/abs/2208.02391v5},
	language = {en},
	urldate = {2025-05-19},
	journal = {arXiv.org},
	author = {Boozer, Allen H.},
	month = aug,
	year = {2022},
	doi = {10.1063/5.0147120},
}

@article{taroni_global_1994,
	title = {Global and local energy confinement properties of simple transport coefficients of the {Bohm} type},
	volume = {36},
	issn = {0741-3335},
	url = {https://dx.doi.org/10.1088/0741-3335/36/10/003},
	doi = {10.1088/0741-3335/36/10/003},
	language = {en},
	number = {10},
	urldate = {2025-05-19},
	journal = {Plasma Physics and Controlled Fusion},
	author = {Taroni, A. and Erba, M. and Springmann, E. and Tibone, F.},
	month = oct,
	year = {1994},
	pages = {1629},
}

@article{kring_situ_2018,
	title = {In situ wavelength calibration system for the {X}-ray {Imaging} {Crystal} {Spectrometer} ({XICS}) on {W7}-{X}},
	volume = {89},
	issn = {0034-6748},
	url = {https://doi.org/10.1063/1.5038809},
	doi = {10.1063/1.5038809},
	number = {10},
	urldate = {2025-05-16},
	journal = {Review of Scientific Instruments},
	author = {Kring, J. and Pablant, N. and Langenberg, A. and Rice, J. and Delgado-Aparicio, L. and Maurer, D. and Traverso, P. and Bitter, M. and Hill, K. and Reinke, M.},
	month = aug,
	year = {2018},
	pages = {10F107},
}

@article{xanthopoulos_controlling_2014,
	title = {Controlling {Turbulence} in {Present} and {Future} {Stellarators}},
	volume = {113},
	url = {https://link.aps.org/doi/10.1103/PhysRevLett.113.155001},
	doi = {10.1103/PhysRevLett.113.155001},
	number = {15},
	urldate = {2025-05-14},
	journal = {Physical Review Letters},
	author = {Xanthopoulos, P. and Mynick, H. E. and Helander, P. and Turkin, Y. and Plunk, G. G. and Jenko, F. and Görler, T. and Told, D. and Bird, T. and Proll, J. H. E.},
	month = oct,
	year = {2014},
	note = {Publisher: American Physical Society},
	pages = {155001},
}

@article{beurskens_ion_2021,
	title = {Ion temperature clamping in {Wendelstein} 7-{X} electron cyclotron heated plasmas},
	volume = {61},
	issn = {0029-5515},
	url = {https://dx.doi.org/10.1088/1741-4326/ac1653},
	doi = {10.1088/1741-4326/ac1653},
	language = {en},
	number = {11},
	urldate = {2025-05-14},
	journal = {Nuclear Fusion},
	author = {Beurskens, M.N.A. and Bozhenkov, S.A. and Ford, O. and Xanthopoulos, P. and Zocco, A. and Turkin, Y. and Alonso, A. and Beidler, C. and Calvo, I. and Carralero, D. and Estrada, T. and Fuchert, G. and Grulke, O. and Hirsch, M. and Ida, K. and Jakubowski, M. and Killer, C. and Krychowiak, M. and Kwak, S. and Lazerson, S. and Langenberg, A. and Lunsford, R. and Pablant, N. and Pasch, E. and Pavone, A. and Reimold, F. and Romba, Th. and von Stechow, A. and Smith, H.M. and Windisch, T. and Yoshinuma, M. and Zhang, D. and Wolf, R.C. and W7-X Team, the},
	month = oct,
	year = {2021},
	note = {Publisher: IOP Publishing},
	pages = {116072},
}

@article{zhang_plasma_2021,
	title = {Plasma radiation behavior approaching high-radiation scenarios in {W7}-{X}},
	volume = {61},
	issn = {0029-5515},
	url = {https://dx.doi.org/10.1088/1741-4326/ac2b75},
	doi = {10.1088/1741-4326/ac2b75},
	language = {en},
	number = {12},
	urldate = {2025-05-09},
	journal = {Nuclear Fusion},
	author = {Zhang, D. and Burhenn, R. and Feng, Y. and König, R. and Buttenschön, B. and Beidler, C.D. and Hacker, P. and Reimold, F. and Thomsen, H. and Laube, R. and Klinger, T. and Giannone, L. and Penzel, F. and Pavone, A. and Krychowiak, M. and Beurskens, M. and Bozhenkov, S. and Brunner, J.K. and Effenberg, F. and Fuchert, G. and Gao, Y. and Geiger, J. and Hirsch, M. and Höfel, U. and Jakubowski, M. and Knauer, J. and Kwak, S. and Laqua, H.P. and Niemann, H. and Otte, M. and Pedersen, T. Sunn and Pasch, E. and Pablant, N. and Rahbarnia, K. and Svensson, J. and Blackwell, B. and Drews, P. and Endler, M. and Rudischhauser, L. and Wang, E. and Weir, G. and Winters, V. and Team, the W7-X.},
	month = oct,
	year = {2021},
	note = {Publisher: IOP Publishing},
	pages = {126002},
}

@article{yamada_impact_2003,
	title = {Impact of heat deposition profile on global confinement of {NBI} heated plasmas in the {LHD}*},
	volume = {43},
	issn = {0029-5515},
	url = {https://dx.doi.org/10.1088/0029-5515/43/8/317},
	doi = {10.1088/0029-5515/43/8/317},
	language = {en},
	number = {8},
	urldate = {2025-05-09},
	journal = {Nuclear Fusion},
	author = {Yamada, H. and Murakami, S. and Yamazaki, K. and Kaneko, O. and Miyazawa, J. and Sakamoto, R. and Watanabe, K. Y. and Narihara, K. and Tanaka, K. and Sakakibara, S. and Osakabe, M. and Peterson, B. J. and Morita, S. and Ida, K. and Inagaki, S. and Masuzaki, S. and Morisaki, T. and Rewoldt, G. and Sugama, H. and Nakajima, N. and Cooper, W. A. and Akiyama, T. and Ashikawa, N. and Emoto, M. and Funaba, H. and Goncharov, P. and Goto, M. and Idei, H. and Ikeda, K. and Isobe, M. and Kawahata, K. and Kawazome, H. and Khlopenkov, K. and Kobuchi, T. and Komori, A. and Kostrioukov, A. and Kubo, S. and Kumazawa, R. and Liang, Y. and Minami, T. and Muto, S. and Mutoh, T. and Nagayama, Y. and Nakamura, Y. and Nakanishi, H. and Narushima, Y. and Nishimura, K. and Noda, N. and Notake, T. and Nozato, H. and Ohdachi, S. and Ohyabu, N. and Oka, Y. and Ozaki, T. and Sagara, A. and Saida, T. and Saito, K. and Sasao, M. and Sato, K. and Sato, M. and Seki, T. and Shimozuma, T. and Shoji, M. and Suzuki, H. and Takeiri, Y. and Takeuchi, N. and Tamura, N. and Toi, K. and Tokuzawa, T. and Torii, Y. and Tsumori, K. and Watanabe, T. and Watari, T. and Xu, Y. and Yamada, I. and Yamamoto, S. and Yamamoto, T. and Yokoyama, M. and Yoshimura, Y. and Yoshinuma, M. and Mito, T. and Itoh, K. and Ohkubo, K. and Ohtake, I. and Satow, T. and Sudo, S. and Uda, T. and Matsuoka, K. and Motojima, O.},
	month = aug,
	year = {2003},
	pages = {749},
}

@article{mandell_laguerrehermite_2018,
	title = {Laguerre–{Hermite} pseudo-spectral velocity formulation of gyrokinetics},
	volume = {84},
	issn = {0022-3778, 1469-7807},
	url = {https://www.cambridge.org/core/journals/journal-of-plasma-physics/article/laguerrehermite-pseudospectral-velocity-formulation-of-gyrokinetics/E0060DECB3071FBD1B09838DB99CA536},
	doi = {10.1017/S0022377818000041},
	language = {en},
	number = {1},
	urldate = {2025-04-30},
	journal = {Journal of Plasma Physics},
	author = {Mandell, N. R. and Dorland, W. and Landreman, M.},
	month = feb,
	year = {2018},
	pages = {905840108},
}

@article{mandell_gx_2024,
	title = {{GX}: a {GPU}-native gyrokinetic turbulence code for tokamak and stellarator design},
	volume = {90},
	issn = {0022-3778, 1469-7807},
	shorttitle = {{GX}},
	url = {https://www.cambridge.org/core/journals/journal-of-plasma-physics/article/gx-a-gpunative-gyrokinetic-turbulence-code-for-tokamak-and-stellarator-design/2C4BB81955E7E749B95B8B8141E997FA#},
	doi = {10.1017/S0022377824000631},
	language = {en},
	number = {4},
	urldate = {2025-04-30},
	journal = {Journal of Plasma Physics},
	author = {Mandell, N. R. and Dorland, W. and Abel, I. and Gaur, R. and Kim, P. and Martin, M. and Qian, T.},
	month = aug,
	year = {2024},
	pages = {905900402},
}

@article{plunk_collisionless_2014,
	title = {Collisionless microinstabilities in stellarators. {III}. {The} ion-temperature-gradient mode},
	volume = {21},
	issn = {1070-664X},
	url = {https://doi.org/10.1063/1.4868412},
	doi = {10.1063/1.4868412},
	number = {3},
	urldate = {2025-04-18},
	journal = {Physics of Plasmas},
	author = {Plunk, G. G. and Helander, P. and Xanthopoulos, P. and Connor, J. W.},
	month = mar,
	year = {2014},
	pages = {032112},
}

@article{pasch_thomson_2016,
	title = {The {Thomson} scattering system at {Wendelstein} 7-{X}},
	volume = {87},
	issn = {0034-6748},
	url = {https://doi.org/10.1063/1.4962248},
	doi = {10.1063/1.4962248},
	number = {11},
	urldate = {2025-04-18},
	journal = {Review of Scientific Instruments},
	author = {Pasch, E. and Beurskens, M. N. A. and Bozhenkov, S. A. and Fuchert, G. and Knauer, J. and Wolf, R. C. and {W7-X Team}},
	month = sep,
	year = {2016},
	pages = {11E729},
}

@article{fuchert_novel_2022,
	title = {A novel technique for an alignment-insensitive density calibration of {Thomson} scattering diagnostics developed at {W7}-{X}},
	volume = {17},
	issn = {1748-0221},
	url = {https://dx.doi.org/10.1088/1748-0221/17/03/C03012},
	doi = {10.1088/1748-0221/17/03/C03012},
	language = {en},
	number = {03},
	urldate = {2025-04-18},
	journal = {Journal of Instrumentation},
	author = {Fuchert, G. and Nelde, P. and Pasch, E. and Beurskens, M.N.A. and Bozhenkov, S.A. and Brunner, K.J. and Meineke, J. and Scott, E.R. and Wolf, R.C. and team, W7-X.},
	month = mar,
	year = {2022},
	note = {Publisher: IOP Publishing},
	pages = {C03012},
}

@misc{plunk_theory_2009,
	title = {The theory of gyrokinetic turbulence: {A} multiple-scales approach},
	shorttitle = {The theory of gyrokinetic turbulence},
	url = {http://arxiv.org/abs/0903.1091},
	doi = {10.48550/arXiv.0903.1091},
	urldate = {2025-04-18},
	publisher = {arXiv},
	author = {Plunk, Gabriel G.},
	month = apr,
	year = {2009},
	note = {arXiv:0903.1091 [physics]},
}

@article{catto_generalized_1981,
	title = {Generalized gyrokinetics},
	volume = {23},
	issn = {0032-1028},
	url = {https://dx.doi.org/10.1088/0032-1028/23/7/005},
	doi = {10.1088/0032-1028/23/7/005},
	language = {en},
	number = {7},
	urldate = {2025-04-18},
	journal = {Plasma Physics},
	author = {Catto, P. J. and Tang, W. M. and Baldwin, D. E.},
	month = jul,
	year = {1981},
	pages = {639},
}

@article{helander_stellarator_2012,
	title = {Stellarator and tokamak plasmas: a comparison},
	volume = {54},
	issn = {0741-3335},
	shorttitle = {Stellarator and tokamak plasmas},
	url = {https://dx.doi.org/10.1088/0741-3335/54/12/124009},
	doi = {10.1088/0741-3335/54/12/124009},
	language = {en},
	number = {12},
	urldate = {2025-04-17},
	journal = {Plasma Physics and Controlled Fusion},
	author = {Helander, P and Beidler, C D and Bird, T M and Drevlak, M and Feng, Y and Hatzky, R and Jenko, F and Kleiber, R and Proll, J H E and Turkin, Yu and Xanthopoulos, P},
	month = nov,
	year = {2012},
	note = {Publisher: IOP Publishing},
	pages = {124009},
}

@phdthesis{jan-peter_bahner_core_2022,
	type = {Phd thesis},
	title = {Core plasma turbulence in {Wendelstein} 7-{X}},
	school = {Universität Greifswald},
	author = {Jan-Peter Bähner},
	month = jan,
	year = {2022},
}

@article{windsor_alpha-particle_1999,
	title = {Alpha-particle physics in tokamaks},
	volume = {357},
	url = {https://royalsocietypublishing.org/doi/10.1098/rsta.1999.0338},
	doi = {10.1098/rsta.1999.0338},
	number = {1752},
	urldate = {2025-04-17},
	journal = {Philosophical Transactions of the Royal Society of London. Series A: Mathematical, Physical and Engineering Sciences},
	author = {Windsor, C. and Keilhacker, M. and Lawson, J. D. and Pert, G. J. and Robinson, D. C. and Putvinski, S. and Heidbrink, W. and Martin, G. and Porcelli, F. and Romanelli, F. and Sadler, G. and Tobita, K. and Dam, J. W. Van and Zweben, S.},
	month = mar,
	year = {1999},
	note = {Publisher: Royal Society},
	pages = {493--513},
}

@misc{nelde_quantification_2023,
	title = {Quantification of systematic errors in the electron density and temperature measured with {Thomson} scattering at {W7}-{X}},
	url = {http://arxiv.org/abs/2111.03562},
	doi = {10.48550/arXiv.2111.03562},
	urldate = {2025-04-17},
	publisher = {arXiv},
	author = {Nelde, Philipp and Fuchert, Golo and Pasch, Ekkehard and Beurskens, Marc N. A. and Bozhenkov, Sergey A. and Brunner, Kai Jakob and Höfel, Udo and Kwak, Sehyun and Meineke, Jens and Scott, Evan R. and Wolf, Robert C. and team, W7-X.},
	month = aug,
	year = {2023},
	note = {arXiv:2111.03562 [physics]},
}

@phdthesis{JuleMLcorrection,
	address = {Max-Planck-Institut für Plasmaphyik Greifswald},
	type = {Bachelor {Thesis}},
	title = {Anwendung von {Machine} {Learning} zur {Kompensation} von {Laserlageschwankungen} bei der {Messung} von {Dichteprofilen} mittels {Thomsonstreuung} an {Wendelstein} 7-{X}},
	language = {German},
	school = {Technische Universität Berlin},
	author = {Jule Muriel Frank},
	month = nov,
	year = {2022},
}

@article{hirshman_steepestdescent_1983,
	title = {Steepest‐descent moment method for three‐dimensional magnetohydrodynamic equilibria},
	volume = {26},
	issn = {0031-9171},
	url = {https://doi.org/10.1063/1.864116},
	doi = {10.1063/1.864116},
	number = {12},
	urldate = {2025-04-17},
	journal = {The Physics of Fluids},
	author = {Hirshman, S. P. and Whitson, J. C.},
	month = dec,
	year = {1983},
	pages = {3553--3568},
}

@book{dwight_r_nicholson_introduction_1983,
	title = {Introduction to {Plasma} {Theory}},
	publisher = {Wiley},
	author = {Dwight R. Nicholson},
	month = mar,
	year = {1983},
}

@misc{perbrunsellFUSIONPHYSICS,
	title = {{FUSION} {PHYSICS} - introduction to the physics behind fusion energy},
	publisher = {Electromagnetic Engineering and Fusion Science KTH},
	author = {Per Brunsell, Jan Scheﬀel},
}

@misc{CollisionalTransportMagnetizedHelander,
	title = {Collisional {Transport} in {Magnetized} {Plasmas} {\textbar} {Plasma} physics and fusion physics},
	url = {https://www.cambridge.org/de/academic/subjects/physics/plasma-physics-and-fusion-physics/collisional-transport-magnetized-plasmas, https://www.cambridge.org/de/academic/subjects/physics/plasma-physics-and-fusion-physics},
	language = {en},
	urldate = {2025-04-16},
	journal = {Cambridge University Press},
}

@incollection{chen_single-particle_2016,
	address = {Cham},
	title = {Single-{Particle} {Motions}},
	isbn = {978-3-319-22309-4},
	url = {https://doi.org/10.1007/978-3-319-22309-4_2},
	language = {en},
	urldate = {2025-04-16},
	booktitle = {Introduction to {Plasma} {Physics} and {Controlled} {Fusion}},
	publisher = {Springer International Publishing},
	author = {Chen, Francis F.},
	editor = {Chen, Francis F.},
	year = {2016},
	doi = {10.1007/978-3-319-22309-4_2},
	pages = {19--49},
}

@misc{W7X_Einfuehrung,
	title = {Einführung},
	url = {https://www.ipp.mpg.de/9296/einfuehrung},
	language = {de},
	urldate = {2025-04-14},
}

@article{sunn_pedersen_key_2017,
	title = {Key results from the first plasma operation phase and outlook for future performance in {Wendelstein} 7-{X}},
	volume = {24},
	issn = {1070-664X},
	url = {https://doi.org/10.1063/1.4983629},
	doi = {10.1063/1.4983629},
	number = {5},
	urldate = {2025-04-14},
	journal = {Physics of Plasmas},
	author = {Sunn Pedersen, Thomas and Dinklage, Andreas and Turkin, Yuriy and Wolf, Robert and Bozhenkov, Sergey and Geiger, Joachim and Fuchert, Golo and Bosch, Hans-Stephan and Rahbarnia, Kian and Thomsen, Henning and Neuner, Ulrich and Klinger, Thomas and Langenberg, Andreas and Trimiño Mora, Humberto and Kornejew, Petra and Knauer, Jens and Hirsch, Matthias and {the W7-X Team} and Pablant, Novimir},
	month = may,
	year = {2017},
	pages = {055503},
}

@misc{ITER2021,
	type = {Text},
	title = {{ITER}: {The} {World}'s {Largest} {Fusion} {Experiment}},
	shorttitle = {{ITER}},
	url = {https://www.iaea.org/bulletin/iter-the-worlds-largest-fusion-experiment},
	language = {en},
	urldate = {2025-04-12},
	month = may,
	year = {2021},
	note = {Publisher: IAEA},
}

@article{beidler_demonstration_2021,
	title = {Demonstration of reduced neoclassical energy transport in {Wendelstein} 7-{X}},
	volume = {596},
	copyright = {2021 The Author(s)},
	issn = {1476-4687},
	url = {https://www.nature.com/articles/s41586-021-03687-w},
	doi = {10.1038/s41586-021-03687-w},
	language = {en},
	number = {7871},
	urldate = {2025-04-12},
	journal = {Nature},
	author = {Beidler, C. D. and Smith, H. M. and Alonso, A. and Andreeva, T. and Baldzuhn, J. and Beurskens, M. N. A. and Borchardt, M. and Bozhenkov, S. A. and Brunner, K. J. and Damm, H. and Drevlak, M. and Ford, O. P. and Fuchert, G. and Geiger, J. and Helander, P. and Hergenhahn, U. and Hirsch, M. and Höfel, U. and Kazakov, Ye O. and Kleiber, R. and Krychowiak, M. and Kwak, S. and Langenberg, A. and Laqua, H. P. and Neuner, U. and Pablant, N. A. and Pasch, E. and Pavone, A. and Pedersen, T. S. and Rahbarnia, K. and Schilling, J. and Scott, E. R. and Stange, T. and Svensson, J. and Thomsen, H. and Turkin, Y. and Warmer, F. and Wolf, R. C. and Zhang, D.},
	month = aug,
	year = {2021},
	note = {Publisher: Nature Publishing Group},
	pages = {221--226},
}

@article{spitzer_stellarator_1958,
	title = {The {Stellarator} {Concept}},
	volume = {1},
	issn = {0031-9171},
	url = {https://doi.org/10.1063/1.1705883},
	doi = {10.1063/1.1705883},
	number = {4},
	urldate = {2025-04-12},
	journal = {The Physics of Fluids},
	author = {Spitzer, Jr., Lyman},
	month = jul,
	year = {1958},
	pages = {253--264},
}

@article{garcia-regana_turbulent_2021,
	title = {Turbulent transport of impurities in {3D} devices},
	volume = {61},
	issn = {0029-5515, 1741-4326},
	url = {http://arxiv.org/abs/2106.05017},
	doi = {10.1088/1741-4326/ac1d84},
	number = {11},
	urldate = {2025-04-12},
	journal = {Nuclear Fusion},
	author = {García-Regaña, J. M. and Barnes, M. and Calvo, I. and González-Jerez, A. and Thienpondt, H. and Sánchez, E. and Parra, F. I. and St. -Onge, D.},
	month = nov,
	year = {2021},
	note = {arXiv:2106.05017 [physics]},
	pages = {116019},
}

@article{HuberLoss,
	title = {Robust {Estimation} of a {Location} {Parameter}},
	volume = {35},
	issn = {0003-4851, 2168-8990},
	url = {https://projecteuclid.org/journals/annals-of-mathematical-statistics/volume-35/issue-1/Robust-Estimation-of-a-Location-Parameter/10.1214/aoms/1177703732.full},
	doi = {10.1214/aoms/1177703732},
	number = {1},
	urldate = {2025-04-10},
	journal = {The Annals of Mathematical Statistics},
	author = {Huber, Peter J.},
	month = mar,
	year = {1964},
	note = {Publisher: Institute of Mathematical Statistics},
	pages = {73--101},
}

@book{rasmussen_gaussian_2005,
	title = {Gaussian {Processes} for {Machine} {Learning}},
	isbn = {978-0-262-25683-4},
	url = {https://direct.mit.edu/books/monograph/2320/Gaussian-Processes-for-Machine-Learning},
	language = {en},
	urldate = {2025-04-10},
	publisher = {The MIT Press},
	author = {Rasmussen, Carl Edward and Williams, Christopher K. I.},
	month = nov,
	year = {2005},
	doi = {10.7551/mitpress/3206.001.0001},
}

@misc{UnivariateSpline,
	title = {{UnivariateSpline} — {SciPy} v1.15.2 {Manual}},
	url = {https://docs.scipy.org/doc/scipy/reference/generated/scipy.interpolate.UnivariateSpline.html},
	urldate = {2025-04-10},
}

@misc{Scitkit_package,
	title = {1. {Supervised} learning},
	url = {https://scikit-learn/stable/supervised_learning.html},
	language = {en},
	urldate = {2025-04-10},
	journal = {scikit-learn},
}

@misc{GaussianProcesses_kernel_scikit,
	title = {1.7. {Gaussian} {Processes}},
	url = {https://scikit-learn.org/stable/modules/gaussian_process.html#radial-basis-function-rbf-kernel},
	language = {en},
	urldate = {2025-04-10},
	journal = {scikit-learn},
}

@article{wappl_web_2024,
	title = {Web apps for profile fitting and power balance analysis at {Wendelstein} 7-{X}},
	volume = {95},
	issn = {0034-6748},
	url = {https://doi.org/10.1063/5.0225315},
	doi = {10.1063/5.0225315},
	number = {9},
	urldate = {2025-04-09},
	journal = {Review of Scientific Instruments},
	author = {Wappl, M. and Bozhenkov, S. A. and Beurskens, M. N. A. and Bannmann, S. and Kuczyński, M. D. and Smith, H. M. and Brunner, K. J. and Ford, O. P. and Fuchert, G. and Knauer, J. P. and Langenberg, A. and Pablant, N. A. and Pasch, E. and Poloskei, P. Zs. and Wolf, R. C. and {W7-X Team}},
	month = sep,
	year = {2024},
	pages = {093529},
}

@article{bozhenkov_thomson_2017,
	title = {The {Thomson} scattering diagnostic at {Wendelstein} 7-{X} and its performance in the first operation phase},
	volume = {12},
	issn = {1748-0221},
	url = {https://dx.doi.org/10.1088/1748-0221/12/10/P10004},
	doi = {10.1088/1748-0221/12/10/P10004},
	language = {en},
	number = {10},
	urldate = {2025-02-06},
	journal = {Journal of Instrumentation},
	author = {Bozhenkov, S.A. and Beurskens, M. and Molin, A. Dal and Fuchert, G. and Pasch, E. and Stoneking, M.R. and Hirsch, M. and Höfel, U. and Knauer, J. and Svensson, J. and Mora, H. Trimino and Wolf, R.C.},
	month = oct,
	year = {2017},
	pages = {P10004},
}

@article{naito_analytic_1993,
	title = {Analytic formula for fully relativistic {Thomson} scattering spectrum},
	volume = {5},
	issn = {0899-8221},
	url = {https://doi.org/10.1063/1.860593},
	doi = {10.1063/1.860593},
	number = {11},
	urldate = {2025-02-06},
	journal = {Physics of Fluids B: Plasma Physics},
	author = {Naito, O. and Yoshida, H. and Matoba, T.},
	month = nov,
	year = {1993},
	pages = {4256--4258},
}

@incollection{hutchinson_scattering_2002,
	address = {Cambridge},
	edition = {2},
	title = {Scattering of electromagnetic radiation},
	isbn = {978-0-521-67574-1},
	url = {https://www.cambridge.org/core/books/principles-of-plasma-diagnostics/scattering-of-electromagnetic-radiation/00CE15871C6F90DB70DFC7CEB3FACE49},
	urldate = {2025-02-06},
	booktitle = {Principles of {Plasma} {Diagnostics}},
	publisher = {Cambridge University Press},
	editor = {Hutchinson, I. H.},
	year = {2002},
	doi = {10.1017/CBO9780511613630.009},
	pages = {273--321},
}

@article{hirshman_three-dimensional_1986,
	title = {Three-dimensional free boundary calculations using a spectral {Green}'s function method},
	volume = {43},
	issn = {0010-4655},
	url = {https://www.sciencedirect.com/science/article/pii/0010465586900585},
	doi = {10.1016/0010-4655(86)90058-5},
	number = {1},
	urldate = {2025-02-03},
	journal = {Computer Physics Communications},
	author = {Hirshman, S. P. and van RIJ, W. I. and Merkel, P.},
	month = dec,
	year = {1986},
	pages = {143--155},
}

@article{yamada_characterization_2005,
	title = {Characterization of energy confinement in net-current free plasmas using the extended {International} {Stellarator} {Database}},
	volume = {45},
	issn = {0029-5515},
	url = {https://dx.doi.org/10.1088/0029-5515/45/12/024},
	doi = {10.1088/0029-5515/45/12/024},
	language = {en},
	number = {12},
	urldate = {2025-01-30},
	journal = {Nuclear Fusion},
	author = {Yamada, H. and Harris, J.H. and Dinklage, A. and Ascasibar, E. and Sano, F. and Okamura, S. and Talmadge, J. and Stroth, U. and Kus, A. and Murakami, S. and Yokoyama, M. and Beidler, C.D. and Tribaldos, V. and Watanabe, K.Y. and Suzuki, Y.},
	month = nov,
	year = {2005},
	pages = {1684},
}

@article{stroth_stellarator-tokamak_2020,
	title = {Stellarator-tokamak energy confinement comparison based on {ASDEX} {Upgrade} and {Wendelstein} 7-{X} hydrogen plasmas},
	volume = {61},
	issn = {0029-5515},
	url = {https://dx.doi.org/10.1088/1741-4326/abbc4a},
	doi = {10.1088/1741-4326/abbc4a},
	language = {en},
	number = {1},
	urldate = {2025-01-30},
	journal = {Nuclear Fusion},
	author = {Stroth, U. and Fuchert, G. and Beurskens, M.N.A. and Birkenmeier, G. and Schneider, P.A. and Scott, E.R. and Brunner, K.J. and Günzkofer, F. and Hacker, P. and Kardaun, O. and Knauer, J.P. and Rahbarnia, K. and Zhang, D. and team, MST1},
	month = nov,
	year = {2020},
	note = {Publisher: IOP Publishing},
	pages = {016003},
}

@article{banon_navarro_assessing_2024,
	title = {Assessing core ion thermal confinement in critical-gradient-optimized stellarators},
	volume = {31},
	issn = {1070-664X},
	url = {https://doi.org/10.1063/5.0204597},
	doi = {10.1063/5.0204597},
	number = {6},
	urldate = {2025-01-28},
	journal = {Physics of Plasmas},
	author = {Bañón Navarro, A. and Roberg-Clark, G. T. and Plunk, G. G. and Fernando, D. and Di Siena, A. and Wilms, F. and Jenko, F.},
	month = jun,
	year = {2024},
	pages = {062508},
}

@article{fuchert_calibration_2024,
	title = {Calibration techniques for {Thomson} scattering diagnostics on large fusion experiments},
	volume = {95},
	issn = {0034-6748},
	url = {https://doi.org/10.1063/5.0219161},
	doi = {10.1063/5.0219161},
	number = {8},
	urldate = {2025-01-28},
	journal = {Review of Scientific Instruments},
	author = {Fuchert, G. and Wagner, J. and Henschke, L. V. and Pasch, E. and Beurskens, M. N. A. and Bozhenkov, S. A. and Brunner, K. J. and Chen, S. and Frank, J. M. and Hirsch, M. and Knauer, J. and Wolf, R. C. and {W7-X Team}},
	month = aug,
	year = {2024},
	pages = {083533},
}




\end{document}